\documentclass[a4paper,12pt]{report}

\newcommand{\auth}{Sam Palmer}   
\newcommand{\thesistitle}{Higher Gauge Theory and M-Theory}
\newcommand{\degree}{Doctor of Philosophy} 
\newcommand{\supdate}{May 2014}            

\usepackage[top=20mm, bottom=20mm, left=40mm, right=20mm]{geometry}
\usepackage{setspace}
\usepackage{fancyhdr}
\usepackage[citecolor=black, urlcolor=black,
  linkcolor=black, colorlinks=true, bookmarksopen=false, bookmarks=true]{hyperref}
\usepackage[nonumberlist]{glossaries}
\usepackage{graphicx}
\usepackage{subfigure}
\usepackage{xspace}
\usepackage{listings}
\usepackage{pdfpages}
\usepackage{alltt}
\usepackage{float} 
\usepackage[notindex, nottoc, notlof, notlot ]{tocbibind} 

\usepackage{stmaryrd}

\usepackage{epsfig,rotating}
\usepackage{amsmath,amssymb}
\usepackage{amsfonts}
\usepackage{mathrsfs}
\usepackage{bbm}
\usepackage{bm}

\def\slasha#1{\setbox0=\hbox{$#1$}#1\hskip-\wd0\hbox to\wd0{\hss\sl/\/\hss}}

\def\periodb#1{\setbox0=\hbox{$#1$}#1\hskip-\wd0\hbox to\wd0{-}}
\newcommand{\nablas}{\slasha{\nabla}}

\newcommand{\binomr}[2]{\binom{\,#1\,}{\,#2\,}}

\newcommand{\delder}[1]{\frac{\delta}{\delta #1}}   		

\newcommand{\lbr}{(\hspace{-0.1cm}(}
\newcommand{\rbr}{)\hspace{-0.1cm})}

\newcommand{\unit}{\mathbbm{1}}   			
\newcommand{\im}{\mathrm{im}}   			
\newcommand{\id}{\mathrm{id}}   			

\newcommand{\CA}{\mathcal{A}}    			

\newcommand{\xb}{\bar{x}}
\newcommand{\xh}{\hat{x}}

\newcommand{\xd}{\dot{x}}

\newcommand{\CC}{\mathcal{C}}
\newcommand{\CCC}{\mathscr{C}}

\newcommand{\CD}{\mathcal{D}}

\newcommand{\CF}{\mathcal{F}}

\newcommand{\CG}{\mathcal{G}}

\newcommand{\CH}{\mathcal{H}}
\newcommand{\CCH}{\mathscr{H}}
\newcommand{\CCI}{\mathscr{I}}
\newcommand{\CI}{\mathcal{I}}

\newcommand{\CL}{\mathcal{L}}

\newcommand{\CN}{\mathcal{N}}
\newcommand{\CO}{\mathcal{O}}

\newcommand{\CR}{\mathcal{R}}

\newcommand{\CT}{\mathcal{T}}
\newcommand{\CCT}{\mathscr{T}}

\newcommand{\CE}{\mathcal{E}}
\newcommand{\fra}{\mathfrak{a}}				
\newcommand{\frg}{\mathfrak{g}}				
\newcommand{\frh}{\mathfrak{h}}				
\newcommand{\frm}{\mathfrak{m}}
\newcommand{\frder}{\mathfrak{der}}				

\newcommand{\fru}{\mathfrak{u}}

\newcommand{\frl}{\mathfrak{l}}

\newcommand{\FR}{\mathbbm{R}}     			
\newcommand{\FC}{\mathbbm{C}}     			
\newcommand{\FH}{\mathbbm{H}}     			
\newcommand{\NN}{\mathbbm{N}}     			
\newcommand{\RZ}{\mathbbm{Z}}     			
\newcommand{\CPP}{{\mathbbm{C}P}}    			

\newcommand{\lambdab}{\bar{\lambda}}

\newcommand{\dd}{\mathrm{d}}     			
\newcommand{\dpar}{\partial}     			
\newcommand{\chib}{{\bar{\chi}}}   	  		
\newcommand{\embd}{{\hookrightarrow}}     		
\newcommand{\diag}{{\mathrm{diag}}}     		
\newcommand{\de}{\mathrm{e}}     			
\newcommand{\di}{\mathrm{i}}     			
\newcommand{\eps}{{\varepsilon}}			
\newcommand{\epsb}{{\bar{\varepsilon}}}			
\renewcommand{\Im}{\mathrm{Im}}     			

\newcommand{\zb}{{\bar{z}}}
\newcommand{\psib}{{\bar{\psi}}}

\newcommand{\sigmab}{{\bar{\sigma}}}

\newcommand{\eand}{{\qquad\mbox{and}\qquad}}     		
\newcommand{\ewith}{{\qquad\mbox{with}\qquad}}

\newcommand{\kernel}{{\mathrm{ker}}}

\newcommand{\der}[1]{\frac{\dpar}{\dpar #1}}   		
\newcommand{\dder}[1]{\frac{\dd}{\dd #1}}   		
\newcommand{\derr}[2]{\frac{\dpar #1}{\dpar #2}}   	
\newcommand{\dderr}[2]{\frac{\dd #1}{\dd #2}}   	
\newcommand{\tr}{\,\mathrm{tr}\,}     			

\newcommand{\agl}{\mathfrak{gl}}     			

\newcommand{\au}{\mathfrak{u}}
\newcommand{\asu}{\mathfrak{su}}
\newcommand{\aso}{\mathfrak{so}}
\newcommand{\aspin}{\mathfrak{spin}}

\newcommand{\sU}{\mathsf{U}}     			

\newcommand{\sSU}{\mathsf{SU}}

\newcommand{\sSO}{\mathsf{SO}}

\newcommand{\sSpin}{\mathsf{Spin}}
\newcommand{\sEnd}{\mathsf{End}\,}
\newcommand{\sHom}{\mathsf{Hom}\,}

\newcommand{\acton}{\vartriangleright}     			
\def\tyng(#1){\hbox{\tiny$\yng(#1)$}}			
\def\tyoung(#1){\hbox{\tiny$\young(#1)$}}			

\newcommand{\beq}{\begin{eqnarray}}
\newcommand{\eeq}{\end{eqnarray}}

\newcommand{\sft}{{\sf t}}
\newcommand{\sfm}{{\sf m}}
\newcommand{\sfg}{{\sf g}}
\newcommand{\sfh}{{\sf h}}
\newcommand{\sfd}{{\sf d}}
\newcommand{\sfb}{{\sf b}}
\newcommand{\sff}{{\sf f}}
\newcommand{\sfk}{{\sf k}}

\newcommand{\dotsp}{\;\cdot\;}

\newcommand{\nablabs}{\slasha{\bar{\nabla}}}
\newcommand{\Zb}{{\bar{Z}}}
\newcommand{\alphab}{{\bar{\alpha}}}
\newcommand{\betab}{{\bar{\beta}}}
\newcommand{\gammab}{{\bar{\gamma}}}

\newcommand{\mt}{\mathsf{t}}
\newcommand{\eor}{{~~~\mbox{or}~~~}}
\newcommand{\actwedge}{\acton}
\newcommand{\pa}{\partial}
\newcommand{\yd}{\dot{y}}
\newcommand{\sMat}{\mathsf{Mat}}
\newcommand{\trng}{t_\circ}
\newcommand{\trngd}{\dot{t}_\circ}
\newcommand{\urng}{u_\circ}
\newcommand{\sG}{\mathsf{G}}
\newenvironment{conditions}{
\vspace{-2mm}\begin{itemize}
\setlength{\itemsep}{-1mm}
}{\vspace{-2mm}\end{itemize}}
\newcommand{\frcs}{\mathfrak{cs}}	
\newcommand{\frb}{\mathfrak{b}}	
\newcommand*{\longhookrightarrow}{\ensuremath{\lhook\joinrel\relbar\joinrel\rightarrow}}
\newcommand{\CEa}{\mathsf{CE}}

\begin{document}
\doublespacing

\pagestyle{empty}
\begin{center}
\begin{spacing}{2}
{\large{\ \\  \vspace{1.5cm}\textbf{\MakeUppercase{\thesistitle}}}}\\
\end{spacing}
\vfill
{\Large\textit{by}}\\\vspace{0.2cm}
{\Large\upshape{\auth}}\\\vspace{1.0cm}
\includegraphics[width=3.5cm]{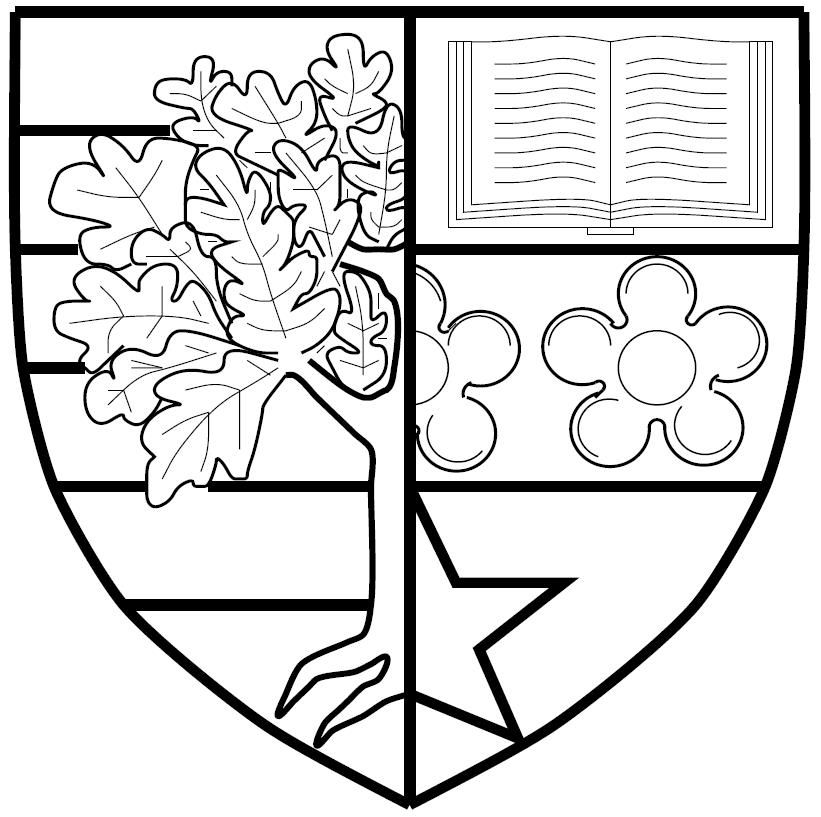}\\
\vspace{1cm}
{\large Submitted for the degree of \\ \degree}\\
\vspace{1cm}
{\large\textsc{Department of Mathematics}\\
\textsc{School of Mathematics and Computer Sciences}\\
\textsc{Heriot-Watt University}}\vfill
{\large{\supdate}}
\end{center}
{\small The copyright in this thesis is owned by the author. Any quotation from the report or use of any of the information contained in it must acknowledge this report as the source of the quotation or information.}
\clearpage
\begin{center}
\LARGE\textbf {Abstract}
\end{center}
\vspace{1cm}

\begin{spacing}{1} 
\noindent

In this thesis, the emerging field of higher gauge theory will be discussed, particularly in relation to problems arising in M-theory, such as selfdual strings and the so-called (2,0) theory. This thesis will begin with a Nahm-like construction for selfdual strings using loop space, the space of loops on spacetime. This construction maps solutions of the Basu-Harvey equation, the BPS equation arising in the description of multiple M2-branes, to solutions of a selfdual string equation on loop space. Furthermore, all ingredients of the construction reduce to those of the ordinary Nahm construction when compactified on a circle with all loops restricted to those wrapping the circle. The rest of this thesis, however, will not involve loop space. We will see a Nahm-like construction for the case of infinitely many selfdual strings, suspended between two M5-branes. This is possible since the limit taken renders the fields describing the M5-branes abelian. This avoids the problem which the rest of this thesis focuses on: What fields describe multiple M5-branes? The answer is likely to involve higher gauge theory, a categorification of gauge theory which describes the parallel transport of extended objects. Any theories which involves 3-algebras, including current M2-brane models and the Lambert-Papageorgakis M5-brane model, are examples of higher gauge theories. Recently, a class of models with $\CN=(1,0)$ supersymmetry have been found, with significant overlap with algebraic structures in higher gauge theory. This overlap suggests that the full $\CN=(2,0)$ theory could involve semistrict $L_\infty$-algebras. Finally, we will see some explicit selfdual string solutions, which may fit into these frameworks.

\end{spacing}

\clearpage
\pagestyle{plain}
\clearpage\pagenumbering{roman}
\noindent
{\LARGE\textbf{Acknowledgements}}
\vspace{1cm}

\begin{spacing}{1} 
\noindent

The main person I should be thanking here is my supervisor, Christian. I've learnt so much over the past three and a half years and it's been almost entirely hugely fun and enjoyable. 

Big hugs for humungous\footnote{Debbietdown and Flint}, Zo\"e, Gilly, Tim, Reuben, Louie, Kitten, Gaga, Orlaith, Ruair\'i, Dionysos, Josh, Terry, Shane, Eoin, Chris Blair, Hussein, Lyonsee and, last but not least, the band King Eider.

\end{spacing}
\tableofcontents
\listoffigures
\clearpage
\pagestyle{fancy}
\pagenumbering{arabic}
\fancyhead{}
\lhead{\slshape \leftmark} 
\cfoot{\thepage}
\renewcommand{\headrulewidth}{0.4pt}
\renewcommand{\footrulewidth}{0.0pt}
\renewcommand{\chaptermark}[1]{\markboth{\chaptername\ \thechapter:\ #1}{}}

\hspace{1cm}\emph{``Who knows anything"}

\begin{flushright}
-Archimedes\\\hfill The sword in the stone
\end{flushright}

\chapter{Introduction}
\label{ch:introduction}

M-theory is arguably the best candidate for a theory of everything, a theory which would explain all repeatable experiments. M-theory's biggest riddle at the moment is the low-energy effective description of M5-branes. This theory will be a six dimensional superconformal field theory with $\CN=(2,0)$ supersymmetry. For supergravity reasons, it should exhibit an $N^3$ scaling of degrees of freedom when describing $N$ M5-branes, as well as reducing to five dimensional super Yang-Mills when one dimension is compactified. 

The theory describing a single M5-brane is well known. It is an abelian theory with a two-form potential $B$, whose field strength $H:=\dd B$ is selfdual. This is an example of a higher gauge theory. Ordinary gauge theory (Yang-Mills) can be thought of as describing the parallel transport of points. Higher gauge theory describes the parallel transport of extended objects (strings, surfaces etc.). This makes sense from a physical point of view since the endpoints of strings on D-branes, giving rise to Yang-Mills theory, are now replaced with the ends of M2-branes on the M5-brane, giving rise to higher gauge theory.

The challenge now is to develop a non-abelian theory which would describe multiple M5-branes. This would involve, at the very least, supersymmetry transformations, gauge transformations and equations of motion for the relevant fields.

Whatever algebraic structures appear, this new field theory will lie at the heart of a web of dualities, it will open up new fields of mathematics and it will fill a gap as the last low-energy effective description of a brane in String/M-theory. Furthermore, it could provide another CFT for an AdS/CFT correspondence.

We will see in this thesis how selfdual strings will play an important role in developing clues relevant to this riddle. Selfdual strings are configurations of M2-branes ending on M5-branes, similar to D1-D3-brane configurations corresponding to magnetic monopoles. When the M2-branes end on the M5-branes along a straight line, the system is half BPS. A thorough understanding of these BPS states will tell us about the algebraic structures relevant to M5-branes as well as part of the fermionic supersymmetry transformation. It may also be possible to generalize the ADHM and ADHMN constructions to M-theory. The existence of these constructions seems plausible given the recent twistor constructions with full $\CN=(2,0)$ supersymmetry in the context of higher gauge theory. Furthermore, selfdual strings should also provide a solution to the problem of quantizing the three-sphere, similarly to the Meyers effect for magnetic monopole configurations, in which quantized two-spheres appear. 

Chapter 2 will begin with an ADHMN-like transform for selfdual strings on loop space, the space of loops in spacetime. This will be based on my first paper with my supervisor \cite{Palmer:2011vx}. The main result is the construction of a transform between solutions of the M2-brane BPS-equation, known as the Basu-Harvey equation, and solutions of the loop space equation
\begin{multline}
 F_{(\mu\sigma)(\nu\tau)}=\big(\eps_{\mu\nu\kappa\lambda}\xd^\kappa(\sigma)D_{(\lambda\tau)}\Phi\big)_{(\sigma\tau)}\\-\Gamma_{\rm ch}\big(\xd_{\mu}(\sigma)D_{(\nu\tau)}\Phi+\xd_{\nu}(\sigma)D_{(\mu\tau)}\Phi-\delta_{\mu\nu}\xd^\kappa(\sigma)D_{(\kappa\tau)}\Phi\big)_{[\sigma\tau]}~,
\end{multline}
where $x\in\CL\FR^4$ denotes a loop on $\FR^4$, the loop space covariant derivative is $D_{(\mu\sigma)}=\delder{x^\mu(\sigma)}+A_{(\mu\sigma)}$, the fields take values in $\frg=\au(N)_+\oplus\au(N)_-$ with $\Gamma_{\rm ch}(\lambda_\pm):=\pm\lambda_\pm$ for $\lambda_\pm\in\au_\pm(N)$ and $(\cdot)_{(\sigma\tau)}$ and $(\cdot)_{[\sigma\tau]}$ denote symmetrization and antisymmetrization in loop parameters $\sigma$ and $\tau$, respectively.

Chapter 3 will then take us out of loop space and we will see an ADHMN-like transform for the special case of infinitely many M2-branes \cite{Harland:2012cj}. That is a transform between solutions of 
\begin{equation}
\frac{\dd t^{\mu}}{\dd s}=~\eps^{\mu\nu\kappa\lambda}\{t^{\nu},t^{\kappa},t^{\lambda}\} ~,
\end{equation}
where $t\in\CC^\infty(M)$ and $M$ is a three-dimensional manifold equipped with a Nambu-Poisson bracket $\{\cdot,\cdot,\cdot\}$, and solutions of 
\begin{equation}
h=*\dd\phi ~,
\end{equation}
on $\FR^4$, where $h,\phi$ take values in $\au(1)$ and satisfy certain conditions. This is a direct generalization of a transform involving infinitely many $\asu(2)$ monopoles which relates solutions of 
\begin{equation}
\frac{\dd t^{i}}{\dd s}=~\eps^{ijk}\{t^{j},t^{k}\} ~,
\end{equation}
where $t\in\CC^\infty(M)$ and $M$ is a two-dimensional manifold equipped with a Poisson bracket $\{\cdot,\cdot\}$, to solutions of 
\begin{equation}
f=*\dd\phi ~,
\end{equation}
on $\FR^3$, where $f,\phi$ take values in $\au(1)$ and satisfy certain conditions.

We will then see in chapter 4 that M5-brane models and also M2-brane models can be regarded as higher gauge theories. This covers the two letters \cite{Palmer:2012ya,Palmer:2013ena}. First we will show that the 3-algebras of M2-brane models are a subset of differential crossed modules and hence Lie 2-algebras, which are two term $L_\infty$-algebras (or strong homotopy Lie algebras). This will allow us to view models involving 3-algebras in the language of higher gauge theory and discuss issues such as fake curvature etc. For this we will need to go one step further in categorification, to Lie 3-algebras, or three term $L_\infty$-algebras.

Higher gauge theories have a significant overlap with a recently developed set of field theories with just $\CN=(1,0)$ supersymmetry, shown in \cite{Palmer:2013pka}. This relationship will be explored in chapter 5. The algebraic structure found in these models are unlike any others found in the literature, however they form a subset of $L_\infty$-algebras. Interestingly, the Lie 2-algebra known as $\mathfrak{string}$ appears as an example for which explicit selfdual string solutions in these models have been constructed\footnote{Note that the fields describing a single M5-brane can also be thought of as living in $\mathfrak{string}$.}.

The final chapter of this thesis, based on \cite{Palmer:2013haa}, will then cover some explicit solutions for selfdual strings based on the 3-algebra $A_4$, as well as so-called higher instanton solutions. Unfortunately, the systems of equations we will consider appear to be under-constrained and admit infinitely many solutions. These solutions, however, provide concrete examples for configurations which, when viewed as principal 2-bundles have non-vanishing fake curvature $\CF\neq0$, yet, when embedded into principal 3-bundle, satisfy the fake curvature condition $\CF=0$.

This thesis explores many open problems concerning higher gauge theory and M-theory, which I believe will be solved in the next few years. This is an exciting time, in which mathematical considerations are applied to physics to reveal potentially fundamental properties of nature and the universe.

\chapter{Loop space selfdual strings}
\label{ch:loop}

In recent years, problems related to finding an effective description of the M2- and M5-branes of M-theory received growing attention. In particular, Bagger-Lambert and independently Gustavsson (BLG) developed an $\CN=8$ supersymmetric Chern-Simons matter theory \cite{Bagger:2007jr,Gustavsson:2007vu}, which is a good candidate for an effective description of stacks of two M2-branes \cite{Mukhi:2008ux}. Soon after, Aharony, Bergman, Jafferis and Maldacena (ABJM) proposed a generalization of this model that is conjectured to provide an effective description of stacks of arbitrarily many M2-branes \cite{Aharony:2008ug}. In favor of this conjecture speak many results, in particular the reproduction of the peculiar $N^{3/2}$ scaling of degrees of freedom with the number $N$ of M2-branes \cite{Drukker:2010nc}.

The corresponding effective description of stacks of M5-branes, however, is much less clear. It is therefore interesting to look at a configuration of M-branes, which exhibits a duality between the M2-brane and the M5-brane theories. Recall that in type IIB superstring theory, there exists such a duality for a configuration of stacks of D1-branes ending on D3-branes. From the point of view of the D1-branes, this configuration is effectively described by the Nahm equation. The description from the perspective of the D3-branes is given by the Bogomolny monopole equation. Both are linked by the so-called Nahm transform, which maps  solutions to the Nahm equation to solutions to the Bogomolny monopole equation and vice versa. The construction of monopole solutions from solutions to the Nahm equation is also known as the Atiyah-Drinfeld-Hitchin-Manin-Nahm (ADHMN) construction \cite{Nahm:1979yw,Nahm:1981nb,Hitchin:1983ay}.

Lifting this D-brane configuration to M-theory, one arrives at a stack of M2-branes ending on a stack of M5-branes. The lift of the Nahm equation yields the Basu-Harvey equation \cite{Basu:2004ed}, while the lift of the Bogomolny monopole equation for gauge group $\sU(1)$ yields the selfdual string equation \cite{Howe:1997ue}. One would therefore expect an ADHMN-like construction linking solutions to the Basu-Harvey equation to selfdual string solitons. For a stack of one or two M2-branes ending on a single M5-brane, this construction was indeed found in \cite{Saemann:2010cp}. 

Interestingly, the lift of the various components in the ADHMN construction very naturally motivates a transition to loop space, in which the selfdual string equation takes the form of a gauge theory equation. It first appears inconvenient to work with an infinite-dimensional base space, but this description has also several advantages. In particular, the selfdual string equation in its original form involves a selfdual three-form and describes only the abelian situation of a single M5-brane. On loop space, however, the corresponding gauge theory equation can be trivially rendered nonabelian and the resulting equation was conjectured in \cite{Saemann:2010cp} to describe M2-branes ending on multiple M5-branes. Further evidence for this was obtained in \cite{Papageorgakis:2011xg}: Here, a set of supersymmetric equations for a 3-Lie algebra (2,0) tensor multiplet \cite{Lambert:2010wm}, which might capture some aspects of M5-brane dynamics, was shown to have a natural interpretation on loop space. The resulting BPS equation was found to be precisely the nonabelian extension of the selfdual string equation on loop space. Moreover, the construction of \cite{Saemann:2010cp} could be straightforwardly extended to the nonabelian case.

The ADHMN-like constructions of \cite{Saemann:2010cp} and \cite{Papageorgakis:2011xg} may be conjectured to capture stacks of $n\leq 2$ M2-branes ending on arbitrarily many M5-branes. The limitation to $n\leq 2$ arises, because the constructions start from the Basu-Harvey equation based on 3-Lie algebras. In this chapter, we will discuss the extension to arbitrary $n$. Correspondingly, we will have to switch to the BPS equation for the ABJM model, that is to a Basu-Harvey equation based on hermitian 3-algebras \cite{Bagger:2008se}. We will also consider the BPS equation of a $\CN=2$ supersymmetric deformation of the BLG model based on real 3-algebras \cite{Cherkis:2008qr}. In both cases, we will demonstrate how solutions to the respective Basu-Harvey equations can be used to construct solutions to the nonabelian selfdual string equation on loop space. We will see various explicit examples of such solutions, as well as corresponding solutions to the Bogomolny monopole equation in the D-brane picture. 

We will also extend the constructions of \cite{Saemann:2010cp,Papageorgakis:2011xg} in another way: These constructions were formulated on the correspondence space of the transgression, which is the Cartesian product of the loop space and $S^1$. Moreover, a reduced differential operator was introduced on correspondence space to guarantee that the transgression was invertible on local abelian gerbes. Here, we will work directly on loop space and use the actual loop space exterior derivative in the construction of the gauge field strength. This leads to a slightly different selfdual string equation on loop space compared to that of \cite{Saemann:2010cp,Papageorgakis:2011xg} and it seems that in the abelian case, the loop space description of selfdual strings is richer than the direct description on space-time.

Interestingly, the fields arising in the construction take values in the gauge algebra $\au(N)_+\oplus\au(N)_-$. This gauge algebra naturally arises as the associated Lie algebra of certain hermitian 3-algebras, cf.\ appendix \ref{app:real3algebras}. The fact that 3-algebras might underly the gauge algebra of an effective description of M5-branes has been used successfully e.g.\ in \cite{Lambert:2010wm}. The gauge algebra we will find fits very well within this picture and its reinterpretation on loop space \cite{Papageorgakis:2011xg}.

There are a few open questions arising from the results. The first one concerns a quantization of $S^3$ by quantizing its loop space, cf.\ e.g.\ \cite{Saemann:2011zq,Saemann:2012ab}: We will employ a Dirac operator containing parameterized loops in the construction. In particular, it contains the expression $\gamma^{\mu\nu}\oint \dd \tau x^\mu(\tau)\xd^\nu(\tau)$, where $x^\mu(\tau)$ with $\tau\in [0,2\pi)$ encodes a parameterized loop and $\xd(\tau)$ is the tangent vector to this loop. A homogeneity argument then suggest that the solutions to the Basu-Harvey equations used in the construction of the Dirac operator should also be dependent on the loop parameter. This would imply that these solutions form coordinates on the quantized loop space of $S^3$. These ideas should be developed in more detail, as they might also yield infinite-dimensional Euclidean 3-Lie algebras, which are not as restrictive as the finite dimensional ones.

Second, recall that by dimensionally reducing the Nahm equation and ``dimensionally oxidizing'' the Bogomolny monopole equation, one obtains\footnote{Up to certain terms in the ADHM equation.} the Nahm-dual pair appearing in the ADHM construction of instantons. It is conceivable that a similar reduction/oxidation procedure could work for the Basu-Harvey equation and the selfdual string equation on loop space, even though the M-brane interpretation is not immediately obvious.

And third, it would be interesting to ``push forward'' the interpretation of the 3-Lie algebra (2,0) tensor multiplet of \cite{Papageorgakis:2011xg} from the correspondence space to loop space. 

\section{Monopoles and selfdual strings}

\subsection{Brane interpretation}

Monopoles of charge $n$ in super Yang-Mills theory with gauge group $\sU(N)$ on $\FR^3$ can be interpreted as stacks of $n$ D1-branes ending on stacks of $N$ D3-branes in type IIB superstring theory as follows  \cite{Diaconescu:1996rk,Tsimpis:1998zh}:
\begin{equation}\label{diag:D1D3}
\begin{tabular}{rcccccccc}
& 0 & 1 & 2 & 3 & 4 & 5 & 6 & \ldots\\
D1 & $\times$ & & & & & & $\vdash$ \\
D3 & $\times$ & $\times$ & $\times$ & $\times$ & & &
\end{tabular}
\end{equation}
An $\times$ indicates a direction that is fully contained in the brane's worldvolume, while a $\vdash$ indicates that the brane's worldvolume is bounded in this direction. We work with Cartesian coordinates $x^0,\ldots,x^6$ on $\FR^{1,6}$ and use the identification $s=x^6$ throughout. The D-brane configuration \eqref{diag:D1D3} is a BPS configuration, and the corresponding time-independent BPS equation in the effective description of the D3-branes is the {\em Bogomolny monopole equation}\footnote{For simplicity the Yang-Mills coupling constant is set to $e=1$, however it will be reintroduced in the following chapter. }
\begin{equation}\label{eq:Bogomolny}
 F:=\dd A+\tfrac{1}{2}[A,A]=*D \Phi~.
\end{equation}
Here, $F$ denotes the $\au(N)$-valued curvature of the connection $D:=\dd+A\acton$, and $\Phi$ is the Higgs field in the adjoint representation of $\au(N)$. The latter describes fluctuations of the D3-branes parallel to the worldvolume of the D1-branes. The time-independent BPS equation on the D1-brane, which gives rise to a dual description, is the {\em Nahm equation}
\begin{equation}\label{eq:Nahm}
 \dder{s}T^i=\tfrac{1}{2}\eps^{ijk}[T^j,T^k]~.
\end{equation}
The $T^i$ are scalar fields taking values in the adjoint of $\au(n)$. They describe the transverse fluctuations of the D1-branes parallel to the worldvolume of the D3-branes. The duality between \eqref{eq:Bogomolny} and \eqref{eq:Nahm} is a special Fourier-Mukai transform, which we will discuss in some detail in section \ref{sec:ADHMNmonopoles}.

The D-brane configuration \eqref{diag:D1D3} can be lifted to M-theory by T-dualizing along the $x^5$-direction and interpreting $x^4$ as the M-theory direction. The resulting configuration is
\begin{equation}\label{diag:M2M5}
\begin{tabular}{rccccccc}
${\rm M}$ & 0 & 1 & 2 & 3 & \phantom{(}4\phantom{)} & 5 & 6 \\
M2 & $\times$ & & & & & $\times$ & $\vdash$ \\
M5 & $\times$ & $\times$ & $\times$ & $\times$ & $\times$ & $\times$ 
\end{tabular}
\end{equation}
This configuration is again a BPS configuration. Contrary to the case of monopoles, the corresponding BPS equation in the effective description of the M5-branes is known only for a single M5-brane, i.e.\ for $N=1$. This is the so-called {\em selfdual string equation} \cite{Howe:1997ue}
\begin{subequations}\label{eq:SelfDualString}
\begin{equation}
 H=*\dd\Phi~
\end{equation}
or in components
\begin{equation}
H_{\mu\nu\kappa}=\eps_{\mu\nu\kappa\lambda}\dpar_\lambda\Phi~,~~~\mu,\nu,\kappa,\lambda=1,\ldots,4~.
\end{equation}
Due to the selfduality of $H$, i.e.\ $H_{\mu\nu\kappa}=\tfrac{1}{3!}\eps_{\mu\nu\kappa\rho\sigma\tau}H^{\rho\sigma\tau}$, it follows that
\begin{equation}
 H_{05\mu}=-\dpar_\mu\Phi~.
\end{equation}
\end{subequations}
As a time-independent BPS equation in the effective description of the M2-branes, Basu and Harvey \cite{Basu:2004ed} suggested the equation
\begin{equation}\label{eq:BasuHarvey}
 \dder{s}T^\mu=\tfrac{1}{3!}\eps^{\mu\nu\kappa\lambda}[T^\nu,T^\kappa,T^\lambda]~,~~~T^\mu\in\CA~,
\end{equation}
which is a natural extension of the $\sSO(3)$-symmetric Nahm equation \eqref{eq:Nahm} describing the $\sSO(3)$-symmetric configuration \eqref{diag:D1D3} to the $\sSO(4)$-symmetric situation \eqref{diag:M2M5}. Here, the $T^\mu$ are scalar fields taking values in the 3-Lie algebra\footnote{See appendix \ref{app:real3algebras} for definitions and conventions related to 3-algebras.} $\CA$. They describe transverse fluctuations of the M2-branes parallel to the worldvolume of the M5-branes. 

\subsection{The ADHMN construction of monopoles}\label{sec:ADHMNmonopoles}

Roughly speaking, the ADHMN construction of monopoles is a Fourier-Mukai transform over a dual pair of degenerate tori $T^4_{\rm D1}$ and $\hat{T}^4_{\rm D3}$ with radii being either infinite or zero. In the D-brane picture \eqref{diag:D1D3}, the degenerate torus $T^4_{\rm D1}=\FR^1$ corresponds to the worldvolume of the D1-branes, while its dual $\hat{T}^4_{\rm D3}=\FR^3$ is to be identified with the D3-branes' worldvolume.

To perform this transform, we start from a special solution to the Nahm equation \eqref{eq:Nahm}. Such a solution is given by a triplet of antihermitian scalar fields $T^i$ over an open interval $\CI\subsetneq \FR$ taking values in the Lie algebra $\au(n)$. Here, $\CI$ is to be identified with the spatial part of the worldvolume of the D1-branes in configuration \eqref{diag:D1D3}. The finite boundaries of $\CI$ correspond to locations of D3-branes. We demand that $T^i$ has simple poles at such finite boundary points of the interval. Moreover, the residues of the solution at these points have to form an irreducible representation of $\mathfrak{su}(2)$ of dimension $n$. 

From this solution, one constructs a Dirac operator $\nablas_{s,x}:W^{1,2}_0(\CI)\otimes\FC^2\otimes \FC^n\rightarrow W^{0,2}(\CI)\otimes\FC^2\otimes \FC^n$. Here, $W^{n,2}$ denotes the Sobolev space of functions on $\CI$, which are square integrable up to their $n$th derivative and the subscript $0$ implies that the functions vanish at finite boundaries of $\CI$, cf.\ \cite{Hitchin:1983ay}. Explicitly, the Dirac operator and its adjoint read as
\begin{equation}
 \nablas_{s,x}=-\unit\dder{s}+\sigma^i\otimes (\di T^i+x^i\unit_n)\ewith \nablabs_{s,x}:=\unit\dder{s}+\sigma^i\otimes (\di T^i+x^i\unit_n)~,
\end{equation}
where the $x^i$ are the Cartesian coordinates on $\FR^3=T^4_{\rm D3}$. Their appearance reflects the twist by the Poincar{\'e} line bundle in the Fourier-Mukai transform \cite{Schenk:1986xe}. The fact that the $T^i$ form a solution to the Nahm equation is equivalent to 
\begin{equation}\label{eq:RelationsNahm}
\Delta_{s,x}:=\nablabs_{s,x}\nablas_{s,x}>0 \eand [\Delta_{s,x},\sigma^i\otimes\unit_n]=0~.
\end{equation}
From the normalized zero modes $\psi^a_{s,x}\in W^{0,2}(\CI)\otimes\FC^2\otimes \FC^n$, $a=1,\ldots,N$, of $\nablabs_{s,x}$ satisfying
\begin{equation}
 \nablabs_{s,x}\psi^a_{s,x}=0~,~~~N=\dim_\FC({\rm ker}\nablabs_{s,x})\eand\delta^{ab}=\int_\CI \dd s\,\psib^a_{s,x}\psi^b_{s,x}~,
\end{equation}
one can construct the following $\au(N)$-valued gauge potential and Higgs field on $T^4_{\rm D3}$:
\begin{equation}\label{eq:BogomolnyFields}
(A_i)^{ab}:=\int_\CI \dd s\,\psib^a_{s,x}\der{x^i}\psi^b_{s,x}\eand\Phi^{ab}:=-\di\int_\CI \dd s\,\psib^a_{s,x}\,s\,\psi^b_{s,x}~.
\end{equation}
Inversely, given fields satisfying the Bogomolny monopole equation \eqref{eq:Bogomolny}, a Dirac operator, zero modes and Nahm data can be constructed. This inverse transform is again a special case of the Fourier-Mukai transform \cite{Schenk:1986xe}.

Using the relations \eqref{eq:RelationsNahm}, it is straightforward to show that the fields \eqref{eq:BogomolnyFields} indeed satisfy the Bogomolny monopole equation \eqref{eq:Bogomolny}. We perform a very similar computation in the case of selfdual strings below. Two explicit examples of this construction are reviewed in section \ref{sec:ADHMNexample}.

\subsection{Examples of solutions to the Nahm equation}

For the simplest case $n=1$, the Nahm data are given by a triplet of constants $T^i\in\FR$ which describe the position of the center of mass of the monopole. In general, the components proportional to $\unit_n$ give this position, which we set to zero in this section, restricting the fields $T^i$ to $\mathfrak{su}(n)$ and fixing the center at the origin.

For $N=1$, the Nahm data live on an interval of the form $(-\infty,v)$ or $(v,\infty)$ with a simple pole at $s=v$. The family of spherically symmetric solutions, corresponding to $n$ coincident D1-branes ending on a single D3-brane, is given by
\begin{equation}\label{eq:sphersymm}
T^i=\frac{e^i}{s-v}~,
\end{equation}
where the $e^i$ form a $n$-dimensional irreducible representation of $\mathfrak{su}(2)$. 

This configuration is known as a \emph{fuzzy funnel} \cite{Constable:1999ac}: Each point of the worldvolume of the D1-brane polarizes into a fuzzy or noncommutative $S^2$ whose radius diverges at $s=v$. The fuzzy funnel describes a transition between D1-branes and D3-branes with a partially noncommutative worldvolume. 

To obtain more general solutions to the Nahm equations, we consider the ansatz $T^i=f_i(s) e^i$, with no sum over $i$. This ansatz was first suggested in \cite{Nahm:1981nb}, and it produces the most general solution for $n\le2$. It reduces the Nahm equations \eqref{eq:Nahm} to 
\begin{equation}\label{eq:spinningtop}
\dder{s}f_1=-f_2 f_3~,~~~
\dder{s}f_2=-f_1 f_3~,~~~ 
\dder{s}f_3=-f_1 f_2~.
\end{equation}
This system of equations is a special case of the Euler-Poinsot equations describing a spinning top in 3 dimensions. There are two constants of motion, related to the mass and energy of the spinning top: $a=f_2^2-f_1^2$ and $b=f_3^2-f_1^2$. The solutions to \eqref{eq:spinningtop} are found by substituting the constants of motion and integrating:
\begin{equation}\label{eq:generalNahm}
f_1=\frac{\sqrt b \ {\rm cn}_{{k}}(\sqrt b \ s)}{{\rm sn}_{{k}} (\sqrt b\ s)}~,~~~ 
f_2=\frac{\sqrt b \ {\rm dn}_{{k}}(\sqrt b \ s)}{{\rm sn}_{{k}} (\sqrt b\ s)}~,~~~  
f_3=\frac{\sqrt b}{{\rm sn}_{{k}} (\sqrt b\ s)}~,
\end{equation}
where $k^2=1-\frac{a}{b}$ and ${\rm cn}_k(s)$, ${\rm dn}_k(s)$ and ${\rm sn}_k(s)$ are the {\em Jacobi elliptic functions} defined in appendix \ref{app:Jacobi}. 

\begin{figure}[h]
\center
\begin{picture}(420,100)
\put(79.0,73.0){\makebox(0,0)[c]{$\dfrac{1}{s}$}}
\put(342.0,73.0){\makebox(0,0)[c]{$f_3$}}
\put(342.0,56.0){\makebox(0,0)[c]{$f_2$}}
\put(342.0,38.0){\makebox(0,0)[c]{$f_1$}}
\includegraphics[width=57mm]{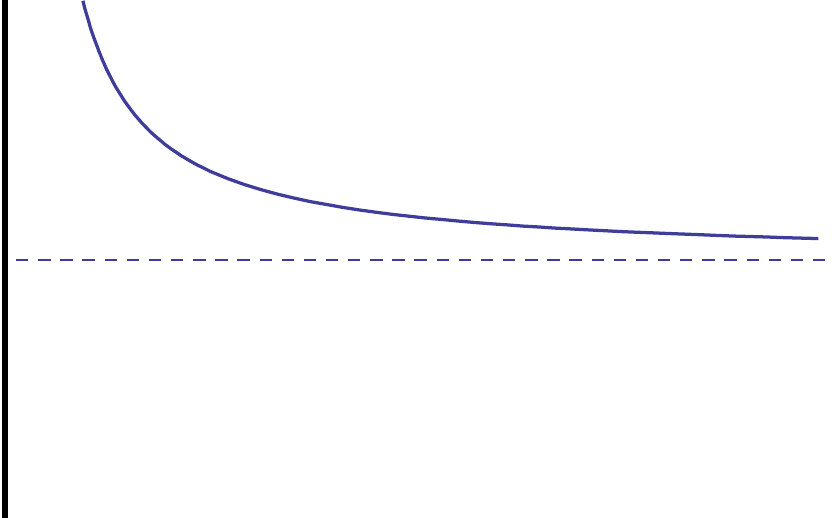}~~~~~~~~~~~~~~~~~~~~~~
\includegraphics[width=57mm]{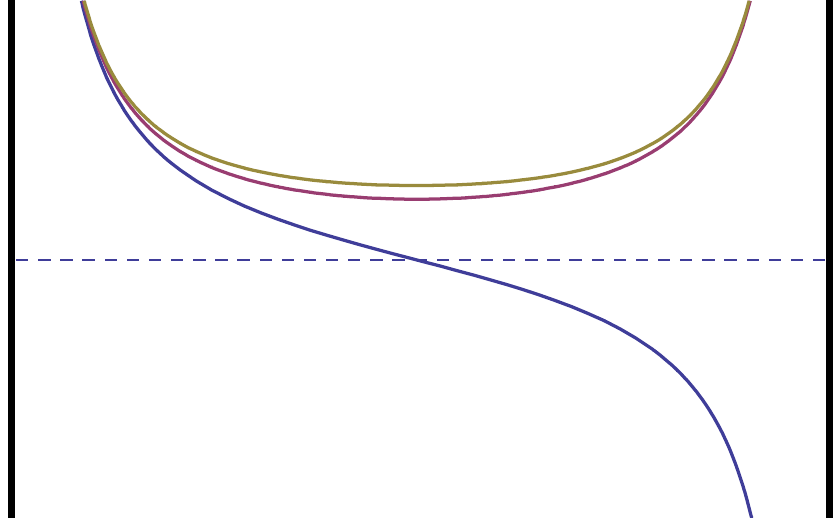}
\end{picture}
\caption{The plot on the left depicts the radial dependence $f=\frac{1}{s}$ in the spherically symmetric configuration \eqref{eq:sphersymm}. The plot on the right shows the corresponding functions $f_1(s),\ f_2(s)$ and $f_3(s)$ in \eqref{eq:generalNahm} for $a=2$, $b=3$. The vertical asymptotes give the positions of D3-branes.}
\end{figure}

The constant of integration is chosen such that one of the poles lies at $s=0$, the other lies at $s=\frac{2}{\sqrt b}{\rm sn}^{-1}_k(1)$. Note that multiplying any two functions by $-1$ gives another solution to the system, although this factor can be absorbed into the $e^i$ to give an equivalent representation of $\asu(2)$. By expanding the solutions \eqref{eq:generalNahm} around the poles, one easily shows that $T^i=\frac{e^i}{s}+$ non-singular terms.

There are two interesting special cases of solution \eqref{eq:generalNahm}. First, there is the axially symmetric case with $a=b$:
\begin{equation}
f_1=\sqrt b/{\rm tan}(\sqrt b \ s)~,~~~f_2=f_3=\sqrt b/{\rm sin}(\sqrt b \ s)~,
\end{equation}
which leads to axially symmetric non-singular monopoles for all charges $n\ge2$, cf.\  \cite{Rossi:1982fq} and references therein. Note that there are no spherically symmetric configurations for $N=2$, $n\ge2$.

Second, there is the case $a=0$, which gives $N=1$ solutions: 
\begin{equation}
f_1=f_2=\sqrt b/{\rm sinh}(\sqrt b \ s)~,~~~f_3=\sqrt b/{\rm tanh}(\sqrt b \ s)~.
\end{equation}
Here, the parameter $b$ corresponds to the separation of the monopoles. Note that the horizontal asymptotes are 0 except for $f_3$, which goes to $\sqrt b$.  Upon taking the limit $b\rightarrow 0$ we recover the spherically symmetric solution \eqref{eq:sphersymm}.

The appearance of elliptic functions is related to the fact that the Nahm equation can be formulated in terms of a Lax pair. This implies that the Nahm equation is linear on the Jacobian variety of its spectral curve \cite{Hitchin:1983ay,Adler1980318}. For the case $n=2$, the spectral curve is a torus, whose doubly-periodic complex coordinate can be identified with the complexification of the variable $s$. The Jacobi elliptic functions form a doubly-periodic basis for functions with maximally simple poles on this torus.

\subsection{Examples of monopole solutions}\label{sec:ADHMNexample}

Consider first the Nahm data \eqref{eq:sphersymm}, corresponding to a stack of $n$ coincident D1-branes ending on $N=1$ D3-branes, which we take to be located at $x^6=s=0$. The spatial part of the worldvolume of the D1-branes is thus $\CI=\FR^{>0}$. The normalized zero mode of $\nablabs_{s,x}$ at the point $\vec x=(0,0,R)^T$ is given by 
\begin{equation}\label{eq:zmMN1}
\psi=\frac{2^{\frac{n}{2}}}{\sqrt{(n-1)!}}R^{\frac{n}{2}} \de^{-s R}s^\frac{n-1}{2} (1,0,\ldots,0)^T~, 
\end{equation}
which yields the Higgs field
\begin{equation}
\Phi=-\frac{\di n}{2R}~.
\end{equation}
For arbitrary $\vec x$, the computation of the zero modes is more difficult. Note that the Higgs field of the charge $n$ monopole is $n$ times that of a charge $1$ monopole, describing $n$ coincident Dirac monopoles. The corresponding field strength is proportional to the volume form on each sphere in the foliation $\FR^3\backslash\{0\}\cong\FR\times S^2$
\begin{equation}\label{eq:Dirac}
\begin{aligned}
F&=-\di n~ \eps_{ijk} \frac{x^k}{2|x|^3}~\dd x^i\wedge\dd x^j=-\frac{\di n}{2|r|^2}~ \mbox{vol}_{S^2}=-\frac{\di n}{2r^2}~ \frac{\dd z_\pm\wedge\dd \zb_\pm}{(1+|z_\pm|^2)^2}~,
\end{aligned}
\end{equation}
where $r$ and $z_\pm$ are the radial and the usual stereographic complex coordinates.

Another nice example is the case of $N=2$ and $n=1$, which gives a non-singular $\sSU(2)$ monopole known as the 't Hooft-Polyakov monopole \cite{Prasad:1975kr}. The Nahm data are constants, taken to be $0$ and the interval is taken to be $(-v,v)$. The normalized zero modes, in matrix notation, are given by
\begin{equation}
\psi=\sqrt{\frac{|x|}{{\rm sinh}(2v|x|)}}({\rm cosh}(|x|s)\unit +{\rm sinh}(|x|s)\frac{x^i\sigma^i}{|x|})~,
\end{equation}
which yields the non-singular fields
\begin{equation}\label{eq:thooft}
\begin{aligned}
\Phi&=\frac{\di \sigma^i x^i}{|x|^2}(v|x|\ {\rm coth}(v|x|)-1)~,\\
A&=\eps_{ijk}\frac{\di \sigma^i x^j}{|x|^2}\left(1-\frac{v|x|}{\sinh(v|x|)}\right)~\dd x^k~.
\end{aligned}
\end{equation}
This is the only spherically symmetric non-singular monopole \cite{Rossi:1982fq} with gauge group $\sSU(2)$. 

\begin{figure}[h]
\center
\begin{picture}(380,100)
~~~~~~\includegraphics[width=50mm]
{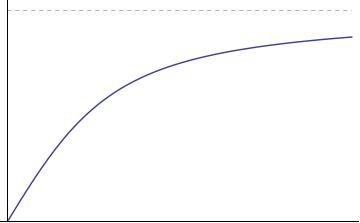}~~~~~~~~\includegraphics[width=50mm]{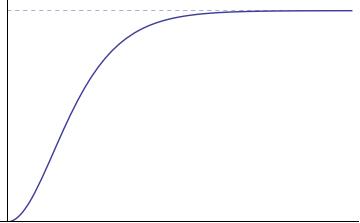}
\put(-187.0,-7.0){\makebox(0,0)[c]{$|x|$}}
\put(-320.0,84.0){\makebox(0,0)[c]{$v$}}
\put(-145.0,84.0){\makebox(0,0)[c]{$1$}}
\put(-13.0,-7.0){\makebox(0,0)[c]{$|x|$}}
\put(-199.0,60.0){\makebox(0,0)[c]{$|\Phi|$}}
\put(-16.0,60.0){\makebox(0,0)[c]{$f_{\rm tHP}$}}
\end{picture}
\caption{The radial dependence of the scalar field $\Phi$ and the function $f_{\rm tHP}=(1-\frac{v|x|}{\sinh(v|x|)})$ appearing in the gauge potential $A$ of the 't Hooft-Polyakov monopole \eqref{eq:thooft}.}
\label{fig:monopoles}
\end{figure}
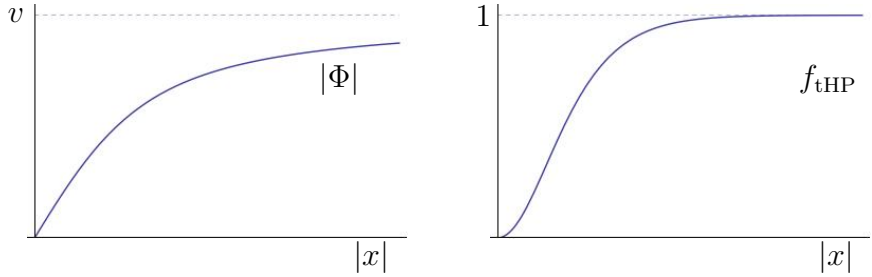

\subsection{Examples of solutions to the Basu-Harvey equation}

The Basu-Harvey equation \eqref{eq:BasuHarvey} also has a unique $\sSO(4)$ invariant, $N=1$ solution given by $T^\mu= \frac{e^\mu}{\sqrt{2(s-v)}}$, where the $e^\mu$ are generators of the 3-Lie algebra $A_4$: $[e^\mu,e^\nu,e^\kappa]=\epsilon^{\mu\nu\kappa\lambda}e^\lambda$. This corresponds to a stack of two coincident M2-branes ending on a single M5-brane and, analogously to the D1-D3-brane configuration, a fuzzy funnel (of one higher dimension) is believed to occur \cite{Basu:2004ed}.

Similarly to the previous ansatz for the Nahm equation, the ansatz $T^\mu=f_\mu (s)e^\mu$ (no sum over $\mu$ implied) reduces the Basu-Harvey equation \eqref{eq:BasuHarvey} to
\begin{equation}\label{eq:3AspinningTop}
\begin{aligned}
\dder{s}f_1&=-f_2 f_3 f_4~,~~~
\dder{s}f_3&=-f_1 f_2f_4~,~~~
\dder{s}f_2&=-f_1 f_3f_4~,~~~
\dder{s}f_4&=-f_1 f_2 f_3~.
\end{aligned}
\end{equation}
The constants of motion for this system are\footnote{As usual in Nambu mechanics \cite{Takhtajan:1993vr}, where the 
Poisson bracket is replaced by a Nambu bracket with 3 arguments, one has an extra Hamiltonian and hence an extra constant of motion.} $a=f_2^2-f_1^2,\ b=f_3^2-f_1^2$ and $c=f_4^2-f_1^2$. The solutions to \eqref{eq:3AspinningTop} were first found in \cite{Nogradi:2005yk}. They are given by \emph{generalized Jacobi elliptic functions}, which are hyperelliptic but can be viewed as single-valued meromorphic functions on a Riemann surface of genus two \cite{Pawellek:2009er}. Using \eqref{genjac}, the solutions can be expressed in terms of Jacobi elliptic functions  
\begin{equation}\label{eq:generalBH}
\begin{aligned}
f_1&=-\frac{\sqrt{a}\ {\rm sn}_{\kappa}( ps)}{\sqrt{1-\frac{a}{c}-{\rm sn}_{\kappa}^2(ps)}}~,~~~
&f_3&=\frac{\sqrt{b(1-\frac{a}{c})}\ {\rm dn}_{\kappa}( ps)}{\sqrt{1-\frac{a}{c}-{\rm sn}_{\kappa}^2(ps)}}~,\\
f_2&=\frac{\sqrt{a(1-\frac{a}{c})}}{\sqrt{1-\frac{a}{c}-{{\rm sn}_{\kappa}^2(ps)}}}~,~~~
&f_4&=\frac{\sqrt{c-a}\ {\rm cn}_{\kappa}( ps)}{\sqrt{1-\frac{a}{c}-{\rm sn}_{\kappa}^2(ps)}}~,
\end{aligned}
\end{equation}
where $p^2=b(c-a)$ and $\kappa^2=\frac{c(b-a)}{b(c-a)}$. This solution exhibits singular behavior at $s=\pm \frac{1}{p}{\rm sn}^{-1}_{\kappa'}(\sqrt{1-\frac{a}{c}}):=\pm v$. Expanding around these points by using the identities \eqref{C1}, we see that $T^\mu\sim \frac{e^\mu}{\sqrt{2(s\pm v)}}+$ non-singular terms.
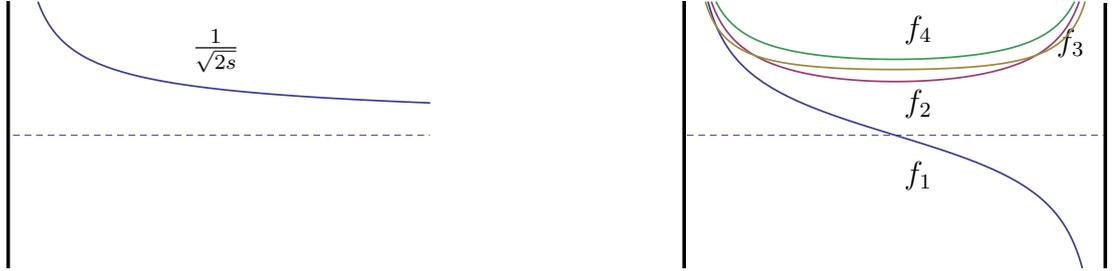
\begin{figure}[h]
\center
\begin{picture}(420,100)
\put(79.0,82.0){\makebox(0,0)[c]{$\frac{1}{\sqrt{2s}}$}}
\put(342.0,90.0){\makebox(0,0)[c]{$f_4$}}
\put(399.0,85.0){\makebox(0,0)[c]{$f_3$}}
\put(342.0,62.0){\makebox(0,0)[c]{$f_2$}}
\put(342.0,35.0){\makebox(0,0)[c]{$f_1$}}
\includegraphics[width=57mm]{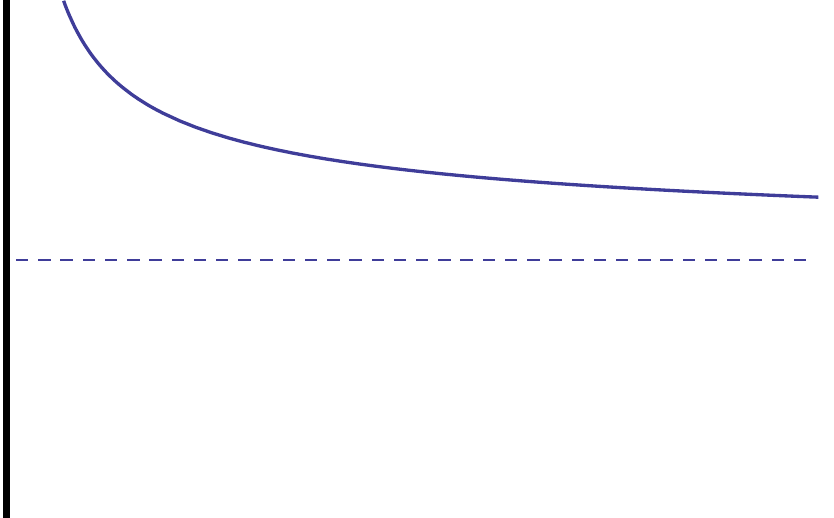}~~~~~~~~~~~~~~~~~~~~~~
\includegraphics[width=57mm]{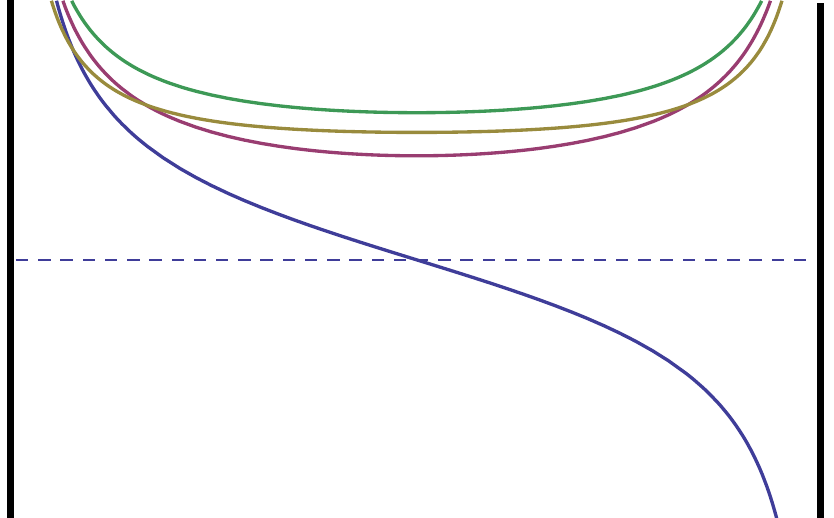}
\end{picture}
\caption{The plot on the left depicts the radial dependence $1/\sqrt{2s}$ of the $N=1$ solution. The plot on the right shows the corresponding functions $f_1(s),\ f_2(s),\ f_3(s)$ and $f_4(s)$ of the solution \eqref{eq:generalBH} for $a=2$, $b=3$, $c=4$. The vertical asymptotes give the positions of the M5-branes.}
\end{figure}

We can again take two interesting limits of the solution \eqref{eq:generalBH}. First, there is the axially symmetric case for $a=b=c$:
\begin{equation}
f_1=-b\ s~\sqrt{\frac{b}{1-b^2s^2}}~,~~~f_2=f_3=f_4=\sqrt{\frac{b}{1-b^2s^2}}~.
\end{equation}
Second, the limit $a\rightarrow0$ takes the period to infinity, giving $N=1$ solutions:
\begin{equation}
\begin{aligned}
f_1=f_2=\frac{ p }{\sqrt{{\rm sinh}(ps)(2p{\rm cosh}(ps)+(b+c){\rm sinh}(ps))}}~,\hspace{2.5cm}&\\
f_3=\frac{ p + b\ {\rm tanh}(ps)}{\sqrt{{\rm tanh}(ps)(2p+(b+c){\rm tanh}(ps))}}~,~
f_4=\frac{ p + c\ {\rm tanh}(ps)}{\sqrt{{\rm tanh}(ps)(2p+(b+c){\rm tanh}(ps))}}~,&
\end{aligned}
\end{equation}
where $p^2=bc$. Taking $b\rightarrow0$ then gives
\begin{equation}\label{eq:sol4}
f_1=f_2=f_3=\frac{1}{\sqrt{s(2+cs)}}~,~~~f_4=\frac{1+cs}{\sqrt{s(2+cs)}}~.
\end{equation}
The horizontal asymptotes are now 0 except for $f_4$, which goes to $\sqrt c$. Taking the parameter $c\rightarrow0$ gives the $\sSO(4)$ symmetric case $T^\mu=\frac{e^\mu}{\sqrt{2s}}$ as expected.

\subsection{Selfdual strings on loop space}

It is not clear how to perform an ADHMN-like construction for selfdual string solitons directly. However, one can reformulate the selfdual string equation \eqref{eq:SelfDualString} on loop space, for which such a construction has been found in \cite{Saemann:2010cp}.

Just as a Dirac monopole is described by the first Chern class $F\in H^2(M,\RZ)$ of a principal $\sU(1)$-bundle over the manifold $M=\FR^3$ or rather\footnote{Dirac monopole solutions on $\FR^3$ are singular at the position of the monopoles, and one should therefore consider the principal $\sU(1)$-bundle on a sphere with the monopole at its center.} $M=S^2$, a selfdual string can be described by the Dixmier-Douady class $H\in H^3(M,\RZ)$ of an abelian $\sU(1)$-gerbe over the manifold $M=S^3$ \cite{Murray:2007ps}. Working with three-form field strengths is rather inconvenient, but there is a trick which allows us to map the Dixmier-Douady class to a first Chern class. This map is called a {\em transgression} \cite{0817647309} and it is defined as follows: Consider $n$ vector fields $v_1,\ldots, v_n$ on the loop space $\CL M$ of $M$. In components, we have
\begin{equation}
v_i=\oint \dd \tau\, v_i^\mu(\tau)\delder{x^\mu(\tau)}~. 
\end{equation}
Any $k+1$-form $\omega\in \Omega^{k+1}(M)$ on $M$ is mapped to an $n$-form $\CT\omega\in\Omega^k(\CL M)$ via
\begin{equation}\label{eq:transgression}
 (\CT\omega)_x(v_1(x),\ldots,v_n(x)):=\oint_{S^1}\dd\tau\,\omega(v_1(\tau),\ldots,v_n(\tau),\dot{x}(\tau))~.
\end{equation}
Here, $x\in \CL M$ denotes a loop and $\xd(\tau)$ is the tangent vector to the loop $x$ at $\tau$. By going to loop space, we thus gain a natural vector, which we can use to fill up one slot of a differential form. Note that the price we have to pay for using the transgression map $\CT$ is that we are now working with an infinite-dimensional base space. One can readily check that $\CT$ is a chain map. This implies that given a three-form field strength $H=\dd B$ of a two-form potential $B$ on $M$, $F=\CT H$ is indeed the field strength for the gauge potential $A=\CT B$ on $\CL M$.

The transgression of the selfdual string equation \eqref{eq:SelfDualString} is given in \cite{Saemann:2010cp} by
\begin{equation}\label{eq:LoopSpaceSDS}
 F_{\mu\nu}(\tau)=\eps_{\mu\nu\kappa\lambda}\xd^\kappa(\tau) \dpar_\lambda\Phi(x(\tau))~,
\end{equation}
where $F$ is a $\au(1)$-valued curvature of some gauge potential, $\Phi$ is a Higgs field and the loop space derivative is
\begin{equation}\label{eq:redloopderivative}
 \dpar_\mu:=\oint_{S^1}\dd \sigma\,\frac{\delta}{\delta x^\mu(\sigma)}~.
\end{equation}

Note that, since the loop parameter $\tau$ appears explicitly in \eqref{eq:LoopSpaceSDS}, this equation does not live on loop space but on the {\it correspondence space} $\CL S^3\times S^1$. In particular, the Higgs field $\Phi(x(\tau))$ is the pullback of the Higgs field $\Phi(x)$ on $S^3$ along the evaluation map $ev:\CL S^3\times S^1\rightarrow S^3: (x(\tau),\tau_0)\mapsto x(\tau_0)$. Here, we intend to perform the construction on loop space itself. That is, we use the loop space exterior derivative 
\begin{equation}
\delta:=\oint \dd \sigma\, \delta x^{\mu}(\sigma)\wedge \delta_{(\mu\sigma)}\ewith \delta_{(\mu\sigma)}:=\delder{x^\mu(\sigma)}~,
\end{equation}
and we consider a Higgs field $\Phi$, which is a $\au(1)$-valued function on $\CL S^3$. Such a function $\Phi$ can be derived from a Higgs field $\Phi_{S^3}$ on $S^3$ by a transgression of functions, i.e.\ via pull-back to the correspondence space and subsequent integration: $\Phi=\oint_{S^1}\dd \tau\, |\xd(\tau)|\,\Phi_{S^3}(x(\tau))$. Moreover, we allow for arbitrary gauge potentials $A$ on $\CL S^3$, which are not necessarily of the form $\CT B$ for some two-form potential $B$ on $S^3$.

Note that a general field strength on loop space is of the form
\begin{equation}
\begin{aligned}
 F:=&\delta A:=\oint \dd \sigma\, \delta x^{\mu}(\sigma)\wedge\delder{x^\mu(\sigma)}\oint \dd \tau\, \delta x^{\nu}(\tau) A_{(\nu\tau)}\\
=&\oint \dd \sigma \oint \dd \tau\, F_{(\mu\sigma)(\nu\tau)}\delta x^{\mu}(\sigma)\wedge \delta x^{\nu}(\tau)~,
\end{aligned}
\end{equation}
where
\begin{equation}
 F_{(\mu\sigma)(\nu\tau)}:=\delder{x^\mu(\sigma)}A_{(\nu\tau)}-\delder{x^\nu(\tau)}A_{(\mu\sigma)}~.
\end{equation}
In equation \eqref{eq:LoopSpaceSDS}, however, only an {\em ultra-local} expression appears. That is, the field strength is of the form
\begin{equation}\label{eq:ultralocal}
 F_{(\mu\sigma)(\nu\tau)}=F_{\mu\nu}(\tau)\delta(\sigma-\tau)~.
\end{equation}

This implies, that we have to extend the selfdual string equation to get both the terms antisymmetric in $\mu\nu$ (and correspondingly symmetric in $\tau\sigma$) as well as the terms symmetric in $\mu\nu$ (and correspondingly antisymmetric in $\tau\sigma$). The extension of \eqref{eq:LoopSpaceSDS} that appears in the construction is given by
\begin{equation}\label{eq:exLoopSpaceSDS}
\begin{aligned}
 F_{(\mu\sigma)(\nu\tau)}=~&\left(\eps_{\mu\nu\kappa\lambda}\xd^\kappa(\sigma)\delder{x^\lambda(\tau)}\Phi\right)_{(\sigma\tau)}\\&-\Gamma_{\rm ch}\left(2\xd_{(\mu}(\sigma)\delder{x^{\nu)}(\tau)}\Phi-\delta_{\mu\nu}\xd^\kappa(\sigma)\delder{x^\kappa(\tau)}\Phi\right)_{[\sigma\tau]}~,
\end{aligned}
\end{equation}
where $(\cdot)_{(\sigma\tau)}$ and $(\cdot)_{[\sigma\tau]}$ denote symmetrization and antisymmetrization in $\sigma$ and $\tau$, respectively. The fields $F_{(\mu\sigma)(\nu\tau)}$ and $\Phi$ now take values in the abelian Lie algebra $\frg=\au(1)_+\oplus\au(1)_-$ and $\Gamma_{\rm ch}$ is a linear involution on $\frg$ with $\Gamma_{\rm ch} (\lambda_\pm)=\pm\lambda_\pm$ for $\lambda_\pm\in\au_\pm(1)$. The obvious nonabelian generalization of the selfdual string equation on loop space \eqref{eq:exLoopSpaceSDS} is:
\begin{multline}\label{eq:NALoopSpaceSDS}
 F_{(\mu\sigma)(\nu\tau)}=\big(\eps_{\mu\nu\kappa\lambda}\xd^\kappa(\sigma)D_{(\lambda\tau)}\Phi\big)_{(\sigma\tau)}\\-\Gamma_{\rm ch}\big(\xd_{\mu}(\sigma)D_{(\nu\tau)}\Phi+\xd_{\nu}(\sigma)D_{(\mu\tau)}\Phi-\delta_{\mu\nu}\xd^\kappa(\sigma)D_{(\kappa\tau)}\Phi\big)_{[\sigma\tau]}~.
\end{multline}
where $D_{(\mu\sigma)}=\delta_{(\mu\sigma)}+A_{(\mu\sigma)}$, the fields take values in $\frg=\au(N)_+\oplus\au(N)_-$ and $\Gamma_{\rm ch}(\lambda_\pm):=\pm\lambda_\pm$ for $\lambda_\pm\in\au_\pm(N)$.

The physical interpretation of this equation is yet unclear: Assuming that a selfdual string is fully described in terms of equation \eqref{eq:SelfDualString}, the components of \eqref{eq:exLoopSpaceSDS} antisymmetric in $\sigma$ and $\tau$ are superfluous, as they cannot be obtained from \eqref{eq:SelfDualString} by a transgression map. Indeed, without the terms antisymmetric in $\sigma\tau$, we have the unextended nonabelian selfdual string equation on loop space \cite{Gustavsson:2008dy,Saemann:2010cp,Papageorgakis:2011xg}.  In \cite{Papageorgakis:2011xg}, this reduced form of equation \eqref{eq:exLoopSpaceSDS} was shown to be the BPS equation to a loop space interpretation of the 3-Lie algebra (2,0) tensor multiplet equations of \cite{Lambert:2010wm}. Note however, that the transgression of \eqref{eq:SelfDualString} is contained in \eqref{eq:exLoopSpaceSDS}. In the abelian case, where the equation is linear, we can therefore project from solutions of \eqref{eq:exLoopSpaceSDS} onto solutions of the transgression of \eqref{eq:SelfDualString}. Moreover, equation \eqref{eq:exLoopSpaceSDS} appears naturally in the Nahm-like construction on loop space, which we develop in the following section. This also motivates the generalization to gauge algebra $\frg$: The Nahm-like construction starts from 3-algebras that often come with an associated Lie algebra of the form $\au(N)_+\oplus\au(N)_-$, which induces a similar splitting onto the constructed fields.

Note that strictly speaking, one should replace $\xd^\rho$ by $R\xd^\rho/|\xd|$, $R\in\FR^{>0}$, in \eqref{eq:NALoopSpaceSDS} and in all of the other equations to arrive at equations invariant under reparameterizations of the loops. To simplify notation, we refrain from doing this but fix the parameterization of all loops by demanding that $\xd^\mu(\tau)\xd^\mu(\tau)=R^2$. 

If the fields take values in $\au(N)_+\oplus\au(N)_-$ with $N>1$, then the Higgs field does not have to diverge and we can extend the considerations from the loop space $\CL S^3$ to the loop space of $\FR^4$.

In the rest of the chapter, we are concerned with constructing various solutions to equations \eqref{eq:NALoopSpaceSDS} by using an ADHMN-like construction.

\section{Selfdual strings from real 3-algebras}

The original construction of selfdual strings developed in \cite{Saemann:2010cp} made use of 3-Lie algebras and the restricted loop space derivative $\dpar_\mu$. Here, we will see the extension involving real 3-algebras and the loop space exterior derivative $\delta$. Recall that all 3-Lie algebras are special cases of real 3-algebras, cf.\ appendix \ref{app:real3algebras}. 

\subsection{The Basu-Harvey equation for real 3-algebras}

The Basu-Harvey equation \eqref{eq:BasuHarvey} is a BPS equation in the BLG model in which the matter fields take values in a 3-Lie algebra and the gauge potential lives in the associated Lie algebra. The problem with using 3-Lie algebras is that they are highly restricted: the only finite-dimensional 3-Lie algebras with positive definite metric are $A_4$ and direct sums thereof. In \cite{Cherkis:2008qr,Cherkis:2008ha}, it was therefore suggested to consider the BLG model with matter fields valued in a real 3-algebra, which preserves at least $\CN=2$ supersymmetries. Another, more interesting generalization of 3-Lie algebras is given by the hermitian 3-algebras, to which we come in section \ref{sec:h3}.

From the supersymmetry transformations given in \cite{Cherkis:2008ha}, it is straightforward to derive the BPS equation corresponding to the Basu-Harvey equation for real 3-algebras. With appropriate normalization, the result is just the ordinary Basu-Harvey equation with the fields $T^\mu$ taking values in a real 3-algebra:
\begin{equation}\label{eq:r3BasuHarvey}
 \dder{s}T^\mu=\tfrac{1}{3!}\eps^{\mu\nu\kappa\lambda}[T^\nu,T^\kappa,T^\lambda]~,~~~T^\mu\in\CA~.
\end{equation}
A class of examples of real 3-algebras is given in the appendix. In particular, the 3-Lie algebra $A_4$ is a sub 3-algebra of the real 3-algebra $\CCC_4$. 

\subsection{The construction}

Analogously to the case of the ADHMN construction, we start from a Dirac operator built from a solution to the Basu-Harvey equation \eqref{eq:r3BasuHarvey}. The solution consists of a quadruplet of real scalar fields over the interval $\CI$ which take values in a metric real 3-algebra $\CA$. Contrary to the case of monopoles, where the solution to the Nahm equation had to have a simple pole at finite boundaries $v$ of $\CI$, we demand here that
\begin{equation}
 T^\mu(s)\sim\frac{e^\mu}{\sqrt{2(s-v)}}+\mbox{regular terms}~.
\end{equation}
The Dirac operator is a map $\nablas_{s,x}:W^{1,2}_0(\CI)\otimes\FC^4\otimes \CA\rightarrow W^{0,2}(\CI)\otimes\FC^4\otimes \CA$ and explicitly, we have
\begin{equation}\label{eq:r3DiracOperators}
\begin{aligned}
 \nablas_{s,x}&=-\gamma_5\dder{s}+\tfrac{1}{2}\gamma^{\mu\nu}\left(D(T^\mu,T^\nu)+\di \oint \dd \tau\, x^\mu(\tau)\xd^\nu(\tau)\right)~,\\
 \nablabs_{s,x}&=+\gamma_5\dder{s}+\tfrac{1}{2}\gamma^{\mu\nu}\left(D(T^\mu,T^\nu)+\di \oint \dd \tau\, x^\mu(\tau)\xd^\nu(\tau)\right)~.
\end{aligned}
\end{equation}
A detailed motivation for the form of this Dirac operator is found in \cite{Saemann:2010cp}. The expressions $x^{\mu\nu}:=\oint \dd \tau\, x^\mu(\tau)\xd^\nu(\tau)$ are also known as the {\em area coordinates} or integrated {\em Pl\"ucker coordinates} of the loop $x$.\footnote{Due to $x^{\mu\nu}=\oint \dd \tau\, x^\mu(\tau)\xd^\nu(\tau)=\oint_C x^\mu\dd x^\nu=\int_V\dd x^\mu\wedge \dd x^\nu$, where $\dpar V=C$, the functions $x^{\mu\nu}$ on $\CL\FR^4$ measure the ``shadow'' of the loop projected onto the coordinate plane $\mu,\nu$.} Since the $T^\mu$ satisfy the Basu-Harvey equation, the Laplace operator $\Delta_{s,x}:=\nablabs_{s,x}\nablas_{s,x}$ is positive and commutes with the generators of $\sSpin(4)$:\footnote{Recall that in the Nahm construction, positivity of the Laplace operator was equivalent to the Dirac operator being constructed from solutions to the Nahm equation. Here, the Laplace operator is positive, if the Dirac operator is constructed from solutions to the Basu-Harvey equation. The inverse statement is only true if the map $D:\CA\wedge \CA\rightarrow \frg_\CA$ is nondegenerate, which is not the case in general.}
\begin{equation}\label{eq:PropDeltaM}
 \Delta_{s,x}>0~,~~~[\Delta_{s,x},\gamma^{\mu\nu}]=0~.
\end{equation}
Note that these properties are preserved, if we shift the Dirac operator by
\begin{equation}\label{eq:shiftDiracReal}
 \nablas_{s,x}\rightarrow \nablas_{s,x}+\gamma^{\mu\nu}\oint \dd \tau\, T^{\mu}_{0}(\tau)\xd^\nu(\tau)\ewith
 \nablabs_{s,x}\rightarrow \nablabs_{s,x}+\gamma^{\mu\nu}\oint \dd \tau\, T^{\mu}_{0}(\tau)\xd^\nu(\tau)~,
\end{equation}
where the field $T^{\mu}_0(\tau)=\di x^\mu_0(\tau)\id_\CA$ with $x^\mu_0\in \CL\FR^4$ allows for a center of mass motion of the selfdual string. For the moment, let us put $x^\mu_0=0$ to simplify the discussion.

We start from the normalized zero modes $\psi^a_{s,x}$ satisfying
\begin{equation}\label{eq:normalizePsiM}
 \nablabs_{s,x}\psi^a_{s,x}=0\eand\delta^{ab}=\int_\CI\dd s\, (\psib^a_{s,x},\psi^b_{s,x})~,
\end{equation}
where $(\dotsp,\dotsp)$ denotes the inner product on $\FC^4\otimes\CA$. We sort the zero modes according to their chirality: We have $N$ zero modes $\psi^a_{s,x}$, $a=1,\ldots,N$, with $\gamma_5\psi^a_{s,x}=\psi^a_{s,x}$ and $N$ zero modes $\psi^a_{s,x}$, $a=N+1,\ldots,2N$, with $\gamma_5\psi^a_{s,x}=-\psi^a_{s,x}$. This is possible because of the block-diagonal structure of the Dirac operator \eqref{eq:r3DiracOperators}.

Analogously to the ADHMN construction, we introduce the following fields:
\begin{equation}\label{eq:MFieldDefinitions}
 A_{(\mu\tau)}^{ab}=\int \dd s\, \left(\psib^a_{s,x}, \delder{x^\mu(\tau)} \psi^b_{s,x}\right)\eand\Phi^{ab}=\di\int \dd s\, \left(\psib^a_{s,x},\,s\,\psi^b_{s,x}\right)~.
\end{equation}
These fields are manifestly anti-hermitian and the sorting of zero modes implies that the fields take values in the gauge algeba $\au(N)_+\oplus\au(N)_-$. Note that the components in $\au(N)_\pm$ depend only on the (anti)-selfdual parts of $D(T^\mu,T^\nu)\pm\tfrac{1}{2}\eps^{\mu\nu\kappa\lambda}D(T^\kappa,T^\lambda)$. 

Let us quickly verify that these fields indeed satisfy the selfdual string equation on loop space \eqref{eq:NALoopSpaceSDS}. For this, we introduce the Green's function $G_{x}(s,t)$ which we can define via
\begin{equation}
\Delta_{s,x} G_{x}(s,t)=-\delta(s-t)
\end{equation}
due to \eqref{eq:PropDeltaM}. We then have the following completeness relation:
\begin{equation}\label{eq:completeness}
 \delta(s-t)=\psi^a_{s,x}\big(\psib^a_{t,x},\dotsp\big)-\nablas_{s,x} G_x(\tau)(s,t)\nablabs_{t,x}~.
\end{equation}
This relation, together with equation \eqref{eq:PropDeltaM} and the identities\footnote{Here and in the following, the sign $\stackrel{[\cdot]}{=}$ means that equality holds after antisymmetrizing the multi-indices $\mu\sigma$ and $\nu\tau$. We include weight factors throughout this thesis in all symmetrizations and antisymmetrizations.}
\begin{equation*}
\begin{aligned}
\gamma^{\mu\kappa}\gamma^{\nu\lambda}\xd^\kappa(\sigma)\xd^\lambda(\tau)\stackrel{[\cdot]}{=}2\gamma^{\mu\lambda}\xd^\nu(\sigma)\xd^\lambda(\tau)-&\delta^{\mu\nu}\gamma^{\kappa\lambda}\xd^\kappa(\sigma)\xd^\lambda(\tau)+\eps_{\mu\nu\kappa\lambda}\gamma^{\kappa\rho}\gamma_5\xd^\lambda(\sigma)\xd^\rho(\tau)~,\\
 \int\dd s\left(\delder{x^\mu(\tau)}\psib^a_{s,x},\psi^b_{s,x}\right)+&\int\dd s\left(\psib^a_{s,x},\delder{x^\mu(\tau)}\psi^b_{s,x}\right)=0~,\\
\left(\delder{x^\mu(\tau)}\nablabs_{s,x}\right) \psi^a_{s,x}+&\nablabs_{s,x}\delder{x^\mu(\tau)}\psi^a_{s,x}=0~,~~~\\
\delder{x^\mu(\tau)}\nablas_{s,x}=\delder{x^\mu(\tau)}\tfrac{\di}{2}\gamma^{\kappa\lambda}&\oint \dd \sigma\,x^\kappa(\sigma)\xd^\lambda(\sigma)=\di\gamma^{\mu\lambda}\xd^\lambda(\tau)~.
\end{aligned}
\end{equation*}
allows us to compute
\begin{equation*}
 \begin{aligned}
  F_{(\mu\sigma)(\nu\tau)}^{ab}&\stackrel{[\cdot]}{=}2\int_\CI \dd s\,\big(\delta_{(\mu\sigma)}\psib^a_{s,x},\delta_{(\nu\tau)}\psi^b_{s,x}\big)+2\int_\CI \dd s\int_\CI \dd t\,\big(\psib^a_{s,x},\delta_{(\mu\sigma)}\psi^c_{s,x}\big)\big(\psib^c_{t,x},\delta_{(\nu\tau)}\psi^b_{t,x}\big)\\
&\stackrel{[\cdot]}{=}-2\int_\CI \dd s\int_\CI \dd t\,\Big(\delta_{(\mu\sigma)}\psib^a_{s,x}\,,\,\left(\nablas_{s,x}G_x(s,t)\nablabs_{t,x}\right)\delta_{(\nu\tau)}\psi^b_{t,x}\Big)\\
&\stackrel{[\cdot]}{=}2\int_\CI \dd s\int_\CI \dd t\,\Big(\psib_{s,x}^a,\left(\gamma^{\mu\kappa}\xd^\kappa(\sigma) G_x(s,t)\gamma^{\nu\lambda}\xd^\lambda(\tau)\right)\psi^b_{t,x}\Big)\\
&\stackrel{[\cdot]}{=}2\eps_{\mu\nu\kappa\lambda}\int_\CI \dd s\int_\CI \dd t\,\Big(\psib_{s,x}^a,\,G_x(s,t)\gamma^{\kappa\rho}\gamma_5\xd^\lambda(\sigma)\xd^\rho(\tau)\psi^b_{t,x}\Big)\\
&\hspace{0.5cm}+\int_\CI \dd s\int_\CI \dd t\,\Big(\psib_{s,x}^a,\,G_x(s,t)\left(4\gamma^{\mu\lambda}\xd^\nu(\sigma)\xd^\lambda(\tau)-2\delta^{\mu\nu}\gamma^{\kappa\lambda}\xd^\kappa(\sigma)\xd^\lambda(\tau)\right)\psi^b_{t,x}\Big).\\
 \end{aligned}
\end{equation*}
It is here that we use the fact that, since the Dirac operator is block diagonal, $\psi^b_{t,x}$ can be arranged into N left and N right-handed zero-modes. Therefore $\psi^b_{t,x}=\gamma_5\Gamma_{\rm ch}{}^b{}_c\psi^c_{t,x}$ where $\Gamma_{\rm ch}$ denotes\footnote{By a slight abuse of notation, we denote the linear involution $\Gamma_{\rm ch}$ on the gauge algebra and the matrix $\diag(\unit_N,-\unit_N)$ leading to it by the same symbol.} the diagonal matrix $\diag(\unit_N,-\unit_N)$.
\begin{equation*}
 \begin{aligned}
&F_{(\mu\sigma)(\nu\tau)}^{ab}\stackrel{[\cdot]}{=}\di\eps_{\mu\nu\kappa\lambda}\xd^\kappa(\sigma)\int_\CI \dd s\,\Big((D_{(\lambda\tau)}\psib_{s,x})^a,\,s\,\psi^b_{s,x}\Big)+\Big(\psib^a_{s,x},\,s\,(D_{(\lambda\tau)}\psi_{s,x})^b\Big)\\
&\hspace{1.2cm}-2\di\xd_{\mu}(\sigma)\int_\CI \dd s\,\Big((D_{(\nu\tau)}\psib_{s,x})^a,\,s\,\psi^c_{s,x}\Big)\Gamma_{\rm ch}{}^b{}_c+\Big(\psib^a_{s,x},\,s\,(D_{(\nu\tau)}\psi^c_{s,x}\Big)\Gamma_{\rm ch}{}^b{}_c\\
&\hspace{1.2cm}-2\di\xd_{\nu}(\sigma)\int_\CI \dd s\,\Big((D_{(\mu\tau)}\psib_{s,x})^a,\,s\,\psi^c_{s,x}\Big)\Gamma_{\rm ch}{}^b{}_c+\Big(\psib^a_{s,x},\,s\,(D_{(\mu\tau)}\psi^c_{s,x}\Big)\Gamma_{\rm ch}{}^b{}_c\\
&\hspace{1.2cm}+\di\delta_{\mu\nu}\xd^\kappa(\sigma)\int_\CI \dd s\,\Big((D_{(\kappa\tau)}\psib_{s,x})^a,\,s\,\psi^c_{s,x}\Big)\Gamma_{\rm ch}{}^b{}_c+\Big(\psib^a_{s,x},\,s\,(D_{(\kappa\tau)}\psi^c_{s,x}\Big)\Gamma_{\rm ch}{}^b{}_c\\
&\stackrel{[\cdot]}{=}\Big(\eps_{\mu\nu\kappa\lambda}\xd^\kappa(\sigma)D_{(\lambda\tau)}\Phi-\Gamma_{\rm ch}(\xd_{\mu}(\sigma)D_{(\nu\tau)}\Phi+\xd_{\nu}(\sigma)D_{(\mu\tau)}\Phi-\delta_{\mu\nu}\xd^\kappa(\sigma)D_{(\kappa\tau)}\Phi)\Big)^{ab}.
 \end{aligned}
\end{equation*}
Thus, the fields \eqref{eq:MFieldDefinitions} indeed satisfy the selfdual string equation on loop space \eqref{eq:NALoopSpaceSDS}.

\subsection{Comments on the reduction to monopoles}\label{sec:r3reduction}

The duality between solutions to the nonabelian selfdual string equation on loop space \eqref{eq:NALoopSpaceSDS} and solutions to the Basu-Harvey equation \eqref{eq:r3BasuHarvey} can be reduced to the duality between solutions to the Bogomolny monopole equation and solutions to the Nahm equation. This reduction has been explained in detail in \cite{Saemann:2010cp} and \cite{Papageorgakis:2011xg} for 3-Lie algebras, and the transition to real 3-algebras is trivially performed. Let us therefore just summarize the key steps in the following.

As usual when going from M-theory to string theory, we have to compactify spacetime along an M-theory direction, which we choose here to be the $x^4$-direction. That is, we arrive at the loop space of $\FR^3\times S^1$ and the radius of the contained $S^1$ is identified with $R=g_{\rm YM}^2=\frac{1}{2\pi}$. We restrict ourselves to loops wrapping this circle by demanding $x^\mu(\tau)=x_0^\mu+R\delta_4^\mu\tau$ and thus $\xd^\mu=R\delta_4^\mu$. In the Dirac operator \eqref{eq:r3DiracOperators}, the generators $\gamma^{\mu\nu}$ of $\sSpin(4)$ are reduced to $\gamma^{i4}$, which generate $\sSU(2)\cong \sSpin(3)\subset \sSpin(4)$. Moreover, because the area coordinates reduce according to
\begin{equation}
 \tfrac{1}{2}\oint \dd \tau\, \gamma^{\mu\nu}x^\mu(\tau)\xd^\nu(\sigma)=\gamma^{i4}x_0^i~,
\end{equation}
the Dirac operator reduces indeed to a Dirac operator for an ADHMN construction for D2-branes ending on D4-branes. As explained in \cite{Saemann:2010cp}, this Dirac operator is a mere doubling of the one appearing in the ordinary ADHMN construction.

Correspondingly, the ultra-local part of the selfdual string equation on loop space \eqref{eq:NALoopSpaceSDS} evidently reduces to the Bogomolny equation \eqref{eq:Bogomolny}. 

In the Basu-Harvey equation, one assumes that the scalar field $T^4$ develops a vacuum expectation value in a 3-algebra direction: $\langle T^4\rangle=v$, $v\in\CA$, cf.\ \cite{Mukhi:2008ux}. To leading order in $v$, the Basu-Harvey equation then reduces to the Nahm equation \cite{Saemann:2010cp,Papageorgakis:2011xg}.

\subsection{Examples}

Let us now give some explicit examples of the above construction. The case of a single M2-brane ending on a single M5-brane corresponds to $n=N=1$ and in this case, the real 3-algebra is abelian. The Nahm data consist of constants and the Dirac operator reduces to 
\begin{equation}\label{eq:Dirack1}
 \nablabs_{s,x(\tau)}=\gamma_5\dder{s}+\tfrac{1}{2}\gamma^{\mu\nu}\oint \dd \tau \left(\di x^\mu(\tau)\xd^\nu(\tau)-T^{\mu}_0(\tau)\xd^\nu(\tau)\right)~.
\end{equation}
As above, we decompose $T^{\mu}_0(\tau)=\di x^\mu_0(\tau)\id_\CA$ and introduce the shifted loop space coordinate $y^\mu(\tau)=x^\mu(\tau)-x^\mu_0(\tau)$ as well as the modified area coordinates $y^{\mu\nu}:=\oint \dd \tau\, y^{[\mu}(\tau)\xd^{\nu]}(\tau)$. The zero modes of the Dirac operator \eqref{eq:Dirack1} are
\begin{equation}\label{eq:zmrSDk1N1}
\begin{aligned}
\psi^+_{s,x(\tau)}&\sim\de^{-r^2_- s}\left(\begin{array}{c}
\di \left(r_-^2+y^{12} -y^{34}\right) \\
y^{13}+y^{24}+\di (y^{23}-y^{14}) \\
0\\ 
0                                
\end{array}
\right)~,\\[0.2cm]
\psi^-_{s,x(\tau)}&\sim\de^{-r^2_+ s}\left(\begin{array}{c} 
0\\ 
0 \\
\di \left(r_+^2+y^{12} +y^{34}\right) \\
y^{13}-y^{24}+\di (y^{23}+y^{14})
\end{array}
\right)~,
\end{aligned}
\end{equation}
where
\begin{equation}
 r_\pm^2:=\tfrac{1}{2}\sqrt{(y^{\mu\nu}\pm\tfrac{1}{2}\eps_{\mu\nu\kappa\lambda}y^{\kappa\lambda})^2}~.
\end{equation}
The resulting Higgs field and gauge potential read as
\begin{equation}\label{eq:Phik1}
\Phi=\begin{pmatrix}
\frac{\di}{2 r_-^{2}}&0\\
0&\frac{\di}{2 r_+^{2}}                      
\end{pmatrix}
\eand
A(\sigma)=\begin{pmatrix}
A^+(\sigma)&0\\
0&A^-(\sigma)                 
\end{pmatrix}~,
\end{equation}
where 
\begin{equation}
A^+(\sigma)=\frac{\di}{2r_-^2(r_-^2+(y^{12}-y^{34}))}\left(\begin{array}{cc}
\xd^{3}(\sigma)(y^{23}-y^{14})+\xd^{4}(\sigma)(y^{13}+y^{24})\\
\xd^{4}(\sigma)(y^{23}-y^{14})-\xd^{3}(\sigma)(y^{13}+y^{24})\\
\xd^{1}(\sigma)(y^{14}-y^{23})+\xd^{2}(\sigma)(y^{13}+y^{24})\\ 
\xd^{2}(\sigma)(y^{14}-y^{23})-\xd^{1}(\sigma)(y^{13}+y^{24})                               
\end{array}
\right)~,
\end{equation}
and $A^-$ is obtained from $A^+$ by substituting $x^4(\sigma)\rightarrow -x^4(\sigma)$. Note that $A^+$ depends only on anti-selfdual combinations of area coordinates, therefore  $A^-$ depends only on selfdual combinations. Altogether, the $\au(1)_+\oplus\au(1)_-$ valued fields are functions of all six linearly independent area coordinates.

Since the field strength of an ordinary Dirac monopole was proportional to the volume form on a two-sphere, one might expect the field strength here to be proportional to the transgression of the volume form on a three-sphere $\CT \mbox{Vol}_{S^3}=\oint \dd\tau \eps^{\mu\nu\kappa\lambda}x^\mu(\tau)\xd^\nu(\tau) \delta x^\kappa(\tau)\wedge\delta x^\lambda(\tau)$ , however this is not the case. Even after looking at the ultralocal part of $F_{(\mu\sigma)(\nu\tau)}$ by setting $\sigma\rightarrow\tau$, they are not the same. Since the form of the field strength is rather lengthy and unilluminating, there is no need to show it here.

One readily checks that these fields satisfy the selfdual string equation on loop space \eqref{eq:exLoopSpaceSDS}. Note that the zero modes \eqref{eq:zmrSDk1N1} reduce to the corresponding zero modes \eqref{eq:zmMN1} in the monopole case for $n=1$, for $x^\mu(\tau)=x_0^\mu+R\delta^\mu_4\tau$ and $s\rightarrow s/r_-$, as expected.

The case $n=1$, $N=2$ has been derived with the reduced loop space derivative \eqref{eq:redloopderivative} in \cite{Papageorgakis:2011xg}. In this case, the Nahm data are trivial and the corresponding Dirac operator directly on loop space is again given by
\begin{equation}\label{eq:Dirack1N2}
 \nablabs_{s,x(\tau)}=\gamma_5\dder{s}+\tfrac{\di}{2}\gamma^{\mu\nu}\oint \dd \tau\,  x^\mu(\tau)\xd^\nu(\tau)~.
\end{equation}
Consider the interval $\CI=(-v,v)$. The zero modes of the Dirac operator \eqref{eq:Dirack1N2} on $\CI$ are
\begin{equation}
 \psi=\eta\left(\left(\begin{array}{cc}\cosh(r_-^2)\unit_2 & 0 \\
             0 & \cosh(r_+^2)\unit_2
            \end{array}\right)
-\frac{\di}{2}\left(\begin{array}{cc}\frac{\sinh(r_-^2)}{r_-^2}\unit_2 & 0 \\
             0 & -\frac{\sinh(r_+^2)}{r_+^2}\unit_2
            \end{array}\right)\gamma^{\mu\nu}y^{\mu\nu}\right)~,
\end{equation}
where the normalization factor $\eta$ reads as
\begin{equation}
 \eta=\left(\begin{array}{cc}\sqrt{\frac{r_-^2}{\sinh(2 v r_-^2)}}\unit_2 & 0 \\
             0 & \sqrt{\frac{r_+^2}{\sinh(2 v r_+^2)}}\unit_2
            \end{array}\right)~.
\end{equation}
The Higgs field resulting from formula \eqref{eq:MFieldDefinitions} is
\begin{equation}
 \Phi=\frac{\di}{2}\left(\begin{array}{cc}\frac{1}{r_-^4}\big(1-2r_-^2 v \coth(2 r_-^2 s0)\big)\unit_2 & 0 \\
             0 & \frac{1}{r_+^4}\big(1-2r_+^2 v \coth(2 r_+^2 s0)\big)\unit_2
            \end{array}\right)~\gamma^{\mu\nu}\gamma_5 y^{\mu\nu}~.
\end{equation}
Note that $\Phi$ takes values in the adjoint representation of $\au(2)_+\oplus\au(2)_-$. It is not clear, what gauge algebra one should expect for a pair of M5-branes. The results of \cite{Papageorgakis:2011xg}, however, suggest that this should be the associated Lie algebra of $A_4$, which is $\frg_{A_4}=\asu(2)\oplus\asu(2)$, in agreement with the result.

For the construction in the case $n=2$, $N=1$, we can use the real 3-algebra $\CCC_4$. As pointed out in appendix \ref{app:real3algebras}, $\CCC_4$ contains $A_4$ as a sub 3-Lie algebra. We can choose the solution of the generalized Basu-Harvey \eqref{eq:r3BasuHarvey} to be 
\begin{equation}
 T^\mu=\frac{e_\mu}{\sqrt{2s}}~,
\end{equation}
where the $e_\mu$ are orthonormal generators of $A_4$ in $\CCC_4$. In the monopole case, we computed for simplicity the Higgs field at $x^3=R$. This was sufficient, as the Higgs field for $n$ coincident monopoles only depends on the radial distance. Here, we expect the Higgs field to depend only on $r_\pm^2$. It is therefore sufficient to compute the Higgs field at $y^{12}=r_-^2=r_+^2=:r^2$. Moreover, the Higgs field just depends on the ``shadow'' of the curve on the 12-plane, not its shape. We can therefore assume that the loop $x$ is a circle:
\begin{equation}
 x(\sigma)=\frac{1}{2\pi}\left(\begin{array}{c}
                  r\sin(\sigma)\\
		  r\cos(\sigma)\\
0 \\ 0
                 \end{array}\right).
\end{equation}
The zero modes of the Dirac operator \eqref{eq:r3DiracOperators} read as\footnote{In the paper \cite{Saemann:2010cp}, compatible representations of $\frg_{A_4}$ were introduced to simplify the reduction to the Nahm equation. Here, we refrain from doing this. Compatible representations could also be used for hermitian 3-algebras in the next section to give the same results.} 
\begin{equation}
\psi=\sqrt{2}r^2\sqrt{s}\de^{-r^2 s}\begin{pmatrix}e_1+\di e_2&&0\\0&&0\\0&&e_1+\di e_2\\0&&0\end{pmatrix}~.
\end{equation}
According to \eqref{eq:MFieldDefinitions}, the Higgs field reads as
\begin{equation}
 \Phi(x)=\frac{\di}{r^2}~\unit_2~,
\end{equation}
which is twice that of \eqref{eq:Phik1}. The charge is thus correctly reproduced.

In principle, we are now able to construct solutions for arbitrary $N$ and $n$ using solutions to the Basu-Harvey equation \eqref{eq:r3BasuHarvey} based on real 3-algebras. As the hermitian 3-algebras are physically more interesting, however, let us continue with these instead.

\section{Selfdual strings from hermitian 3-algebras}\label{sec:h3}

The extension of the construction of selfdual strings developed in \cite{Saemann:2010cp} to a construction involving hermitian 3-algebras is particularly interesting: Hermitian 3-algebras underlie the ABJM model, which has good chances of effectively describing stacks of multiple M2-branes. Therefore, the duality between the two effective descriptions of the configuration \eqref{diag:M2M5} from the perspective of the M2- and the M5-brane, respectively, should make use of hermitian 3-algebras.

\subsection{The Basu-Harvey equation for hermitian 3-algebras}

We start again from the configuration \eqref{diag:M2M5} of M2-branes ending on M5-branes, but we switch from a real description of this configuration to a complex one. Explicitly, we replace the four real coordinates $x^\mu$, $\mu=1,\ldots,4$, transverse to the M2-branes by two complex coordinates $z^1=x^1+\di x^2$ and $z^2=-x^3-\di x^4$. Correspondingly, the real fields $T^\mu$ appearing in the Basu-Harvey equation \eqref{eq:BasuHarvey} are replaced by two complex fields $Z^1:=T^1+\di T^2$ and $Z^2:=-T^3-\di T^4$. If we extend the range of these fields from a 3-Lie algebra to a hermitian 3-algebra, we obtain the analogue of the Basu-Harvey equation in the ABJM model. 

Recall that the BLG model has $\CN=8$ supersymmetry and correspondingly R-sym\-metry group $\sSO(8)$. In going from a real description to a complex one, we break the manifest R-symmetry group from $\sSO(8)$ to $\sSU(4)\simeq \sSO(6)$. The ABJM model is then obtained by generalizing the BLG action such that the matter fields can take values in a hermitian 3-algebra, upon which supersymmetry is indeed reduced from $\CN=8$ to $\CN=6$ in general.

Recall that the metric hermitian 3-algebra appearing in the ABJM model is $\CA={\rm Mat}_\FC(n)$ with a 3-bracket and inner product given respectively by\footnote{We use the notation $\bar{a}=a^\dagger$ as well as $\bar Z_\beta := (Z^\beta)^\dagger$ to avoid overdecorating symbols.}
\begin{equation}
 [a,b;c]:=a\bar c b-b \bar c a\eand(a,b):=\tr(\bar a b)~,~~~a,b,c\in\CA~.
\end{equation}
The metric 3-Lie algebra $A_4$ is reproduced in this way by choosing the basis
\begin{equation}
\left(\tfrac{\di}{\sqrt{2}}\sigma^1, \tfrac{\di}{\sqrt{2}}\sigma^2,\tfrac{\di}{\sqrt{2}}\sigma^3,\tfrac{1}{\sqrt{2}}\unit_2\right)~,
\end{equation}
where the $\sigma^i$, $i=1,2,3$, are the standard Pauli matrices. Using this case, we can adjust the normalization of the fields such that they match the normalization for the real 3-algebras.

The analogue of the Basu-Harvey equation in the ABJM model was previously derived in \cite{Gomis:2008vc,Terashima:2008sy,Hanaki:2008cu} and reads in our conventions as\footnote{We rescaled the fields and thus dropped the Chern-Simons level appearing in \cite{Gomis:2008vc,Terashima:2008sy,Hanaki:2008cu}.}
\begin{equation}\
\dder{s}Z^\alpha=\tfrac{1}{2}(Z^\alpha\Zb_\beta Z^\beta-Z^\beta \Zb_\beta Z^\alpha)~,~~~\alpha,\beta=1,2~.
\end{equation}
Written in the abstract 3-bracket notation explained in appendix \ref{app:real3algebras}, we have
\begin{equation}\label{eq:h3Basu-Harvey}
\dder{s}Z^\alpha=\tfrac{1}{2}[Z^\alpha,Z^\beta;Z^\beta]=-\tfrac{\di}{2}D(\di Z^\beta,Z^\beta)\acton Z^\alpha~,
\end{equation}
and it is this equation that we use as a Basu-Harvey equation for hermitian 3-algebras. We inserted the factors of $\di$ in \eqref{eq:h3Basu-Harvey}, as we choose to work with antihermitian generators of $\frg_\CA$. The unusual contraction over two upper indices of $\sSU(2)$ is due to the antilinearity of the 3-bracket and the map $D(\dotsp,\dotsp)$.

\subsection{The construction}

Here we wish to rewrite the Dirac operator \eqref{eq:r3DiracOperators}  in terms of complex fields and coordinates, however to get both selfdual and anti-selfdual combinations of coordinates that appear in the lower-right and upper-left blocks, respectively, we need to introduce coordinates $\hat z_1:=z^1=x^1+\di x^2~,~~\hat z_2:=\zb^2=-x^3+\di x^4$. Now we can use 
\begin{equation}\label{eq:sigmaident}
 \gamma^{\mu\nu} x^\mu\otimes x^\nu=\tfrac{1}{4}\gamma^{\mu\nu}\left((\sigma^{\mu\nu}{}_\alpha{}^\beta (z^\alpha\otimes \zb_\beta-\zb_\beta\otimes z^\alpha)+\sigmab^{\mu\nu}{}^\alpha{}_\beta (\hat z_\alpha\otimes \hat \zb^\beta-\hat \zb^\beta\otimes\hat z_\alpha)\right)~,
\end{equation}
where we used
\begin{equation}
 \sigma^{\mu\nu}=\tfrac{1}{4}(\sigma^\mu\sigmab^\nu-\sigma^\nu\sigmab^\mu)~,~~~\sigma^\mu=(-\di\sigma^i,\unit)~,~~~\sigmab^\mu=(\di\sigma^i,\unit)~.
\end{equation}
Recall that the $\sigma^{\mu\nu}$ satisfy the identities
\begin{equation}\label{eq:identitiessigmamunu}
\begin{aligned}
{}[\sigma^{\mu\nu},\sigma^{\kappa\lambda}]&=\delta^{\nu\kappa}\sigma^{\mu\lambda}-\delta^{\mu\kappa}\sigma^{\nu\lambda}+\delta^{\mu\lambda}\sigma^{\nu\kappa}-\delta^{\nu\lambda}\sigma^{\mu\kappa}~,\\
\{\sigma^{\mu\nu},\sigma^{\kappa\lambda}\}&=\tfrac{1}{4}\left(\delta^{\nu\kappa}\delta^{\mu\lambda}-\delta^{\mu\kappa}\delta^{\nu\lambda}+\delta^{\mu\lambda}\delta^{\nu\kappa}-\delta^{\nu\lambda}\delta^{\mu\kappa}+2\eps^{\mu\nu\kappa\lambda}\right)\unit_2~,\\
\sigma^{\mu\nu}{}_\alpha{}^\beta\sigma^{\mu\nu}{}_{\gamma}{}^\delta&=\delta_\alpha^\beta\delta_\gamma^\delta-2\delta_\alpha^\delta\delta_\gamma^\beta~,~~~
\sigma^{[\mu\kappa}{}_\alpha{}^\beta\sigma^{\kappa\nu]}{}_\gamma{}^\delta=\tfrac{1}{2}(\sigma^{\mu\nu}{}_\alpha{}^\delta\delta_\beta^\gamma-\sigma^{\mu\nu}{}_\gamma{}^\beta\delta_\alpha^\delta)~.
\end{aligned}
\end{equation}
So using \eqref{eq:sigmaident} we can write the upper-left block of the Dirac operator 
\begin{equation}
 \nablas_{s,z}:=\left(\begin{array}{cc} 
\nablas^+_{s,z} & 0 \\ 
0 & \nablas^-_{s,z}
\end{array}\right)
\end{equation}
as
\begin{equation}
\begin{aligned}
 \nablas^+_{s,z}&=-\unit_2\dder{s}-\tfrac{\di}{4}\sigma^{\mu\nu}\sigma^{\mu\nu}{}_\alpha{}^\beta\left(D(\di Z^\alpha,Z^\beta)-\oint\dd \tau\,z^{\alpha}(\tau)  \dot{\zb}_\beta(\tau) - \dot z^\alpha(\tau) \bar z_\beta(\tau)\right)~,\\
 \nablabs^+_{s,z}&=+\unit_2\dder{s}-\tfrac{\di}{4}\sigma^{\mu\nu}\sigma^{\mu\nu}{}_\alpha{}^\beta\left(D(\di Z^\alpha,Z^\beta)-\oint\dd \tau\,z^{\alpha}(\tau)  \dot{\bar z}_\beta(\tau) - \dot z^\alpha(\tau) \bar z_\beta(\tau)\right)~,
\end{aligned}
\end{equation}
where $Z^\alpha\in\CA$ and $\CA$ is a metric hermitian 3-algebra. The lower-right block $\nablas^-_{s,z}$ can be written in a similar way using $\hat z_\alpha$ and $\hat Z_1:=Z^1=T^1+\di T^2~,~~\hat Z_2:=\Zb^2=-T^3+\di T^4$.

Note that as done in the real case in \eqref{eq:shiftDiracReal}, one could include an additional central part in the above Dirac operator to allow for center of mass motion of the selfdual strings. 

The first step in the construction is to verify that the Laplace operator $\Delta^+_{s,z}:=\nablabs^+_{s,z}\nablas^+_{s,z}$ is positive and central in $\sU(2)$, if the $Z^\alpha$ satisfy the Basu-Harvey equation \eqref{eq:h3Basu-Harvey}. One readily computes the non-central part of the Laplace operator to be
\begin{equation}
\sigma^{\mu\nu}\sigma^{\mu\nu}{}_\alpha{}^\beta\left(-\tfrac{\di}{2}\right)\dder{s}D(\di Z^\alpha,Z^\beta)-\tfrac{1}{4}\sigma^{\mu\nu} \sigma^{\mu\kappa}{}_\alpha{}^\beta\sigma^{\kappa\nu}{}_\gamma^\delta[D(\di Z^\alpha,Z^\beta),D(\di Z^\gamma,Z^\delta)]~.
\end{equation}
Using the fundamental identity \eqref{eq:GenFundIdent} and the identities \eqref{eq:identitiessigmamunu} simplifies this further to
\begin{multline}
\sigma^{\mu\nu}\sigma^{\mu\nu}{}_\alpha{}^\beta\tfrac{1}{2}\dder{s}D(Z^\alpha,Z^\beta)\\+\tfrac{1}{8}\sigma^{\mu\nu}(\sigma^{\mu\nu}{}_\alpha{}^\delta\delta_\beta^\gamma-\sigma^{\mu\nu}{}_\gamma{}^\beta\delta_\alpha^\delta)\big(D([Z^\gamma,Z^\alpha;Z^\beta],Z^\delta)-D(Z^\gamma,[Z^\delta,Z^\beta;Z^\alpha])\big)~.
\end{multline}
Due to $\sigma^{\mu\nu}{}_\alpha{}^\delta\eps^{\beta\alpha}=\sigma^{\mu\nu}{}_\alpha{}^\beta\eps^{\delta\alpha}$, we have
\begin{equation}
\begin{aligned}
 -\sigma^{\mu\nu}{}_\gamma{}^\beta D([Z^\gamma,Z^\alpha;Z^\beta],Z^\alpha)&=\sigma^{\mu\nu}{}_\alpha{}^\delta D([Z^\beta,Z^\alpha;Z^\beta],Z^\delta)~,\\
 -\sigma^{\mu\nu}{}_\alpha{}^\delta D(Z^\beta,[Z^\delta,Z^\beta;Z^\alpha])&=\sigma^{\mu\nu}{}_\gamma{}^\beta D(Z^\gamma,[Z^\alpha,Z^\beta;Z^\alpha])~,
\end{aligned}
\end{equation}
and the non-central part of the Laplace operator becomes proportional to
\begin{equation}
 \sigma^{\mu\nu}\sigma^{\mu\nu}{}_\alpha{}^\beta\left(\dder{s}D(Z^\alpha,Z^\beta)+\tfrac{1}{2}D([Z^\gamma,Z^\alpha;Z^\gamma],Z^\beta)+\tfrac{1}{2}D(Z^\alpha,[Z^\gamma,Z^\beta;Z^\gamma]\right)~.
\end{equation}
This expression vanishes, if the Basu-Harvey equation \eqref{eq:h3Basu-Harvey} is satisfied. In this case, the Laplace operator $\Delta^-_{s,z}:=\nablabs^-_{s,z}\nablas^-_{s,z}$ and thus $\Delta_{s,z}:=\nablabs_{s,z}\nablas_{s,z}$ are positive and central in $\sU(2)$, too. Note that the inverse statement is not necessarily true, as the map $D:\CA\times\CA\rightarrow \frg_\CA$ could be degenerate. 

As in the case of real 3-algebras, we again have $2N$ zero modes $\psi_{s,z}^a\in W^{0,2}(\CI)\otimes \FC^2\otimes\FC^N\otimes \CA$, $a=1,\ldots,2N$, of the Dirac operator $\nablabs_{s,z}$. We sort them according to their chirality and normalize them such that
\begin{equation}\label{eq:abjmnormalizePsiM}
 \delta^{ab}=\int_\CI\dd s\, (\psib^{a}_{s,z},\psi^{b}_{s,z})~,
\end{equation}
where $(\dotsp,\dotsp)$ denotes the inner product on $\FC^4\otimes \CA$. Contrary to the real case, we now define a complex gauge potential,
\begin{subequations}\label{eq:h3FieldDefinitions}
\begin{equation}
\begin{aligned}
\big(A_{(\alpha\tau)}\big)^{ab}&=\int \dd s\, \left(\psib^a_{s,z}, \delder{z^\alpha(\tau)}\psi^b_{s,z}\right)~,\\
\big(A^{(\alphab\tau)}\big)^{ab}&=\int \dd s\, \left(\psib^a_{s,z}, \delder{\zb_\alpha(\tau)}\psi^b_{s,z}\right)~,
\end{aligned}
\end{equation}
and a scalar field
\begin{equation}
 \Phi^{ab}=\di \int \dd s\, \left(\psib^a_{s,z}\,,s\,\psi^b_{s,z}\right)~.
\end{equation}
\end{subequations}
These fields take values in the gauge algebra $\au(N)_+\oplus\au(N)_-$. The selfdual string equation on loop space \eqref{eq:NALoopSpaceSDS} for the $\au(N)_+$-components of the complex gauge potential and the Higgs field reads as
\begin{equation}\label{eq:h3NASDSeq}
\begin{aligned}
 F_{(\alpha\sigma)(\beta\tau)}=[D_{(\alpha\sigma)},D_{(\beta\tau)}]&=\tfrac{1}{2}(\dot{\zb}_{\beta}(\sigma)D_{(\alpha\tau)}\Phi-\dot{\zb}_{\alpha}(\tau)D_{(\beta\sigma)}\Phi)~,\\
 F^{(\alphab\sigma)(\betab\tau)}=[D^{(\alphab\sigma},D^{(\betab\tau)}]&=\tfrac{1}{2}(\dot{z}^\beta(\sigma)D^{(\alphab\tau)}\Phi-\dot{z}^\alpha(\tau)D^{(\betab\sigma)}\Phi)~,\\
 F_{(\alpha\sigma)}{}^{(\betab\tau)}=[D_{(\alpha\sigma)},D^{(\betab\tau)}]&=\tfrac{1}{2}\eps_{\alpha\gamma}\eps^{\beta\delta}(\dot{z}^\gamma(\tau) D_{(\delta\sigma)}\Phi-\dot{\zb}_\delta(\sigma)D^{(\gammab\tau)}\Phi)~,
\end{aligned}
\end{equation}
where $D_{(\alpha\sigma)}:=\delder{z^\alpha(\sigma)}+A_{(\alpha\sigma)}$, $D^{(\alphab\sigma)}:=\delder{\zb_\alpha(\sigma)}+A^{(\alphab\sigma)}$ and $\eps^{12}=-\eps_{12}:=1$. The corresponding equations for the $\au(N)_-$ components are obtained from \eqref{eq:h3NASDSeq} by substituting $z\rightarrow\hat z$.

The proof that the fields \eqref{eq:h3FieldDefinitions} indeed satisfy these equations closely follows the real case. For simplicity, we restrict to the $\au(N)_+$ components. The proof for the $\au(N)_-$ components is completely analogous. We start by introducing the Green's function $G_{z}(s,t)$ of the Laplace operator $\Delta^+_{s,z}$ leading again to the completeness relation
\begin{equation}\label{eq:abjmcompleteness}
 \delta(s-t)=\psi^a_{s,z}\big(\psib^a_{t,z},\dotsp\big)-\nablas^+_{s,z} G_z(s,t)\nablabs^+_{t,z}~.
\end{equation}
We then compute
\begin{equation}
\begin{aligned}
(F_{(\alpha\sigma)(\beta\tau)})^{ab}=~&2\int_\CI \dd s\,(\delta_{[(\alpha\sigma)}\psib^a_{s,z},\delta_{(\beta\tau)]}\psi^b_{s,z})\\&+2\!\int_\CI \dd s\!\int_\CI \dd t\,(\psib^a_{s,z},\delta_{[(\alpha\sigma)}\psi^c_{s,z})(\psib^c_{t,z},\delta_{(\beta\tau)]}\psi^b_{t,z})\\
=~&-2\int_\CI \dd s\int_\CI \dd t\,\Big(\delta_{[(\alpha\sigma)}\psib^a_{s,z}\,,\,\left(\nablas^+_{s,z}G_z(s,t)\nablabs^+_{t,z}\right)\delta_{(\beta \tau)]}\psi^b_{t,z}\Big)
\end{aligned}
\end{equation}
and
\begin{equation}
(F_{(\alpha\sigma)}{}^{(\betab\tau)})^{ab}=-2\int_\CI \dd s\int_\CI \dd t\,\Big(\delta_{[(\alpha\sigma)}\psib^a_{s,z}\,,\,\left(\nablas^+_{s,z}G_z(s,t)\nablabs^+_{t,z}\right)\delta^{(\betab \tau)]}\psi^b_{t,z}\Big)~.
\end{equation}
Here, we need the identities
\begin{equation}
\begin{aligned}
 \sigma^{\mu\nu}\sigma^{\kappa\lambda}\big(\sigma^{\mu\nu}{}_{\alpha}{}^\gamma\sigma^{\kappa\lambda}{}_{\beta}{}^\delta\dot{\zb}_\gamma(\sigma)\dot{\zb}_\delta(\tau)\big)\stackrel{[\cdot]}{=}
2\sigma^{\mu\nu}\sigma^{\mu\nu}{}_{\alpha}{}^\gamma\dot{\zb}_\gamma(\tau)\dot{\zb}_{\beta}(\sigma)~,\hspace{2cm}\\
\sigma^{\mu\nu}\sigma^{\kappa\lambda}\big(\sigma^{\mu\nu}{}_\alpha{}^\gamma\sigma^{\kappa\lambda}{}_\delta{}^\beta \dot{\zb}_\gamma(\sigma)\dot{z}^\delta(\tau)-\sigma^{\mu\nu}{}_\delta{}^\beta\sigma^{\kappa\lambda}{}_\alpha{}^\gamma\dot{z}^\delta(\tau)\dot{\zb}_\gamma(\sigma)\big)\hspace{4cm}\\
=-2\epsilon_{\alpha \gamma}\epsilon^{\beta\delta}\sigma^{\mu\nu}(\sigma^{\mu\nu}{}_\kappa{}^\gamma \dot z^\kappa(\tau)\dot{\zb}_\delta(\sigma)+\sigma^{\mu\nu}{}_\delta{}^\kappa \dot z^\gamma(\tau)\dot{\zb}_\kappa(\sigma))~,
\end{aligned}
\end{equation}
where $\stackrel{[\cdot]}{=}$ denotes weighted antisymmetrization under $(\alpha\sigma)\leftrightarrow (\beta\tau)$. The identities lead to
\begin{equation}
\begin{aligned}
(F_{(\alpha\sigma)(\beta\tau)})^{ab}&\stackrel{[\cdot]}{=}\int_\CI \dd s\int_\CI \dd t\,\Big(\psib_{s,z}^a,\,\left(\sigma^{\mu\nu}\sigma^{\mu\nu}{}_{\alpha}{}^\gamma\dot{\zb}_{(\gamma\tau)}\dot{\zb}_{(\beta\sigma)} G_z(s,t)\right)\psi^b_{t,z}\Big)\\
&=\di\dot{\zb}_{[(\beta\sigma)}\int_\CI \dd s\,\Big(D_{(\alpha\tau)]}\psib^a_{s,z},\,s\,\psi^b_{s,z}\Big)+\Big(\psib^a_{s,z},\,s\,D_{(\alpha\tau)]}\psi^b_{s,z}\Big)\\
&=\tfrac{1}{2}(\dot{\zb}_{\beta}(\sigma)D_{(\alpha\tau)}\Phi^{ab}-\dot{\zb}_{\alpha}(\tau)D_{(\beta\sigma)}\Phi^{ab})~,
\end{aligned}
\end{equation}
and
\begin{equation}
 F_{(\alpha\sigma)}{}^{(\betab\tau)}=\tfrac{1}{2}\eps_{\alpha\gamma}\eps^{\beta\delta}(\dot{z}^\gamma(\tau) D_{(\delta\sigma)}\Phi-\dot{\zb}_\delta(\sigma)D^{(\gammab\tau)}\Phi)~.
\end{equation}

\subsection{Comment on the reduction to monopoles}

In the complex description of selfdual strings we work with loops wrapping the $x^4$-direction by imposing the condition $\dot{\zb}_\alpha=-\di R\delta_\alpha^2$, cf.\ section \ref{sec:r3reduction}. Then the whole reduction procedure for hermitian 3-algebras works fully analogously to the case of real 3-algebras. We therefore refrain from going into further details.

\subsection{Examples}

We now will see a few simple examples of the construction. We start with the simplest case $n=N=1$, which is a mere rewriting of the same case for real 3-algebras in complex notation. We can rewrite $r_{-}^2=\sqrt{\frac{1}{4}z^\alpha{}_\alpha z^\beta{}_\beta-\frac{1}{2}z^\alpha{}_\beta z^\beta{}_\alpha}$, where we've used complex area coordinates: $z^\alpha{}_\beta:=\frac{1}{2}\int\dd\tau(z^{\alpha}(\tau)\dot{\zb}_\beta(\tau)-\dot z^\alpha(\tau)\zb_\beta(\tau))$. As in the real case, the Nahm data are trivial: $Z^\alpha=0$ and the zero mode reads before normalization as
\begin{equation}
\psi^+\sim\de^{-r_{-}^2 s}\begin{pmatrix} \di r_{-}^2 +z^1{}_1-{z}^2{}_2\\ 2z^1{}_2\\ 0 \\ 0\end{pmatrix}~,~~~
\psi^-\sim\de^{-r_{+}^2 s}\begin{pmatrix} 0 \\ 0 \\ \di r_{+}^2 +\hat{z}^1{}_1-\hat{z}^2{}_2\\ 2\hat{z}^1{}_2\end{pmatrix}~,
\end{equation}
and leads to the expected Higgs field
\begin{equation}
\Phi=\left(\begin{array}{cc}
\frac{\di}{2r_{-}^2} & 0 \\ 0 &
\frac{\di}{2r_{+}^2}
           \end{array}\right)~.
\end{equation}

Next, let us consider the case $N=1$, $n$ arbitrary. Note that for $n>2$, this case could not have been treated using 3-Lie algebras. The corresponding solution to the Basu-Harvey equation has been found in \cite{Gomis:2008vc}. In our conventions, it reads as
\begin{equation*}
\begin{aligned}
 Z^1=\frac{1}{\sqrt{s}}&\left(\begin{array}{ccccc} 
	    0 &  0 & 0 & \ldots & 0 \\ 
	    0 & \sqrt{1} & 0 & & \vdots \\ 
	    0  & 0 & \sqrt{2} & & \\
	    \vdots  & & & \ddots & \\
	    0 & \ldots & & & \sqrt{n-1}
	  \end{array}\right)~,~\\
 Z^2=\frac{1}{\sqrt{s}}&\left(\begin{array}{ccccc} 
	    0 &  0 & 0 & \ldots & 0 \\ 
	    \sqrt{n-1} & 0 & 0 & & \vdots \\ 
	    0  & \sqrt{n-2} & 0 & & \\
	    \vdots  & & \ddots &  & \\
	    0 & \ldots & 0 & 1 & 0
	  \end{array}\right)~.
\end{aligned}
\end{equation*}
As before, we consider the zero modes only at $y^{41}=r_{\pm}^2=\di z^1{}_2=\di z^2{}_1=:r^2$ and extract the Higgs field as a consistency check. The zero modes of the Dirac operator $\nablabs_{s,z}$ with this restriction are given by 
\begin{equation*}
 \psi^+\sim\de^{-r^2 s}s^{\frac{n-1}{2}}\left(\begin{array}{c}\zeta\\ \zeta\\ 0 \\ 0\end{array}\right)\eand 
 \psi^-\sim\de^{-r^2 s}s^{\frac{n-1}{2}}\left(\begin{array}{c}0 \\ 0\\ \zeta\\ \zeta\end{array}\right)
\end{equation*}
with
\begin{equation}
 \zeta=\left(\begin{array}{ccccc}\sqrt{\binomr{n-1}{0}} & \sqrt{\binomr{n-1}{1}}\ & \sqrt{\binomr{n-1}{2}}\ & \ldots & \sqrt{\binomr{n-1}{n-1}}\\ 0 & 0 & 0 & 0 & 0\\
             &&\vdots&&\\
0 & 0 & 0 & 0 & 0
\end{array}\right)~.
\end{equation}
One readily computes the Higgs field 
\begin{equation}
 \Phi=\frac{\di n}{2r^2}\,\unit_2~.
\end{equation}
and we indeed recovered a selfdual string of charge $n$.

\chapter{Magnetic Domains}
\label{ch:background}

In this chapter we will break away from loop space and construct a transform for selfdual strings which does not involve loop space. This transform only works for the special limiting case of infinitely many M2-branes. In \cite{Harland:2011tm}, a Nahm transform for infinitely many monopoles (magnetic bags) was found, which is the basis for the selfdual string transform in this chapter. This can be generalized easily to objects in higher dimensions. We will call all of these objects, including magnetic bags, magnetic domains.

The extension of the Nahm transform to certain configurations of infinitely many D1-branes was developed in  \cite{Harland:2011tm}. The crucial observation is that the Lie algebra $\au(n)$ can be viewed as the algebra of functions on the fuzzy sphere, with $1/n$ playing the role of the non-commutativity parameter. The fields describing the transverse fluctuations of the D1-brane are $\au(n)$-valued functions on an interval $\CI$. In the limit $n\rightarrow\infty$ they become functions on $S^2 \times \CI$. These fields are then put together into a map
\begin{equation}
t:S^2\times \CI \rightarrow\FR^3~,
\end{equation}
from which the fields on $\FR^3$ can easily be constructed.

The resulting configurations are known as magnetic bags.  Magnetic bags are abelian configurations that were introduced in \cite{Bolognesi:2005rk}.
They are widely believed to describe the large $n$ limit of $n$-monopoles in non-abelian gauge theory. This is known as Bolognesi's conjecture. 

In this chapter we investigate various extensions of the Nahm transform, in particular also to bags of selfdual strings. We will begin in section \ref{sec:Bags} with a discussion of the 3-dimensional situation.  The notion of magnetic bags is generalized to that of {\em magnetic domains in three dimensions}; the latter may appear as limits not only of monopoles, but also of monopole walls, monopole chains, and probably other configurations.  We will state and prove a Nahm transform for magnetic domains which generalizes that given in \cite{Harland:2011tm}.  We will also see a partial proof of Bolognesi's conjecture for the case of magnetic discs, which are flattened magnetic bags.

In section \ref{sec:branes} we present a D-brane interpretation of magnetic bags and their Nahm transform.  The surfaces of magnetic bags are junctions of D-branes which are related by T- and S-duality to junctions of $(p,q)$ 5-branes.  These junctions appear in the Nahm data as defects.  The D-brane picture is valuable not only as further support for the magnetic bag conjecture, but also as a guide in generalizing the magnetic bag conjecture to M-theory.  Indeed, it seems very likely that $n$ M2-branes stretching between two M5-branes will form a bag as $n\to\infty$.  A striking feature here is that the bags are abelian, and thus evade the usual difficulties associated with writing down non-abelian higher gauge theories.

In sections \ref{sect:SDS}--\ref{sect:LSSDS} we investigate in detail bags and more general domains formed by selfdual strings.  A precise definition of these domains is formulated in section \ref{sect:SDS}, and we state and prove the Nahm transform for them.  The Nahm-dual picture for a selfdual string bag consists of solutions of the Basu-Harvey equation based on the algebra of functions on the 3-sphere.  These can be combined into a map
\begin{equation}
t:S^3\times \CI\rightarrow\FR^4~,
\end{equation}
from which the bag can be recovered.  This substantially improves a result of Ho and Matsuo \cite{Ho:2008nn}, who showed that the Bagger-Lambert-Gustavsson action based on the algebra of functions on a 3-manifold at least has the correct low energy degrees of freedom to describe M5-branes.

We go on to show in section \ref{sect:HermitianBH} that this Basu-Harvey equation is the large $n$ limit of the equation introduced in \cite{Terashima:2008sy,Gomis:2008vc,Hanaki:2008cu} for describing $n$ M2-branes.  The bags obtained in this large $n$ limit are quite constrained: they are necessarily invariant under a certain action of $\sU(1)$.  In fact, they can be identified with ordinary magnetic bags using the Hopf fibration.

We show in section \ref{sect:LSSDS} that our Nahm equation for selfdual string bags also has a natural loop space formulation and re-interpret the Nahm transform from that perspective.

Finally, we provide in section \ref{sec:higher} a construction for bags in higher gauge theories.  An interesting feature here is that the Nahm equation can be written as a Maurer-Cartan equation for an element of an $L_\infty$-algebra.

\section{Magnetic domains in three dimensions}\label{sec:Bags}

\subsection{From magnetic monopoles to magnetic domains}

We will reintroduce the coupling constant $e$ since a double-scaling limit will need to be taken to regulate the size of the magnetic bag.

To get to magnetic bags we begin with $\sSU(2)$ Yang-Mills-Higgs theory: an $\sSU(2)$ principal bundle over $\FR^3$ with connection 1-form $A$, curvature 2-form $F$ and an adjoint Higgs field $\Phi$. We define
\begin{equation}
F=\dd A +e A\wedge A~,~~~D\Phi=\dd\Phi+e[A,\Phi]~,
\end{equation}
where $e$ is the Yang-Mills coupling constant. The Yang-Mills-Higgs energy functional 
\begin{equation}
E = \tfrac{1}{2}\int_{\FR^3} \tr\left(F\wedge\ast F + D\Phi\wedge * D \Phi\right)
\end{equation}
admits a Bogomolny bound
\begin{equation}
E = \int_{\FR^3} \tr\left(\tfrac{1}{2}|D\Phi-*F|^2+D\Phi\wedge F\right) \ge \int_{S^2_\infty} \tr(F\Phi)~.
\end{equation}

This bound is saturated (and the Yang-Mills-Higgs equations of motion are satisfied) for BPS monopoles, which are defined as solutions $(A,\Phi)$ to the Bogomolny monopole equation 
\begin{equation}\label{eq:BogomolnyMonopole}
F=*D\Phi~,
\end{equation}
together with the asymptotic condition $||\Phi||:=\sqrt{\frac{1}{2}\tr(\Phi^\dagger\Phi)}\rightarrow v>0$ as $r\rightarrow\infty$.  This asymptotic condition on $\Phi$ breaks the gauge symmetry to $\sU(1)$, and therefore it makes sense to talk about the magnetic charge $q$ of a monopole.  It is well-known that the magnetic charge is quantized:
\begin{equation}\label{magnetic charge}
q := -\tfrac{1}{2}\int_{S^2_\infty} \frac{\tr(F\Phi)}{\|\Phi\|} = \frac{2\pi n}{e}~.
\end{equation}
Here $n\in \RZ$ is a topological charge which counts the number of monopoles. The Bogomolny bound can now be written as $E\ge vq$.

In this section, we are interested in monopole configurations that arise in the limit $n\rightarrow \infty$.  For example, consider a BPS configuration of an odd number $n$ of monopoles in $\FR^3$ located on a one-dimensional lattice at $\vec{x}=(i,0,0)$, $i\in \RZ$, $|i|\leq (n-1)/2$, where we use the usual Cartesian coordinates on $\FR^3$.  Such configurations of monopoles are known to exist, and in a certain limit $n,v\to\infty$ one obtains a solution of the Bogomolny equation invariant under a translation group $\RZ$ \cite{Dunne:2005pr,Harland:2009yh}. This is an example of a {\em monopole chain} \cite{Cherkis:2000cj,Ward:0505254}.

Similarly, one can consider doubly-periodic monopoles invariant under the action of $\RZ^2$, given by $(x^1,x^2,x^3)\mapsto(x^1+i,x^2+j,x^3)$ for $i,j\in\RZ^2$.  One has the freedom to impose different boundary conditions as $z\to\pm\infty$, and configurations satisfying $\|\Phi\|\to A$ as $z\to-\infty$ and $\|\Phi\|\sim Bz$ as $z\to\infty$ for constants $A,B$ are know as {\em monopole walls}\footnote{Configurations for which $\|\Phi\|\sim B|z|$ as $z\to\pm\infty$ are called {\em monopole sheets}.} \cite{Lee:1998isa,Ward:0505254,Ward:2006wt}. Monopole walls are thought to be related to the boundary of magnetic bags. If a monopole wall has non-zero charge per unit period, then the total charge $n$ is again infinite.

Inductive reasoning might lead one to consider triply-periodic monopoles, but the following argument shows that there are no non-trivial examples of these.  Any triply-periodic monopole would correspond to a monopole on the compact manifold $T^3$.  The equation of motion $\square_A \Phi=0$ would then imply that $0=\int_{T^3}\tr(\Phi~D\ast D\Phi)=-\int_{T^3} \tr(D\Phi\wedge\ast D\Phi)$, and hence that $D\Phi$ vanishes\footnote{Note that although $\Phi$ is a section of an associated vector bundle, the expression under the integral is globally defined.}.

Our final examples of monopoles with $n\to\infty$ are {\em magnetic bags} \cite{Bolognesi:2005rk}.  Heuristically, a magnetic bag with finite charge $n$ consists of a finite-area segment of a monopole wall, folded around to form a closed surface.  The existence of such monopoles is an open question (which we discuss further in section \ref{commentsConjecture}), however, five examples are known with $n=3,4,5,7,11$ \cite{Lee:2008ze}.  These magnetic bags are roughly spherical in shape, and the lattice structures on their surfaces resemble the five Platonic solids.\footnote{There are actually two types of magnetic bag, termed ``abelian'' and ``non-abelian'' in \cite{Lee:2008ze}, but this distinction becomes irrelevant in the limit $n\to\infty$ that we consider.}  The size of the Platonic monopoles has been shown to be in good agreement with predictions of the bag model \cite{Manton:2011vm}.  Constructing further examples of magnetic bags on $\FR^3$ is difficult, because there are no further Platonic solids 
whose symmetries can be exploited.  The situation is much better on AdS space, where numerical methods can be used to construct magnetic bags with a large range of values of $n$ \cite{Sutcliffe:2011sr}.  Thus it is widely believed that magnetic bags exist for infinitely many values of $n$, and that they are the most tightly-packed configurations of monopoles.

The magnetic charge $q=4\pi n/e$ of a magnetic bag remains finite in the limit $n\to\infty$ provided one takes a double-scaling limit $e\to\infty$ such that $n/e$ remains finite.  In this limit the BPS energy $E=vq$ and also the size of the bag remain finite.  The double scaling limit causes two of the three $\asu(2)$ components of the fields to be exponentially suppressed. This can be seen from the D-brane interpretation discussed in section \ref{dbraneint}, where the `W-boson' strings stretching between different D-branes have masses which diverge as $\sim e||\Phi||$. With just one generator of $\asu(2)$ left, we have $\au(1)$ valued fields, which we denote $\phi$ and $f$. Explicitly, we have 
\begin{equation}
\Phi\rightarrow~\di\begin{pmatrix}\phi&&0\\0&&-\phi\end{pmatrix}\eand F\rightarrow~\di\begin{pmatrix}f&&0\\0&&-f\end{pmatrix}
\end{equation}
in local gauges as $n\to\infty$.  The surface of a bag becomes infinitely thin as $n,e\to\infty$, and can be represented by a surface $S\subset\FR^3$.  One has $f=0$ inside the bag, and hence that $\phi$ is constant; $\phi$ is continuous on $S$, but $f$ is not.  We will assume that $\phi=0$ inside the bag.

We will be concerned with magnetic bags only in this abelian double-scaling limit.  One could take similar limits of walls or chains: here one sends the topological charge per unit area (or length) and the coupling constant $e$ to infinity, in such a way that the magnetic charge per unit area (or length) stays finite.  The limiting configuration for walls could have a discontinuity in $f$ along a plane, while for chains one could perhaps arrange for a singularity along a line or a discontinuity on a cylinder.

All of these abelian limiting configurations are examples of what we will refer to as {\em magnetic domains} $\Omega$ in three dimensions: These are monopole configurations characterized by continuous $\au(1)$-valued fields $(f,\phi)$ satisfying the following properties:
\begin{itemize}
 \setlength{\itemsep}{-1mm}
 \item $f$ is closed, and therefore we have locally a gauge potential $a$ with $f=\dd a$,
 \item $f$ and $\phi$ satisfy the Bogomolny monopole equation $f=*\dd \phi$ in the region $\Omega\subset \FR^3$,
 \item $\dd \phi\neq 0$ in $\Omega$ and
 \item depending on the shape and dimensionality of the boundary of the domain $\Omega$, $\phi$ satisfies certain boundary conditions.
\end{itemize}

\subsection{Nahm transform and the fuzzy funnel}\label{sect:FuzzyFunnel}

Instead of regarding the fields appearing in the Nahm equation as functions on the interval $\CI=(-v,v)$ taking values in $\au(n)$, we can interpret them as functions on $S^2_F\times \CI$, where $S^2_F$ is a fuzzy sphere at level $n$. To understand this statement, let us briefly recall the Berezin-Toeplitz quantization of the 2-sphere \cite{Berezin:1974du}, see also \cite{Balachandran:2005ew,IuliuLazaroiu:2008pk} and references therein. We start from the round sphere $S^2\cong\CPP^1$ endowed with its Fubini-Study metric and the corresponding K\"ahler form $\omega$. As usual in geometric quantization, we have to pick an ample line bundle (the prequantum line bundle), from whose global sections we derive a Hilbert space $\CCH_n$. We choose the line bundle $L_n=\CO(n-1)$ with first Chern number $c_1=n-1$ and we will moreover work with K\"ahler polarization. This means that the Hilbert space is given by the global holomorphic sections of $L_n$: \begin{equation}
\CCH_n=H^0(\CPP^1,L_n)\cong \FC^n~.
\end{equation}
Using the volume form $\omega$, one can construct an inner product on $\CCH_n$ via
\begin{equation}
 \langle s_1|s_2\rangle:=\int_{\CPP^1} \frac{\omega(z,\zb)}{(1+z\zb)^n}~ \overline{s_1(z)}s_2(z)~,
\end{equation}
where $z\in \FC\cup\{\infty\}$ denotes a point on $\CPP^1$. Moreover, we can construct an overcomplete set of coherent states $|z\rangle\in \CCH_n$ for each $z$. These are used in the definition of the coherent state projector $P_{z,\zb}:=\frac{|z\rangle\langle z|}{\langle z| z \rangle}$, which provides a bridge between the classical and the quantum world, as $P_{z,\zb}\in\CC^\infty(\CPP^1)\otimes\sEnd(\CCH_n)$. We define the {\em Berezin symbol map}
\begin{equation}
 \sigma_n: \sEnd(\CCH_n)\rightarrow \CC^\infty_n(\CPP^1)\subset \CC^\infty(\CPP^1)\ewith \sigma_n(A):=\tr(P_{z,\zb} A)~,
\end{equation}
and the {\em Toeplitz quantization map}
\begin{equation}
 \CCT_n: \CC^\infty(\CPP^1)\rightarrow \sEnd(\CCH_n)\ewith \CCT_n(f):=\int_{\CPP^1}\omega(z,\zb) f(z,\zb)P_{z,\zb}~.
\end{equation}
The set $\CC^\infty_n(\CPP^1)$ is called the {\em set of quantizable functions at level $n$}. Both the above maps combine to the Berezin transform $\beta_n:\CC^\infty(\CPP^1)\rightarrow \CC^\infty_n(\CPP^1)$, where $\beta_n(f)=\sigma_n(\CCT_n(f))$. We now have the following results in the large $n$ limit \cite{Bordemann:1993zv}, see also \cite{Schlichenmaier:1998mv}:
\begin{equation}\label{eq:approximations}
 \|\di n[\CCT_n(f),\CCT_n(g)]-\CCT_n(\{f,g\})\|=\CO\left(\frac{1}{n}\right)\hspace{-0.25cm}\eand\hspace{-0.25cm}
 \beta_n(f)(z,\zb)=f(z,\zb)+\CO\left(\frac{1}{n}\right).
\end{equation}

On the set of quantizable functions $\CC^\infty_n(\CPP^1)$, we can invert $\sigma_n$ to obtain a quantization map $\sigma^{-1}_n$ from real functions in $\CC^\infty_n(\CPP^1)$ to $\au(n)$, the set of real endomorphisms on $\CCH_n$. In this quantization procedure, $n$ plays essentialy the role of $1/\hbar$. The fuzzy sphere is now defined via its algebra of functions $\sEnd(\CCH_n)\cong \CC^\infty_n(\CPP^1)$. Note that the operator product on $\sEnd(\CCH_n)$ induces a ``star product'' on $\CC^\infty_n(\CPP^1)$ by $f\star g=\sigma_n^{-1}(\sigma_n(f)\sigma_n(g))$.

Explicitly, the coordinate functions $x^i$ describing the embedding $S^2\subset \FR^3$ are mapped to the operators $X^i:=\frac{2\di J^i}{n}\in \au(n)$, where $J^i$ form an $n$-dimensional irreducible representation of $\asu(2)$. For these, we have the identities
\begin{eqnarray}
\label{eq:4.1}
X^iX^j-X^jX^i &=& \frac{2\di}{n}\eps_{ijk}X^k~,\\
\label{eq:4.2}
(X^1)^2+(X^2)^2+(X^3)^2 &=& 1 - \frac{1}{n^2}~,
\end{eqnarray}
which makes the limit $S^2_F\rightarrow S^2\subset \FR^3$ as $n\rightarrow\infty$ clear. 

General functions in $\CC^\infty_n(S^2)$ split up into representations of the rotation group $\sSO(3)\simeq \sSU(2)$. These representations are given by the spherical harmonics $Y_{\ell m}$, labeled by integers $0\leq \ell< n$, $m\in \RZ$ with $|m|\leq \ell$:
\begin{equation}
\CC_n^\infty(S^2)= \bigoplus_{\ell=0}^{n-1} \bigoplus_{m=-\ell}^{\ell}Y_{\ell m}\cong \bigoplus_{i=1}^n ({\bf 2i-1})_\FR~.
\end{equation}
Here, ${\bf i}$ is the $i$-dimensional irreducible representation of $\asu(2)$. A subscript $\FR$ denotes projection onto its real part under the obvious antilinear involution. Note that the functions $Y_{1m}$, $m=-1,0,1$ are linear combinations of the coordinate functions $x^1,x^2,x^3$. Under quantization, elements of $\CC_n^\infty(S^2)$ are mapped to general elements of $\au(n)$, which form the same sums of representations of $\asu(2)\cong \aso(3)$:
\begin{equation}
(\sEnd(\FC^n))_\FR\cong\au(n)\cong(\overline{{\bf n}}\otimes{\bf n})_\FR= \bigoplus_{i=1}^n ({\bf 2i-1})_\FR~,
\end{equation}

In the limit $n\rightarrow\infty$, the fuzzy sphere $S_F^2$ becomes the ordinary sphere $S^2$ and $\overline{\CC^\infty_n(S^2)}\rightarrow \CC^\infty(S^2)$. As implied by \eqref{eq:approximations}, the Lie bracket on $\au(n)$ goes over to the Poisson bracket on $\CC^\infty(S^2)$. All this suggests that in the case of magnetic bags, for which $n\rightarrow \infty$, the Nahm data should be extended from functions on an interval $\CI$ taking values in $\au(n)$ to functions on $S^2\times\CI$. A Nahm construction using this point of view has been developed in \cite{Harland:2011tm}. In the following, we will also allow for other 2-manifolds such as $\FR^2$ and $\FR\times S^1$ to replace $S^2$ and thus extend this construction to a large class of magnetic domains.

\subsection{Nahm transform for magnetic domains}\label{sect:NahmTrans}

We start from a real two-dimensional manifold $M$ without boundary. A volume form $\omega$ on $M$ induces a symplectic structure, which in turn leads to a Poisson bracket on $\CC^\infty(M)$. This Poisson bracket can be trivially extended to a Poisson bracket $\{\cdot,\cdot\}_\omega$ on $\CC^\infty(M\times \CI)$, where $\CI$ is the union of finitely many intervals on the positive real line. We denote the resulting Poisson algebra by $\Pi_\omega$. 

By {\em $\Pi_\omega$-valued Nahm data} or {\em $\Pi_\omega$-Nahm data} for short, we understand a triple of functions $t^i\in \CC^\infty(M\times \CI)$, which satisfy the $\Pi_\omega$-Nahm equation
\begin{equation}\label{eq:NahmInfinity}
 \frac{\partial t^i}{\partial s} = \frac{4\pi}{q}\frac{1}{2}\eps_{ijk} \{t^j,t^k\}_\omega~.
\end{equation}

Below we will state and prove a theorem which shows how solutions of the $\Pi_\omega$-Nahm equation can be used to construct magnetic domains.  However, before doing so we need to introduce the concept of the volume type of a volume form on a 2-manifold.  In general a non-compact manifold $M$ may be written as a union of a compact subset $K$ and a collection of open sets $U$, called ends.  For example $\FR^2\backslash\{(0,0)\}$ has two ends, one near $r=\infty$ and one near $r=0$:
\begin{equation}
\begin{aligned}
\FR^2\backslash\{(0,0)\} &= U_0\cup K\cup U_\infty~,\\
U_0 = \{x^ix^i < 1 \}~,~~~K &= \{x^ix^i = 1 \}~,~~~U_\infty = \{x^ix^i > 1 \}~.
\end{aligned}
\end{equation}
Given any volume form $\omega$ on $M$ one may measure the volume of each of its ends, and this could be either infinite or finite.  We say that two volume forms $\omega_1$, $\omega_2$ have the same {\em volume type} if every end $U$ has either infinite $\omega_1$-volume and infinite $\omega_2$-volume, or finite $\omega_1$-volume and finite $\omega_2$-volume.  The notion of volume type is independent of the choice of compact set $K$ provided that $K$ is big enough -- see appendix \ref{app:A} or reference \cite{Greene:1979aa} for more details. Note that two volume forms on a compact manifold are trivially of the same volume type.

As a simple example, consider the following two volume forms on $\FR^2\backslash\{(0,0)\}$:
\begin{equation}
\omega_1 = \dd x^1 \wedge \dd x^2~,~~~\omega_2 = \frac{\dd x^1 \wedge \dd x^2}{x^ix^i}~.
\end{equation}
These do not have the same volume type, since the volume of $U_0$ is finite when measured with $\omega_1$ but infinite when measured with $\omega_2$.  Note however that both volume forms give $M$ infinite volume.

It is not hard to show that two volume forms related by a diffeomorphism that does not permute the ends have the same volume type.  The converse statement was proven in \cite{Greene:1979aa}: if $\omega_1,\omega_2$ have the same volume type then there exists a diffeomorphism $u$ of $M$ such that $u^*\omega_2=\omega_1$.

Our theorem states: up to gauge equivalence, there is a one-to-one correspondence between
\vspace{-0.2cm}
\begin{itemize}\label{thm:2}
 \setlength{\itemsep}{-1mm}
 \item sets of $\Pi_\omega$-Nahm data with the property that the map from $M\times \CI$ to $\Omega\subset\FR^3$ defined by the $t^i$ is a diffeomorphism $t:M\times \CI\rightarrow \Omega$, and
 \item magnetic domains $\Omega$ that are diffeomorphic to $M\times \CI$, where the restriction of $f$ to any slice $M\times\{s_0\}$ has the same volume type as $\omega$ and $\CI$ is the range of $\phi$. Explicitly, there is a diffeomorphism $u:\Omega\rightarrow M\times \CI$ such that $f=\frac{q}{4\pi}u^*\omega$ and $\phi=s\circ u$ on $\Omega$. 
\end{itemize}

The proof follows closely that given in \cite{Harland:2011tm} for magnetic bags, and is similar to one given in \cite{Dunajski:2003ep}: The $\Pi_\omega$-Nahm data provide us with a diffeomorphism $t$ and its inverse $u$,
\begin{equation}\label{eq:diff}
M\times \CI ~\overset{t}{\underset{u}\rightleftarrows}~\Omega\subset\FR^3~.
\end{equation}
We will use local coordinates $\theta^{1,2}$ on $M$, $s$ on $\CI$ and Cartesian coordinates $y^i$ on $\Omega\subset \FR^3$. By definition of the Poisson bracket, the Nahm equation \eqref{eq:NahmInfinity} is equivalent to 
\begin{equation}
\dd t^i\wedge\omega=\frac{4\pi}{q}\frac{1}{2}~\eps_{ijk}~\dd t^j \wedge \dd t^k\wedge \dd s ~,
\end{equation}
where $\omega$ is the volume form on $M$. This is an equation on $M\times \CI$, which we want to pull back along $u$ to an equation on $\Omega\subset\FR^3$, identifying $y^i=u^* t^i$:
\begin{equation}
\begin{aligned}
\dd y^i\wedge u^*\omega=\frac{4\pi}{q}\frac{1}{2}\eps_{ijk}\dd y^j \wedge \dd y^k\wedge u^*\dd s&=\frac{4\pi}{q}* \dd y^i\wedge u^*\dd s=\frac{4\pi}{q} \dd y^i\wedge *~u^*\dd s\\
\iff~\frac{q}{4\pi}u^*\omega&=*\dd~ u^*s~,
\end{aligned}
\end{equation}
and therefore $f=*\dd \phi$. Note that $\omega$ is a volume form on $M$ and therefore closed. This means that locally, there exists a gauge potential $a$ such that $f=\dd a$.

Alternatively, we can start from the fields $f:=\frac{q}{4\pi}u^*\omega$ and $\phi:=u^*s$ and determine the conditions necessary for the Bogomolny equation $f=*\dd \phi$ to hold. For this, we pull back $f=*\dd \phi$ to $M\times \CI$ to get 
\begin{equation}
\frac{q}{4\pi}~\omega= t^**\dd\phi~. 
\end{equation}
We compute 
\begin{equation}
 t^**\dd\phi=\eps^{ijk}  \frac{1}{2}\frac{\partial s}{\partial t^i}\left(\frac{\partial t^j}{\partial \theta^a}\frac{\partial t^k}{\partial  s}\dd\theta^a\wedge\dd s+\frac{\partial t^j}{\partial \theta^a}\frac{\partial t^k}{\partial  \theta^b}\dd\theta^a\wedge\dd\theta^b\right)~.
\end{equation}
When the $\Pi_\omega$-Nahm equation holds, the unwanted term $\eps^{ijk}\frac{\partial s}{\partial t^i}\frac{\partial t^j}{\partial \theta^a}\frac{\partial t^k}{\partial  s}$ vanishes since\linebreak $\frac{\partial s}{\partial t^i}\frac{\partial t^i}{\partial \theta^a}=\frac{\partial s}{\partial \theta^a}=0$ and the remaining term gives $\frac{q}{4\pi}\omega$. With a little more work, it can be shown that the Nahm equation is in fact equivalent to $f=*\dd \phi$. We will use a similar argument when discussing the loop space selfdual string bags in section \ref{sect:LSSDS}.

The inverse construction is done for each connected component in $\Omega$ separately. Let us therefore restrict to one connected component $\Omega^c$ of $\Omega$, on which the range of $\phi$ is given by some interval $\CI$. By assumption, the magnetic domain $\Omega^c$ is diffeomorphic to $M\times\CI$ with $\dd\phi\neq0$ everywhere. The direct product structure $M\times\CI$ translates into a foliation of $\Omega^c$ by two-dimensional surfaces $\Sigma_\phi$ that are diffeomorphic to $M$. These surfaces are formed by the level sets of $\phi$. We pick an element $\phi_0=s_0\in \CI$ and the corresponding level set $\Sigma_{\phi_0}=\{p\in \Omega^c|\phi(p)=\phi_0\}$ together with the embedding $i: \Sigma_{\phi_0}\embd \Omega^c$. Now $f$ and $\omega$ have the same volume type, so the non-compact version of Moser's theorem \cite{Greene:1979aa} implies that there is a diffeomorphism $w:M\rightarrow \Sigma_{\phi_0}$ and a constant $q\in\FR$ such that $w^*i^*f=\frac{q}{4\pi}\omega$ (see also \cite{
0198504519}).  We will now extend the map $i$ to a diffeomorphism $t:\CI\times M\rightarrow \Omega^c$ as done in \cite{
Harland:2011tm}: The vector field $\der{s}$ has the properties
\begin{equation}
 \CL_{\der{s}}s=1\eand \iota_{\der{s}}\omega=0~,
\end{equation}
where $\CL$ denotes the Lie derivative. On $\Omega^c$, we have analogously the normalized gradient of $\phi$, i.e.\ the vector field
\begin{equation}
 Y=\left(\derr{\phi}{y^j}\derr{\phi}{y_j}\right)^{-1}\derr{\phi}{y^i}\der{y_i}~,
\end{equation}
which satisfies
\begin{equation}
 \CL_Y\phi=1\eand \iota_Y f=0~.
\end{equation}
We now solve the differential equations
\begin{equation}
 \dderr{y^i}{s}=Y^i(y(s))
\end{equation}
with the boundary condition $y(s_0)=w\circ i$ at $s_0=\phi_0$. The map $y$ yields a diffeomorphism between $\tilde{\CI}\times M$ and $\Omega^c$, where $\tilde{\CI}$ is some interval in $\FR$ containing $s_0$. Because of
\begin{equation}
 \dderr{\phi(y^i(s))}{s}=\dderr{y^i}{s}\derr{\phi}{y^i}=\CL_Y\phi=1~,
\end{equation}
$\tilde{\CI}$ is identical to the range of $\phi$ and therefore to $\CI$, and we can identify $t$ with $y$. The one-to-one correspondence is then shown by composing the transform with the inverse transform to get the identity. This completes the proof.

\

The fact that we can find a prescription for the explicit construction of magnetic domains reflects that they are described by integrable equations. They therefore come with an infinite number of conserved charges as shown in \cite{Harland:2011tm} for magnetic bags. 

The boundary conditions imposed on the $\Pi_\omega$-Nahm data at the edges of the intervals contained in $\CI$ are in direct correspondence to the boundary conditions of the fields describing the magnetic domain, as we will show in detail for various examples in the next section.

\subsection{Examples}\label{sec:3DExamples}

For magnetic bags, the boundary $S$ of the domain $\Omega$ is diffeomorphic to a sphere. The domain $\Omega$ itself is then 
$\FR^3$ with the interior of $S$ excluded. The boundary conditions imposed are that $\phi=0$ on $S$ and $\phi$ tends to some positive constant $v$ as $r\to\infty$.  Due to the Bogomolny equation \eqref{eq:BogomolnyMonopole}, the Higgs field $\phi$ is a harmonic function on $\Omega$ and furthermore, because of the definition of $q$ in \eqref{magnetic charge}, $\phi$ has the following asymptotic expansion:
\begin{equation}\label{eq:HiggsB}
 \phi\sim v - \frac{q}{4\pi r}+\CO(1)~~~\mbox{for}~~~r\rightarrow \infty~.
\end{equation}
The $\Pi_\omega$-Nahm data are now functions of $S^2\times\CI$, where $\CI=[0,v)$. The lower bound of $\CI$ corresponds to the surface $S$, while the upper bound of $\CI$ corresponds to $S^2_\infty$, the boundary of $\FR^3$ at infinity. This asymptotic behavior of the Higgs field \eqref{eq:HiggsB} induces the following boundary condition on the $\Pi_\omega$-Nahm data:
\begin{equation}
t^i(x,s) = \frac{q}{4\pi} \frac{ x^i}{v-s}+\CO(1)~~~\mbox{as}~~~s\rightarrow v~.
\end{equation}

The simplest example for a magnetic bag is the spherical one. It has $\Pi_\omega$-Nahm data \cite{Harland:2011tm}
\begin{equation}\label{eq:NDsphbag}
t^i(x,s) = \frac{q}{4\pi} \frac{ x^i}{v-s} ~~,
\end{equation}
where $ x \in S^2\subset \FR^3$. The inverse map $u:\Omega\rightarrow S^2\times I$ is 
\begin{equation}
u(\vec y)=\left(\frac{\vec y}{r},v-\frac{q}{4\pi r}\right)~,
\end{equation}
from which we compute
\begin{equation}
\phi=u^* s =\begin{cases} v-\frac{q}{4\pi r} & r\geq \frac{q}{4\pi v} \\ 0 & r < \frac{q}{4\pi v} \end{cases} ~~. 
\end{equation}
Thus, $\Omega$ is given by $\{\vec{y}~|~|\vec{y}|\geq \frac{q}{4\pi v}\}\subset \FR^3$. Now on $S^2$ we have $\omega=\sin\theta^1\dd\theta^1\wedge\dd\theta^2=\frac{1}{4}\eps_{ijk} x^i \dd  x^j\wedge\dd  x^k$ and so
\begin{equation}
\begin{aligned}
 f=&\frac{q}{4\pi}u^*\omega=\begin{cases} \frac{q}{8\pi r^3}\eps_{ijk}y^i\dd y^j\wedge\dd y^k & r\geq \frac{q}{4\pi v} \\ 0 & r < \frac{q}{4\pi v} \end{cases}~~.
\end{aligned}
\end{equation}
These fields satisfy $f=*\dd\phi$ on $\Omega$.

A generalization of this example is the ellipsoidal bag, stretched in the $y^3$ direction, for which the $\Pi_\omega$-Nahm data reads as 
\begin{equation}
 t(x,s)=\frac{q\lambda}{4\pi} \left(\frac{ x^{1}}{\sinh(\lambda(v-s))},\frac{ x^{2}}{\sinh(\lambda(v-s))},\frac{ x^3}{\tanh(\lambda(v-s))} \right)~.
\end{equation}
In the limit $\lambda\rightarrow 0$, the $\Pi_\omega$-Nahm data reduce to the spherical case \eqref{eq:NDsphbag}. Let us now restrict to $\lambda=1$ for simplicity. 

The inverse map $u:\Omega\rightarrow S^2\times \CI$ is 
\begin{equation}
u(\vec y)=\left(\left(\frac{y^1}{\alpha},\frac{y^2}{\alpha},y^3\sqrt\frac{ p}{p\alpha^2+1}\right),v-\sinh^{-1}\left(\frac{1}{\sqrt{p}\alpha}\right)\right)~,
\end{equation}
where $p=(4\pi/q)^2$ and 
\begin{equation}
\alpha^2=\frac{p r^2-1+\sqrt{4p((y^1)^2+(y^2)^2)+(pr^2-1)^2}}{2p}~.
\end{equation}
Therefore $\phi(\vec y)=v-\sinh^{-1}(\frac{q}{4\pi\alpha(\vec y)})$ and $f=\eps_{ijk} \frac{1}{\alpha\sqrt{p\alpha^2+1}}\frac{\pa\alpha}{\pa y^i}\dd y^j\wedge\dd y^k$ on
\begin{equation}
 \Omega := \left\{ \vec{y}~\Big|~(y^1)^2+(y^2)^2+\frac{1}{\cosh^2 v} (y^3)^2\geq \frac{1}{p\sinh^2 v}\right\}\subset \FR^3~.
\end{equation}

We can also consider a circular disc, i.e.\ a degenerate magnetic bag completely squashed in the $y^3$ direction, with $\Pi_\omega$-Nahm data 
\begin{equation}
 t(x,s)=\frac{q}{8v} \left(\frac{ x^{1}}{\sin(\pi(v-s)/2v)},\frac{ x^{2}}{\sin(\pi(v-s)/2v)},\frac{ x^3}{\tan(\pi(v-s)/2v)} \right)
\end{equation}
and inverse 
\begin{equation}
u(\vec y)=\left(\left(\frac{y^1}{\alpha},\frac{y^2}{\alpha},y^3\sqrt\frac{ p}{p\alpha^2-1}\right),v-\frac{2v}{\pi}\sin^{-1}\left(\frac{1}{\sqrt{p}\alpha}\right)\right)~,
\end{equation}
where $p=(8v/q)^2$ and 
\begin{equation}
\alpha^2=\frac{p r^2+1+\sqrt{-4p((y^1)^2+(y^2)^2)+(pr^2+1)^2}}{2p}~.
\end{equation}
Therefore $\phi=v(1-\frac{2}{\pi}\sin^{-1}(\frac{q}{8v\alpha(\vec y)}))$ and $f=\eps_{ijk} \frac{1}{\alpha\sqrt{p\alpha^2-1}}\frac{\pa\alpha}{\pa y^k}\dd y^j\wedge\dd y^k$ on $\Omega=\FR^3\backslash D$, where $D$ is a disc in the $y^1$-$y^2$-plane with radius $q/8v$. The Higgs field $\phi$ (and $-\phi$) are used in the plots in Figure 1. These plots will find a natural interpretation in terms of D3-branes as explained in section \ref{sec:branes}.
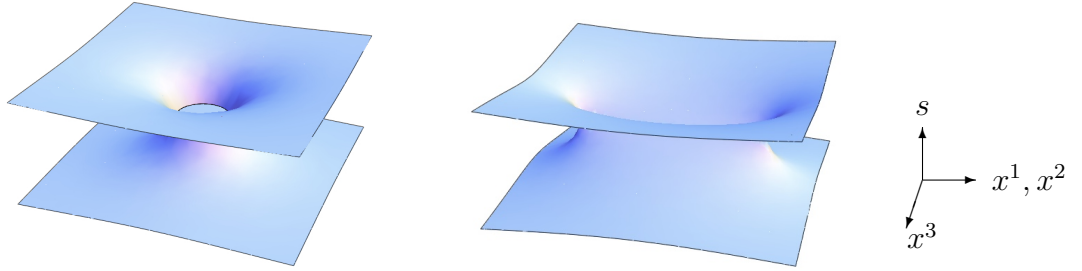
\begin{figure}[h]
\center
\begin{picture}(420,100)
~~~~\includegraphics[width=50mm]
{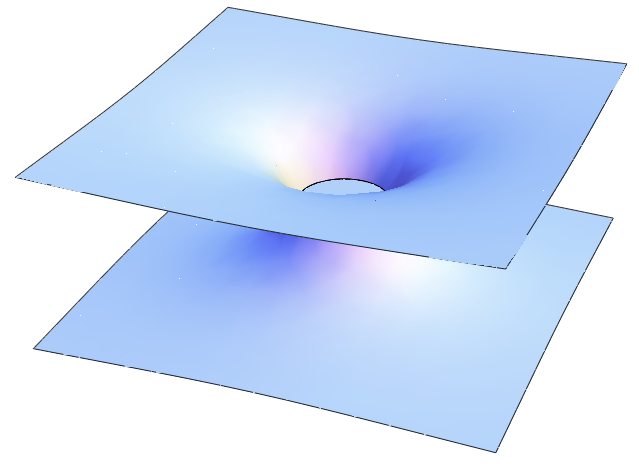}~~~~~~~~\includegraphics[width=50mm]{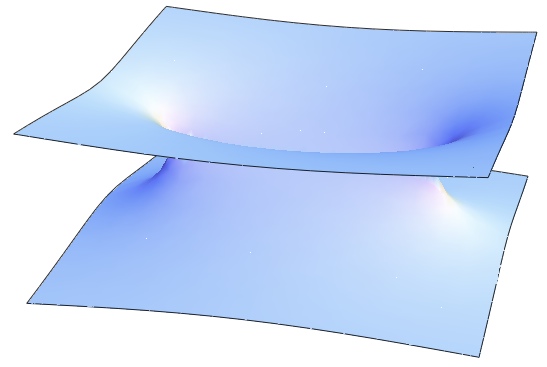}
\put(30.0,35.0){\vector(1,0){20}}
\put(30.0,35.0){\vector(0,1){20}}
\put(30.0,35.0){\vector(-1,-3){6}}
\put(70.0,35.0){\makebox(0,0)[c]{$x^1,x^2$}}
\put(30.0,62.0){\makebox(0,0)[c]{$s$}}
\put(30.0,14.0){\makebox(0,0)[c]{$x^3$}}
\end{picture}
\caption{Plots of the Higgs field $\phi$ (and $-\phi$) for the spherical magnetic bag and the circular magnetic disc. The vertical axis is the s-direction and one of the circular symmetric directions is suppressed.}
\end{figure}

The flat magnetic wall \cite{Lee:1998isa} arises from a map $t:\FR^2\times \CI\rightarrow \FR^3_{y^3> 0}$, where $\CI=[0,\infty)$. The Poisson bracket on $\FR^2$, arising from the symplectic form $\omega=\dd x^1\wedge\dd x^2$, is just 
\begin{equation}
\{x^a,x^b\}=\eps^{ab}~,~~~a,b=1,2~.
\end{equation}
The $\Pi_\omega$-Nahm data for the flat wall \cite{Harland:2011tm} are
\begin{equation}
t^a(x,s)=x^a~,~~t^3(x,s)= \frac{4\pi}{q} s~,
\end{equation}
and the inverse map is
\begin{equation}
u(\vec y)=\left( (y^1,y^2),\frac{q}{4\pi} y^3\right)~.
\end{equation}
This gives solutions to the Bogomolny equation 
\begin{equation}
\phi=u^* s= \frac{q}{4\pi} y^3 ~,~~f=\frac{q}{4\pi} u^* \omega =\frac{q}{4\pi} \dd y^1\wedge \dd y^2~.
\end{equation}

Note that the Higgs field is a harmonic function on $\Omega$, which is independent of $y^1$ and $y^2$ as expected. More general magnetic walls, correspondingly, would still have $\Pi_\omega$-Nahm data $t:\FR^2\times \CI\rightarrow \FR^3_{y^3> 0}$ satisfying the boundary condition
\begin{equation}
t^3(x,s)\sim \frac{4\pi}{q} s ~~\mbox{as }s\rightarrow\infty~.
\end{equation}

Finally, we can also consider a magnetic tube along the $y^3$ axis. This arises from a map $t:S^1\times\FR\times \CI\rightarrow \Omega\subset\FR^3$. The Poisson bracket on $S^1\times\FR$, induced by the symplectic form $\omega=\eps^{ab}x^a\dd x^b\wedge\dd z$, is
\begin{equation}
\{x^1,x^2\}=0~,~~\{z,x^a\}=\eps^{ab}x^b~,
\end{equation}
where $(x^1,x^2)\in S^1\subset\FR^2$. The $\Pi_\omega$-Nahm data are given by
\begin{equation}
t^a(x,z,s)=\de^{\frac{4\pi}{q}(s-v)}x^a~,~~t^3(x,z,s)=z~,~~~a=1,2~,
\end{equation}
and the inverse map is
\begin{equation}
u(\vec y)=\left( \left(\frac{y^1}{r},\frac{y^2}{r}\right),y^3,\frac{q}{4\pi}\ln(r)+v\right)~,
\end{equation}
where $r^2:=(y^1)^2+(y^2)^2$. From here we can see that the bag surface is a cylinder along the $y^3$-axis with radius $r=\de^{-\frac{4\pi}{q}v}$ and $\Omega$ is the exterior of this cylinder in $\FR^3$.

This gives solutions to the Bogomolny equation 
\begin{equation}
\phi=u^* s= \frac{q}{4\pi}\ln(r)+v ~,~~f=\frac{q}{4\pi} u^* \omega =\eps^{ab}\frac{q}{4\pi}\frac{y^a}{r^2}\dd y^b\wedge \dd y^3~.
\end{equation}

General magnetic tubes would have $\Pi_\omega$-Nahm data with the boundary condition 
\begin{equation}
t^a(x,z,s)\sim\de^{\frac{4\pi}{q}(s-v)}x^a~~~\mbox{for }~a=1,2~~\mbox{as }s\rightarrow\infty~.
\end{equation}
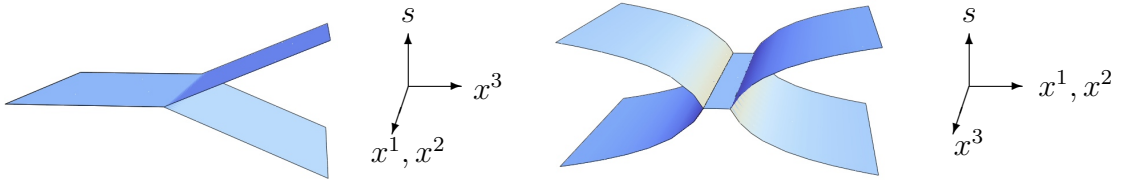
\begin{figure}[h]
\center
\begin{picture}(420,70)
\includegraphics[width=45mm]{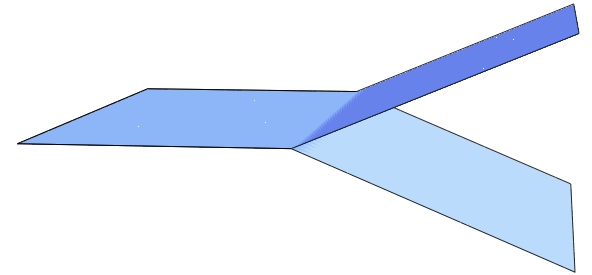}~~~~~~~~~~~~~~~~~~~~\includegraphics[width=45mm]{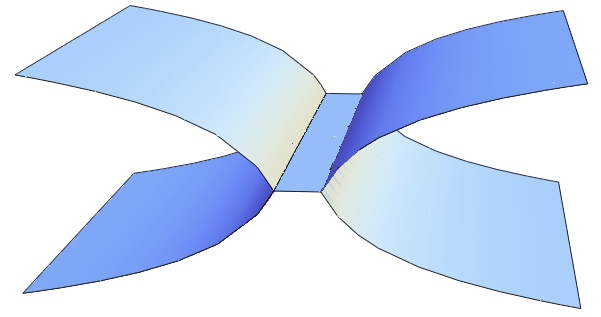}
\put(30.0,35.0){\vector(1,0){20}}
\put(30.0,35.0){\vector(0,1){20}}
\put(30.0,35.0){\vector(-1,-3){6}}
\put(70.0,35.0){\makebox(0,0)[c]{$x^1,x^2$}}
\put(30.0,62.0){\makebox(0,0)[c]{$s$}}
\put(30.0,14.0){\makebox(0,0)[c]{$x^3$}}
\put(-180.0,35.0){\vector(1,0){20}}
\put(-180.0,35.0){\vector(0,1){20}}
\put(-180.0,35.0){\vector(-1,-3){6}}
\put(-150.0,35.0){\makebox(0,0)[c]{$x^3$}}
\put(-180.0,62.0){\makebox(0,0)[c]{$s$}}
\put(-180.0,10.0){\makebox(0,0)[c]{$x^1,x^2$}}
\end{picture}
\caption{Magnetic wall and magnetic tube. The vertical axis is the s-direction and one of the symmetric directions of $\FR^3$ is suppressed.}
\end{figure}

\subsection{Magnetic domains as limits of monopole configurations}\label{commentsConjecture}

As stated above, abelian magnetic bags are expected to correspond to the large $n$ limits of non-abelian magnetic monopoles.  More precisely, Bolognesi has made the following conjecture \cite{Bolognesi:2005rk}, cf.\ \cite{Harland:2011tm}:

 For any magnetic bag $(f,\phi)$, there is a sequence $(A^{(n)},\Phi^{(n)})$ of charge $n$ solutions to the Bogomolny monopole equations $F^{(n)}=\dd_{A^{(n)}} A^{(n)}=e_n\,\star \dd \Phi^{(n)}$ with coupling constant $e_n\in\FR$ and gauge group $\sSU(2)$, such that in the limit $n\rightarrow \infty$:
 \begin{equation}\label{eq:limits}
  2\pi \, \frac{n}{e_n}\rightarrow q~,~~~\|\Phi^{(n)}\|\rightarrow\phi\eand-\frac{\tr(F^{(n)}\Phi^{(n)})}{2\|\Phi^{(n)}\|}\rightarrow f~.
 \end{equation}

Recall that the ADHMN construction gives a one-to-one correspondence between gauge equivalence classes of sets of Nahm data and gauge equivalence classes of solutions to the Bogomolny monopole equations. Note that the limits \eqref{eq:limits} in the conjecture are gauge invariant. This suggests that if the conjecture is true, then for each set of $\Pi_\omega$-Nahm data $(t^i(s))$ corresponding to a magnetic bag $(f,\phi)$, one can find a sequence of Nahm data $(T^i_{(n)}(s))$ for finite-charge monopoles that converges towards $t^i$ in the large $n$ limit. Moreover, the solutions $(T^i_{(n)}(s))$ can be extended to the full interval $\CI_2=\CI\cup -\CI$. Let us be more precise, we conjecture:

 For each solution $(t^i)$, $t^i\in\CC^\infty(S^2\times \CI)$ of the infinite-charge Nahm equation \eqref{eq:NahmInfinity} corresponding to a magnetic bag, there is a sequence of solutions $(T^i_{(n)})$, $T^i_{(n)}\in \au(n)\otimes \CC^\infty(\CI_2)$ of the finite-charge Nahm equation such that in the limit $n\rightarrow \infty$: $\sigma_n(T^i_{(n)}(s))\rightarrow t^i(s)$ on $\CI$. Here, $\sigma_n$ is the Berezin symbol map $\sigma_n:\au(n)\rightarrow \CC_n^\infty(S^2)$ introduced above.

To find a sequence of sets of Nahm data $T^i_{(n)}(s)$ converging towards a set of $\Pi_\omega$-Nahm data $t^i(s)$ for a magnetic bag, one would ideally like a non-trivial Lie algebra homomorphism from the Poisson algebra $\CC^\infty(S^2)$ to the Lie algebra $\au(n)$. However, such a map does not exist. The best one can do is to use an approximate Lie algebra homomorphisms, just as the Toeplitz quantization map, cf.\ \eqref{eq:approximations}. 

First, it is necessary to extend the $\Pi_\omega$-Nahm data for magnetic bags from the half-interval $\CI=[0,v)$ to the full interval $\CI_2=(-v,v)$.  The operation of transposition on a matrix can be interpreted as the operation of a reflection $R\in \mathrm{O}(3)$ on the fuzzy sphere \cite{Harland:2011tm}, so the reality condition $T^i(-s)=T^i(s)^t$ for monopole Nahm data should be replaced by the condition $t^i(x,-s)=t^i(Rx,s)$ for bag $\Pi_\omega$-Nahm data.  Thus $\Pi_\omega$-Nahm data on $\CI$ can be extended to $\CI_2$, but doing so may introduce a discontinuity at $s=0$.

The discontinuity is not present if the $\Pi_\omega$-Nahm data satisfy $t^i(Rx,0)=t^i(x,0)$.  If this is the case the corresponding magnetic bag will be degenerate, in the sense that the volume contained inside the magnetic bag will vanish.  It is not hard to convince oneself that Bolognesi's conjecture is true for these degenerate bags, at least in the form of the new conjecture: to obtain Nahm data for a monopole corresponding to a degenerate bag, one only needs to take $T^i_{(n)}(0):=\CCT_n(t^i(0))$ as initial conditions and solve the Nahm equation.  The Nahm equation implies that the condition $T^i(-s)_{(n)}=T^i(s)_{(n)}^t$ is automatically satisfied, because the matrices $T^i_{(n)}(0)$ are by construction symmetric. As the failure of the Toeplitz quantization map $\CCT_n$ to be a Lie algebra homomorphism is of order $\CO(1/n)$, in the limit, the deviation of $T^i_{(n)}(s)$ from $\CCT_n(t^i(s))$ vanishes:
\begin{equation}
 \int_{\CI^{(n)}_2}\dd s~ ||T^i_{(n)}(s)-\CCT_n(t^i(s))||^2\ \rightarrow \ 0~~~\mbox{as}~~~n\rightarrow \infty~.
\end{equation}
Here, $\CI^{(n)}_2$ is the maximal interval on which both the $T^i_{(n)}(s)$ and the $t^i(s)$ are defined. As the functions $t^i(s)$ diverge at $s=v$, the same should hold for the matrix-valued functions $T^i_{(n)}$ in a neighborhood of $v$ that becomes smaller with $n$, i.e.\ $\CI^{(n)}_2\rightarrow \CI_2$. The $T^i_{(n)}(s)$ thus indeed describe a sequence of Nahm data that encodes monopole solution and converges to the $\Pi_\omega$-Nahm data $t^i(s)$ of a magnetic bag. These arguments suggest that Bolognesi's conjecture is true at least for degenerate bags like e.g.\ magnetic discs.

For non-degenerate bags, the situation is more subtle: the extension of the $\Pi_\omega$-Nahm data to $\CI_2$ via $t^i(x,-s):=t^i(Rx,s)$ has a discontinuity at $s=0$ as $t^i(Rx,0)\neq t^i(x,0)$ for some $i=1,2,3$. Thus the limiting configuration $t^i(s)$ must satisfy a modified Nahm equation.  Because the $t^i(x,s)$ satisfy the Nahm equation on $\CI_2\backslash\{0\}$, we are led to
\begin{equation}\label{eq:NahmJumpBolognesi}
 \dderr{t^i}{s}=\frac{2\pi}{q}\eps_{ijk}\{t^j,t^k\}+\zeta^i \delta(s)~,
\end{equation}
where $\zeta^i\in\CC^\infty(S^2)$ determines the size of the jump at $s=0$.  Solutions of this modified Nahm equation are expected to be good approximations to solutions of the usual Nahm equation in the large $n$ limit.  To understand this modification in more detail, let us turn to the brane interpretation of magnetic domains in string theory.

\section{Brane interpretation}\label{sec:branes}

\subsection{Brane interpretation of magnetic walls and bags}\label{dbraneint}

We can consider the configuration of $n$ D1-branes ending on $N$ D3-branes at positions $x^6=s_i$, $i=1,\ldots,N$:
\begin{equation}
\begin{tabular}{rcccccccc}
& 0 & 1 & 2 & 3 & 4 & 5 & 6 & \ldots\\
D1 & $\times$ & & & & & & $\vdash$ \\
D3 & $\times$ & $\times$ & $\times$ & $\times$ & & & $s_i$
\end{tabular}
\end{equation}
To compare with Chalmers-Hanany-Witten configurations \cite{Chalmers:1996xh,Hanany:1996ie}, we T-dualize along the $x^4$- and $x^5$-direc\-tions, S-dualize and obtain
\begin{equation}\label{diag:D3NS5}
\begin{tabular}{rcccccccc}
& 0 & 1 & 2 & 3 & 4 & 5 & 6 & \ldots\\
D3 & $\times$ & & & & $\times$ & $\times$ & $\vdash$ \\
NS5 & $\times$ & $\times$ & $\times$ & $\times$ & $\times$ & $\times$ & $s_i$
\end{tabular}
\end{equation}
Dirac monopoles correspond to a single $N=1$ NS5-brane at e.g.\ $s=0$. The usual $\sSU(2)$-monopoles yield $N=2$ NS5-branes at positions $s_1=-v$ and $s_2=v$ and D3-branes suspended between them, where $\CI=(-v,v)$ is the interval over which the Nahm data is supported. The BPS equations in the gauge theory description of configuration \eqref{diag:D3NS5} are just the ordinary Nahm equations, cf.\ e.g.\ \cite{Hanany:1996ie,Cherkis:2008ip,Cherkis:2011ee}. 	

As a first nontrivial configuration, let us consider a so-called monopole wall \cite{Ward:2006wt}, i.e.\ a doubly periodic monopole. A brane interpretation of such a monopole wall has been recently discussed in \cite{Cherkis:2012qs}. Here,  we consider an NS5-brane at $s=0$ and D3-branes whose endpoints form a two-dimensional lattice in the $\FR^2_{12}$-directions\footnote{Subscripts on manifolds denote the directions in which these spaces extend into the target space $\FR^{1,9}$.}. Alternatively, we can replace the subspace $\FR^3_{123}$  with $T^2_{12}\times \FR_3$ and consider a single monopole on this space at $x^1=x^2=0$. Let us assume that the radii of the torus $T^2_{12}$ are sufficiently small and therefore the Higgs field $\Phi$ is effectively constant in the compactified directions. It therefore satisfies the Laplace equation in the $x^3$-direction:
\begin{equation}
 \der{x^3}\der{x^3} \phi(x_3)=\tan\theta~\delta(0)~,
\end{equation}
where $x^3=0$ is the position of the endpoint of the D3-brane on the NS5-brane and the angle $\theta$ is related to the lattice spacing or, equivalently, the radii of the torus $T^2_{12}$, cf.\ e.g.\ \cite{Aharony:1997ju}. The solution of this equation is
\begin{equation}
 \phi(x_3)=\frac{\tan\theta}{2}|x^3|+b x^3+c~,~~~b,c\in\FR~.
\end{equation}
The constants can be fixed by demanding that $x^6=\phi(x^3)=0$ for $x^3\leq 0$, which yields $b=\tfrac{\tan\theta}{2}$ and $c=0$. This configuration is in fact related to a bound state between D5- and NS5-branes. To see this, let us T-dualize along $T^2_{12}$, and we arrive at the configuration
\begin{equation*}\label{diag:pq}
\begin{tabular}{rcccccccc}
& 0 & 1 & 2 & 3 & 4 & 5 & 6 & \ldots\\
D5 & $\times$ & $\times$ & $\times$ & 0 & $\times$ & $\times$ & $\vdash$ \\
NS5 & $\times$ & $\times$ & $\times$ & $\vdash$ & $\times$ & $\times$ & 0\\
(1,1) & $\times$ & $\times$ & $\times$ & \rotatebox[origin=c]{45}{$\vdash$} & $\times$ & $\times$ &  \rotatebox[origin=c]{45}{$\vdash$}
\end{tabular}
\begin{picture}(290,55)(0,0)
\put(30.0,20.0){\line(1,0){50}}
\put(40.0,12.0){\makebox(0,0)[c]{NS5}}
\put(80.0,20.0){\line(0,-1){50}}
\put(90.0,-22.0){\makebox(0,0)[c]{D5}}
\put(80.0,20.0){\line(1,1){25}}
\put(130.0,36.0){\makebox(0,0)[c]{$(1,1)$-brane}}
\put(130.0,-25.0){\vector(1,0){20}}
\put(130.0,-25.0){\vector(0,1){20}}
\put(160.0,-25.0){\makebox(0,0)[c]{$x^3$}}
\put(130.0,2.0){\makebox(0,0)[c]{$x^6$}}
\end{picture}\vspace*{0.4cm}
\end{equation*}
The NS5-brane ends at $x^3=0$ and turns into a $(p,q)$-fivebrane with $p=q=1$ which extends diagonally in $\FR^2_{36}$ as indicated by the symbol \rotatebox[origin=c]{45}{$\vdash$}. A $(p,q)$-brane \cite{Aharony:1997ju} is a bound state of $p$ NS5-branes and $q$ D5-branes, fused together at a junction like the one above. The angle $\theta$ is restricted by $\tan\theta=g_s \frac{p}{q}$, $p,q\in\NN$, where $g_s$ is the string coupling.

Combining two such monopole walls and tuning the length of the connecting D5-branes to zero, we obtain the following picture:
\begin{equation*}
\begin{picture}(130,55)
\put(30.0,20.0){\line(1,0){50}}
\put(40.0,12.0){\makebox(0,0)[c]{NS5}}
\put(30.0,-10.0){\line(1,0){50}}
\put(40.0,-18.0){\makebox(0,0)[c]{NS5}}
\put(80.0,20.0){\line(0,-1){30}}
\put(90.0,6.0){\makebox(0,0)[c]{D5}}
\put(80.0,20.0){\line(1,1){25}}
\put(130.0,36.0){\makebox(0,0)[c]{$(1,1)$-brane}}
\put(80.0,-10.0){\line(1,-1){25}}
\put(130.0,-26.0){\makebox(0,0)[c]{$(1,1)$-brane}}
\end{picture}~~~~\longrightarrow
\begin{picture}(200,45)
\put(30.0,6.0){\line(1,0){50}}
\put(30.0,4.0){\line(1,0){50}}
\put(40.0,-5.0){\makebox(0,0)[c]{NS5}}
\put(80.0,6.0){\line(0,-1){2}}
\put(80.0,6.0){\line(1,1){25}}
\put(130.0,38.0){\makebox(0,0)[c]{$(1,1)$-brane}}
\put(80.0,4.0){\line(1,-1){25}}
\put(130.0,-28.0){\makebox(0,0)[c]{$(1,1)$-brane}}
\end{picture}\vspace*{1.2cm}
\end{equation*}
The right configuration is a useful picture for the neighborhood of the edge of a magnetic bag. Let us now try to model a complete spherical magnetic bag. For this, consider two NS5-branes as above, which extend into $\FR^6_{012345}$ at $s_1=-v$ and $s_2=v$, together with D5-branes extending into $\FR_0\times \FR^3_{456}$, wrapping a 2-sphere $S^2$ in $\FR^3_{123}$, and ending on the NS5-branes at $s_1$ and $s_2$. The boundary of the D5-branes in $\FR^3_{123}$ is given by the 2-sphere, which is identified with the surface of the magnetic bag. We now perform again the analysis of the Higgs field as above. The Higgs field now has to satisfy the Laplace equation in three dimensions, which yields $\phi\sim v-\frac{1}{r}$, where $r$ is the radial distance from the center of the 2-spheres $S^2$. As a boundary condition, we demand that the NS5-branes are flat in the interior of the bag. This deforms them to $(1,1)$-branes on the outside of the bag:
\begin{equation}\label{diag:pq2}
\begin{tabular}{rc}
D5 & $\FR_0\times S^2_{123}\times \FR^3_{456}$\\ 
NS5 & $\FR_0\times B^3_{123}\times \FR^2_{45}$\\ 
(1,1) & $\FR_0\times S^2_{123}\times \FR^+_{1236}\times \FR^2_{45}$
\end{tabular}
\begin{picture}(190,35)
\put(15.0,15.0){\line(1,1){10}}
\put(165.0,15.0){\line(1,1){10}}
\put(15.0,-25.0){\line(1,1){10}}
\put(165.0,-25.0){\line(1,1){10}}
\put(15.0,15.0){\line(1,0){150}}
\put(25.0,25.0){\line(1,0){150}}
\put(25.0,25.0){\line(1,0){150}}
\put(15.0,-25.0){\line(1,0){150}}
\put(25.0,-15.0){\line(1,0){150}}
\qbezier(95, -10)(120, -10)(120, 0)
\qbezier(95, 10)(120, 10)(120, 0)
\qbezier(95, 10)(70, 10)(70, 0)
\qbezier(95, -10)(70, -10)(70, 0)
\qbezier(120, 0)(120, 20)(170, 20)
\qbezier(120, 0)(120, -20)(170, -20)
\qbezier(70, 0)(70, -20)(20, -20)
\qbezier(70, 0)(70, 20)(20, 20)
\put(95.0,0.0){\makebox(0,0)[c]{NS5}}
\put(190.0,20.0){\makebox(0,0)[c]{$(1,1)$}}
\put(190.0,-20.0){\makebox(0,0)[c]{$(1,1)$}}
\end{picture}\vspace*{0.6cm}
\end{equation}
After taking the length of the D5-branes in the $x^6$ direction to zero, the Higgs field has the profile of that of the spherical magnetic bag.

\subsection{Approximating the Nahm data for magnetic bags}

We now return to equation \eqref{eq:NahmJumpBolognesi} and its interpretation in terms of branes. We start again from two NS5 branes at $s_1=-v$ and $s_2=v$ and $n$ D3-branes suspended between them. The source at $s=0$ in equation \eqref{eq:NahmJumpBolognesi} signals that there is an `impurity' in the worldvolume of the D3-branes. Such impurity theories have been extensively studied, see e.g.\ \cite{Cherkis:2012qs} and references therein. In string theory, the impurities can be modeled by inserting fivebranes whose worldvolumes are orthogonal to the direction $x^6$. These fivebranes are assumed to be heavy compared to the D3-branes, and therefore they are considered as static. Moreover, the distribution-like source induces a jump in the Nahm datum $T^1$ at $s=0$, signaling a breaking of the D3-branes in the $x^1$-direction:
\begin{equation}\label{diag:pq3}
\begin{picture}(170,60)
\put(0.0,0.0){\line(0,1){50}}
\put(0.0,10.0){\line(1,0){50}}
\put(0.0,12.0){\line(1,0){50}}
\linethickness{1pt}
\put(50.0,0.0){\line(0,1){50}}
\linethickness{0.4pt}
\put(50.0,40.0){\line(1,0){50}}
\put(50.0,38.0){\line(1,0){50}}
\put(100.0,0.0){\line(0,1){50}}
\put(13.0,45.0){\makebox(0,0)[c]{NS5}}
\put(113.0,45.0){\makebox(0,0)[c]{NS5}}
\put(66.0,5.0){\makebox(0,0)[c]{defect}}
\put(25.0,17.0){\makebox(0,0)[c]{D3s}}
\put(75.0,32.0){\makebox(0,0)[c]{D3s}}
\put(140.0,10.0){\vector(1,0){20}}
\put(140.0,10.0){\vector(0,1){20}}
\put(170.0,10.0){\makebox(0,0)[c]{$x^6$}}
\put(140.0,37.0){\makebox(0,0)[c]{$x^1$}}
\end{picture}\vspace*{0.0cm}
\end{equation}
If we insert a D5-brane at $s=0$ parallel to the NS5-branes and such that the D3-branes can intersect it, the strings connecting the D3- and D5-branes yield an additional fundamental hypermultiplet \cite{Hanany:1996ie}. Giving a vacuum expectation value (vev) to this hypermultiplet, we obtain additional source terms to the Nahm equation, which are of the form\footnote{As remarked in \cite{Tsimpis:1998zh}, it is expected that stringy effects will regulate the $\delta(s)$-term to an exponential approximation.}
\begin{equation}\label{eq:NahmJumpFinite}
 \dderr{T^i_{(n)}}{s}=\frac{2\pi}{q}\eps_{ijk}[T^j_{(n)},T^k_{(n)}]+h_a\otimes h_b^*\sigma^i_{ab}\delta(s)~,
\end{equation}
where $h_a\in \FC^k\otimes \FC^2$. This is the Nahm equation appearing in the construction of an $\sSU(3)$ monopole \cite{Hurtubise:1989qy,Tsimpis:1998zh}. Note that the expression $h_a\otimes h_b^*\in\sEnd(\FC^2)\otimes \sEnd(\FC^k)$ is of rank one in the gauge part $\sEnd(\FC^k)$. This amounts to the fact that only one of the D3-branes suspended between the two NS5-branes can break up on the D5-brane\footnote{This is also related to the s-rule \cite{Hanany:1996ie}, which states that only one D3-brane can be supersymmetrically suspended between any given pair of NS5- and D5-branes.}. Here, however, we want all the D3-branes to break in the $x^1$ direction. 

The alternative is to insert an NS5-brane. This generates an additional bifundamental hypermultiplet at $s=0$ arising from strings connecting the D3-branes to the left and the right of the NS5-brane \cite{Hanany:1996ie,Cherkis:2012qs}. Giving a vev to this hypermultiplet, we obtain the Nahm equation
\begin{equation}\label{eq:NahmJumpFinite2}
 \dderr{T^i_{(n)}}{s}=\frac{2\pi}{q}\eps_{ijk}[T^j_{(n)},T^k_{(n)}]+\zeta^i\delta(s)~,
\end{equation}
which is the finite $n$ version of \eqref{eq:NahmJumpBolognesi}. Here, $\zeta^i\in\au(n)$ is determined by the vev of the hypermultiplet. This configuration, however, does not describe an $\sSU(2)$ monopole. In fact, the configuration we arrived at is S-dual to a sequence of D5-branes at $s=-v$, $s=0$ and $s=v$, which is the usual description of an $\sSU(3)$ monopole, except for the fact that all the D3-branes break on the D-brane in the middle. To obtain a brane configuration corresponding to an $\sSU(2)$-monopole, we compactify the direction $x^6$ on a circle and identify the NS5-branes at $s=-v$ and $s=+v$. On the latter NS5-brane, the D3-branes end with the usual Nahm boundary condition, 
while on the NS5-brane at $s=0$, they break up and their worldvolume becomes discontinuous in the $x^1$-direction. Inverting the process of T- and S-dualizing, we recover a D-brane configuration with two D3-branes and $2n$ D1-branes, which describes an $\sSU(2)$-monopole configuration. While this string theory interpretation is certainly no proof of the Bolognesi conjecture, it gives at least strong evidence for its validity.

Finally, let us try to connect configuration \eqref{diag:pq2} to \eqref{diag:pq3}. While \eqref{diag:pq3} is the na\"ive, classical picture, configuration \eqref{diag:pq2} incorporates quantum corrections bending the branes. We know that each point of the worldvolume of the D3-branes in \eqref{diag:pq3} polarizes into a fuzzy sphere due to the Myers effect \cite{Myers:1999ps,Constable:1999ac}. In the limit $n\rightarrow \infty$, the D3-branes therefore turn into D5-branes wrapping a sphere $S^2_{123}$. Moreover, if we assume that all the D3-branes come in pairs such that the configuration \eqref{diag:pq3} is symmetric with respect to the $x^6$-coordinate axes, we arrive at the following quantum corrected picture:
\begin{equation*}
\begin{picture}(190,60)
\put(0.0,10.0){\line(-1,-1){15}}
\put(0.0,40.0){\line(-1,1){15}}
\put(100.0,10.0){\line(1,-1){15}}
\put(100.0,40.0){\line(1,1){15}}
\put(0.0,10.0){\line(0,1){30}}
\put(0.0,10.0){\line(1,0){50}}
\put(0.0,12.0){\line(1,0){50}}
\linethickness{1pt}
\put(50.0,0.0){\line(0,1){50}}
\linethickness{0.4pt}
\put(50.0,40.0){\line(1,0){50}}
\put(50.0,38.0){\line(1,0){50}}
\put(0.0,40.0){\line(1,0){50}}
\put(0.0,38.0){\line(1,0){50}}
\put(50.0,10.0){\line(1,0){50}}
\put(50.0,12.0){\line(1,0){50}}
\put(100.0,10.0){\line(0,1){30}}
\put(-23.0,45.0){\makebox(0,0)[c]{$(p,q)$}}
\put(-23.0,5.0){\makebox(0,0)[c]{$(p,q)$}}
\put(125.0,45.0){\makebox(0,0)[c]{$(p,q)$}}
\put(125.0,5.0){\makebox(0,0)[c]{$(p,q)$}}
\put(-13.0,25.0){\makebox(0,0)[c]{NS5}}
\put(113.0,25.0){\makebox(0,0)[c]{NS5}}
\put(66.0,3.0){\makebox(0,0)[c]{defect}}
\put(25.0,17.0){\makebox(0,0)[c]{D5s}}
\put(75.0,32.0){\makebox(0,0)[c]{D5s}}
\put(75.0,17.0){\makebox(0,0)[c]{D5s}}
\put(25.0,32.0){\makebox(0,0)[c]{D5s}}
\put(160.0,10.0){\vector(1,0){20}}
\put(160.0,10.0){\vector(0,1){20}}
\put(190.0,10.0){\makebox(0,0)[c]{$x^6$}}
\put(160.0,37.0){\makebox(0,0)[c]{$x^1$}}
\end{picture}\vspace*{0.2cm}
\end{equation*}
Up to the defect at $s=0$, this configuration is identical to \eqref{diag:pq2}. Note that to obtain a magnetic bag, we have to tune the distance between the NS5-branes to zero. To our knowledge, it is still unclear how to describe Chalmers-Hanany-Witten configurations with stacks of multiple NS5-branes as impurities. Studying our example of a magnetic bag in more detail might provide some new insights into this issue. In particular, it might explain the appearance of the additional defect at $s=0$. 

It is clear that the D-brane configurations we considered in this section all have lifts to M-theory. In particular, the M-brane configuration obtained from lifting \eqref{diag:pq2} describes a bag of selfdual strings, which are bounded by three-dimensional surfaces diffeomorphic to $S^3$. We will present the corresponding Nahm constructions in the following.

\section{Magnetic domains in four dimensions}\label{sect:SDS}

\subsection{From selfdual strings to magnetic domains}

We saw in sections \ref{sec:Bags} and \ref{sec:branes} that magnetic domains, obeying an abelian equation, can appear in the $n\to\infty$ limit of $n$ D1-branes stretched between two D3-branes. We expect something similar to happen to selfdual strings here: If we consider the limit of infinitely many M2-branes stretched between two M5-branes, the theory should become abelian. We will refer to the resulting configurations again as {\em magnetic domains}. To stress that the field strength $H$ is abelian, we will denote it by $h$ for the rest of this chapter. These domains in four dimensions are described by a Higgs field $\phi$ and a closed 3-form $h$ in a domain $\Omega\subset\FR^4$, both taking values in $\au(1)$ and having the following properties:
\begin{itemize}
 \setlength{\itemsep}{-1mm}
 \item $h$ is closed, and therefore we have locally a 2-form potential $b$ with $h=\dd b$,
 \item $h$ and $\phi$ satisfy the selfdual string equation $h=*\dd \phi$ in the region $\Omega\subset \FR^4$,
 \item $\dd \phi\neq 0$ in $\Omega$ and
 \item depending on the shape and dimensionality of the boundary of the domain $\Omega$, $\phi$ satisfies certain boundary conditions.
\end{itemize}
We clearly expect to find four-dimensional generalizations of the magnetic domains we know from three dimensions, in particular magnetic bags, magnetic tubes and magnetic walls. Analogously to the name monopole bags, we will refer to magnetic domains in four dimensions as selfdual string bags.

It is interesting to note that, similar to the Yang-Mills-Higgs energy functional, we can define a functional
\begin{equation}
E=\tfrac{1}{2}\int_\Omega h\wedge *h+\dd\phi\wedge *\dd\phi~,
\end{equation}
which has a Bogomolny bound 
\begin{equation}
E=\int_\Omega \tfrac{1}{2}|\dd\phi-*h|^2+d\phi\wedge h~\ge vq~,~~~q:=\int_{S^3_\infty}h~,
\end{equation}
saturated by solutions to the selfdual string equation.

\subsection{Nambu-Poisson structure and the Basu-Harvey equation}

The Basu-Harvey equation 
\begin{equation}\label{eq:BasuFinite}
\frac{\dd T^\mu}{\dd s} = \frac{e}{3!}\,\eps_{\mu\nu\rho\sigma}[T^\nu,T^\rho,T^\sigma]~.
\end{equation}
involves fields living in a 3-Lie algebra. The only finite dimensional non-trivial normed 3-Lie algebra is $A_4$. In contrast, there are many examples of infinite-dimensional normed 3-Lie algebras.  Let $M$ be any 3-manifold equipped with a non-vanishing volume form $\omega$.  The space $\CC^{\infty}(M)$ of smooth functions forms a 3-Lie algebra, with 3-bracket defined by the equation
\begin{equation}
 \{f,g,h\}\omega = \dd f\wedge \dd g\wedge\dd h~.
\end{equation}
In addition to the fundamental identity \eqref{eq:GenFundIdent}, the 3-bracket satisfies the Leibniz rule
\begin{equation}\label{eq:Leibniz}
 \{f_1f_2,g,h\} = f_1\{f_2,g,h\} + \{f_1,g,h\}f_2 ~.
\end{equation}
This implies that for any $g,h\in \CC^\infty(M)$ the map $D(g,h):f\to\{g,h,f\}$ is a derivation, which means that $D(g,h)$ is a vector field.  In general, a 3-Lie algebra structure on the algebra of functions over a manifold obeying the Leibniz rule is called a {\em Nambu-Poisson structure} \cite{Nambu:1973qe,Takhtajan:1993vr}.

Solutions to the Basu-Harvey equation based on the 3-Lie algebra $\CA_4$ are conjectured to describe two M2-branes stretching between M5-branes.  We will show below that the appropriate 3-Lie algebra for describing selfdual string bags is $\CC^\infty(S^3)$ equipped with the Nambu-Poisson 3-Lie bracket induced by the $\sSO(4)$-invariant volume form $\omega$.  In standard polar coordinates $0\le\theta^1,\theta^2\le\pi$,  $0\le\theta^3\le2\pi$ the volume form is
\begin{equation}
\omega=\sin^2\theta^1 \sin \theta^2~\dd\theta^1\wedge\dd\theta^2\wedge\dd\theta^3~,
\end{equation}
and the Nambu-Poisson 3-bracket is given by
\begin{equation}
\{f,g,h\}=\frac{1}{\sin^2\theta^1 \sin \theta^2} \eps^{ijk}\frac{\pa f}{\pa \theta^{i}}\frac{\pa g}{\pa \theta^j}\frac{\pa h}{\pa \theta^{k}}~~.
\end{equation}
It will be convenient to denote by $x^1,x^2,x^3,x^4$ the functions on $S^3$ obtained by restricting coordinate functions from $\FR^4$.  These of course satisfy $x^\mu x^\mu=1$, and their 3-brackets with each other are
\begin{equation}
 \{ x^\mu, x^\nu, x^\rho \} = \eps^{\mu\nu\rho\sigma} x^\sigma ~.
\end{equation}
Thus the $x^\mu$ span a sub-algebra of $\CC^\infty(S^3)$ isomorphic to $\CA_4$.

\subsection{Nahm transform and its inverse}\label{sect:SDStransform}

In general, we will denote the Nambu-Poisson structure on a three-dimensional manifold $M$ induced by its volume form $\omega$ by $\Pi_\omega$. Under {\em $\Pi_\omega$-Basu-Harvey data} for magnetic domains in four dimensions, we understand a set of four functions $t^\mu$ on $M\times \CI$, where $\CI$ is a union of finitely many intervals, satisfying the $\Pi_\omega$-Basu-Harvey equation 
\begin{equation}\label{eq:BHInfinite}
\frac{\dd t^{\mu}}{\dd s}=\frac{2\pi^2}{ 3! q}~\eps^{\mu\nu\kappa\lambda}\{t^{\nu},t^{\kappa},t^{\lambda}\}_\omega ~.
\end{equation}
We will discuss in section \ref{sect:HermitianBH} how this Basu-Harvey equation emerges as the large $n$ limit of Basu-Harvey equations based on hermitian 3-algebras. Analogously to the case of magnetic domains in $\FR^3$, we have the following theorem, which refines a result of Dunajski \cite{Dunajski:2003ep}: 

Up to gauge equivalence, there is a one-to-one correspondence between
\vspace{-0.2cm}
\begin{itemize}\label{thm:4}
 \setlength{\itemsep}{-1mm}
 \item sets of $\Pi_\omega$-Basu-Harvey data with the property that the map from $M\times \CI$ to $\Omega\subset\FR^4$ defined by the $t^\mu$ is a diffeomorphism $t:M\times \CI\rightarrow \Omega$, and
 \item magnetic domains $\Omega$ that are diffeomorphic to $M\times \CI$, where the restriction of the 3-form curvature $h$ to any slice $M\times\{s_0\}$ has the same volume type as $\omega$ and $\CI$ is the range of $\phi$. Explicitly, there is a diffeomorphism $u:\Omega\rightarrow M\times \CI$ such that $h=\frac{q}{2\pi^2}u^*\omega$ and $\phi=s\circ u$ on $\Omega$. 
\end{itemize}

The proof is a minor generalization of that of theorem \ref{thm:2}. The Basu-Harvey data defines a diffeomorphism $t$ from $M\times \CI$ to a subset $\Omega\subset\FR^4$ with inverse $u$.  By definition of the Nambu 3-bracket, the infinite-charge Basu-Harvey equation \eqref{eq:BHInfinite} is equivalent to
 \begin{equation}
\dd t^{\mu}\wedge \omega=\frac{2\pi^2}{3! q}~\eps_{\mu\nu\kappa\lambda}\dd t^{\nu}\wedge\dd t^{\kappa}\wedge \dd t^{\lambda}\wedge\dd s~.
\end{equation}
This implies the following equation on $\FR^4$:
\begin{equation}
 \dd y^{\mu}\wedge u^*\omega=\frac{2\pi^2}{q}\dd y^\mu\wedge*\dd (u^* s)~.
\end{equation}
Thus $\phi=u^* s$ and $h=\frac{q }{2\pi^2}~u^*\omega$ solve the selfdual string equation.

To define the inverse transform we restrict ourselves again to a connected component. We choose a value $\phi_0=s_0\in\CI$, which yields the level surface $\Sigma_{\phi_0}$, which is embedded in $\Omega$ via the map $i:\Sigma_{\phi_0}\embd \Omega$. Because the restriction of $h$ and $\omega$ have the same volume type, there is a diffeomorphism $w:M\rightarrow \Sigma_{\phi_0}$ such that $w^*i^*h=\frac{q}{2\pi^2}\omega$. The diffeomorphism $w\circ i$ can be extended to all of $M\times \CI$ by solving the differential equation 
\begin{equation}
 \frac{\dd y^\mu}{\dd s} = Y^\mu(y(s))~,~~~Y=\left(\derr{\phi}{y^\nu}\derr{\phi}{y_\nu}\right)^{-1}\derr{\phi}{y^\mu}\der{y_\mu}
\end{equation}
with boundary condition $y(s_0)=w\circ i$. Here, the solution $y$ can again be identified with the diffeomorphism $t:M\times \CI\rightarrow \Omega$. It can be readily checked that this construction inverts the Nahm transform.

\subsection{Examples}

First we consider what we will call {\em selfdual string bags}: magnetic domains in four dimensions for which $\Omega$ is the exterior of a hypersurface $\Sigma\subset\FR^4$ diffeomorphic to $S^3$. On the interior of this hypersurface $\Sigma$, we have $\phi=0$, and on the exterior, $\phi\sim v-\frac{1}{r^2}$ as $r\rightarrow \infty$.  The corresponding $\Pi_\omega$-Basu-Harvey data consists of four functions on $S^3\times [0,v)$ satisfying
\begin{equation}
t^{\mu}\sim\frac{ x^\mu}{2\pi}\left(\frac{q}{v-s}\right)^\frac{1}{2}~\mbox{ as }s\rightarrow v~.
\end{equation}

The simplest example of $\Pi_\omega$-Basu-Harvey data is
\begin{equation}\label{nahmboundary}
t^{\mu}=\frac{ x^\mu}{2\pi}\left(\frac{q}{v-s}\right)^\frac{1}{2}~.
\end{equation}
The image of the map $t:S^3\times [0,v)\to\FR^4$ is the set $\Omega=\{r^2\geq q/4\pi^2v\}$, and the inverse map $u:\Omega\rightarrow S^3\times \CI$ is 
\begin{equation}
u(\vec y)=\left(\frac{\vec y}{r},v-\frac{q}{4\pi^2 r^2}\right).
\end{equation}
Thus the corresponding selfdual string bag is the following spherically-symmetry configuration:
\begin{equation}
\begin{aligned}
\phi =& \begin{cases} v-\frac{q}{4\pi^2 r^2} & r^2\geq \frac{q}{4\pi^2v} \\ 0 & r^2 < \frac{q}{4\pi^2v} \end{cases}~,\\
h =& \begin{cases} \frac{q}{3!2\pi^2r^4} \eps_{\mu\nu\rho\sigma}y^\mu\dd y^\nu\wedge\dd y^\rho\wedge\dd y^\sigma & r^2\geq \frac{q}{4\pi^2v} \\ 0 & r^2 < \frac{q}{4\pi^2v} \end{cases} ~.
\end{aligned}
\end{equation}

Another example of Basu-Harvey data, this time describing an ellipsoidal bag extended in the $y^4$ direction, cf.\ \eqref{eq:sol4}, is given by
\begin{equation}
t\left( x^\mu,s\right)= \sqrt \frac{q}{2\pi^2} \left(\frac{ x^{i}}{\sqrt{(v-s)(2+v-s)}},\frac{ x^{4}(1+v-s)}{\sqrt{(v-s)(2+v-s)}} \right)~,~~i=1,\dots,3~.
\end{equation}
The inverse map $u:\Omega\rightarrow S^3\times \CI$ is then
\begin{equation}
u(y)=\left(\left(\frac{y^i}{\alpha},\frac{y^4}{\alpha\sqrt{1+\frac{q}{2\pi^2\alpha^2}}}\right),v+1-\sqrt{1+\frac{q}{2\pi^2\alpha^2}}\right)~,
\end{equation}
where 
\begin{equation}
\alpha^2=\tfrac{1}{2}\left(r^2-\frac{q}{2\pi^2}+\sqrt{\frac{2q}{\pi^2}(r^2-(y^4)^2)+(r^2-\frac{q}{2\pi^2})^2}\right)~.
\end{equation}
This gives the magnetic domain
\begin{equation}
\begin{aligned}
&\phi=v+1-\sqrt{1+\frac{q}{2\pi^2\alpha^2}}~,~~ h=\eps_{\mu\nu\rho\sigma} \frac{q}{2\pi^2\alpha^3\sqrt{1+\frac{q}{2\pi^2\alpha^2}}}\frac{\pa\alpha}{\pa y^\mu}\dd y^\nu\wedge\dd y^\rho\wedge\dd y^\sigma~\\
&\mbox{on}~~~ \Omega = \left\{ y~\Big|~(y^1)^2+(y^2)^2+(y^3)^2+\frac{1}{(1+ v)^2} (y^4)^2\geq \frac{q}{2\pi^2(2v+v^2)}\right\}\subset \FR^4~.
\end{aligned}
\end{equation}

Analogously to the 3-dimensional examples presented in section \ref{sec:3DExamples}, one can also construct 4-dimensional magnetic domains from manifolds $M=\FR^3$, $M=\FR^2\times S^1$ and $M=\FR\times S^2$ endowed with a volume form. The boundary conditions for the scalar field $\phi$ can be fixed by demanding that $\phi$ asymptotes to a harmonic function with appropriate symmetries, and these induce boundary conditions on the Basu-Harvey data.

\subsection{Conserved charges}\label{sect:cons}

It is interesting to note that the Basu-Harvey equation is integrable.  Rather than a Lax pair, the integrability manifests itself through a Lax triple $(t,A,B)$ with spectral parameters $\eta,\zeta\in\CPP^1$:
\begin{equation}
\begin{aligned}
 t(\eta,\zeta) &= (t^1+\mathrm{i} t^2) + \zeta (t^3+\mathrm{i} t^4) + \eta (t^3-\mathrm{i} t^4) + \zeta\eta (-t^1+\mathrm{i} t^2)~, \\
 A(\eta) &=  (t^3+\mathrm{i} t^4) + \eta (-t^1+\mathrm{i} t^2)~, \\
 B(\zeta) &=  (t^3-\mathrm{i} t^4) + \zeta (-t^1+\mathrm{i} t^2)~.
\end{aligned}
\end{equation}
Here, $t^i\in \CC^\infty(\CI)\otimes \CA$, where $\CA$ is a 3-Lie algebra. Using the anti-symmetry of the 3-bracket, it can be shown that the Basu-Harvey equation is equivalent to
\begin{equation}
\label{Lax equation}
 \frac{\mathrm{d}}{\mathrm{d} s} t(\eta,\zeta) = [A(\eta),B(\zeta),t(\eta,\zeta)]~.
\end{equation}

Specializing now to the 3-Lie algebra $\CC^\infty(M)$, we define an infinite tower of conserved charges by taking the coefficients of the polynomials,
\begin{equation}
\int_{M\times s_0} \omega~ t(\eta,\zeta)^n~, \quad n\in\NN~,
\end{equation}
where $\omega$ is again the volume form on $M$.  We assume that these integrals converge, which is certainly the case when $M$ is compact.  The fact that these quantities are conserved follows from the Lax equation \eqref{Lax equation} and the observation that the integral of the 3-bracket of any three functions is zero.

The conserved charges can equivalently be defined in the Nahm dual picture as the integrals over level sets $\{\phi=s_0\}$. That these integrals are independent of $\phi_0$ follows from repeated applications of Stokes theorem.

For the Basu-Harvey equation based on $\CA_4$, one can construct conserved charges $(t(\zeta,\eta),t(\zeta,\eta))$ using the positive definite norm $(\cdot,\cdot)$.

\section{Hermitian bags}\label{sect:HermitianBH}

The Basu-Harvey equation based on the trivial 3-algebra $\FR$ and the 3-algebra $\CA_4$ describe one or two parallel M2-branes ending on M5-branes.  For $n>2$ M2-branes, we need the generalization of this equation based on hermitian 3-algebras. In this section we will show that our proposed equation \eqref{eq:BHInfinite} for an infinite number of M2-branes arises in the large $n$ limit of this equation.

In order to do this, we first show that the hermitian 3-algebras converge to a sub-algebra of $\CC^\infty(S^3)$ as $n\to\infty$.  The fact that the limit yields a sub-algebra, rather than the whole of $\CC^\infty(S^3)$, places constraints on the bag obtained via the Nahm transform.  We discuss the implications of these constraints at the end of the section: essentially, the bag obtained is invariant under an action of $\sU(1)$, and can be identified with a magnetic bag on $\FR^3$.

\subsection{Equivariant fuzzy 3-sphere}

The ABJM model is built from the hermitian 3-algebra $\sMat_{n\times n}(\FC)$ with bracket 
\begin{equation}
 [C,A;B] = -2n(A\bar BC - C\bar B A) = D(A,B)\acton C~,
\end{equation}
where the bar denotes matrix transposition combined with complex conjugation. Here, we will focus on the sub 3-algebra $\CH_n$ of $(n-1\times n)$-dimensional matrices, which is relevant for the hermitian Basu-Harvey equation.

Note that it is also possible to construct hermitian 3-algebras from 3-Lie algebras.  Given a 3-Lie algebra $\CA$, one defines $\CH=\FC\otimes\CA$ and for the hermitian 3-bracket chooses
\begin{equation}
 [a,b;c] = [a,b,\bar c]~.
\end{equation}
This means for example that the space $\FC\otimes \CC^\infty(S^3)$ of complex functions on $S^3$ forms a hermitian 3-algebra.  We will show below that the large $n$ limit of $\CH_n$ can be identified with a sub-algebra $\CH_\infty$ of $\FC\otimes \CC^\infty(S^3)$.

Consider the following two distinguished elements $W^1,W^2\in\CH_n$:
\begin{equation}
\begin{aligned}
 W^1 &= \frac{1}{\sqrt{n}}\left( \begin{array}{ccccc} 
 0 & \sqrt{1} & 0 & & \vdots \\
 0 & 0 & \sqrt{2} & & \\
 \vdots & & & \ddots & 0 \\
 0 & \cdots & & 0 & \sqrt{n-1}
 \end{array} \right)~, \\
 W^2 &= \frac{1}{\sqrt{n}}\left( \begin{array}{ccccc}
 \sqrt{n-1} & 0 & 0 & & \vdots \\
 0 & \sqrt{n-2} & 0 & & \\
 \vdots & & & \ddots & \\
 0 & \cdots & 0 & \sqrt{1} & 0
 \end{array} \right)~.
\end{aligned}
\end{equation}
These special elements were introduced in \cite{Gomis:2008vc}, where it was noted that they satisfy
\begin{eqnarray}
\label{Ak radius 1}
 \bar W^1 W^1 + \bar W^2 W^2 &=& \frac{n-1}{n} \mathbf{1}_n~, \\
\label{Ak radius 2}
 W^1 \bar W^1 + W^2 \bar W^2 &=& \mathbf{1}_{n-1}~,
\end{eqnarray}
and
\begin{equation}
\label{Ak 3-bracket}
 [W^\alpha,W^\beta;W^\gamma] = 2\eps^{\alpha\beta}\eps^{\gamma\delta}W^\delta~.
\end{equation}
It follows from \eqref{Ak 3-bracket} that the Lie algebra of derivations spanned by $D(-\frac{\mathrm{i}}{2}W^\alpha,W^\beta)$ is $\mathfrak{u}(2)$, and that $W^1,W^2$ transform in the fundamental representation of this Lie algebra.  Thus there is a natural action of $\mathfrak{u}(2)$ on $\CH_n$.

The diagonal sub-algebra $\mathfrak{u}(1)\subset\mathfrak{u}(2)$ is generated by
\begin{equation}
\label{Ak xi}
\Xi = D\left(-\frac{\mathrm{i}}{2} W^\alpha,W^\alpha\right)~,
\end{equation}
and this $\mathfrak{u}(1)$ sub-algebra acts in the following way:
\begin{equation}
\label{eq:Xi action}
\Xi\acton A = \mathrm{i}A
\end{equation}
for all $A\in \CH_n$. The action of the Lie sub-algebra $\mathfrak{su}(2)\subset\mathfrak{u}(2)$ can be summarized by saying that $\CH_n$ transforms in the following representation of $\mathfrak{su}(2)$:
\begin{equation}
\label{Ak functions}
 \CH_n = \overline{\mathbf{n-1}}\otimes\mathbf{n} = \bigoplus_{i=1}^{n-1} \mathbf{2i}~.
\end{equation}

As was noted in \cite{Gomis:2008vc}, the algebraic identities \eqref{Ak radius 1}, \eqref{Ak radius 2} suggest an interpretation of $\CH_n$ as a fuzzy 3-sphere. As we will see now, this interpretation is problematic. It is natural to try to identify the elements $W^1,W^2\in\CH_n$ with the complex functions $w^1=x^1+\mathrm{i} x^2,w^2=x^3+\mathrm{i} x^4$, which satisfy
\begin{equation}
\label{Ainfty radius}
 \bar w^1 w^1 + \bar w^2 w^2 = 1~.
\end{equation}
The hermitian 3-brackets of these functions satisfy the same relations as the 3-brackets of the $W^\alpha$:
\begin{equation}
\label{Ainfty 3-bracket}
 [w^\alpha,w^\beta;w^\gamma]:=\{w^\alpha,w^\beta,\bar{w}^\gamma\}= 2\eps^{\alpha\beta}\eps^{\gamma\delta}w^\delta~,
\end{equation}
where $\{\cdot,\cdot,\cdot\}$ denotes the Nambu-Poisson bracket induced by the canonical volume form on $S^3$. The derivations $D(\frac{\mathrm{i}}{2}w^\alpha,w^\beta)$ therefore span the Lie algebra $\mathfrak{u}(2)$, which can be identified with a Lie sub-algebra of the rotation Lie algebra $\mathfrak{so}(4)\cong \asu(2)\oplus\asu(2)$.

We now explain how this Lie algebra acts on $\FC\otimes \CC^\infty(S^3)$.  We start with the derivation
\begin{equation}
\label{Ainfty xi}
\xi=D\left(-\frac{\mathrm{i}}{2}w^\alpha,w^\alpha\right)=\mathrm{i} \left(w^\alpha\frac{\pa}{\pa w^\alpha}-\bar w^\alpha\frac{\pa}{\pa \bar w^\alpha}\right)~,
\end{equation}
which generates the diagonal $\mathfrak{u}(1)$.  The set of eigenvalues of $\xi$ is $\RZ$, and the fundamental identity \eqref{eq:GenFundIdent} implies that the eigenspaces of $\xi$ are closed under the hermitian 3-bracket.  In view of \eqref{eq:Xi action} it seems reasonable to identify the large $n$ limit of $\CH_n$ with the hermitian 3-algebra,
\begin{equation}
 \CH_\infty := \{ f:S^3\to\FC\,|\, \xi\acton f = \CL_\xi f = \mathrm{i} f \}~.
\end{equation}
The vector space $\CH_\infty$ may be identified with the space of chiral spinors on the 2-sphere. That is, $\CH_\infty$ is the closure of the span of the polynomials of the form $w^{\alpha_1}\ldots w^{\alpha_{\ell+1}}\bar{w}^{\beta_1}\ldots \bar{w}^{\beta_\ell}$, $\alpha_i,\beta_i=1,2$, $\ell\in\NN$. The space $\CH_\infty$ therefore does {\em not} contain all functions on $S^3$, as would be required by an interpretation of $\CH_n$ as a fuzzy 3-sphere.

Now we consider the action of $\mathfrak{su}(2)$.  It is well-known that $\CH_\infty$ transforms in the following representation of $\mathfrak{su}(2)$:
\begin{equation}
\label{Ainfty functions}
 \CH_\infty = \bigoplus_{i=1}^\infty \mathbf{2i}~.
\end{equation}
This clearly coincides with the $n\to\infty$ limit of \eqref{Ak functions}.  Due to the similarities between equations \eqref{Ak radius 1}, \eqref{Ak radius 2}, \eqref{Ak 3-bracket} and \eqref{Ainfty radius}, \eqref{Ainfty 3-bracket}, it is clear that $\CH_\infty$ is the formal $n\to\infty$ limit of $\CH_n$.

There are obvious parallels to be drawn with the discussion in section \ref{sect:FuzzyFunnel} of the Berezin-Toeplitz quantization of $S^2$: just as $\CCH_n\otimes\overline{\CCH_{n}}$ quantizes the Poisson bracket functions on $S^2$, we have shown that $\CCH_n\otimes\overline{\CCH_{n+1}}$ quantizes the 3-bracket structure on the space of chiral spinors on $S^2$.  It would be interesting to investigate this idea from an analytical point of view, i.e.\ to find analogous formulas to \eqref{eq:approximations} involving Nambu-Poisson and hermitian 3-algebra brackets.

\subsection{The hermitian Basu-Harvey equation}

The \emph{hermitian Basu-Harvey equation}
\begin{equation}
\label{eq:HBH}
\frac{\dd}{\dd s} Z^\alpha = \frac{\pi^2}{q}[Z^\alpha,Z^\beta;Z^\beta]~
\end{equation}
has fields taking values in a hermitian 3-algebra. The ordinary hermitian 3-algebra chosen in \cite{Gomis:2008vc} was the hermitian 3-algebra of $n\times n$ matrices, however, we saw in the previous section that in order to obtain a reasonable large $n$ limit, it is sensible to restrict attention to the sub-algebra $\CH_n$ of $n-1\times n$ matrices. All irreducible solutions of \eqref{eq:HBH} can be restricted to this sub-algebra \cite{Gomis:2008vc}.

Thus in the large $n$ limit, we obtain $\CH_\infty$-valued functions $z^1(s)$, $z^2(s)$ obeying
\begin{equation}
\label{eq:HBHinf}
\frac{\dd}{\dd s} z^\alpha = \frac{\pi^2}{q}[z^\alpha,z^\beta;z^\beta]~.
\end{equation}
This equation is equivalent to the Basu-Harvey equation \eqref{eq:BHInfinite} if we identify $z^1=t^1+\mathrm{i}t^2$, $z^2=t^3+\mathrm{i} t^4$.   The natural range for the variable $s$ is here $[0,v)$, and the boundary condition \eqref{nahmboundary} can be rewritten as
\begin{equation}
\label{eq:HBHBC}
z^{\alpha}=\frac{ w^\alpha}{2\pi}\sqrt{\frac{q}{v-s}}+\CO((v-s)^{\frac{1}{2}}) \mbox{ as } s\to v ~.
\end{equation}
A selfdual string bag on $\FR^4$ can be obtained by applying the Nahm transform to any solution of \eqref{eq:HBHinf}, \eqref{eq:HBHBC} as in subsection \ref{sect:SDStransform}.

However, the fact that $z^1,z^2$ take values in $\CH_\infty$ and not the full function space $\CC^\infty(S^3)$ imposes constraints on the bag obtained.  We will now show that bags resulting from solutions to \eqref{eq:HBHinf}, \eqref{eq:HBHBC} are invariant under a certain $\sU(1)$-action, and moreover that they are equivalent to magnetic bags on $\FR^3$.

Let $\eta$ be the following vector field on $\FR^4$:
\begin{equation}
\eta = y^1\frac{\pa}{\pa y^2} - y^2\frac{\pa}{\pa y^1} + y^3\frac{\pa}{\pa y^4} - y^4\frac{\pa}{\pa y^3}~.
\end{equation}
The fact that $\CL_\xi z^\alpha=\mathrm{i}z^\alpha$ implies that
\begin{equation}
 \CL_\xi t^1=-t^2~,\quad
 \CL_\xi t^2=t^1~,\quad 
 \CL_\xi t^3=-t^4~,\quad
 \CL_\xi t^4=t^3~.
\end{equation}
It follows that the push-forward of $\xi$ under the map $t:S^3\times[0,v)\to\Omega\subset\FR^4$ is $\eta$: $t_\ast\xi=\eta$.  Now the coordinate function $s$ and the 3-form $\omega$ on $S^3\times[0,v)$ satisfy $\CL_\xi s=0$ and $\CL_\xi \omega=0$; therefore the function $\phi$ and 3-form $h$ obtained under the Nahm transform satisfy $\CL_\eta\phi=0$, $\CL_\eta h=0$.  This means that $\phi$ and $h$ are invariant under the action of $\sU(1)$ generated by $\eta$, and similarly the bag surface $\Sigma=\dpar \Omega$ is $\sU(1)$-invariant.

\subsection{Magnetic bags from selfdual string bags}

Since the bag on $\FR^4$ obtained from a solution to \eqref{eq:HBHinf}, \eqref{eq:HBHBC} is $\sU(1)$-invariant, it is natural to try to identify it with some configuration on the quotient space.  It is well-known that $\FR^4/\sU(1)\cong\FR^3$; standard coordinates on $\FR^3$ are defined by the $\sU(1)$-invariant functions,
\begin{equation}
\label{R4 to R3}
r^i := \left( \begin{array}{cc} y^1-\mathrm{i}y^2 & y^3-\mathrm{i} y^4 \end{array} \right) \sigma^i 
\left( \begin{array}{cc} y^1+\mathrm{i}y^2 \\ y^3+\mathrm{i} y^4 \end{array} \right)~.
\end{equation}
We will denote this projection from $\FR^4$ to $\FR^3$ by $\pi$. When restricted to $S^3\embd \FR^4$, the projection $\pi$ is nothing but the Hopf fibration $S^1\rightarrow S^3\stackrel{\pi}{\rightarrow} S^2$. 

Let us return to the solution $(h,\phi)$ constructed in the previous subsection. The function $\phi$ is $\sU(1)$-invariant, so it must be the pull-back of some function $\psi$ on (a subset of) $\FR^3$.  The 3-form $h$ cannot be the pull-back of a 3-form on $\FR^3$, because $\iota_\eta h\neq 0$.  However, the 2-form $\iota_\eta h$ satisfies $\iota_\eta(\iota_\eta h)=0$ and $\CL_\eta(\iota_\eta h)=0$, so it is the pull-back of some 2-form $f$ on $\FR^3$.  This 2-form $f$ is closed, because
\begin{equation}
\pi^\ast\dd f = \dd \pi^\ast f = \dd \iota_\eta h = \CL_\eta h + \iota_\eta\dd h = 0~.
\end{equation}

Now we will determine what equation $(f,\psi)$ must satisfy.  It can be shown that, for any 1-form $u$ on $\FR^3$,
\begin{equation}
\ast_4 \pi^* u = \theta\wedge\pi^\ast(\ast_3 u)~,
\end{equation}
where $\pi:\FR^4\to\FR^3$ is the projection, $\ast_4$ and $\ast_3$ are the Hodge star operators on $\FR^4$ and $\FR^3$ with respect to the standard flat metrics, and
\begin{equation}
\theta := \frac{1}{y^\mu y^\mu}\left( -y^2\dd y^1 + y^1\dd y^2 - y^4\dd y^3 + y^3\dd y^4 \right)~.
\end{equation}
Since $\iota_\eta \theta = 1$, it follows that
\begin{equation}
\pi^\ast f = \iota_\eta h = \iota_\eta (\ast_4\dd\phi) = \iota_\eta (\ast_4 \pi^\ast \dd\psi) = \iota_\eta(\theta\wedge \pi^\ast (\ast_3\dd\psi)) = \pi^\ast (\ast_3\dd\psi)~,
\end{equation}
Therefore $(f,\psi)$ satisfy $f=\ast_3\dd\psi$ and define a magnetic bag on $\FR^3$.

Conversely, given any magnetic bag $(f,\psi)$ on $\FR^3$, a selfdual string bag on $\FR^4$ can be obtained by setting $\phi=\pi^\ast\psi$, $h=\theta\wedge\pi^\ast f$.  One can check that $(h,\phi)$ satisfy the selfdual string equation:
\begin{equation}
\ast_4\dd\phi = \ast_4\pi^\ast\dd\psi = \theta\wedge\pi^\ast(\ast_3\dd\psi) = \theta\wedge\pi^\ast f = h~.
\end{equation}
Moreover, $h$ is closed, because
\begin{equation}
\iota_\eta \dd h = \iota_\eta(\dd\theta\wedge\pi^\ast f -\theta\wedge\dd\pi^\ast f) = \iota_\eta(\dd\theta\wedge\pi^\ast f) = 0~,
\end{equation}
where in the last equality we have used the facts that $\iota_\eta\pi^\ast f=0$ and $\iota_\eta \dd\theta=0$.  Any 4-form whose inner derivative with $\eta$ vanishes must be zero, so it must be the case that $\dd h=0$.  Thus we have established a bijective correspondence between $\sU(1)$-invariant selfdual string bags and magnetic bags.

This correspondence can also be seen at the level of $\Pi_\omega$-Nahm data.  Viewed as functions on $S^3\times [0,v)$, $t^i=\bar{z}^\alpha\sigma^i_{\alpha\beta} z^\beta$ are invariant under the group $\sU(1)$ generated by $\xi$. The space of $\sU(1)$-invariant functions on $S^3$ can be identified with the space of functions on $S^2$, via the Hopf fibration.  This function space is equipped with a Poisson bracket, defined via
\begin{equation}
4\{f,g\}_{S^3}~ \iota_\xi\omega = \dd f\wedge\dd g,
\end{equation}
where $f,g$ are any $\sU(1)$-invariant functions on $S^3$.  The Poisson bracket can be lifted to $S^3\times[0,v)$ by wedging both sides of this equation with $\dd s$.  It is straightforward (but tedious) to check that the hermitian Basu-Harvey equation \eqref{eq:HBHinf} implies that $t^i$ satisfy the Nahm equation,
\begin{equation}
\frac{\dd }{\dd s} t^i = \frac{4\pi^2}{q}\frac{1}{2}\eps_{ijk}\{t^j,t^k\}_{S^3}~,
\end{equation}
Altogether, we have proved the following theorem:

Up to gauge equivalence, we have one-to-one correspondences between the following sets:
\begin{equation}
\label{commutative diagram}
\begin{array}{ccc}
\mbox{$\CH_\infty$ hermitian Basu-Harvey data} & \longleftrightarrow & \mbox{$\sU(1)$-invariant bags on }\FR^4 \\
\updownarrow & & \updownarrow \\
\mbox{$\Pi_\omega$-Nahm data for magnetic bags} & \longleftrightarrow & \mbox{magnetic bags on }\FR^3
\end{array}
\end{equation}

\section{Loop space selfdual string bags}\label{sect:LSSDS}

We will see in this section how the Nahm transform for selfdual string bags has a formulation in loop space; this sets the transform in a wider context.  This formulation makes essential use of naturally defined Poisson-like brackets on 1-forms and loop space, so we begin by reviewing these constructions.

\subsection{Poisson-like structures on 1-forms}\label{subsec:PoissonForms}

To any 1-form $\alpha$ on $S^3$, a vector field $X_\alpha$ can be associated via the equation
\begin{equation}
 \dd \alpha = \iota_{X_\alpha}\omega~.
\end{equation}
It follows directly that $\CL_{X_\alpha}\omega=0$, so the vector field $X_\alpha$ is volume-preserving or divergence-free. This generalizes the relationship between functions and vector fields on a symplectic manifold.  The 1-form $\alpha$ is called a Hamiltonian 1-form and $X_\alpha$ is the corresponding Hamiltonian vector field, while the volume form $\omega$ is sometimes called a 2-plectic form.\footnote{More generally, a closed non-degenerate $p+1$-form $\omega$ on a manifold is called a $p$-plectic form, and one can speak of Hamiltonian $p-1$-forms and vector fields.  It is not true in general that every $p-1$-form is Hamiltonian, however, on $S^3$ every 1-form is Hamiltonian.}

There are two obvious generalizations of the Poisson bracket on 1-forms \cite{Baez:2008bu}: the {\em hemi-bracket} is defined as
\begin{equation}
 \{\alpha,\beta\}_h:=\CL_{X_\alpha}\beta~,
\end{equation}
and the {\em semi-bracket} is given by
\begin{equation}
  \{\alpha,\beta\}_s:=\iota_{X_\alpha}\iota_{X_\beta}\omega~.
\end{equation}
The hemi-bracket satisfies the Jacobi-identity but it is not antisymmetric, while the semi-bracket is anti-symmetric but does not satisfy the Jacobi-identity.\footnote{Note that for so-called exact multisymplectic manifolds, which $S^3$ is not, a further bracket can be constructed that is both antisymmetric and satisfies the Jacobi identity \cite{Forger:2002aa}.}  The difference between the hemi- and semi-brackets is an exact 1-form: 
\begin{equation}\label{eq:diffHemiSemi}
 \{\alpha,\beta\}_h-\{\alpha,\beta\}_s=\dd \iota_{X_\alpha} \beta~.
\end{equation}
It follows that $\{\alpha,\beta\}_h$ and $\{\alpha,\beta\}_s$ induce the same vector field on $S^3$.  In fact, one has that
\begin{equation}
 X_{\{\alpha,\beta\}_h} = X_{\{\alpha,\beta\}_s} = [X_\alpha,X_\beta]~.
\end{equation}

On $S^3$, we may write $\alpha=\dd\theta^i\,\alpha_i$ and $\beta=\dd\theta^i\,\beta_i$, where $\theta^i$, $i=1,2,3$, denote again the canonical angles. Then the semi-bracket explicitly reads as
\begin{equation}
\{\alpha,\beta\}_s=\iota_{X_\alpha}\iota_{X_\beta}\omega=\dd\theta^i~\frac{\eps^{jkl}}{\sin^2\theta^1 \sin\theta^2}~\der{\theta^{[j}}\alpha_{i]}~\der{\theta^k} \beta_l~.
\end{equation}

\subsection{Poisson structures on loop space}

Consider now the free loop space $\CL S^3$ of $S^3$, whose elements are given by loops $\theta:S^1\to S^3$. The tangent space at a loop $\theta$ is given by
\begin{equation}
 T_\theta\CL S^3=\CC^\infty(S^1,\theta^* TS^3)~.
\end{equation}
Thus, we will write tangent vectors as 
\begin{equation}
 \xi=\oint \dd \tau~ \xi^i(\theta,\tau)\,\delder{\theta^i(\tau)}=\oint \dd \tau~ \xi^{i\tau}(\theta)\,\delder{\theta^{i\tau}}~,
\end{equation}
and dual 1-forms as 
\begin{equation}
 \chi=\oint \dd \tau~ \chi_{i\tau}(x)\,\delta\theta^{i\tau}~,
\end{equation}
with $\langle \delta \theta^{i\tau},\delder{\theta^{j\sigma}}\rangle=\delta^i_j\delta(\tau-\sigma)$.  The total differential is
\begin{equation}
 \delta=\oint \dd \tau~\delta \theta^{i\tau}\delder{\theta^{i\tau}}~.
\end{equation}

Reparameterizations of a loop $\theta(\tau)$ are generated by the vector fields
\begin{equation}
\label{eq:trivial tangent vector}
\Gamma= \oint \dd \tau~ \gamma(\tau)\,\dot\theta^{i}(\tau)\,\delder{\theta^{i\tau}}~,
\end{equation}
where $\gamma$ is a function of $\tau$, transforming appropriately under reparameterizations. The quotient of the free loop space by this action is the space of unparameterized loops\footnote{Strictly speaking, we restrict ourselves to the space of singular knots, see \cite{0817647309} for details.}, which we denote by $\CL S^3$. We will still describe these loops by maps $\theta:S^1\rightarrow S^3$, but we will ensure that all our formulas are reparameterization invariant. Moreover, we impose the relations
\begin{equation}
 \dot{\theta}^{i}(\tau)\delder{\theta^{i\tau}}=\dot{\theta}_{i}(\tau)\delta\theta^{i\tau}=0\quad\forall\tau\in S^1~.
\end{equation}

The \emph{transgression map} \cite{0817647309} sends $p$-forms on a manifold $M$ to $p-1$-forms on its loop space $\CL M$: One of the $p$-form's indices can be contracted with the tangent vector to the loop under consideration. For a $p$-form $\omega=\frac{1}{p!}\omega_{i_1\cdots i_p}(\theta)\dd \theta^{i_1}\wedge\cdots\wedge \dd \theta^{i_p}$ we have explicitly the following local expression:
\begin{equation}\label{eq:Transgression}
 (\CT \omega)(\theta)=\oint_x \dd \tau~ \tfrac{1}{(p-1)!}\,\omega(\theta(\tau))_{i_1\cdots i_{p}}~\dot{\theta}^{i_p}~\delta \theta^{i_1\tau}\wedge \cdots \wedge \delta \theta^{i_{p-1}\tau}~.
\end{equation}
Note that $\CT\omega$ is reparameterization invariant. Furthermore, the transgression map is a chain map, which means that closed forms are mapped to closed forms and exact forms are mapped to exact forms.  In particular, the transgression of an exact 1-form is zero: 
\begin{equation}
 \CT(\dd f)=\oint \dd \tau~ \dot\theta^i(\tau)\left.\pa_if\right|_{\theta(\tau)}=\oint \dd \tau~ \dder{\tau} f(\theta(\tau))=0~.
\end{equation}
Note that the transgression map is not surjective. We will call forms on $\CL S^3$ which are in the image of $\CT$ {\em ultralocal}. Moreover, we will call forms on $\CL S^3$ that can be written in terms of a single loop integral {\em local}.

Consider now the standard volume form $\omega$ on $S^3$. The 2-form $\CT\omega$ is closed and non-degenerate, and therefore the volume form (or 2-plectic structure) on $S^3$ is lifted by the transgression map to a symplectic structure on $\CL S^3$ \cite{0817647309}.  Thus to any function $f$ on loop space, one can associate a Hamiltonian vector field $X_f$ in the usual way, and a Poisson bracket can be defined on loop space by the formula $\{f,g\}=\iota_{X_f}\iota_{X_g}\CT\omega$.

Interestingly, the components of the Hamiltonian vector field of a 1-form $\alpha\in \Omega^1(S^3)$ with respect to a 2-plectic form $\omega$ are identical to those of the Hamiltonian vector field of $\CT\alpha$ with respect to $\CT \omega$:
\begin{equation}
 X_\alpha= X^i_\alpha\der{\theta^i}~~~\Rightarrow~~~X_{\CT\alpha}=\oint \dd \tau~ X^i_{\alpha}(\theta(\tau))\,\delder{\theta^i(\tau)}~.
\end{equation}
This implies that the transgression maps both semi- and hemi-brackets on $(M,\omega)$ to the Poisson bracket on $(\CL S^3,\CT\omega)$:
\begin{equation}
 \CT\{\alpha,\beta\}_h=\CT\{\alpha,\beta\}_s=\{\CT\alpha,\CT\beta\}_{\CT\omega}~.
\end{equation}
Note that the transgressions of the hemi- and semi-brackets agree because their difference is an exact 1-form, and the transgression of exact 1-forms is zero.

\subsection{The Basu-Harvey equation in loop space}

We have now all the preliminaries covered to discuss the Basu-Harvey equation on loop space.  If $t^\mu(s)$ solve the Basu-Harvey equation \eqref{eq:BHInfinite} then the vector fields $D(t^\mu,t^\nu)$ solve
\begin{equation}
\label{eq:VFinfBH}
\eps_{\mu\nu\kappa\lambda}\frac{\dd}{\dd s}D(t^\kappa,t^\lambda)=\frac{4\pi^2}{ q}~[D(t^{\mu},t^{\kappa}),D(t^{\nu},t^{\kappa})]~.
\end{equation}
Consider the following 1-forms on $S^3$:
\begin{equation}
 t^{\mu\nu}:=t^{[\mu}\dd_{S^3} t^{\nu]}~,
\end{equation}
where $\dd_{S^3}$ denotes the exterior derivative on $S^3$, i.e.\ $\dd_{S^3} t^\mu(\theta^i,s)=\partial_i t^\mu\,\dd \theta^i$. The vector fields $D(t^\mu, t^\nu)$ are Hamiltonian vector fields associated to these 1-forms,
\begin{equation}
 X_{t^{\mu\nu}}=D(t^{\mu},t^{\nu})~.
\end{equation}
Since two Hamiltonian one-forms yielding the same Hamiltonian vector field on $S^3$ can differ only by an exact form $\gamma$, the Basu-Harvey equation implies the following equation for the 1-forms $t^{\mu\nu}$:
\begin{equation}
\eps_{\mu\nu\rho\lambda}\frac{\dd}{\dd s}t^{\rho\lambda}=\frac{4\pi^2}{ q}~\{t^{\mu\kappa},t^{\nu\kappa}\}_s+\gamma~,
\end{equation}
where $\dd \gamma=0$. Let us now switch to loop space via the transgression map \eqref{eq:Transgression}. The 1-forms $t^{\mu\nu}$ are mapped to the following functions on $\CL S^3\times[0,v)$:
\begin{equation}
 \trng^{\mu\nu}(\theta,s):=\CT t^{\mu\nu}=\oint\dd\tau\, t^\mu(\theta^a(\tau),s)\frac{\dd}{\dd\tau}t^\nu(\theta^a(\tau),s)~.
\end{equation}
These functions satisfy the loop space Basu-Harvey equation,
\begin{equation}\label{eq:LSinfBH}
\eps_{\mu\nu\kappa\lambda}\frac{\dd}{\dd s}\trng^{\kappa\lambda}=\frac{4\pi^2}{ q}~\{\trng^{\mu\kappa},\trng^{\nu\kappa}\}_{\CT\omega}~,
\end{equation}
where $\{\cdot,\cdot\}_{\CT\omega}$ denotes the natural Poisson structure on $\CL S^3$ induced by the transgressed volume form on $S^3$.

We would like to point out that while the Basu-Harvey equation implies equation \eqref{eq:VFinfBH}, the converse is not true. For example, the solution $t^1=\frac{e}{s}$, $t^2=t^3=t^4=0$, $e\in \CA$, to equation \eqref{eq:VFinfBH} does not satisfy the Basu-Harvey equation. Furthermore, one can exploit Gustavsson's observation \cite{Gustavsson:2008dy} that \eqref{eq:VFinfBH} is equivalent to two copies of the Nahm equation; these are obtained by projecting out the selfdual and anti-selfdual parts of the antisymmetric tensors $D(t^\kappa,t^\lambda)$ using the 't Hooft tensors.  Ashtekar et al.\ have shown \cite{Ashtekar:1987qx} that selfdual and anti-selfdual Einstein metrics can be constructed from any solution of the Nahm equation based on a Lie algebra of volume-preserving diffeomorphisms.  This means in particular that there are two Einstein metrics naturally associated to solutions of \eqref{eq:VFinfBH}, one selfdual and the other anti-selfdual.  A short calculation shows that these metrics both 
coincide with the flat metric on $\FR^4$ for all solutions of \eqref{eq:VFinfBH} obtained from the Basu-Harvey equation \eqref{eq:BHInfinite}.  Thus the space of solutions of the Basu-Harvey equation forms a special subspace of the space of solutions of \eqref{eq:VFinfBH}.

\subsection{Constructing loop space selfdual strings}

Let us now come to the loop space version of the selfdual string equation. Recall that the diffeomorphism $t:S^3\times \CI\rightarrow \Omega$ induces a foliation of $\Omega$ with leaves $\Sigma_{\phi}\cong S^3$, and for the interval $\CI=[0,v)$, $v\in\FR^+$, this foliation can be considered to be a fibration $\Omega\rightarrow\FR^+$ with fiber $\Sigma_{\phi(r)}$. Replacing $S^3$ with its loop space, we have the following induced map
\begin{equation}
 \trng:\CL S^3\times \CI\rightarrow \CL\Omega~,~~~\trng(\theta,s,\tau):=t(\theta(\tau),s)~,
\end{equation}
where $\theta$, $s$ and $\tau$ are coordinates on $S^3$, $\CI$ and $S^1$, respectively. This map is a diffeomorphism only between $\CL S^3\times \CI$ and its image in $\CL\Omega$. The latter space consists of loops that lie entirely in the fibers of $\Omega\rightarrow \FR^+$, and we effectively have a diffeomorphism $\trng: \CL S^3\times \CI\rightarrow \CL \Sigma\times \FR^+$ together with its inverse $\urng$. 

As before, we would like to construct a field strength $\CF$ together with a Higgs field $\varPhi$ by pulling back the transgressed volume form $\CT\omega$ and the coordinate function $s$ along $\urng$. The pull-back 2-form $\CF:=\frac{q}{2\pi^2}\urng^* \CT \omega$ has components 
\begin{equation}
\CF=\tfrac{1}{2}\oint \dd\tau~\CF_{(\mu\tau)(\nu\tau)}\delta y^{\mu\tau}\wedge\delta y^{\nu\tau}~.
\end{equation}
Note that the transgression $\CT \omega$ is by definition ultralocal, and so is the pull-back $\CF=\urng^*\CT\omega$ along inverses $\urng$ of induced maps $\trng$.

Recall from chapter \ref{ch:loop}, the abelian local loop space selfdual string equation is
\begin{equation}\label{eq:ex1LoopSpaceSDS}
 \CF_{(\mu\tau)(\nu\tau)}=\eps_{\mu\nu\kappa\lambda}~\frac{\yd^{\kappa\tau}}{|\yd^\tau|}\delder{y^{\lambda\tau}}\varPhi~,
\end{equation}
where $\varPhi$ is a $\au(1)$-valued function on $\CL\Omega$. Note that in the case of the non-abelian version of this equation, a formulation in terms of local forms is no longer gauge invariant and therefore not very useful. In the abelian case, however, gauge transformations act trivially on $\CF$ and $\varPhi$.

The right-hand side of \eqref{eq:ex1LoopSpaceSDS} can be understood as a Hodge-star for certain local forms generalized to loop space\footnote{Recall that there is no Hodge-star operation on general forms, because loop space is infinite dimensional.}. We define on $\CL\Omega\subset \CL\FR^4$:
\begin{equation}
\begin{aligned}
 * \oint \dd \tau~\alpha_{\mu_1\ldots \mu_p,\tau}~\delta y^{\mu_1\tau}&\wedge \cdots \wedge \delta y^{\mu_p\tau}:=\\
&\oint \dd \tau~\frac{(-1)^{p+1}}{p!}\alpha_{\mu_1\ldots \mu_p,\tau}\eps^{\mu_1\ldots \mu_{4}}\frac{\yd_{\mu_{p+1}\tau}}{|\yd^\tau|}\delta y^{\mu_{p+2}\tau}\wedge \cdots\wedge \delta y^{\mu_4\tau}~,
\end{aligned}
\end{equation}
where $0\leq p\leq 3$. One easily verifies that $*^2=\id$. The loop space selfdual string equation \eqref{eq:ex1LoopSpaceSDS} then reduces to $\CF=*\delta \varPhi$. 

We now restrict to $\CL\Sigma\times \FR^+$, which is diffeomorphic to $\CL S^3\times \CI$. As before, the $\Sigma_\varPhi$ are the level sets of the Higgs field $\varPhi$, and $\varPhi(\theta,r)=\varPhi(r)$. The total differential $\delta$ on $\CL\Sigma\times \FR^+$ reduces such that the equation $\CF=*\delta \varPhi$ becomes
\begin{equation}\label{eq:6.27}
\CF=\oint \dd\sigma~\eps_{\mu\nu\kappa\lambda}\yd^{\kappa\sigma}\left. \frac{\partial \varPhi(x)}{\partial x^\lambda}\right|_{x=y(\sigma)}\delta y^{\mu\sigma}\wedge\delta y^{\nu\sigma}~.
\end{equation}
Let us now verify that the fields $\CF=\frac{q}{2\pi^2}\urng^*\CT \omega$ and $\varPhi=s\circ \urng$ indeed solve this equation, where $u_\circ$ is the inverse of the diffeomorphism $t_\circ$.  Equation \eqref{eq:6.27} is equivalent to the following equation on $\CL S^3\times\CI$:
\begin{multline}
\frac{q}{2\pi^2}\CT \omega = \oint \dd\sigma~\eps_{\mu\nu\kappa\lambda}\trngd^{\kappa}(\sigma)\left. \frac{\partial s}{\partial x^\lambda}\right|_{x=t(\theta(\sigma),s)} \times \\
\left[ \frac{1}{2}\oint \dd \tau'\oint \dd \tau'' \frac{\delta \trng^{\mu\sigma}}{\delta \theta^{i\tau'}}\frac{\delta \trng^{\nu\sigma}}{\delta\theta^{j\tau''}} \delta\theta^{i\tau'}\wedge\delta \theta^{j\tau''} +\oint \dd \tau \frac{\partial \trng^{\mu\sigma}}{\partial s} \frac{\delta \trng^{\nu\sigma}}{\delta \theta^{i\tau}} \dd s\wedge\delta\theta^{i\tau}\right]~.
\end{multline}
The right hand side of this equation can be simplified using the identities,
\begin{equation}
 \frac{\delta \trng^{\mu\sigma}}{\delta \theta^{i\tau}}= \left. \frac{\partial t^{\mu}(\chi,s)}{\partial \chi^{i}}\right|_{\chi=\theta(\sigma)}\delta(\sigma-\tau)~~,~~~\trngd^{\mu}(\sigma)=\left. \frac{\partial t^{\mu}(\chi,s)}{\partial \chi^{i}}\right|_{\chi=\theta(\sigma)}\dot \theta^{i}(\sigma)~,
\end{equation}
yielding
\begin{equation}
\oint \dd\sigma\,\eps_{\mu\nu\kappa\lambda}\, \partial_\lambda s\, \dot\theta^k(\sigma)\,\partial_k t^\kappa
\left[ \frac{1}{2} \partial_i t^\mu \partial_j t^\nu \delta\theta^{i\sigma}\wedge\delta \theta^{j\sigma} + \partial_s t^{\mu} \partial_i t^\nu \dd s\wedge\delta\theta^{i\sigma}\right].
\end{equation}
On substituting for $\pa_s t^\mu$ using the Basu-Harvey equation \eqref{eq:BHInfinite}, the second term in the square bracket becomes an expression which vanishes due to $\frac{\partial s}{\partial t^\mu}\frac{\partial t^\mu}{\partial \theta^a}=0$.  The first term in the square bracket can be rearranged using the Basu-Harvey equation \eqref{eq:BHInfinite} to give
\begin{equation}
\frac{q}{2\pi^2} \oint\dd\tau \frac{1}{2}\sin^2\theta^1(\tau)\sin\theta^2(\tau)\eps_{ijk} \dot\theta^i(\tau)\delta\theta^{j\tau}\delta\theta^{k\tau}.
\end{equation}
This expression is clearly equal to $\frac{q}{2\pi^2}\CT \omega$, so the Basu-Harvey equation \eqref{eq:BHInfinite} implies the loop space selfdual string equation \eqref{eq:6.27}.

\section{Magnetic domains in higher dimensions}\label{sec:higher}

Although string- and M-theory only motivate the study of magnetic domains in three and four dimensions, it is still interesting to consider higher-dimensional generalizations of these objects. 

\subsection{From higher BPS equations to magnetic domains}

Recall that the curvature 2-form $f$ of a magnetic domain $\Omega$ in three dimensions defines topologically a vector bundle over $\Omega$. Using repeatedly the Poincar\'e lemma, we obtain gauge potentials on patches of a covering of $\Omega$ and transition functions on overlaps of patches. Analogously, the curvature 3-form $h$ of a magnetic domain in four dimensions defines a gerbe, with 2-form potentials on patches etc.\ Vector bundles and gerbes are examples of so-called {\em $k$-gerbes} with $k=0$ and $k=1$, respectively. In general, $k$-gerbes are defined in terms of curvature $k+2$-forms, with associated $k+1$-form potential etc.

Let us now generalize the previous discussion to magnetic domains in $k+3$-dimensions, which are described in terms of an abelian Higgs field together with the curvature $k+2$-form $g$ of a $k$-gerbe. As before, we can obtain a Bogomolny bound from the energy functional 
\begin{equation}
E=\tfrac{1}{2}\int_\Omega~ g\wedge * g+\dd\phi\wedge *\dd\phi~,
\end{equation}
where $\Omega\subset\FR^{k+3}$ and $g=\dd c$ for some $k+1$-form potential $c$. The Bogomolny bound then becomes
\begin{equation}
E=\int_\Omega~ \tfrac{1}{2}|\dd\phi-*g|^2+\dd\phi\wedge g\ge vq~,
\end{equation}
where $q:=\int_{S^{k+2}_\infty}g$, and the bound is saturated if 
\begin{equation}
g=*\dd\phi~.
\end{equation}

On a $k+2$-dimensional orientable manifold $M$, a volume form $\omega$ yields a $k+1$-plectic structure, i.e.\ a non-degenerate and closed $k+2$-form. This form can be inverted to a multivector field, which defines a Nambu-Poisson structure on $\CC^\infty(M)$. That is, we have a $k+2$-ary bracket $\{\cdot,\,\cdots,\cdot\}$ which is linear in each argument, totally antisymmetric and satisfies the obvious generalizations of the fundamental identity \eqref{eq:GenFundIdent} and the Leibniz rule \eqref{eq:Leibniz}. 

The $\Pi_\omega$-Nahm data is here given by a $k+3$-tuple of functions $t^m$, $m=1,\ldots,k+3$ on a $k+1$-plectic manifold $M$, which solve the higher Nahm equation
 \begin{equation}\label{eq:higherNahm}
\frac{\dd t^{m_1}}{\dd s}=\frac{{\rm vol}(S^{k+2})}{(k+2)! q}~\eps_{m_1\dots m_{k+3}}\{t^{m_2},\dots,t^{m_{k+3}}\}~,
\end{equation}
where the volume of the unit $S^{k+2}$-sphere is
\begin{equation}
{\rm vol}(S^{k+2})=\frac{2\pi^\frac{k+3}{2}}{\Gamma(\frac{k+3}{2})}~~.
\end{equation}
The generalization of theorems \ref{thm:2} and \ref{thm:4} now reads as
Up to gauge equivalence, there is a one-to-one correspondence between
\vspace{-0.2cm}
\begin{itemize}\label{thm:6}
 \setlength{\itemsep}{-1mm}
 \item sets of $\Pi_\omega$-Nahm data with the property that the map from $M\times \CI$ to $\Omega\subset\FR^{k+3}$ defined by the $t^m$ is a diffeomorphism $t:M\times \CI\rightarrow \Omega$, and
 \item magnetic domains $\Omega$ that are diffeomorphic to $M\times \CI$, where the restriction of the $k+2$-form curvature $g$ to any slice $M\times\{s_0\}$ has the same volume type as $\omega$ and $\CI$ is the range of $\phi$. Explicitly, there is a diffeomorphism $u:\Omega\rightarrow M\times \CI$ such that $g=\frac{q}{{\rm vol}(S^{k+2})}u^*\omega$ and $\phi=s\circ u$ on $\Omega$. 
\end{itemize}

The proof is an obvious generalization of the proofs of theorems \ref{thm:2} and \ref{thm:4}. 

\

For a $k+2$-dimensional magnetic bag in $\FR^{k+3}$, the boundary condition for the Higgs field reads as
\begin{equation}
\phi=v-\frac{q}{{\rm vol}(S^{k+2}) r^{k-1}(k-1)}+\CO(r^{-k})
\end{equation}
for $r\rightarrow \infty$. The corresponding boundary condition for the $\Pi_\omega$-Nahm data is
 \begin{equation}
t^{i}= x^i\left(\frac{q}{{\rm vol}(S^{k+2})(k-1)(v-s)}\right)^{\frac{1}{k-1}}+\CO(s^{\frac{k}{1-k}})
\end{equation}
as $s\rightarrow v$. Instead of discussing examples of such magnetic domains in more detail, let us comment on the relation of our equations to $L_\infty$-algebras.

\subsection{Comments on the relation to strong homotopy Lie algebras}\label{ssec:SHalgebras}

The appearances of higher brackets in our equations suggests to look for a relationship to strong homotopy Lie algebras or $L_\infty$-algebras for short. Roughly speaking, an $L_\infty$-algebra is a graded vector space together with brackets with arbitrarily many arguments that satisfy homotopy Jacobi identities. They are the most natural generalization of Lie algebras to vector spaces endowed with higher brackets. The definition is found in appendix \ref{app:SH_Lie_algebras}.

If the graded vector space underlying an $L_\infty$-algebra $L$ is concentrated in degrees $k=0,\ldots,n-1$, i.e.\ $L=\oplus_{k\in\RZ} L_k$ with $L_k=0$ unless $0\leq k\leq n-1$, then $L$ is a {\em Lie $n$-algebra}. In a Lie $n$-algebra, we have $\mu_i=0$ for $i>n+1$. Note that the homotopy Jacobi identity \eqref{eq:homotopyJacobi} implies $\mu_1^2=0$, such that $\mu_1$ is a differential. Therefore, a Lie $1$-algebra is an ordinary Lie algebra. A Lie 2-algebra, or 2-term $L_\infty$-algebra, consists of two vector spaces $V_0$ and $V_1$ with differential $\mu_1:V_1\rightarrow V_0$, a binary map $\mu_2:V_i\times V_j\rightarrow V_{i+j}$, $i,j,i+j=0,1$ and a ternary map $\mu_3:V_0\times V_0\times V_0\rightarrow V_1$, all satisfying \eqref{eq:homotopyJacobi}.

Strong homotopy Lie algebras appear in modern deformation theory. Here, the definition of the deformation functor involves the so-called {\em homotopy Maurer-Cartan equations} \cite{Merkulov:1999aa,Lazaroiu:2001nm} on an element $\phi$ of an $L_\infty$-algebra $L$:
\begin{equation}\label{eq:MCeqs}
 \sum_{i=1}^\infty \frac{(-1)^{i(i+1)/2}}{i!}\mu_i(\phi,\cdots,\phi)=0~.
\end{equation}
These equations are invariant under the infinitesimal gauge transformations
\begin{equation}\label{eq:MCgaugetrafos}
 \delta \phi=-\sum_i \frac{(-1)^{i(i-1)/2}}{(i-1)!}\mu_i(\alpha,\phi,\cdots,\phi)~,
\end{equation}
where $\alpha$ is an element of $L$ of degree 0. The classical Maurer-Cartan equations $\dd \phi+\frac{1}{2}[\phi,\phi]=0$ appear as a special case for Lie 1-algebras. 

We see three ways in which $L_\infty$-algebras are concealed in our previous discussion. First of all, the semi-bracket on 1-forms introduced in section \ref{subsec:PoissonForms} yields a semi-strict Lie 2-algebra \cite{Baez:2008bu}. As Lie 2-algebras are 2-term $L_\infty$-algebras, the 1-form description on $S^3$, which transgresses to the loop space description yields an $L_\infty$-algebra. Second, as we will see in the next chapter, 3-Lie algebras are special cases of differential crossed modules. The category of the latter is equivalent to that of strict Lie 2-algebras \cite{Baez:2003aa}, and we arrive again at an $L_\infty$-algebra. Let us 
stress, however, that the 3-bracket of 3-Lie algebras cannot be interpreted as a ternary product $\mu_3$ in an $L_\infty$-algebra unless one gives up the grading \cite{IuliuLazaroiu:2009wz}.

Both the above appearances of $L_\infty$-algebras do not seem to provide any further insights into our discussion. We therefore give up the grading and turn to another interpretation advocated for $n$-Lie algebras in \cite{IuliuLazaroiu:2009wz}. There, it was shown that both the Nahm and the Basu-Harvey equations correspond to Maurer-Cartan equations in certain $n$-term $L_\infty$-algebras \cite{IuliuLazaroiu:2009wz}. We now briefly review these structures and demonstrate that the higher $\Pi_\omega$-Nahm equations also fit into this picture.

We start from the gauge covariant form of the Nahm equation with coupling constants put to 1,
\begin{equation}
 \dder{s}T^{m_1}+[A_s,T^{m_1}]= \eps^{m_1\ldots m_{p+1}}[T^{m_2},\ldots,T^{m_{j+1}}]~,
\end{equation}
where the $T^m$ are functions on $\CI$ with values in the $p$-Lie algebra $\CA$ and $A_s$ is the gauge potential with values in $\frg_\CA$, the Lie algebra of inner derivations of $\CA$. We choose $\CA$ to be the $p$-Lie algebra $\CA_{p+1}\cong\FR^{j}$ with generators $e_i$, $i=1,\ldots,p$ and bracket
\begin{equation}
 [e_{i_1},\cdots,e_{i_p}]=\eps_{i_1\cdots i_pj}e_j~.
\end{equation}
Note that the algebra of inner derivations of $\CA_{p+1}$ is $\frg_{\CA_{p+1}}\cong \sSO(p)$. Consider now the $L_\infty$-algebra $L=L_0+L_1+L_2$ with
\begin{equation}\label{eq:Lcomponents}
\begin{aligned}
 L_0&=\Omega^0(\CI)\otimes \frg_{\CA}~,\\
 L_1&=\left(\Omega^0(\CI)\otimes \CC\ell(\FR^{n+1})\otimes \CA\right)~\oplus~\left(\frg_{\CA}\otimes\Omega^1(\CI)\right)~,\\
 L_2&=\CC\ell(\FR^{n+1})\otimes \CA\otimes\Omega^1(\CI)~,
\end{aligned}
\end{equation}
where $\CC\ell(\FR^{n+1})$ is the Clifford algebra with $n+1$ generators $\gamma_\mu$, $\mu=1,\ldots,n+1$, satisfying $\gamma_\mu\gamma_\nu+\gamma_\nu\gamma_\mu=2\delta_{\mu\nu}$. An element $\phi$ of $L$ decomposes as 
\begin{equation}
 \phi=\underbrace{\lambda}_{L_0}+\underbrace{T^\mu \gamma_\mu+A\,\dd s}_{L_1}+\underbrace{S^\mu\gamma_\mu \dd s}_{L_2}~,
\end{equation}
where $\lambda\in L_0$, $T^\mu, S^\mu\in \Omega^0(\CI)$ and $A\in \Omega^0(\CI)\otimes\frg_{A_n}$. We define the following products:
\begin{equation}
\begin{aligned}
\mu_1(\phi)&=\dder{s}\lambda\,\dd s+ \left(\dder{s} T^\mu\right)\,\gamma_\mu\,\dd s~,\\
\mu_2(\phi_1,\phi_2)&=[\lambda_1,A_2]\,\dd s+[A_1,\lambda_2]\,\dd s+A_1\acton T^\mu_2\,\gamma_\mu\,\dd s-A_2\acton T^\mu_1\,\gamma_\mu\,\dd s\\
&~~~+\lambda_1\acton T^\mu_2\,\gamma_\mu-\lambda_2\acton T^\mu_1\,\gamma_\mu~,\\
\mu_p(\phi_1,\ldots,\phi_p)&=[T^{\mu_1}_1,\ldots,T^{\mu_n}_n]\,\eps_{\mu_1\cdots\mu_p\nu}\gamma_\nu\,\dd s~.
\end{aligned}
\end{equation}
For $p=2$, the two products defined above have to be added. The Maurer-Cartan equation \eqref{eq:MCeqs} for $\phi\in L$ with grading 1 correspond to the Nahm equation for $p=2$, the Basu-Harvey equation for $p=3$ and corresponding higher Nahm equations for $p>3$. Note that also the gauge transformations of the (higher) Nahm equations in gauge covariant form are given by the corresponding gauge transformations \eqref{eq:MCgaugetrafos} for an $\alpha\in L$ with grading 0.

In the rest of this thesis we will not pursue this use of $L_\infty$-algebras. Instead we will keep the grading on the vector spaces and relate $L_\infty$-algebras to differential crossed modules and hence 3-algebras. This observation makes M-theory models involving 3-algebras into \emph{higher gauge theories}.

\chapter{Higher gauge theory}
\label{ch:high}

In this chapter we will see how models involving 3-algebras can be reformulated in terms of differential crossed modules. This turns these models into \emph{higher gauge theories}, putting them in the same framework as that of M5-brane models. We will see a caveat in which the so-called \emph{fake curvature condition} will require us to reformulate the models in terms of differential 2-crossed modules, in such a way that the fake curvature condition is satisfied.

M5-branes interact via M2-branes ending on them. An effective description of M5-branes should therefore be a gauge theory describing the parallel transport of the one-dimensional boundaries of these M2-branes in the worldvolume of the M5-branes. This is where higher gauge theory \cite{Baez:2004in,Baez:2010ya} enters the picture. In general, higher gauge theory with principal $n$-bundles captures the parallel transport of $(n-1)$-dimensional objects. 

It is known that the effective dynamics of a single M5-brane involves an $\CN=(2,0)$ tensor multiplet in six dimensions, which contains a 2-form potential $B$. Higher gauge theory naturally contains this 2-form potential, even in a non-abelian generalization: it is the gauge potential for the parallel transport of a one-dimensional object along a surface.

In the first chapter we saw a loop space Nahm-like transform, connecting solutions of the Basu-Harvey equation to solutions of a loop space version of the selfdual string equation. This is somewhat surprising, as the original Nahm transform is a duality between identical equations: instanton solutions on a four-torus are mapped to instantons on the corresponding dual four-torus. Taking certain infinite-radius limits, one then arrives e.g.\ at the ADHMN construction of solutions to the Bogomolny monopole equation from solutions to the Nahm equation. The Basu-Harvey equation and the selfdual string equation, however, seem very different. If a full Nahm transform is to exist, one would need a reformulation of both equations in a common language.

We have seen a non-abelian version of the selfdual string equation on loop space. However, it seems clear that a direct formulation on space-time is very likely to involve the non-abelian gerbes defined in \cite{Breen:math0106083} and equivalently in \cite{Aschieri:2003mw}. These non-abelian gerbes can be described in terms of higher gauge theories involving Lie 2-groups as gauge groups. The category of (strict) Lie 2-groups is equivalent to the category of Lie crossed modules and the gauge algebra in the higher gauge theories is therefore given by differential crossed modules. In this chapter, we make the observation that 3-algebras relevant to M-brane models are special cases of differential crossed modules. Therefore, these models can be regarded as higher gauge theories.

The existence of this reformulation can also be expected from the following point of view: the fuzzy funnel described by the Basu-Harvey equation should contain a non-commutative 3-sphere \cite{Basu:2004ed}. One would expect that the corresponding Hilbert space is a categorification of an ordinary Hilbert space and therefore based on a 2-vector space. The fields in the Basu-Harvey equation should correspond to endomorphisms of this 2-Hilbert space, which are organized in the structure of a Lie 2-algebra.

Besides making the existence of a full generalized Nahm transform more conceivable, our reformulation also has other advantages. In particular, it yields more than the 3-algebra based M-brane models already known, and therefore we can use it as a framework\footnote{Another framework for generalizing the Basu-Harvey equation and M2-brane models in general has been proposed in \cite{IuliuLazaroiu:2009wz} in the form $L_\infty$-algebras. The $L_\infty$-structures identified there are different from the ones found here.} for generalizing these models. Within this larger class of models, one might overcome some of the problems of current M-brane models. For example, one might hope to find Chern-Simons matter theories with $\CN=8$ supersymmetry beyond the Bagger-Lambert-Gustavsson model based on the 3-Lie algebra $A_4$. Encouraging is that we do find new $\CN=(2,0)$ supersymmetric tensor multiplet equations beyond a recently proposed set based on 3-Lie algebras.

We start our discussion by reviewing Lie 2-groups and crossed modules as well as the derivation of 3-algebras from these. We give a few non-trivial examples that go beyond the usual picture of 3-algebras. We then present the interpretation of the 3-Lie algebra valued tensor multiplet equations of \cite{Lambert:2010wm} as well as the M2-brane models of \cite{Bagger:2007jr,Gustavsson:2007vu,Aharony:2008ug,Bagger:2008se} in the framework of higher gauge theories. For the latter we will need to cover the machinery of principal 3-bundles which come equipped with differential 2-crossed modules. This will allow the fake curvature condition to be satisfied.

\section{Lie 2-algebras and 3-algebras}

In this section, we make the connection between Lie 2-algebras and 3-algebras using an extension of the so-called Faulkner construction. For a detailed account of the Faulkner construction for 3-algebras, see \cite{deMedeiros:2008zh}. For a motivation and a more extensive discussion of categorified gauge structures, see \cite{Baez:2002jn,Baez:0511710,Baez:2010ya}.

\subsection{Lie 2-groups, Lie 2-algebras and crossed modules}

While the parallel transport of a point particle along a path assigns a group element to each path, the parallel transport of a string along a surface leads naturally to the concept of a Lie 2-group\footnote{In this letter, we will restrict ourselves to strict Lie 2-groups.}. A Lie 2-group is a categorification of the notion of a Lie group. Recall that a group is a (small) category with one object in which each morphism is invertible. A Lie 2-group is analogously built from a corresponding 2-category, i.e.\ a category with additional ``morphisms between morphisms''. Furthermore, the category of Lie 2-groups can be shown to be equivalent to the category of Lie crossed modules and it is this language that we will use.

Recall that a {\em crossed module} is a pair of groups $\CG$ and $\CH$ together with an automorphism action $\acton$ of $\CG$ onto $\CH$ and a group homomorphisms $\mt:\CH\rightarrow \CG$, which satisfy the following conditions:
\begin{itemize}
\item[$i)$] $\mt$ is equivariant with respect to conjugation,
\begin{subequations}\label{eq:LCMIdentities}
\begin{equation}
 \mt(g\acton h)=g \mt(h) g^{-1}~,
\end{equation}
\item[$ii)$] and the so-called {\em Pfeiffer identity} holds:
\begin{equation}
 \mt(h_1)\acton h_2=h_1h_2h_1^{-1}~,
\end{equation}
\end{subequations}
\end{itemize}
for all $g\in\CG$ and $h,h_1,h_2\in\CH$. A {\em Lie crossed module} is a crossed module $(\mt:\CH\rightarrow\CG,\acton)$, where $\CG$ and $\CH$ are Lie groups. A simple example of a Lie crossed module is $\CG=\CH=\sU(N)$ with $\mt$ the identity map and $\acton$ the adjoint action.

Just as a Lie algebra can be obtained by linearizing a Lie group at the identity element, so can a Lie 2-algebra be obtained by linearizing a Lie 2-group. These Lie 2-algebras correspond to differential crossed modules.

A {\em differential crossed module} $(\sft:\frh\rightarrow \frg,\acton)$ is a pair of Lie algebras $\frg$, $\frh$ together with an action $\acton$ of elements of $\frg$ as derivations of $\frh$ and a Lie algebra homomorphism between $\frh$ and $\frg$, which we will also denote by $\mt$, slightly abusing notation. We demand that $\acton$ and $\mt$ satisfy the linearized versions of the identities \eqref{eq:LCMIdentities}:
\begin{equation}\label{eq:Pfeif}
 \mt(x\acton y)=[x,\mt(y)]\eand \mt(y_1)\acton y_2=[y_1,y_2]
\end{equation}
for all $x\in \frg$ and $y,y_1,y_2\in \frh$. The differential version of our simple example from above is evidently $\frg=\frh=\au(N)$ with $\acton$ being the adjoint action and $\sft$ the identity map.

Note that Lie 2-algebras are 2-term $L_\infty$-algebras \cite{Baez:2003aa}. These have a 3-bracket called the Jacobiator, which is different from the 3-bracket we define later. These 2-term $L_\infty$-algebras are only equivalent to differential crossed modules when the Jacobiator vanishes. In this case, they are called \emph{strict} Lie 2-algebras.

To write down action functionals, we need to extend the above notion to that of a {\em metric differential crossed module}. The additional metric structure on a differential crossed module $(\mt:\frh\rightarrow\frg,\acton)$ is given by non-degenerate hermitian forms $\lbr\cdot,\cdot\rbr$ on $\frg$ and $(\cdot,\cdot)$ on $\frh$, which are invariant under the obvious Lie algebra actions:
\begin{equation}
\begin{aligned}
 \lbr [x_1,x_2],x_3\rbr+\lbr x_2,[\bar x_1,x_3]\rbr&=0~,\\
 (x\acton y_1,y_2)+(y_1,\bar x\acton y_2)&=0~.
\end{aligned}
\end{equation}
The last equation also implies that $(\cdot,\cdot)$ is $\frh$-invariant: $ ([y_1,y_2],y_3)+(y_2,[\bar y_1,y_3])=0$.

Note that the introduction of the metric structure allows us to define a map $\mt^*:\mathfrak g\rightarrow\mathfrak h$ implicitly by
\begin{equation}
 (\mt^*(x),y):=\lbr x,\mt(y)\rbr~.
\end{equation}
One readily verifies useful identities, e.g.\
\begin{equation}\label{eq:tstaridentity}
 \sft^*([x_1,x_2])=x_1\acton \sft^*(x_2)=-x_2\acton \sft^*(x_1)~.
\end{equation}

To avoid the appearance of ghosts from matter fields in our M-brane models, we will always choose the metric on $\frh$ to be positive definite. For $\frg$, however, we would like to allow split signature. The reason for this is that all the 3-algebra M2-brane models are given by Chern-Simons matter theories, which are a priori not parity invariant. Having a gauge algebra $\frg$ of the form $\frg_L\oplus \frg_R$ with split signature yields a pair of Chern-Simons terms with opposite Chern-Simons levels. These are then mapped into each other under a parity flip.
\subsection{Deriving 3-algebras from differential crossed modules}\label{ssec:3fromDCM}

It is possible to construct all 3-algebras from metric Lie algebras together with certain faithful representations via the Faulkner construction \cite{Faulkner:1973aa,deMedeiros:2008zh}. These pairs of Lie algebras and representations correspond to metric differential crossed modules $(\mt:\frh\rightarrow\frg,\acton)$ with abelian $\frh$ and trivial $\mt$. Thus, all real and hermitian 3-algebras are obtained by applying the Faulkner construction to such differential crossed modules whose Lie algebras $\frh$ are real or complex, respectively. However, we can extend this construction to arbitrary metric differential crossed modules: Allowing $\frh$ to be non-abelian and $\mt$ non-trivial still gives structures with 3-brackets which satisfy the fundamental identity \eqref{eq:fun}.

Starting from a metric differential crossed module $(\mt:\frh\rightarrow\frg,\acton)$, there is a unique linear map $D: \frh\otimes \frh \rightarrow \frg$ such that\footnote{The usual definition, in e.g.\ \cite{Baez:2002jn}, is $\lbr x,D(y_1,y_2)\rbr = (x\acton y_1,y_2)$. This agrees with our definition in the real case, but in the complex case our definition gives antilinearity in the second argument, which is the convention chosen for the hermitian 3-algebras in \eqref{eq:3bra}.}
\begin{equation}
 \lbr x,D(y_1,y_2)\rbr=-(x\acton y_2,y_1)
\end{equation}
for all $x\in \frg$ and $y_1,y_2\in \frh$. The map $D$ is skew-hermitian since
\begin{equation}
\begin{aligned}
 \lbr x,D(y_1,y_2)\rbr&=-(x\acton y_2,y_1)=(y_2,\bar x\acton y_1)\\&=\overline{(\bar x\acton y_1,y_2)}=-\overline{\lbr \bar x,D(y_2,y_1)\rbr}=\lbr x,-\overline{D(y_2,y_1)}\rbr~,
\end{aligned}
\end{equation}
and satisfies the identity \cite{Martins:2010ry}
\begin{equation}
[x,D(y_1,y_2)]=D(x\acton y_1,y_2)+D( y_1,\bar x\acton y_2)~,
\end{equation}
which implies the fundamental identity \eqref{eq:GenFundIdent}. Therefore, we can define 3-brackets according to
\begin{equation}
[y_1,y_2,y_3] :=D(y_1,y_2)\acton y_3\eand [y_3,y_1;y_2] :=D(y_1,y_2)\acton y_3
\end{equation}
for real and hermitian 3-algebras, respectively.

\subsection{3-algebra examples}

Let us now reconstruct the familiar examples of 3-algebras. The simplest way of realizing the 3-algebra $A_4$ as a differential crossed module is to take $(\mt:\FR^4\rightarrow\aso(4),\acton)$, where $\acton$ is the ordinary action of $\aso(4)$ on the fundamental representation, $\sft=0$ is the trivial map and the metric on $\aso(4)\cong\asu(2)\times\asu(2)$ is of split signature. $\FR^4$ is viewed here as an abelian Lie algebra with trivial Lie bracket and Euclidean metric. This gives the completely anti-symmetric 3-bracket $[e_\mu,e_\nu,e_\kappa]:=D(e_\mu,e_\nu)\acton e_\kappa=\eps_{\mu\nu\kappa\lambda}e_\lambda$ on the standard basis vectors $e_\mu\in\FR^4$.

The hermitian 3-algebras occurring in the ABJM model \cite{Aharony:2008ug,Bagger:2008se} are equivalent to crossed modules of the form $(\mt:\agl(N,\FC)\rightarrow \agl(N,\FC)\times \agl(N,\FC),\acton)$, where $\mt=0$ and $\frh=\agl(N,\FC)$ is regarded as an (additive) abelian Lie algebra. The action of $\frg$ on $\frh$ is given by
\begin{equation}
(x_1,x_2)\acton y := x_1 y  - y  x_2~,
\end{equation}
which yields the following Lie bracket:
\begin{equation}
[(x_1,x_2), (x_3,x_4)]=([x_1,x_3],[x_2,x_4])~.
\end{equation}
The metric structures on $\frh$ and $\frg$ are given by
\begin{equation}
(y_1,y_2):=\tr(y_1 y_2^\dagger)~,~~\lbr (x_1,x_2), (x_3,x_4)\rbr:=\tr(x_1 x_3^\dagger-x_2 x_4^\dagger)~,
\end{equation}
and from these we derive the derivations
\begin{equation}
D(y_1,y_2)= (y_1  y_2^\dagger, y_2^\dagger y_1)~,
\end{equation}
which yield the 3-bracket
\begin{equation}\label{eq:ABJM3bracket}
[y_1,y_3;y_2]:=D(y_1,y_2)\acton y_3= y_1  y_2^\dagger y_3-y_3 y_2^\dagger y_1~.
\end{equation}
This is the 3-bracket used for the matter fields in the ABJM model. The gauge fields however are required to live in the real Lie algebra $\au(N)\times\au(N)$. We therefore define the differential crossed module $\frm_{\rm ABJM}(N)$ as $(\mt:\agl(N,\FC)\rightarrow\au(N)\times\au(N),\acton)$, where $\mt=0$ and the action of $\frg$ on $\frh$ reads as
\begin{equation}
(x_1,x_2)\acton y =  x_1 y  - y  x_2~.
\end{equation}
The metrics are
\begin{equation}
(y_1,y_2):=\tr(y_1 y_2^\dagger+ y_1^\dagger y_2)\eand\lbr (x_1,x_2), (x_3,x_4)\rbr:=-\tr(x_1 x_3-x_2 x_4)~,
\end{equation}
from which we derive
\begin{equation}
D(y_1,y_2)= (y_1  y_2^\dagger  - y_2  y_1^\dagger, y_2^\dagger y_1 -  y_1^\dagger y_2)~.
\end{equation}
In the case $N=2$, the bracket is totally anti-symmetric and the 3-algebra becomes $A_4$.

The 3-algebras $\CC_{2N}$ used in \cite{Cherkis:2008qr} involve only the real Lie algebras $(\mt:\agl(N,\FR)\rightarrow\aso(N)\times\aso(N),\acton)$. Similarly, one can obtain all 3-algebras, in particular those appearing in the classification of \cite{Cherkis:2008ha}, from differential crossed modules with $\sft=0$.

\subsection{Nontrivial examples of differential crossed modules}\label{sec:NontrivialExamples}

The non-abelian gerbes of Breen and Messing \cite{Breen:math0106083} use automorphism Lie 2-groups, whose differential crossed modules are of the form $(\mt:\frh\rightarrow \mathsf{Der}(\frh),\acton)$, where $\sft$ is the obvious map from the Lie algebra $\frh$ to its derivations $\mathsf{Der}(\frh)$ and $\acton$ is the action of these derivations. The simplest example is $(\mt:\au(N)\rightarrow\au(N),\acton)$, with $\mt$ being the identity and $\acton$ the adjoint action. With Hilbert-Schmidt metrics, this non-abelian gerbe has a 3-bracket
\begin{equation}
[y_1,y_2,y_3]:=D(y_1,y_2)\acton y_3= [[y_1 , y_2], y_3]~.
\end{equation}
This example can be trivially reduced to the differential crossed module $(\mt:\au(N)\rightarrow\asu(N),\acton)$, where $\sft(\unit):=0$. It is this differential crossed module that we will encounter in the M5-brane model.

Finally, we will consider an example from \cite{Martins:2010ry}. Let $\frh$ be the Lie algebra of complex block matrices with blocks of sizes
\begin{equation}
\begin{pmatrix} m\times m & m\times p\\n\times m & n\times p\end{pmatrix}
\end{equation}
endowed with the Lie bracket (which is not the ordinary matrix commutator)
\begin{equation}
\left[   \begin{pmatrix} A & B \\C  & D\end{pmatrix} ,\begin{pmatrix} A' & B' \\C'  & D'\end{pmatrix} \right]=\begin{pmatrix} [A,A'] &  AB'-A'B \\ CA'-C'A & CB'-C'B\end{pmatrix} ~.
\end{equation}
The Lie algebra $\frg$ consists of pairs of these matrices of the form
\begin{equation}
\left( \begin{pmatrix} A & 0\\C & D\end{pmatrix},\begin{pmatrix} A & B'\\0 & D'\end{pmatrix}  \right)~,
\end{equation}
where the Lie bracket is the usual matrix commutator. Now the map  $\mt:\frh\rightarrow\frg$ is given by
\begin{equation}
\mt\begin{pmatrix} A & B \\C  & D\end{pmatrix}:=\left ( \begin{pmatrix} A & 0\\ C & 0 \end{pmatrix}, \begin{pmatrix} A & B\\  0 & 0 \end{pmatrix}\right)~,
\end{equation}
and the action of $\frg$ on $\frh$ is the usual combination of left and right actions
\begin{equation}
\begin{aligned}
\left  ( \begin{pmatrix} A & 0\\C & D\end{pmatrix},\begin{pmatrix} A & B'\\0 & D'\end{pmatrix}  \right) &\acton \begin{pmatrix} A_1& B_1\\C_1& D_1\end{pmatrix}\\&:= \begin{pmatrix} A & 0\\C & D\end{pmatrix} \begin{pmatrix}  A_1& B_1\\C_1& D_1\end{pmatrix} -\begin{pmatrix}  A_1& B_1\\C_1& D_1\end{pmatrix}\begin{pmatrix} A & B'\\0 & D'\end{pmatrix}~.
\end{aligned}
\end{equation}
We can endow the Lie algebras $\frh$ and $\frg$ with Hilbert-Schmidt metrics, which we choose to be positive definite on $\frh$ and of split signature on $\frg$. Then we find
\begin{equation}
\begin{aligned}
&D\left( \begin{pmatrix} A_1 & B_1 \\C_1  & D_1\end{pmatrix} ,\begin{pmatrix} A_2 & B_2 \\C_2  & D_2\end{pmatrix}\right)\\
&=\left( \begin{pmatrix} A_1 A_2^\dagger+(B_1 B_2^\dagger+C_1 C_2^\dagger)/2 & 0 \\C_1 A_2^\dagger +B_1A_2^\dagger  & C_1C_2^\dagger+D_1D_2^\dagger\end{pmatrix} ,\right.\\&\hspace{4cm}\left.\begin{pmatrix} A_1 A_2^\dagger+(B_1 B_2^\dagger+C_1 C_2^\dagger)/2& C_2^\dagger D_1 +A_2^\dagger B_1  \\0 & B_2^\dagger B_1+D_2^\dagger D_1\end{pmatrix}\right)~,
\end{aligned}
\end{equation}
from which one can derive a corresponding 3-bracket as $[x,y,z]:=D(x,y)\acton z$, where $x,y,z\in\frh$.

\section{M5-brane models}

Let us now apply our observation that 3-Lie algebras are special cases of differential crossed modules. After briefly reviewing higher gauge theories, we rewrite a recently proposed set of supersymmetric equations of motion for the non-abelian (2,0) tensor multiplet in this language. We then consider the corresponding re-interpretation of the BLG model.

\subsection{Higher gauge theory with differential crossed modules}

In this letter, we will restrict ourselves to trivial principal 2-bundles over $\FR^n$, such that there is no distinction between local and global objects. Similar to trivial principal bundles, all \v{C}ech cocycles defining the bundle are trivial, and all non-trivial information is contained in the connection. Moreover, all potentials defining this connection are given in terms of Lie algebra valued differential forms.

Consider a (trivial) principal 2-bundle $\CE$ over $\FR^n$. Let the structure Lie 2-group of $\CE$ be given in terms of the Lie crossed module $(\sft:\CH\rightarrow \CG,\acton)$ with corresponding differential crossed module $(\sft:\frh\rightarrow \frg,\acton)$. A {\em connection} on $\CE$ is a pair $(A,B)$, where $A$ is a $\frg$-valued 1-form and $B$ is an $\frh$-valued 2-form, cf.\ e.g.\ \cite{Baez:2002jn}. We also introduce the corresponding {\em curvatures} as a pair $(F,H)$, where $F$ takes values in $\frg$ and $H$ takes values in $\frh$, according to
\begin{equation}
 \begin{aligned}
  F:=\dd A+\tfrac{1}{2}[A, A]\eand H:=D B:=\dd B+A \actwedge B~.
 \end{aligned}
\end{equation}
The wedge products of Lie algebra valued differential forms are defined in the obvious way: Consider $\frg$-valued forms $X_{1,2}=X^a_{1,2} \tau_a$, where $X^a_{1,2}\in \Omega^\bullet(\FR^n)$ and the $\tau_a$ are generators of $\frg$ and an $\frh$-valued form $Y=Y^a\rho_a$, where $Y^a\in \Omega^\bullet(\FR^n)$ and the $\rho_a$ are generators of $\frh$. Then
\begin{equation}
 X_1\wedge X_2:=(X^a_1\wedge X^b_2)\otimes [\tau_a,\tau_b]\eand X_1\actwedge Y:=(X^a_1\wedge Y^b)\otimes (\tau_a\acton \rho_b)~.
\end{equation}
We evidently have
\begin{equation}\label{eq:Bianchi}
 D F=0\eand D H=F\actwedge B~.
\end{equation}

It can be shown \cite{Baez:0511710}, see also \cite{Girelli:2003ev}, that a connection $(A,B)$ gives rise to well-defined parallel transport over surfaces if the so-called {\em fake curvature} vanishes:
\begin{equation}\label{eq:FakeCurvature}
 \CF:=F-\sft(B)=0~.
\end{equation}
Note that this, together with \eqref{eq:Bianchi}, implies
\begin{equation}
 \sft(H)=0\eand D H=0~.
\end{equation}

Finite gauge transformations are specified by a pair $(g,\Lambda)$ of a $\CG$-valued function $g$ and an $\frh$-valued 1-form $\Lambda$. They act according to
\begin{equation}
\begin{aligned}
 A&\rightarrow \tilde{A}:=g^{-1} A g+g^{-1} \dd g-\sft(\Lambda)~,\\
 B&\rightarrow \tilde{B}:=g^{-1}\acton B -\tilde{A}\actwedge \Lambda-\dd \Lambda-\tfrac{1}{2}[ \Lambda, \Lambda]~.
\end{aligned}
\end{equation}
This implies
\begin{equation}
\begin{aligned}
 F&\rightarrow \tilde{F}=g^{-1} F g+\sft(\dd \Lambda)-\sft(-\dd \Lambda+\tfrac{1}{2}[ \Lambda, \Lambda])-\tilde{A}\acton \sft(\Lambda)~,\\
 H&\rightarrow \tilde{H}=g^{-1}\acton H-(F-\sft(B))\actwedge \Lambda~,\\
 \CF&\rightarrow \tilde{\CF}=g^{-1}\CF g~.
\end{aligned}
\end{equation}
We will follow the nomenclature of e.g.\ \cite{Martins:2010ry} and refer to gauge transformations parameterized by $(g,0)$ as {\em thin} and those parameterized by $(0,\Lambda)$ as {\em fat}. In addition, we will call gauge transformations $(g,\Lambda)$ with $\sft(\Lambda)=0$ {\em ample}.

A few remarks are in order. First, as stated above, the non-abelian gerbes of Breen and Messing \cite{Breen:math0106083} are obtained when we use automorphism Lie 2-groups. Therefore, our discussion contains non-abelian gerbes, but it is more general. Second, if $\CH$ is abelian and $\acton$ and $\sft$ are trivial, we obtain the usual picture of abelian gerbes. Third, we can always use a fat gauge transformation to remove the part of $A$ that lies in the image of $\sft$. This poses the problem that, for finite dimensional differential crossed modules where the compliment of the image of $\sft$ is a Lie algebra, the fake curvature condition implies $F$ can be gauged away by a fat gauge transformation. This problem can be avoided by, for example, using a $\RZ_2$-graded Lie algebra. Finally, note that an M5-brane model has been recently
proposed \cite{Ho:2011ni} that uses the above language. In the following, however, we will discuss a different model built from 3-Lie algebras.

\subsection{Tensor multiplet equations of motion}\label{ssec:tensormultiplet}

In \cite{Lambert:2010wm}, a set of equations for the fields in the non-abelian tensor multiplet in six dimensions was proposed, which are invariant under $\CN=(2,0)$ supersymmetry. The field content of the tensor multiplet, i.e.\ the selfdual 3-form field strength $h_{\mu\nu\kappa}$, the scalars $X^I$ and superpartners $\Psi$, were all assumed to take values in a 3-Lie algebra $\CA$. It was found that for the closure of the supersymmetry algebra, it was necessary to introduce an additional gauge potential taking values in the associated Lie algebra $\frg_\CA$. Moreover, a covariantly constant, $\CA$-valued vector field $C^\mu$ had to be introduced. Altogether, the proposed equations of motion read as
\begin{equation}\label{eq:eomTensor}
\begin{aligned}
 D^2 X^I-\tfrac{\di}{2}[\bar{\Psi},\Gamma_\nu\Gamma^I\Psi,C^\nu]+[X^J,C^\nu,[X^J,C_\nu,X^I]]&=0~,\\
 \Gamma^\mu D_\mu\Psi-[X^I,C^\nu,\Gamma_\nu\Gamma^I\Psi]&=0~,\\
 D_{[\mu}h_{\nu\kappa\lambda]}+\tfrac{1}{4}\eps_{\mu\nu\kappa\lambda\sigma\tau}[X^I,D^\tau X^I,C^\sigma]+\tfrac{\di}{8}\eps_{\mu\nu\kappa\lambda\sigma\tau}[\bar{\Psi},\Gamma^\tau\Psi,C^\sigma]&=0~,\\
F_{\mu\nu}-D(C^\lambda,h_{\mu\nu\lambda})&=0~,\\
D_\mu C^\nu=D(C^\mu,C^\nu)&=0~,\\
D(C^\rho,D_\rho X^I)=D(C^\rho,D_\rho\Psi)=D(C^\rho,D_\rho h_{\mu\nu\lambda})&=0~,
\end{aligned}
\end{equation}
and the supersymmetry transformations leaving these equations invariant are given by
\begin{equation}\label{eq:SUSYTensor}
 \begin{aligned}
  \delta X^I&=\di \epsb \Gamma^I\Psi~,\\
  \delta \Psi&=\Gamma^\mu\Gamma^ID_\mu X^I\eps+\tfrac{1}{2\times 3!}\Gamma_{\mu\nu\lambda}h^{\mu\nu\lambda}\eps-\tfrac{1}{2}\Gamma^{IJ}\Gamma_\lambda[X^I,X^J,C^\lambda]\eps~,\\
  \delta h_{\mu\nu\lambda}&=3\di \epsb\Gamma_{[\mu\nu}D_{\lambda]}\Psi+\di\epsb\Gamma^I\Gamma_{\mu\nu\lambda\kappa}[X^I,\Psi,C^\kappa]~,\\
  \delta A_\mu&=\di\epsb\Gamma_{\mu\lambda} D(C^\lambda,\Psi)~,\\
  \delta C^\mu&=0~.
 \end{aligned}
\end{equation}
Here, $(\Gamma_\mu,\Gamma_I)$, $\mu=0,\ldots,5$, $I=1,\ldots,5$, form the generators of the Clifford algebra of $\FR^{1,10}$. 

Let us note in passing that, as far as a unification of M2- and M5-brane models is concerned, the right-hand side of one of the BLG equations of motion
\begin{equation}\label{eq:BLGFeom}
F_{\mu\nu}=\eps_{\mu\nu\lambda}(D(X^I,D^\lambda X^I)+\tfrac{\di}{2}D(\bar \Psi,\Gamma^\lambda\Psi))~.
\end{equation}
also appears in the third equation in \eqref{eq:eomTensor}. 

One of the major problems of this model is that it seems impossible to consistently introduce a potential 2-form field $B$ for $h$. In \cite{Papageorgakis:2011xg}, the equations \eqref{eq:eomTensor} found a natural interpretation on loop space: The constraints on $C^\mu$ imply a factorization, $C^\mu=c^\mu C$, where $C$ is a constant element of $\CA$, and the remaining covariantly constant vector $c^\mu$ can be identified with the tangent vector to the loop. This implies that the equation $F_{\mu\nu}-D(C^\lambda,h_{\mu\nu\lambda})=0$ is very similar to the transgression map \eqref{eq:transgression}.

Here, however, we want to reformulate equations \eqref{eq:eomTensor} in terms of a differential crossed module $(\sft:\frh\rightarrow\frg,\acton)$. That is, we replace $\CA$ and $\frg_\CA$ by $\frh$ and $\frg$, respectively. Instead of having an extra element $C\in\CA$, we substitute all expressions $D(y,C)$, $y\in\CA$, by $\sft(y)$. Correspondingly, all 3-brackets containing $C$, i.e.\ $[y_1,C,y_2]=D(y_1,C)\acton y_2$, $y_1,y_2\in \CA$, become $\sft(y_1)\acton y_2=[y_1,y_2]$. Note that in equations \eqref{eq:eomTensor} and \eqref{eq:SUSYTensor}, $C$ appears in every 3-bracket and in every expression containing the map $D$. We will therefore obtain equations containing only the Lie structures on $\frh$ and $\frg$.

We cannot work with differential crossed modules yielding 3-Lie algebras, because in these cases, the map $\sft$ is trivial. However, we find that the equations \eqref{eq:eomTensor} e.g.\ with 3-Lie algebra $\CA=A_4$ correspond to equations using the differential crossed module $(\mt:\au(2)\rightarrow\asu(2),\acton)$ defined in section \ref{sec:NontrivialExamples}.

While the equation $F_{\mu\nu}-D(C^\lambda,h_{\mu\nu\lambda})=0$ looks like a transgression in the loop space picture, in the context of differential crossed modules it is a candidate for the fake curvature constraint \eqref{eq:FakeCurvature}. Consequently, we are led to identify $B_{\mu\nu}=h_{\mu\nu\lambda}c^\lambda$. For simplicity, we will assume $|c|>0$. Given a $B_{\mu\nu}$ satisfying $B_{\mu\nu}c^\nu=0$, we can then write
\begin{equation}\label{eq:Defh}
 h_{\mu\nu\kappa}=\frac{1}{|c|^2}\left(B_{[\mu\nu}c_{\kappa]}+\tfrac{1}{3!}\eps_{\mu\nu\kappa\lambda\rho\sigma}B^{[\lambda\rho}c^{\sigma]}\right)~,
\end{equation}
where $[\cdots]$ denotes antisymmetrization of $n$ indices with weight $1/n!$ . Note that locally and before taking gauge invariance into account, a selfdual 3-form in six dimensions has just as many components as a 2-form satisfying $B_{\mu\nu}c^\nu=0$. Such a 2-form has non-trivial components only in the five dimensional space perpendicular to $c$.

Let us now rewrite \eqref{eq:eomTensor} in the language of differential crossed modules:
\begin{equation}\label{eq:NewEomTensor}
\begin{aligned}
 D^2 X^I-\tfrac{\di}{2}[\bar{\Psi},\Gamma\Gamma^I\Psi]+|c|^2[X^J,[X^J,X^I]]&=0~,\\
 \Gamma^\mu D_\mu\Psi+[X^I,\Gamma\Gamma^I\Psi]&=0~,\\
 D_{[\mu}h_{\nu\kappa\lambda]}+\tfrac{1}{4}\eps_{\mu\nu\kappa\lambda\sigma\tau}c^\sigma \left([X^I,D^\tau X^I]+\tfrac{\di}{2}[\bar{\Psi},\Gamma^\tau\Psi]\right)&=0~,\\
H_{\mu\nu\kappa}-\tfrac{1}{3!}\eps_{\mu\nu\kappa\rho\sigma\tau}H^{\rho\sigma\tau}&=0~,\\
F_{\mu\nu}-\sft(B_{\mu\nu})&=0~,\\
\dpar_\mu c^\nu=\sft(D_c X^I)=\sft(D_c \Psi)=\sft(D_c B_{\mu\nu})&=0~,
\end{aligned}
\end{equation}
where $\Gamma:=c^\nu\Gamma_\nu$, $D_c:=c^\nu D_\nu$ and $h$ is given in \eqref{eq:Defh}. Note that the commutators of spinors are to be read as commutators of the gauge structure only.

From the third equation in \eqref{eq:NewEomTensor}, we find
\begin{equation}
 c^\lambda (D_{[\mu}h_{\nu\kappa\lambda]})=0~.
\end{equation}
Using this, we compute
\begin{equation}
\begin{aligned}
 H&:=D B=c^\lambda D_\lambda h_{\mu\nu\kappa}\dd x^\mu\wedge \dd x^\nu\wedge \dd x^\kappa~,\\
 *H&=\tfrac{1}{3!}\eps_{\mu\nu\kappa\rho\sigma\tau}c_\lambda D^\lambda h^{\rho \sigma\tau}\dd x^\mu\wedge \dd x^\nu\wedge \dd x^\kappa~,
\end{aligned}
\end{equation}
from which (together with the selfduality of $h$) we conclude that
\begin{equation}
 H=*H\eand \sft(H)=0~~~\Rightarrow~~~\sft(D_c B_{\mu\nu})=0~.
\end{equation}
Thus, our definition of $B$ yields indeed a selfdual curvature 3-form. Moreover, it also answers the question why there is no potential for $h$: The field $h$ encodes the potential. And finally, note that the degrees of freedom in the gauge potential are completely determined by the 2-form potential $B$ via the fake curvature condition $F-\sft(B)=0$. Therefore, there are no additional degrees of freedom in the supermultiplet.

As we merely rewrote the equations of motion, it is clear that for certain differential crossed modules $(\sft:\frh\rightarrow\frg,\acton)$, equations \eqref{eq:NewEomTensor} are invariant under the maximal $\CN=(2,0)$ supersymmetry transformations
\begin{equation}\label{eq:newSUSY}
 \begin{aligned}
  \delta X^I&=\di \epsb \Gamma^I\Psi~,\\
  \delta \Psi&=\Gamma^\mu\Gamma^ID_\mu X^I\eps+\tfrac{1}{2\times 3!}\Gamma_{\mu\nu\lambda}h^{\mu\nu\lambda}\eps-\tfrac{1}{2}\Gamma^{IJ}\Gamma [X^I,X^J]\eps~,\\
  \delta B_{\mu\nu}&=3\di \epsb\Gamma_{[\mu\nu}c^\lambda D_{\lambda]}\Psi~,\\
  \delta A_\mu&=\di\epsb\Gamma_{\mu\lambda}c^\lambda \sft(\Psi)~,\\
  \delta c^\mu&=0~.
 \end{aligned}
\end{equation}
Recall that equations \eqref{eq:eomTensor} are maximally supersymmetric if the contained 3-brackets are totally antisymmetric and satisfy the fundamental identity \cite{Lambert:2010wm}. The consequences of these properties in equations \eqref{eq:eomTensor} are preserved under the rewriting $D(y,C)\rightarrow \sft(y)$, as is readily verfied. One would therefore expect that equations \eqref{eq:NewEomTensor} are invariant under the supersymmetry transformations \eqref{eq:newSUSY} for any differential crossed module $(\sft:\frh\rightarrow\frg,\acton)$. An explicit computation along the lines of \cite{Lambert:2010wm} confirms this expectation. One interesting result of the calculation is that the fake curvature condition is not necessary for the closure of the supersymmetry algebra, see appendix \ref{app:m5susy} for comments on the calculation. 

We have therefore significantly extended the previously known examples of $\CN=(2,0)$ tensor multiplet equations.

\subsection{Comments on the tensor multiplet equations}

First of all, it is not clear to us how to make the above equations invariant under general fat gauge transformations. The equations \eqref{eq:NewEomTensor} are only invariant under thin gauge transformations $(g,0)$ with
\begin{equation}
 X^I\rightarrow \tilde{X}^I:=g\acton X^I\eand \Psi\rightarrow \tilde{\Psi}:=g\acton \Psi~.
\end{equation}
We thus recover the gauge symmetry already suggested in \cite{Lambert:2010wm}.

Second, it is nice that for $\sft$ trivial, i.e.\ the case of an abelian gerbe, $\frh$ must be abelian and the field strength $F$ necessarily vanishes. We can therefore gauge away the gauge potential and obtain the known free theory:
\begin{equation}
 \dpar^2X^I=\Gamma^\mu\dpar_\mu \Psi= H-(*H)=0~.
\end{equation}

Third, we can follow \cite{Lambert:2010wm} and reduce equations \eqref{eq:NewEomTensor} to five-dimensional maximally supersymmetric Yang-Mills (mSYM) theory. For this, we dimensionally reduce along $x^5$ by imposing $\der{x^5}=0$ and fixing $c^\mu=\delta^{\mu5}g^2_{\rm YM}$. Due to $B_{\mu\nu}=h_{\mu\nu\kappa}c^\kappa$, we conclude that $B_{\mu5}=0$. This implies that $F_{\mu5}=0$ and we can therefore partially gauge fix $A_5=0$. The relation $B_{\mu 5}=0$ together with $\der{x^5}=0$ and the selfduality of $H$ also yields $H=0$. We are therefore left with the field content of mSYM theory in five dimensions. If we use the differential crossed module $(\sft:\au(N)\rightarrow\au(N),\acton)$, equations \eqref{eq:NewEomTensor} reduce to the mSYM equations with gauge algebra $\au(N)$.

As a final test, let us briefly derive the BPS equation corresponding to a (non-abelian) selfdual string. That is, we dimensionally reduce the above equations along the $x^0$- and $x^5$-directions and put $\Phi:=X^6\neq 0=X^7,\ldots,X^{10}$ as well as $H_{0ij} = H_{5ij} = 0$. Then the supersymmetry transformation of the spinors reduces to
\begin{equation}\label{eq:BPS}
 \Gamma^i\Gamma^6D_i \Phi\eps+\tfrac{1}{2\times 3!}\Gamma_{ijk}h^{ijk}\eps=0~,~~~i,j,k=1,\ldots,4~.
\end{equation}
To break half of the supersymmetry, as expected for the BPS equation, we impose $\Gamma^{05}\eps=\Gamma^6\eps$ and arrive at
\begin{equation}\label{eq:sdseqn}
 h_{ijk}=\eps_{ijk\ell}D^\ell \Phi\eor B_{ij}=\eps_{ijk\ell}c^kD^\ell\Phi~.
\end{equation}
The fact that this equation is close but not identical to the desired $H=*D \Phi$ indicates that the equations \eqref{eq:NewEomTensor} need further generalization. Note that after applying $\sft$ to both sides of equation \eqref{eq:sdseqn} and using the fake curvature constraint \eqref{eq:FakeCurvature}, we obtain
\begin{equation}
 F_{ij}=\eps_{ijk\ell}c^kD^\ell\sft(\Phi)~.
\end{equation}
This should be interpreted as the Bogomolny monopole equation obtained by dimensionally reducing a selfdual string along the direction $c^k$.

Altogether, we can conclude that the 3-Lie algebra tensor multiplet equations proposed in \cite{Lambert:2010wm} can be naturally reformulated in the language of differential crossed modules while preserving $\CN=(2,0)$ supersymmetry. However, the BPS equation and issues with fat gauge transformations suggest that the obtained equations \eqref{eq:NewEomTensor} are not the final answer.

\section{M2-brane models}

The BLG and ABJM M2-brane models can be trivially rewritten in terms of differential crossed modules with $\sft$ trivial. The fake curvature condition \eqref{eq:FakeCurvature} however would imply $F=\sft(B)=0$. This problem can be circumvented by introducing the machinery of differential 2-crossed modules, which appear as the structure algebras of principal 3-bundles. This is effectively using the observation that a higher gauge theory on a principal 2-bundle with non-vanishing $\CF$ can be reformulated as a higher gauge theory on a principal 3-bundle for which the fake curvature does vanish \cite{Schreiber:N01,Baez:2010ya}. Additional motivation for the use of principal 3-bundles comes from the fact that M2-branes couple to a 3-form potential, which suggests an underlying picture involving principal 3-bundles.

The machinery needed can be constructed from considering the inner derivations of a differential crossed module, as we will see in the following.

\subsection{Inner derivation 2-crossed modules}\label{ssec:2crossmod}

Just as a Lie algebra comes with a differential crossed module governing the action of inner derivations, a differential crossed module (or strict Lie 2-algebra) comes with a differential 2-crossed module of inner derivations as implied e.g.\ by the results of \cite{Roberts:0708.1741}. In higher category theoretical terms, differential 2-crossed modules are certain Lie 3-algebras, which must not be confused with 3-Lie algebras. 

Note that more generally, the Chevalley-Eilenberg algebra of the inner derivations of an $L_\infty$-algebra $\frg_\infty$ is known as the {\em Weil algebra} of $\frg_\infty$.

The definition of a differential 2-crossed module \cite{Conduche:1984:155} is a triple of Lie algebras $\frl,\frh,\frg$ arranged in a normal complex
\begin{equation}
 \frl\ \xrightarrow{~\sft~}\ \frh\ \xrightarrow{~\sft~}\ \frg~.
\end{equation}
There are $\frg$-actions $\acton$ onto $\frh$ and $\frl$ by derivations. The Peiffer identity $\sft(h_1)\acton h_2=[h_1,h_2]$ is now lifted by a $\frg$-equivariant bilinear map, called {\em Peiffer lifting} and denoted by $\{-,-\}: \frh\times \frh\rightarrow \frl$. These maps satisfy the following axioms for all $g\in\frg$, $h,h_1,h_2,h_3\in\frh$ and $\ell,\ell_1,\ell_2\in\frl$:
\begin{itemize}
 \setlength{\itemsep}{-1mm}
 \item[(i)] $\sft(g\acton \ell)=g\acton\sft(\ell)$ and $\sft(g\acton h)=[g,\sft(h)]$.
 \item[(ii)] $\sft(\{h_1,h_2\})=[h_1,h_2]-\sft(h_1)\acton h_2$.
 \item[(iii)] $\{\sft(\ell_1),\sft(\ell_2)\}=[\ell_1,\ell_2]$.
 \item[(iv)] $\{[h_1,h_2],h_3\}=\sft(h_1)\acton\{h_2,h_3\}+\{h_1,[h_2,h_3]\}-\sft(h_2)\acton\{h_1,h_3\}-\{h_2,[h_1,h_3]\}$.
 \item[(v)] $\{h_1,[h_2,h_3]\}=\{\sft(\{h_1,h_2\}),h_3\}-\{\sft(\{h_1,h_3\}),h_2\}$.
 \item[(vi)] $\{\sft(\ell),h\}+\{h,\sft(\ell)\}=-\sft(h)\acton \ell$.
\end{itemize}

Given a differential crossed module $\frh\xrightarrow{~\tilde{\sft}~}\frg$ with action $\tilde{\acton}:\frg\times \frh\rightarrow \frh$, the corresponding differential 2-crossed module of inner derivations, denoted $ \frder\big(\frh\xrightarrow{~\tilde{\sft}~}\frg\big)$, has the underlying normal complex \cite{Roberts:0708.1741}
\begin{equation}
\frh\ \xrightarrow{~\sft~}\ \frg \ltimes \frh\ \xrightarrow{~\sft~}\ \frg~.
\end{equation}
Recall that the Lie bracket on $\frg \ltimes \frh$ reads as 
\begin{equation}
[(g_1,h_1),(g_2,h_2)]:= ([g_1,g_2],[h_1,h_2]+g_1\tilde{\acton}h_2-g_2\tilde{\acton}h_1)~.
\end{equation}
The maps $\sft$ are defined as
\begin{equation}
 \sft(h):=(\tilde{\sft}(h),-h)\eand \sft(g,h):=\tilde{\sft}(h)+g~,
\end{equation}
the $\frg$-actions and the Lie bracket on $\frh$ are given by
\begin{equation}
g\acton h := g~\tilde{\acton}~h\eand g_1 \acton (g_2,h):=([g_1,g_2],g_1~\tilde{\acton}~h)
\end{equation}
and the Peiffer lifting reads as
\begin{equation}
 \{(g_1,h_1),(g_2,h_2)\}:=g_2\tilde{\acton} h_1
\end{equation}
for all $g,g_1,g_2\in\frg$, $h,h_1,h_2\in \frh$. One readily checks that this structure satisfies the axioms of a differential 2-crossed module.

\subsection{Inner derivations of $\frm_{\rm ABJM}(N)$}

The inner derivations of $\frm_{\rm ABJM}(N)$ are captured by a differential 2-crossed module that is constructed from $\frm_{\rm ABJM}(N)$ as described in the previous section. To simplify the discussion, let us use the following picture: We consider a chain complex of block matrices
\begin{equation}\label{eq:rep_d2cm}
\frh=\left(\begin{array}{cc} 0 & \agl(N,\FC)\\ 0 & 0\end{array}\right)\xrightarrow{~\sft~}\frg\ltimes \frh=\left(\begin{array}{cc} \au(N) & \agl(N,\FC)\\ 0 & \au(N)\end{array}\right)\xrightarrow{~\sft~}\frg=\left(\begin{array}{cc} \au(N) & 0\\ 0 & \au(N)\end{array}\right),
\end{equation}
where the two maps $\sft:\frh \rightarrow \frg\ltimes \frh$ and $\sft:\frg\ltimes \frh\rightarrow\frg$ read as
\begin{equation}
\sft:\left(\begin{array}{cc} 0 & h\\ 0 & 0\end{array}\right)\mapsto \left(\begin{array}{cc} 0 & -h\\ 0 & 0\end{array}\right) \eand 
\sft:\left(\begin{array}{cc} g_L & h\\ 0 & g_R\end{array}\right)\mapsto \left(\begin{array}{cc} g_L & 0\\ 0 & g_R\end{array}\right)
\end{equation}
respectively, for $g_{L,R}\in\au(N)$ and $h\in\agl(N,\FC)$. All $\frg$-actions as well as the Lie algebra commutators are given by the corresponding matrix commutators. The Peiffer lifting is defined as
\begin{equation}
\left\{\left(\begin{array}{cc} g_{L1} & h_1\\ 0 & g_{R1}\end{array}\right),\left(\begin{array}{cc} g_{L2} & h_2\\ 0 & g_{R2}\end{array}\right)\right\}:=\left(\begin{array}{cc} 0 & g_{L2} h_1-h_1 g_{R2}\\ 0 & 0\end{array}\right)~,
\end{equation}
where $g_{L1,2},g_{R1,2}\in \au(N)$ and $h_{1,2}\in\agl(N,\FC)$. As a consistency check, one can easily verify that this Peiffer lifting indeed captures the failure of the Peiffer identity according to
\begin{equation}
 \sft(\{(g_1,h_1),(g_2,h_2)\})=[(g_1,h_1),(g_2,h_2)]-\sft(g_1,h_1)\acton (g_2,h_2)~.
\end{equation}
We will denote this differential 2-crossed module by $\frder(\frm_{\rm ABJM}(N))$.

\subsection{Higher gauge theory with differential 2-crossed modules}

We will need the basics of the local description of higher gauge theory by a connective structure on a trivial principal 3-bundle over $M=\FR^{1,2}$. The detailed picture for gauge theory on principal 3-bundles was developed in \cite{Saemann:2013pca}, see \cite{Martins:2009aa} for a partial earlier account. Let us work for the moment with a general differential 2-crossed module $\frl\stackrel{\sft}{\rightarrow}\frh\stackrel{\sft}{\rightarrow}\frg$, we will restrict ourselves to the case $\frder(\frm_{\rm ABJM}(N))$ in the next section.

Consider 1-, 2- and 3-form potentials $A\in \Omega^1(M,\frg)$, $B\in \Omega^2(M,\frh)$ and $C\in \Omega^3(M,\frl)$. From these, we construct the corresponding field strengths
\begin{equation}
  F:=\dd A+\tfrac{1}{2}[A,A]~,~~~H:=\dd B+A\acton B~,~~~G:=\dd C+A\acton C+ \{B,B\}~.
\end{equation}

The gauge transformations of the gauge potentials are given by \cite{Saemann:2013pca}
\begin{equation}\label{eq:gauge_transformations}
 \begin{aligned}
\tilde C&=g^{-1}\acton C-\tilde D^0\big(\Sigma-\tfrac12\{\Lambda,\Lambda\}\big)+\{\tilde B,\Lambda\}+\{\Lambda,\tilde B\}-\{\Lambda,\tilde{D}\Lambda+\tfrac12[\Lambda,\Lambda]\}~,\\
 \tilde B&=g^{-1}\acton B-\tilde{D}^0\Lambda-\tfrac12\sft(\Lambda)\acton \Lambda-\sft(\Sigma)~,\\
 \tilde A&=g^{-1}A g+g^{-1}\dd g-\sft(\Lambda)~,
 \end{aligned}
\end{equation}
where $g$ is a function on $M$ taking values in a Lie group $\mathsf{G}$ with $\frg=\mathsf{Lie}(\mathsf{G})$, $\Lambda\in \Omega^1(M,\frh)$ and $\Sigma\in \Omega^2(M,\frl)$. Moreover, we used abbreviations $\tilde D\ :=\ \dd+\tilde A\acton$ and $\tilde D^0\ :=\ \dd+\big(\tilde A+\sft(\Lambda)\big)\acton$.

For the higher gauge theory to describe a parallel transport of membranes along three-dimensional volumes that is invariant under reparameterizations of the volume, the so-called {\em fake curvatures} have to vanish:
\begin{equation}\label{eq:fake_curvature_conditions}
 \CF:=F-\sft(B)=0\eand \CH:=H-\sft(C)=0~.
\end{equation}
This implies $\sft(H)=0$.

\subsection{Higher gauge theory formulation of the ABJM model}

The ABJM model describes a stack of $N$ flat M2-branes with a $\FC^4/\RZ_k$ orbifold in the transverse directions. These eight transverse directions of the M2-branes are thus packaged into four complex fields $Z^A$, $A=1,\dots,4$, which have spinors $\psi^A$ as their superpartners. These matter fields take values in $\frh:=\agl(N,\FC)$. The gauge fields $A^\mu$, $\mu=0,1,2$, live in $\frg:=\au(N)\times\au(N)$. We use the representation \eqref{eq:rep_d2cm} of the differential 2-crossed module $\frder(\frm_{\rm ABJM}(N))$, where the action of the gauge potentials on matter fields corresponds to the matrix commutator. Besides this, there is also the ordinary matrix product between matter fields and their adjoints, which we will need for the potential terms in the ABJM model.

The ABJM action can then be written in the following way:
\begin{equation}
\CL_{\rm ABJM} = \int_{\FR^{1,2}} \tr\left(\tfrac{k}{4\pi
}\eta~A\wedge(\dd A
+\tfrac{1}{3}[A,A])-D Z_A^{\dag}\wedge \star D Z^A-\star\di\bar\psi^{A}\wedge \slasha{D} \psi_A\right)+V~,
\end{equation}
where $D=\dd+A\acton$ and $\eta=-\sigma_3\otimes \unit_N$ yields a metric of split signature on the gauge algebra $\au(N)\times \au(N)$. By $\tr(-)$, we mean the trace in the matrix representation \eqref{eq:rep_d2cm}. The potential is given by 
\begin{equation}
\begin{aligned}
V=& \int_{\FR^{1,2}}~\star{\rm tr}\Big(-\di\bar\psi^{A\dag} \psi_{A} Z^\dag_B
Z^B-\di\bar\psi^{A\dag} Z^B Z^\dag_B\psi_{A}+2\di\bar\psi^{A\dag}\psi_{B} Z_A^\dag Z^B-2\di\bar\psi^{A\dag} Z^B Z_A^\dag\psi_{B}\\
&\hspace{2.5cm}+\di\varepsilon_{ABCD}\bar\psi^{A\dag}
Z^C\psi^{B\dag} Z^D   -\di\varepsilon^{ABCD}Z_D^\dag\bar \psi_A Z_C^\dag\psi_B-\tfrac{2}{3}\Upsilon^{CD}_B\Upsilon^{\dagger B}_{CD}\Big)~,\\
\Upsilon^{CD}_B &:=
  Z^CZ^\dagger_B Z^D-\frac{1}{2}\delta^C_BZ^EZ^\dagger_E Z^D+\frac{1}{2}\delta^D_BZ^EZ^\dagger_E Z^C~.
\end{aligned}
\end{equation}
This theory exhibits $\CN=6$ supersymmetry and it has passed some highly non-trivial tests as an effective description of M2-branes. 

Next, we extend this action to implement the fake curvature conditions \eqref{eq:fake_curvature_conditions}, introducing 2- and 3-form potential $B\in \Omega^2(\FR^{1,2},\frg\ltimes \frh)$ and $C\in \Omega^2(\FR^{1,2},\frh)$. In the matrix representation \eqref{eq:rep_d2cm} of $\frder(\frm_{\rm ABJM}(N))$, the fake curvature conditions amount to 
\begin{equation}\label{eq:fake_curvature_conds_matrix}
\begin{aligned}
B=\left(\begin{array}{cc} F_L & b\\ 0 & F_R\end{array}\right)~,~~H= \left(\begin{array}{cc} 0 & \dd b +A_L b-bA_R\\ 0 & 0\end{array}\right)=\sft(C)=\left(\begin{array}{cc} 0 & -c\\ 0 & 0\end{array}\right)
\end{aligned}
\end{equation}
for some $b,c\in\agl(N,\FC)$, where $A_L$ and $A_R$ are the first and second block diagonal entries of $A$ and $F_{L,R}=\dd A_{L,R}+\tfrac{1}{2}[A_{L,R},A_{L,R}]$. Note that because of $\sft(H)=0$, $H$ has no block diagonal entries. 

To enforce \eqref{eq:fake_curvature_conds_matrix}, we introduce Lagrange multipliers $\lambda_{1}\in \Omega^1(\FR^{1,2},\frg)$, $\lambda_{2}\in \Omega^0(\FR^{1,2},\frg\ltimes \frh)$ and $\lambda_{3}\in \Omega^3(\FR^{1,2},\frg)$, adding the following terms to the action\footnote{As it stands, this action is not real. However, one can either impose reality conditions on $H$ and $\lambda_2$ or add complex conjugate terms to correct for this in a straightforward manner. Again we suppress these technical details.}:
\begin{equation}
 S_{\rm HGT}=S_{\rm ABJM}+ \int_{\FR^{1,2}}\tr\left(\lambda^\dagger_1\wedge(F-\sft(B))+\lambda_2^\dagger(H-\sft(C))+\lambda_3^\dagger\sft(\lambda_2)\right)~.
\end{equation}

Varying with respect to $\lambda_1$ and $\lambda_2$, we obtain
\begin{equation}\label{eq:variation_l1_l2}
 F-\sft(B)=0~,~~~H-\sft(C)+\sft^*(\lambda_3)=0~,
\end{equation}
where $\sft^*$ is the adjoint to $\sft$. This map is the trivial embedding of $\frg$ into $\frg\ltimes \frh$. Because $H-\sft(C)$ is a block off-diagonal in $\frg\ltimes \frh$, \eqref{eq:variation_l1_l2} reduces to
\begin{equation}
 F-\sft(B)=0~,~~~H-\sft(C)=0~,~~~\lambda_3=0~.
\end{equation}
Varying $S_{\rm HGT}$ with respect to $\lambda_3$ and $C$, we have
\begin{equation}
 \sft(\lambda_2)=\sft^*(\lambda_2)=0~~~\Leftrightarrow~~~\lambda_2=0~,
\end{equation}
where $\sft^*$ is here the obvious projection of $\frg\ltimes \frh$ onto $\frh$. Finally, varying the action with respect to $B$ yields
\begin{equation}
 \sft^*(\lambda_1)+D\lambda_2=0~,
\end{equation}
which implies $\lambda_1=0$ due to $\lambda_2=0$.

Varying $S_{\rm HGT}$ with respect to the gauge potential, we obtain the usual equation of motion of the ABJM model plus terms containing the Lagrange multipliers $\lambda_1$ and $\lambda_2$. Since both vanish on-shell, we recover
\begin{equation}\label{eq:eom_F}
F_{\mu\nu}=\eps_{\mu\nu\kappa}\left(D^\kappa Z^A Z^\dagger_A-Z^AD^\kappa  Z^{\dagger}_{A}+Z^\dagger_AD^\kappa Z^A -D^\kappa Z^\dagger_A Z^A  -\di \bar\psi^A \gamma^\kappa\psi^\dagger_A-\di \bar\psi^{\dagger A} \gamma^\kappa\psi_A\right)~.
\end{equation}
The equations of motion for the matter fields remain obviously those of the ABJM model. Note that the four-form curvature $G$ trivially vanishes, as our trivial principal 3-bundle lives over $\FR^{1,2}$.

Altogether, the action $S_{\rm HGT}$ yields the equations of motion of the ABJM model, together with the fake curvature conditions \eqref{eq:fake_curvature_conds_matrix}. We therefore reformulated the ABJM model as a higher gauge theory. 

Supersymmetry and gauge symmetry of the ABJM model are trivially preserved, if we demand that $\lambda_{1,2,3}$ transform appropriately. Explicitly, we can demand that the fields $B$ and $C$ transform in the same way as $\sft^*(F)$ and $\sft^*(H)$, which renders the fake curvature conditions invariant under supersymmetry. The Lagrange multipliers can then be chosen to be invariant under supersymmetry, too. 

Gauge transformations should act on the Lagrange multipliers as
\begin{equation}
      \lambda_1\rightarrow \tilde{\lambda}_1=\gamma\lambda_1\gamma^{-1}+\gamma[\lambda_2,\Lambda^\dagger]\gamma^{-1}~,~~~\lambda_{2,3}\rightarrow \tilde{\lambda}_{2,3}=\gamma\lambda_{2,3}\gamma^{-1}~,
\end{equation}
where $\gamma\in \Omega^0(M,\mathsf{G})$ and $\Lambda\in \Omega^1(M,\frg\ltimes\frh)$ are the gauge parameters. The second term in the $\lambda_1$ transformation renders the action gauge invariant off-shell. The 2- and 3-form potentials $B$ and $C$ transform as specified in \eqref{eq:gauge_transformations}. 

Note however, that the ABJM model is {\em not} invariant under the general tensor transformations parametrized by $\Lambda$ in \eqref{eq:gauge_transformations}. In particular, the equation of motion for the 2-form curvature \eqref{eq:eom_F} breaks this symmetry. We are therefore left with the ample gauge transformations, which are parametrized by a $\Lambda$ with $\sft(\Lambda)=0$. This solves a common problem when working with higher gauge theories: In many cases, e.g.\ if $\sft:\frh\rightarrow \frg$ is surjective, the potential 1-form $A$ can be gauged away by a tensor transformation, leaving an abelian theory. This is not possible if these transformations are broken down to the ample ones. 

The same observation was made in \cite{Baez:2012bn}, where teleparallel gravity was reformulated as a higher gauge theory. Here, all field configurations can be gauge transformed away by tensor transformations. However, the action of the theory is not invariant under these symmetries, leaving only the usual group-valued gauge transformations.

The $\Sigma$-transformations in \eqref{eq:gauge_transformations}, affect only the new terms added to $S_{\rm ABJM}$, which contain the Lagrange multipliers. All these terms are invariant under these transformations.

\subsection{ABJ-model}

The ABJ model \cite{Aharony:2008gk} is a Chern-Simons matter theory closely related to the ABJM model and also invariant under $\CN=6$ supersymmetry. We follow precisely the same formulation as above, merely replacing $\frm_{\rm ABJM}(N)$ by $\frm_{\rm ABJ}(N_1,N_2)$, which is the differential crossed module $\sHom(\FC^{N_2},\FC^{N_1})\stackrel{\sft}{\rightarrow}\au(N_1)\times \au(N_2)$. We then obtain a differential 2-crossed module of inner derivations, which we can represent in terms of matrices as
 \begin{equation}
\begin{aligned}
\left(\begin{array}{cc} 0 & \sHom(\FC^{N_2},\FC^{N_1})\\ 0 & 0\end{array}\right)\rightarrow \left(\begin{array}{cc} \au(N_1) & \sHom(\FC^{N_2},\FC^{N_1})\\ 0 & \au(N_2)\end{array}\right)\rightarrow \left(\begin{array}{cc} \au(N_1) & 0\\ 0 & \au(N_2)\end{array}\right)~.
\end{aligned}
\end{equation}

It does not seem possible to use more general types of differential crossed modules to obtain $\CN=6$ Chern-Simons matter theories. The hermitian 3-Lie algebras underlying such models seem to be very rigid. Note in particular that, as shown in \cite{Cherkis:2008ha}, the only hermitian 3-Lie brackets that can be written as products of matrices and their adjoints are of the form of the ABJM 3-bracket \eqref{eq:ABJM3bracket}.

\subsection{BLG-model}

The BLG Lagrangian reads, in terms of 3-brackets, as
\begin{equation}\label{eq:BLGLagrangian}
\begin{aligned}
\CL_{\rm BLG}=&\tfrac{1}{2}\lbr A, \dd A+\tfrac{1}{3} A \wedge A \rbr-\tfrac{1}{2}(D_\mu X^I,D^\mu X^I)+\tfrac{\di}{2}(\bar\Psi,\Gamma^\mu D_\mu\Psi)\\&-\tfrac{\di}{4}(\bar \Psi,\Gamma_{IJ}[X^I,X^J,\Psi])-\tfrac{1}{6}([X^I,X^J,X^K],[X^I,X^J,X^K])~.
\end{aligned}
\end{equation}

We can restrict the ABJM model to the Bagger-Lambert-Gustavsson model by restricting to $\frm_{\rm ABJM}(2)$, splitting the four complex matter fields into eight real ones, reducing $(\agl(2,\FC),\au(2)\times \au(2))$ to $(\asu(2)\oplus \di\,\au(1),\asu(2)\times \asu(2))$. This turns the hermitian 3-Lie algebra into the (real) 3-Lie algebra $A_4$, which is a real four dimensional vector space with totally antisymmetric 3-bracket
\begin{equation}\label{eq:A4}
[e_\mu,e_\nu,e_\rho]=\eps_{\mu\nu\rho\sigma}e_\sigma~,
\end{equation}
on the basis elements $e_\mu\in A_4$. The Lie algebra of inner derivations is represented by the matrices
\begin{equation}
\begin{aligned}
\left(\begin{array}{cc} 0 & \asu(2)\oplus \di\,\au(1)\\ 0 & 0\end{array}\right)\rightarrow \left(\begin{array}{cc} \asu(2) & \asu(2)\oplus \di\,\au(1)\\ 0 & \asu(2)\end{array}\right)\rightarrow \left(\begin{array}{cc} \asu(2) & 0\\ 0 & \asu(2)\end{array}\right)~.
\end{aligned}
\end{equation}
The resulting action $S_{\rm HGT}$ will have enhanced $\CN=8$ supersymmetry. In the hopes that it will be helpful to someone in the future, the supersymmetry closure calculation is presented in detail in appendix \ref{app:blgsusy}, as well as some useful identities for M5-brane supersymmetry calculations in appendix \ref{app:m5susy}.

\chapter{(1,0) superconformal models}
\label{ch:design}

We will now look at a different avenue of approach for the problem of finding the non-abelian six-dimensional (2,0) superconformal field theory, or (2,0)-theory for short. In this chapter, we will relate the gauge structure appearing in an approach based on tensor hierarchies in supergravity \cite{Samtleben:2011fj} to various algebraic structures appearing in the context of categorification, such as Courant algebroids, Courant-Dorfman algebras, differential crossed modules, differential 2-crossed modules, strong homotopy Lie algebras and string Lie 2-algebras.

The six-dimensional model of \cite{Samtleben:2011fj} exhibits $\CN=(1,0)$ superconformal invariance, and its field content comprises, besides the usual gauge potential one-form $A$, also gauge potential 2- and 3-forms $B$ and $C$, all taking values in a priori different vector spaces. A non-abelian action of $A$ onto $B$ and $C$ is defined, together with various other algebraic structures on the three vector spaces. The analysis of \cite{Samtleben:2011fj} led to a list of constraints on these algebraic structures necessary for closure of the (1,0) supersymmetry algebra and, in some cases, for an action to be formulated, see also \cite{Samtleben:2012mi,Samtleben:2012fb,Bandos:2013jva}. These constraints can be regarded as generalizations of the familiar Jacobi identity of Lie algebras. A special case of these theories contains the $\sG\times \sG$-model proposed in \cite{Chu:2011fd}, to which an action and interesting solutions have been constructed in \cite{Chu:2012um,Chu:2013hja,Chu:2013joa}. For solutions, such as solitons, in the general (1,0) model, see \cite{Akyol:2012cq}.

We start our analysis of the (1,0) gauge structure by noting that it forms a differential graded Leibniz algebra. Restricting the (1,0) gauge structure to an interesting class of examples, we find exact agreement of the resulting structure with Courant-Dorfman algebras \cite{Roytenberg:0902.4862}. Moreover, a general (1,0) gauge structure is a weak Courant-Dorfman algebra in the sense of \cite{Ekstrand:2009qz}. We investigate the possibility that these arise from Voronov's derived bracket construction \cite{Voronov:math0304038}, unfortunately this does not seem to be the case.

Weak Courant-Dorfman algebras, and in particular (1,0) gauge structures have a large overlap with strong homotopy Lie algebras or semistrict Lie $n$-algebras that replace gauge algebras in the context of higher gauge theory. We find that (1,0) gauge structures corresponding to Courant-Dorfman algebras form Lie 2-algebras, while many another interesting classes form Lie 3-algebras or can be extended to Lie 4-algebras. This establishes, at least in part, the desired relation to higher gauge theory. 

To strengthen the link between the (1,0) model and higher gauge theory further, we continue by studying a number of examples. The connective structure of an abelian gerbe, which underlies abelian higher gauge theory, is easily identified as a special case of the gauge potentials of the (1,0) model. Similarly, we discover the gauge algebraic structures as well as the field content and the gauge transformations of special classes of principal 2- and principal 3-bundles in the (1,0) model, establishing an overlap of the (1,0) model with strict higher gauge theory. We thus have to conclude that (1,0) models do not allow for general differential crossed and 2-crossed modules as higher gauge algebras.

We briefly comment on a number of further examples. First, we show how to recover both the gauge algebra as well as the action of gauge transformations of the $\sG\times \sG$-model proposed in \cite{Chu:2011fd} from the (1,0) model. Then we show that two canonical examples in higher gauge theory, the string Lie algebra of a simple Lie algebra and the Chern-Simons Lie 3-algebra of $\au(1)$ both form (1,0) gauge structures. Finally, we consider the two extreme examples of Courant-Dorfman algebras.

An interesting open question remaining is the comparison of the equations of motion of the (1,0) model to the superconformal (2,0) equations that can be obtained from a twistor construction, cf.\ \cite{Saemann:2011nb,Saemann:2012uq,Mason:2012va,Saemann:2013pca}. However, the fact that the (1,0) model makes use of structures that are only accessible in the semistrict case suggests that the twistor constructions should first be extended to principal 2-bundles with semistrict gauge 2-algebras.

\section{The (1,0) model}

In this section, we will briefly review the recently derived superconformal field theories in six dimensions with $\CN=(1,0)$ supersymmetry \cite{Samtleben:2011fj}. We will focus on the gauge structure, but we will also see the field content, gauge transformations as well as the equations of motion. 

\subsection{(1,0) gauge structures}

Consider two vector spaces $\frg$ and $\frh$ together with two linear maps $\sfg:\frg^*\rightarrow \frh$ and $\sfh: \frh\rightarrow \frg$, where $\frg^*$ denotes the dual of $\frg$. Demanding that $\sfh\circ \sfg=0$, we obtain the chain complex
\begin{equation}\label{eq:chain_complex}
 \frg^*\ \stackrel{\sfg}{\longrightarrow}\ \frh\ \stackrel{\sfh}{\longrightarrow}\ \frg~.
\end{equation}
We will denote elements of $\frg^*$, $\frh$ and $\frg$ by $\lambda$, $\chi$ and $\gamma$, respectively. Assume that we have further bilinear maps 
\begin{equation}\label{eq:maps_on_g_h}
 \sff:\frg\wedge \frg \rightarrow \frg~,~~~\sfd: \frg\odot \frg\rightarrow \frh~,~~~\sfb:\frh\otimes \frg\rightarrow \frg^*~.
\end{equation}
We also have the dual maps 
\begin{equation}
 \sfg^*: \frh^*\rightarrow \frg~,~~~\sfh^*: \frg^*\rightarrow \frh^*~,
\end{equation}
and, by considering one of the arguments as a parameter,
\begin{equation}
  \sff^*: \frg\times\frg^*\rightarrow \frg^*~,~~~\sfd^*:\frh^*\times \frg\rightarrow \frg^*~.
\end{equation}

We demand that all these maps satisfy the following equations \cite{Samtleben:2011fj}:
\begin{subequations}\label{eq:algebra_relations}
\begin{equation}
2(\sfd(\sfh(\sfd(\gamma_1,\gamma_{(2})),\gamma_{3)})-\sfd(\sfh(\sfd(\gamma_2,\gamma_3)),\gamma_1))=2\sfd(\sff(\gamma_1,\gamma_{(2}),\gamma_{3)})-\sfg(\sfb(\sfd(\gamma_2,\gamma_3),\gamma_1))~,\label{eq:algebra_relation_a}
\end{equation}
\begin{equation}
 \begin{aligned}
    \sfd^*(\sfh^*(\sfb(\chi,\gamma_2)),\gamma_1)\,+\,&\sfb(\chi,\sfh(\sfd(\gamma_1,\gamma_2)))+2\sfb(\sfd(\gamma_1,\sfh(\chi)),\gamma_2)=\\&\sff^*(\gamma_1,\sfb(\chi,\gamma_2))+\sfb(\chi,\sff(\gamma_1,\gamma_2))+\sfb(\sfg(\sfb(\chi,\gamma_1)),\gamma_2)\label{eq:algebra_relation_b}
 \end{aligned}
\end{equation}
and
\begin{eqnarray}
 \sfh(\sfg(\lambda))&=&0~,\label{eq:algebra_relation_c}\\
 \sff(\sfh(\chi),\gamma)-\sfh(\sfd(\sfh(\chi),\gamma))&=&0~,\label{eq:algebra_relation_d}\\
 \sff(\gamma_{[1},\sff(\gamma_2,\gamma_{3]}))-\tfrac{1}{3}\sfh(\sfd(\sff(\gamma_{[1},\gamma_2),\gamma_{3]}))&=&0~,\label{eq:algebra_relation_e}\\
 \sfg(\sfb(\chi_1,\sfh(\chi_2)))-2\sfd(\sfh(\chi_1),\sfh(\chi_2))&=&0~,\label{eq:algebra_relation_f}\\
 \sfg(\sff^*(\gamma,\lambda)-\sfd^*(\sfh^*(\lambda),\gamma)+\sfb(\sfg(\lambda),\gamma))&=&0~.\label{eq:algebra_relation_g}
\end{eqnarray}
\end{subequations}
We will refer to such a structure, i.e.\ a chain complex \eqref{eq:chain_complex} together with maps \eqref{eq:maps_on_g_h} satisfying \eqref{eq:algebra_relations} as a {\em (1,0) gauge structure}.

As an initial remark, note that the map $ \sff:\frg\wedge \frg \rightarrow \frg$ is very similar to a Lie bracket on $\frg$, with \eqref{eq:algebra_relation_e} showing the failure of the Jacobi identity to hold.

Equations \eqref{eq:algebra_relations} guarantee that there is a Lie algebra $\CA$ isomorphic to $\frg$ as a vector space that has the following two representations on $\frg$ and $\frh$:
\begin{subequations}\label{eq:reps_A}
\begin{equation}
 \rho(X)\acton \gamma=-\sff(X,\gamma)+\sfh(\sfd(X,\gamma))~,
\end{equation}
and 
\begin{equation}
 \rho(X)\acton \chi:=2\sfd(X,\sfh(\chi))-\sfg(\sfb(\chi,X))
\end{equation}
for $X\in\CA$. The representation on $\frg$ also induces a representation on $\frg^*$,
\begin{equation}
 \rho(X)\acton \lambda=\sff^*(X,\lambda)-\sfd^*(\sfh^*(\lambda),X)~.
\end{equation}
\end{subequations}
All the representations satisfy the relation\footnote{Note that equations \eqref{eq:reps_A} and \eqref{eq:rep_commutator} define the Lie algebra $\CA$ only up to representations. Unless one of them is faithful, there is no unique Lie algebra structure on $\CA$ that could be reconstructed.}
\begin{equation}\label{eq:rep_commutator}
 [\rho(X_1),\rho(X_2)]=\rho(-\sff(X_1,X_2)+\sfh(\sfd(X_1,X_2)))=\rho(-\sff(X_1,X_2))~.
\end{equation}
Finally, all the maps introduced above are invariant under the action of $\CA$ because equations
\begin{subequations}
\begin{eqnarray}
 \rho(X)\acton \sfd(\gamma_1,\gamma_2)&=&\sfd(\rho(X)\acton\gamma_1,\gamma_2)+\sfd(\gamma_1,\rho(X)\acton\gamma_2)~,\\
 \rho(X)\acton \sfb(\chi,\gamma)&=&\sfb(\rho(X)\acton\chi,\gamma)+\sfb(\chi,\rho(X)\acton\gamma)~,\\
 \rho(X)\acton \sfh(\chi)&=&\sfh(\rho(X)\acton\chi)~,\label{eq:rep_rel_c}\\
 \rho(X)\acton \sfg(\lambda)&=&\sfg(\rho(X)\acton\lambda)\label{eq:rep_rel_d}
\end{eqnarray}
\end{subequations}
are equivalent to \eqref{eq:algebra_relation_a}, \eqref{eq:algebra_relation_b}, \eqref{eq:algebra_relation_d} and \eqref{eq:algebra_relation_g}, respectively. Furthermore, the invariance of $\sff$ implies \eqref{eq:algebra_relation_e}.

To analyze the above equations further, one can choose a convenient basis for $\frg$ and $\frh$, in which either the map $\sfg$ or $\sfh$ is diagonal as was done in \cite{Samtleben:2012mi}.

If one demands that the (1,0) model allows for an action principle, one has to require in addition that there is a nondegenerate bilinear form $(\cdot,\cdot)_\frh$ on $\frh$, which induces a linear nondegenerate map $\sfm:\frh\rightarrow \frh^*$ with $\sfm\circ \sfm^*=\sfm^*\circ\sfm=\mathsf{id}$. Furthermore, the following conditions have to be satisfied:
\begin{subequations}\label{eq:algebra_relations_action}
\begin{eqnarray}
  \sfg(\lambda)&=&\sfm^*(\sfh^*(\lambda))~,\label{eq:algebra_relation_action_a}\\
  \sfb(\chi,\gamma)&=&2\sfd^*(\sfm(\chi),\gamma)~,\label{eq:algebra_relation_action_b}\\
  \big(\sfd(\gamma_1,\gamma_{(2}),\sfd(\gamma_2,\gamma_{3)})\big)_\frh&=&0~.\label{eq:algebra_relation_action_c}
\end{eqnarray}
\end{subequations}
Below, we will impose the additional relations \eqref{eq:algebra_relations_action} only if explicitly stated.

\subsection{Field content}

The field content of the superconformal (1,0) theory is given by a gauge potential one-form $A$ taking values in $\frg$, a two-form potential $B$ taking values in $\frh$ and a three-form potential $C$ with values in $\frg^*$. Their curvatures read as 
\begin{subequations}\label{eq:curvatures}
\begin{eqnarray}
\CF&=&\dpar  A-\tfrac{1}{2}\sff(A,A)+\sfh(B)~,\label{eq:curvatureF10}\\
\CH&=&DB+\sfd(A,\dpar  A-\tfrac{1}{3}\sff(A,A))+\sfg(C)\nonumber\\
&=&\dpar  B+ 2\sfd(A,\sfh(B))-\sfg(\sfb(B,A))+\sfd(A,\dpar  A-\tfrac{1}{3}\sff(A,A))+\sfg(C)~\label{eq:curvatureH10},
\end{eqnarray}
\end{subequations}
where, to avoid confusion with the map $\sfd:\frg\odot \frg\rightarrow \frh$ , we will use $\dpar$ for the exterior derivative for this chapter only, e.g.
\begin{equation}
\dpar  A:=\dpar_{[\mu}A_{\nu]}\dd x^\mu\wedge\dd x^\nu~.
\end{equation} 
The covariant derivative acts by $D=\dpar +\rho(A)\acton$ and, in our notation, maps acting on the (1,0) gauge structure do not act on the form part of the fields, e.g.
\begin{equation}
 \sff(A,A):=\sff(A_\mu,A_\nu)\dd x^\mu\wedge\dd x^\nu~.
\end{equation}

Infinitesimal gauge transformations are parametrized by a function $\alpha$ taking values in $\frg$, as well as 1- and 2-forms $\Lambda$ and $\Xi$ with values in $\frh$ and $\frg^*$, respectively. Their action on the potential forms are
\begin{equation}\label{eq:qauge}
\begin{aligned}
\delta A&=D\alpha-\sfh(\Lambda)~,\\
\delta B&=D \Lambda+\sfd(A,D \alpha -\sfh(\Lambda))-2\sfd(\alpha,\CF)-\sfg(\Xi)~,\\
\delta  C&=D \Xi+\sfb(B,D\alpha-\sfh(\Lambda))-\tfrac{1}{3}\sfb(\sfd(D\alpha-\sfh(\Lambda),A),A)+\sfb(\Lambda,\CF)+\sfb(\CH,\alpha)+\dots~,
\end{aligned}
\end{equation}
where $\dots$ represents further terms in the kernel of $\sfg$. Later, we will find it useful to use a shifted version of these gauge transformations. Taking the shifted parameters $(\alpha,\Lambda,\Xi)\rightarrow(\alpha,\Lambda+\sfd(\alpha,A), \Xi-\sfb(B,\alpha)+\tfrac{1}{3}\sfb(\sfd(\alpha,A),A))$ we obtain
\begin{equation}\label{eq:shiftqauge}
\begin{aligned}
\delta A=&~\dpar  \alpha-\sff(A,\alpha)-\sfh(\Lambda)~,\\
\delta B=&~\dpar  \Lambda+\sfd(A,\sfh(\Lambda))+\sfg(\sfb(\Lambda,A))-\sfd(\alpha,\sfh(B))+\sfg(\sfb(B,\alpha))-\sfg(\Xi)\\
&-\sfd(\alpha,\CF)+\tfrac{1}{6}(\sfd(\sff(A,A),\alpha)+2\sfd(\sff(A,\alpha),A))~,\\
\delta  C=&~\dpar  \Xi-\sfb(\dpar  B,\alpha)+\tfrac{1}{3}(\sfb(\sfd(\alpha,\dpar  A),A)-\sfb(\sfd(\alpha,A),\dpar  A))\\&-\sfb(\sfg(\Xi-\sfb(B,\alpha)+\tfrac{1}{3}\sfb(\sfd(\alpha,A),A))),A)\\&+\sfb(B,-\sff(A,\alpha)-\sfh(\Lambda))-\tfrac{1}{3}\sfb(\sfd(-\sff(A,\alpha)-\sfh(\Lambda),A),A)\\&+\sfb(\Lambda+\sfd(\alpha,A),\CF)+\sfb(\CH,\alpha)+\dots~,
\end{aligned}
\end{equation}
where we used \eqref{eq:algebra_relation_g} and \eqref{eq:algebra_relation_a} in the form of
\begin{equation}\label{eq:forma}
\begin{aligned}
\sfd(A,\sff(A,\alpha)-3\sfh(\sfd(A,\alpha))-\sfd(\alpha,\sff(A,A))=\sfg(\sfb(\sfd(\alpha,A),A))~.
\end{aligned}
\end{equation}

\subsection{Bianchi identities and extended complexes}\label{ssec:Bianchi_Extension}

By construction, the field strengths satisfy the Bianchi identity
\begin{equation}
\begin{aligned}
D \CF=\sfh(\CH)~.
\end{aligned}
\end{equation}
Furthermore, demanding that 
\begin{equation}
\begin{aligned}
D \CH=\sfd(\CF,\CF)+\sfg(\CH^{(4)})~,
\end{aligned}
\end{equation}
for some four-form $\CH^{(4)}$, defined up to terms in the kernel of $\sfg$, leads to 
\begin{equation}
\begin{aligned}
D \CH^{(4)}=\sfb(\CH,\CF)+\dots~,
\end{aligned}
\end{equation}
where $\dots$ again represents terms in the kernel of $\sfg$. 

This process can be continued by extending the complex\footnote{Such an extension can always be found; for example, we could put $\frl=\ker(\sfg)$ and $\sfk$ is its embedding into $\frg$.}
\begin{equation}
\frl\overset{\sfk}{\longrightarrow} \frg^*\longrightarrow \frh\longrightarrow \frg ~,
\end{equation}
and defining a five-form $\CH^{(5)}\in\frl$ such that 
\begin{equation}
\begin{aligned}
D \CH^{(4)}=\sfb(\CH,\CF)+\sfk(\CH^{(5)})~,
\end{aligned}
\end{equation}
and such that $\CH^{(5)}$ satisfies its own Bianchi identity involving new maps into $\frl$ which satisfy additional constraints. These are found in \cite{Samtleben:2011fj} and \cite{Bandos:2013jva}. In the latter paper this extended model was used to write down a PST-like action. This extension is very similar to that of higher gauge theory with iterated categorifications of principal bundles. In the following, however, we will restrict ourselves to the non-extended case.

\subsection{Supersymmetry and field equations}

For this section we will introduce the notation 
\begin{equation}
\begin{aligned}
\gamma&=\gamma_\mu\dd x^\mu~,~~&\gamma^{(2)}&=\tfrac{1}{2}\gamma_{\mu\nu}\dd x^\mu\wedge\dd x^\nu~,~~&\gamma^{(3)}&=\tfrac{1}{6}\gamma_{\mu\nu\rho}\dd x^\mu\wedge\dd x^\nu\wedge\dd x^\rho~,\\
\slasha{D}&=\gamma^\mu D_\mu~,~&\slasha{\CF}&=*(\CF\wedge*\gamma^{(2)})=\tfrac{1}{2}\gamma^{\mu\nu}\CF_{\mu\nu}~,&\slasha{\CH}&=*(\CH\wedge*\gamma^{(3)})=\tfrac{1}{6}\gamma^{\mu\nu\rho}\CH_{\mu\nu\rho}
\end{aligned}
\end{equation}
where $*$ is the Hodge star operation. The fields above belong to the (1,0) vector and tensor supermultiplets $(A,\lambda^i,Y^{ij})$ and $(\phi,\chi^i,B)$, for $i,j=1,2$, taking values in $\frg$ and $\frh$, respectively. In \cite{Samtleben:2011fj}, it was found that the supersymmetry transformations 
\begin{equation}\label{eq:susy}
\begin{aligned}
\delta A&=-\epsb \gamma \lambda~,~~~&\delta B&=-\sfd(A,\epsb \gamma \lambda)-\epsb\gamma^{(2)}\chi~,\\
\delta \lambda^i&=\tfrac{1}{4}\slasha{\CF}\eps^i-\tfrac{1}{2}Y^{ij}\eps_j+\tfrac{1}{4}\sfh(\phi)\eps^i~,~~~&\delta \chi^i&=\tfrac{1}{8}\slasha{\CH}\eps^i+\tfrac{1}{4}\slasha{D}\phi~\eps^i-*\tfrac{1}{2}\sfd(\gamma\lambda^i,*\epsb\gamma\lambda)~,\\
\delta  Y^{ij}&=-\epsb^{(i}\slasha{D} \lambda^{j)}+2 \epsb^{(i}\sfh(\chi^{j)})~,~~~&\delta \phi&=\epsb\chi~,\\
&&&\hspace{-4.3cm}\delta C=-\sfb(B,\epsb \gamma\lambda)-\tfrac{1}{3}\sfb(\sfd(A,\epsb \gamma \lambda), A)-\sfb(\phi,\epsb\gamma^{(3)}\lambda)~,
\end{aligned}
\end{equation}
close up to translations, gauge transformations and the equations of motion
\begin{equation}\label{eq:eom}
\begin{aligned}
\CH^-&=-\sfd(\lambdab,\gamma^{(3)}\lambda)~,\\
\slasha{D} \chi^i&=\sfd(\slasha{\CF},\lambda^i)+2\sfd(Y^{ij},\lambda_j)+\sfd(\sfh(\phi),\lambda^i)-2\sfg(\sfb(\phi,\lambda^i))~,\\
D^2 \phi&=2\sfd(Y^{ij},Y_{ij})-*2\sfd(\CF,*\CF)-4\sfd(\lambdab,\slasha{D} \lambda)\\&~~~~-2\sfg(\sfb(\chib,\lambda))+16\sfd(\lambdab,\sfh(\chi))-3\sfd(\sfh(\phi),\sfh(\phi))~,
\end{aligned}
\end{equation}
where $\CH=\CH^++\CH^-$ is split into selfdual and anti-selfdual parts: $\CH^\pm=\pm*\CH^\pm$. These tensor multiplet equations \eqref{eq:eom} are connected by supersymmetry to the following vector multiplet equations 
\begin{equation}\label{eq:eom2}
\begin{aligned}
\sfg(\sfb(\phi,Y_{ij})+2\sfb(\chib_{(i},\lambda_{j)}))&=0~,\\
\sfg(\sfb(\phi,\CF)-2\sfb(\chib,\gamma^{(2)}\lambda))&=\tfrac{1}{2}\sfg(*\CH^{(4)})~,\\
\sfg(\sfb(\phi,\slasha{D} \lambda_i)+\tfrac{1}{2}\sfb(\slasha{D}\phi, \lambda_i))&=\sfg(*\tfrac{1}{2}\sfb(\gamma^{(2)}\chi_i,*\CF)+\tfrac{1}{4}\sfb(\slasha{\CH},\lambda_i)-\sfb(\chi^j,Y_{ij})\\&~~~~+\tfrac{3}{2}\sfb(\phi,\sfh(\chi))+*\tfrac{1}{3}\sfb(\sfd(\gamma\lambda_i,\lambdab),*\gamma\lambda))~.
\end{aligned}
\end{equation}

\section{(1,0) gauge structures and\\ weak Courant-Dorfman algebras}

\subsection{Differential graded Leibniz algebra}

We now come to the analysis of the gauge structure that is defined by the maps \eqref{eq:maps_on_g_h} together with equations \eqref{eq:algebra_relations}. The fact that underlying the (1,0) gauge structure is the chain complex \eqref{eq:chain_complex} suggests that we are working with some differential graded algebra\footnote{For a detailed analysis of the general tensor hierarchy algebra from the perspective of Lie superalgebras, see \cite{Palmkvist:2013vya}.}. We first focus on the representations of the Lie algebra $\CA$ \eqref{eq:reps_A} on the vector spaces $\frg$, $\frh$ and $\frg^*$. As they satisfy the Jacobi identity, we arrive at a Leibniz algebra.

Recall that a {\em differential graded Leibniz algebra}\footnote{or a {\em differential graded Loday algebra}} $(L,\CD,\acton)$ is a ($\RZ$-)graded vector space $L$ equipped with a degree $1$ linear map $\CD$ and a degree $0$ bilinear map $\acton$ such that 
\begin{conditions}
 \item[(i)] $\CD$ is a differential: $\CD^2=0$ and $\CD(\ell_1\acton \ell_2)=(\CD \ell_1)\acton \ell_2+(-1)^{|\ell_1|}\ell_1\acton (\CD \ell_2)$~,
 \item[(ii)] a Leibniz identity holds: $\ell_1\acton(\ell_2\acton \ell_3)=(\ell_1\acton \ell_2)\acton \ell_3+(-1)^{|\ell_1||\ell_2|}\ell_2\acton(\ell_1\acton \ell_3)$~,
\end{conditions}
where $\ell_1,\ell_2,\ell_3\in L$ and $|\ell_i|$ denotes the grading of $\ell_i$. 

In the case of a (1,0) gauge structure, we have\footnote{Recall that $V[-n]$ denotes the vector space $V$ shifted by $-n$ degrees in the grading. In particular, $\frg^*[-2]$ consists of elements in $\frg^*$, and each element has homogeneous grading -2.}
\begin{equation}
 L=\frg^*[-2]\oplus \frh[-1]\oplus \frg~,~~~\CD|_{\frg^*}=\sfg~,~~~\CD|_{\frh}=\sfh~,
\end{equation}
and the only nontrivial actions $\acton$ are given by \eqref{eq:reps_A}:
\begin{equation}
 \gamma_1\acton \gamma_2:=\rho(\gamma_1)\acton \gamma_2~,~~~\gamma_1\acton \chi:=\rho(\gamma_1)\acton \chi~,~~~\gamma_1\acton \lambda:=\rho(\gamma_1)\acton \lambda
\end{equation}
for all $\gamma_1,\gamma_2\in\frg$, $\chi\in\frh$ and $\lambda\in\frg^*$. Conditions (i) and (ii) are readily verified: (i) follows from \eqref{eq:algebra_relation_c} together with \eqref{eq:rep_rel_c} and \eqref{eq:rep_rel_d}, while (ii) follows from the fact that $\rho$ forms a representation of $\CA$.

The characterization of (1,0) gauge algebras in terms of Leibniz algebras is certainly too general. In particular, we would like to identify a structure in which the maps $\sff$, $\sfd$ and $\sfb$ are given an intrinsic meaning. Clearly, considering separately the antisymmetrization and the symmetrization of 
\begin{equation}
 \gamma_1\acton\gamma_2:=\rho(\gamma_1)\acton\gamma_2=-\sff(\gamma_1,\gamma_2)+\sfh(\sfd(\gamma_1,\gamma_2))
\end{equation}
would allow us to extract $\sff$ as well as $\sfd$ up to terms in the kernel of $\sfh$. Note, however, that these new maps cannot be expected to satisfy the Leibniz identity anymore. The transition between a product satisfying a Leibniz identity and its antisymmetrization that violates the Leibniz rule (which here amounts to the Jacobi identity) is in fact a very common one in the context of Courant algebroids. We therefore turn our attention to those in the following.

\subsection{Courant algebroids}

A particularly nice class of examples of (1,0) gauge structures is obtained from Courant algebroids. Recall that a Courant algebroid is a symplectic Lie 2-algebroid, or, equivalently, a symplectic NQ-manifold\footnote{a Q-manifold with non-negatively integer grading which is endowed with a symplectic form}, cf.\ \cite{Roytenberg:0203110}. Here, we define it as a Euclidean vector bundle $(E, \langle \cdot,\cdot\rangle)$ over a smooth manifold $M$ that is endowed with a bilinear operation $\acton$ on sections of $E$ and a bundle map $\varrho:E\rightarrow TM$ called the {\em anchor} satisfying the following axioms for all $e,e_1,e_2\in \Gamma(E)$ and $f\in\CC^\infty(M)$:
\begin{conditions}
 \item[(i)] $e\acton(e_1\acton e_2)=(e\acton e_1)\acton e_2+e_1\acton(e\acton e_2)$,
 \item[(ii)] $e_1\acton e_2+e_2\acton e_1=\CD\langle e_1,e_2\rangle$,
 \item[(iii)] $\varrho(e_1\acton e_2)=[\varrho(e_1),\varrho(e_2)]$,
 \item[(iv)] $e_1\acton (fe_2)=f(e_1\acton e_2)+(\varrho(e_1)\cdot f)e_2$,
 \item[(v)] $\varrho(e)\cdot \langle e_1,e_2\rangle=\langle e\acton e_1,e_2\rangle+\langle e_1,e\acton e_2\rangle$.
\end{conditions}
Here $\varrho(e)\cdot f$ denotes the action of the vector field $\varrho(e)$ onto $f$, $[\cdot,\cdot]$ denotes the Lie bracket of vector fields and $\CD$ is the pullback of the exterior derivative $\dpar$ on $M$ via the adjoint map $\varrho^*$:
\begin{equation}
 \langle \CD f,e \rangle:=\tfrac{1}{2}\varrho(e)\cdot f~.
\end{equation}

A Courant algebroid contains a differential graded Leibniz algebra, and one can show that it forms a (1,0) gauge structure with trivial maps $\sfg$ and $\sfb$. Instead of doing this using the above definition, which stems from \cite{Roytenberg:0203110}, we can switch to the original and equivalent definition from \cite{Liu:1997aa}. For this, we introduce the antisymmetric {\em Courant bracket}
\begin{equation}\label{eq:switch_brackets}
\llbracket e_1,e_2\rrbracket:=\tfrac{1}{2}(e_1\acton e_2-e_2\acton e_1)=e_1\acton e_2-\tfrac{1}{2}\CD\langle e_1,e_2\rangle~.
\end{equation}
In this context, the action $\acton$ is often called a {\em Dorfman bracket}. For the Courant bracket, the axioms in the definition of a Courant algebroid become
\begin{conditions}
 \item[(i')] $\llbracket\llbracket e_1,e_2\rrbracket,e_3\rrbracket+\llbracket\llbracket e_2,e_3\rrbracket, e_1\rrbracket+\llbracket\llbracket e_3,e_1\rrbracket,e_2\rrbracket+\tfrac{1}{2}\CD\big\langle\llbracket e_{[1},e_2\rrbracket, e_{3]}\big\rangle=0$,
 \item[(iii')] $\varrho(\llbracket e_1,e_2\rrbracket)=[\varrho(e_1),\varrho(e_2)]$,
 \item[(iv')] $\llbracket e_1,f e_2\rrbracket=f\llbracket e_1,e_2\rrbracket+(\varrho(e_1)\cdot f)e_2-\langle e_1,e_2\rangle\CD f$,
 \item[(v')] $\varrho(e)\cdot\langle e_1,e_2\rangle =\big\langle\llbracket e,e_1\rrbracket+\CD\langle e,e_1\rangle,e_2 \big\rangle+ \big\langle e_1 ,\llbracket e,e_2\rrbracket+\CD\langle e ,e_2\rangle \big\rangle$,
 \item[(vi')] $\langle \CD f,\CD g\rangle=0$,
\end{conditions}
where again $e,e_1,e_2\in \Gamma(E)$ and $f,g\in \CC^\infty(M)$.

Given a Courant algebroid, we can define a (1,0) gauge structure by putting
\begin{equation}
 \begin{aligned}
    \frg:=\Gamma(E)~,~~\frh:=\CC^\infty(M)~,~~\sfh:=\CD~,~~\sff:=-\llbracket\cdot,\cdot\rrbracket~,~~\sfd:=\tfrac{1}{2}\langle \cdot,\cdot \rangle~,~~ \sfg:=0~,~~~\sfb:=0.
 \end{aligned}
\end{equation}
The relations \eqref{eq:algebra_relation_b}, \eqref{eq:algebra_relation_c}, \eqref{eq:algebra_relation_g} are trivially satisfied. Moreover, the relations \eqref{eq:algebra_relation_a}, \eqref{eq:algebra_relation_e} and \eqref{eq:algebra_relation_f} are equivalent to the axioms (v'), (i') and (vi'), respectively. Finally, equation \eqref{eq:algebra_relation_d} has been shown to hold for Courant algebroids \cite[Prop. 4.2]{Roytenberg:1998vn}.

To capture finite dimensional (1,0) gauge structures, we need to reformulate the notion of a Courant algebroid in purely algebraic terms. This leads to the concept of a Courant-Dorfman algebra.

\subsection{Courant-Dorfman algebras}\label{ssec:Courant_Dorfman_algebras}

A {\em Courant-Dorfman algebra} \cite{Roytenberg:0902.4862}, see also \cite{Keller:0807.0584}, consists of a commutative $\mathbb{K}$-algebra $\CR$ together with an $\CR$-module $\CE$ endowed with a derivation $\CD:\CR\rightarrow \CE$, a symmetric bilinear form (not necessarily non-degenerate) $\langle\cdot,\cdot\rangle: \CE\otimes_\CR\CE\rightarrow \CR$ and a {\em Dorfman bracket} $\acton:\CE\otimes \CE\rightarrow \CE$, which satisfy the following axioms:
\begin{conditions}
 \item[(i)] $e_1\acton (e_2\acton e_3)=(e_1\acton e_2)\acton e_3+e_2\acton(e_1\acton e_3)$,
 \item[(ii)] $e_1\acton e_2+e_2\acton e_1=\CD\langle e_1,e_2\rangle$,
 \item[(iii)] $(\CD r)\acton e=0$,
 \item[(iv)] $e_1\acton r e_2=r(e_1\acton e_2)+\langle e_1,\CD r\rangle e_2$,
 \item[(v)] $\langle e_1,\CD\langle e_2,e_3\rangle\rangle=\langle e_1\acton e_2,e_3\rangle+\langle e_2,e_1\acton e_3\rangle$,
 \item[(vi)] $\langle \CD r_1,\CD r_2\rangle=0$,
\end{conditions}
where $e,e_1,e_2,e_3\in\CE$ and $r,r_1,r_2\in\CR$. Note that if the bilinear form $\langle\cdot,\cdot\rangle$ is non-degenerate, axioms (iii), (iv) and (vi) are redundant. Moreover, if we consider a Euclidean vector bundle $E\rightarrow M$ with a fiber metric which we identify with $\langle \cdot,\cdot\rangle$, if we put $\CE=\Gamma(E)$ and $\CR=\CC^\infty(M)$ and define $\CD$ as the pullback of the exterior derivative on $M$, then we recover the notion of a Courant algebroid.

As before, we can reformulate these axioms by switching from the Dorfman bracket $\acton$ to the Courant bracket via relation \eqref{eq:switch_brackets}, and we are left with
\begin{conditions}
 \item[ (i')] $\llbracket e_{[ 1},\llbracket e_2,e_{3]}\rrbracket \rrbracket +\tfrac{1}{6}\CD\langle e_{[ 1},\llbracket e_2,e_{3]}\rrbracket \rangle=0$,
 \item[ (iii')] $\llbracket \CD r,e\rrbracket +\tfrac{1}{2}\CD\langle \CD r,e\rangle=0$,
 \item[ (iv')] $\llbracket e_1,r e_2\rrbracket =r\llbracket e_1,e_2\rrbracket +\langle e_1,\CD r\rangle e_2+\tfrac{1}{2}r(\CD\langle e_1,e_2\rangle)-\tfrac{1}{2}\CD\langle e_1,r e_2\rangle$,
 \item[ (v')] $\langle \CD\langle e_1,e_{(2}\rangle,e_{3)}\rangle-\langle \CD\langle e_2,e_3\rangle,e_1\rangle+2\langle\llbracket e_1,e_{(2}\rrbracket ,e_{3)}\rangle=0$,
 \item[ (vi')] $\langle \CD r_1,\CD r_2\rangle=0$.
\end{conditions}

Given a Courant-Dorfman algebra, we can construct a (1,0) gauge structure by putting 
\vspace*{-0.6cm}
\begin{equation}
 \begin{aligned}
    \frg:=\CE~,~~~\frh:=\CR~,~~~\sfh:=\CD~,~~~\sff:=-\llbracket\cdot,\cdot\rrbracket~,~~~\sfd:=\tfrac{1}{2}\langle \cdot,\cdot \rangle~,~~~ \sfg:=0~,~~~\sfb:=0~.
 \end{aligned}
\end{equation}
Axioms \eqref{eq:algebra_relation_a}, \eqref{eq:algebra_relation_d}, \eqref{eq:algebra_relation_e} and \eqref{eq:algebra_relation_f} of the (1,0) gauge structure correspond to the axioms (v'), (iii'), (i') and (vi') of the Courant-Dorfman algebra, respectively.

Inversely, a (1,0) gauge structures with $\sfg$ and $\sfb$ trivial gives rise to a Courant-Dorfman algebra, where the action of $\CR=\frh$ onto $\CE=\frg$ is given by
\begin{equation}
r e:=\CD\langle e,\CD r\rangle=\sfh(\rho(e)\acton r)~. 
\end{equation}
Axiom (iv') holds then by definition, the other axioms are related to those of the (1,0) gauge structure as before.

\subsection{Weak Courant-Dorfman algebras}

To extend this correspondence to the case of (1,0) gauge structures with non-trivial maps $\sfg$ and $\sfb$, we have to allow for some more general Courant-Dorfman algebras. In particular, we have to weaken axioms (v') and (vi'), which correspond to \eqref{eq:algebra_relation_a} and \eqref{eq:algebra_relation_f} only for trivial $\sfg$ and $\sfb$. Interestingly, this generalization has already been introduced in \cite{Ekstrand:2009qz} by dropping axioms (iv), (v) and (vi) (or, equivalently, (iv'), (v') and (vi')) of a Courant-Dorfman algebra:

A {\em weak Courant-Dorfman algebra} consists of two vector spaces $\CR$ and $\CE$ together with a symmetric bilinear form $\langle \cdot,\cdot\rangle:\CE\otimes \CE\rightarrow \CR$, a map $\CD:\CR\rightarrow \CE$ and a Dorfman bracket $\acton:\CE\otimes \CE\rightarrow \CE$. These satisfy the following axioms:
\begin{conditions}
 \item[(i'')] $e_1\acton (e_2\acton e_3)=(e_1\acton e_2)\acton e_3+e_2\acton(e_1\acton e_3)$,
 \item[(ii'')] $e_1\acton e_2+e_2\acton e_1=\CD\langle e_1,e_2\rangle$,
 \item[(iii'')] $(\CD r)\acton e=0$.
\end{conditions}

An important class of examples is given by the higher generalizations of exact Courant algebroids $TM\oplus \wedge^p T^*M$ together with the standard Courant brackets. Since these do not seem to be related to our discussion, we refrain from going into further details.

Note that the above axioms imply the following weaker form of (v) and (vi) \cite{Ekstrand:2009qz}:
\begin{equation}
 \begin{aligned}
 \CD\big(\langle e_1,\CD\langle e_2,e_3\rangle\rangle-\langle e_1\acton e_2,e_3\rangle-\langle e_2,e_1\acton e_3\rangle\big)&=0~,\\
 \CD\langle \CD e_1,\CD e_2\rangle&=0~.
 \end{aligned}
\end{equation}
These equations are precisely the generalizations necessary to accommodate a (1,0) gauge structure with non-trivial $\sfg$ and $\sfb$, as axioms \eqref{eq:algebra_relation_a} and \eqref{eq:algebra_relation_f} are modified by terms in the image of $\sfg$, which vanishes under $\CD$ due to $\sfh\circ\sfg=0$. We therefore conclude that (1,0) gauge structures are special cases of weak Courant-Dorfman algebras.

\subsection{Comments on derived brackets}

To construct weak Courant-Dorfman algebras, one is quickly led to the notion of derived brackets: Courant algebroids are symplectic NQ-manifolds \cite{Severa:2001aa,Roytenberg:0203110}, see also \cite{Kotov:2010wr}, and $\CD$, as well as the Courant bracket $\llbracket\cdot,\cdot\rrbracket$ on sections, are derived from the symplectic structure on an NQ-manifold \cite{Roytenberg:0203110} via a derived bracket construction \cite{Kosmann-Schwarzbach:0312524,Voronov:math0304038}. Unfortunately, this approach to (1,0) gauge structures seems too restrictive, at least if one uses the superextension due to \cite{Voronov:math0304038}, as we demonstrate in the following.

We start from a Lie superalgebra $L$ with Lie bracket $\{\cdot,\cdot\}$ together with a projector $P\in \sEnd L$ onto an abelian subalgebra of $L$ such that 
\begin{equation}
 P^2=P~,~~~\{P\ell_1,P\ell_2\}=0\eand P\{\ell_1,\ell_2\}=P\{P\ell_1,\ell_2\}+P\{\ell_1,P\ell_2\}~.
\end{equation}
Given an odd element $Q\in L$ (with appropriate $\RZ$-grading) such that $Q^2=\tfrac{1}{2}\{Q,Q\}=0$, we can define the brackets
\begin{equation}
 \mu_i(\ell_1,\ell_2,\ldots,\ell_i):=P\{\ldots\{\{Q,\ell_1\},\ell_2\},\ldots,\ell_i\}~,
\end{equation}
which turn $L$ into an $L_\infty$-algebra \cite[Cor.\ 1]{Voronov:math0304038}. In particular, the condition $Q^2=0$ is equivalent to the higher homotopy relations \eqref{eq:homotopyJacobi}. Note that the grading of the $L_\infty$-algebra is again that of the Lie superalgebra shifted by one.

We now wish to identify the additional structure maps $\sfd$ and $\sfb$ with (parts of) a Poisson bracket. For this, note that equation \eqref{eq:algebra_relation_a} implies
\begin{equation}
 \sfg(\sfb(\sfd(\gamma_{(1},\gamma_2),\gamma_{3)}))=0~.
\end{equation}
If we impose either the additional constraint \eqref{eq:algebra_relation_action_c} or consider the extended tensor hierarchy (cf.\ \cite{Hartong:2009vc,Samtleben:2011fj}), one has the stronger relation
\begin{equation}\label{eq:angular_graded_jacobi}
 \sfb(\sfd(\gamma_{(1},\gamma_2),\gamma_{3)})=0~.
\end{equation}
This relation is in fact the graded Jacobi identity we require, assuming a parity shift of $\frg$ by one to odd grading. We are thus led to identify
\begin{equation}
 \{\gamma_1,\gamma_2\}=\sfd(\gamma_1,\gamma_2)\eand \{\gamma,\chi\}=\sfb(\chi,\gamma)~.
\end{equation}
If we demand in addition that
\begin{equation}
 P\{\gamma_1,\gamma_2\}=\{\gamma_1,\gamma_2\}~,
\end{equation}
then relations \eqref{eq:algebra_relation_c}, \eqref{eq:algebra_relation_d} and \eqref{eq:algebra_relation_e} are automatically satisfies, as one readily verifies. Equation \eqref{eq:algebra_relation_f} leads to a constraint:
\begin{equation}
\sfd(\sfh(\chi_1),\sfh(\chi_2))=\{P\{Q,\chi_1\},P\{Q,\chi_2\}\}=0\stackrel{!}{=}\tfrac{1}{2}\sfg(\sfb(\chi_1,\sfh(\chi_2)))~.
\end{equation}
A similar constraint is derived from \eqref{eq:algebra_relation_a}. More importantly, however, we have
\begin{equation}
 \{\mu_2(\gamma_1,\gamma_2),\mu_2(\gamma_3,\gamma_4)\}=\{P\{\{Q,\gamma_1\},\gamma_2\},P\{\{Q,\gamma_3\},\gamma_4\}\}=0~.
\end{equation}
All these constraints impose severe restrictions on the maps $\sff$, $\sfd$ and $\sfb$, which renders this approach essentially uninteresting for the construction of (1,0) gauge algebras.

\section{(1,0) gauge structures as Lie 3-algebra}

Having identified (1,0) gauge structures with weak Courant-Dorfman algebras, we would now like to make contact with higher or categorified gauge theory. As a first step towards this goal, we need to identify categorified Lie algebras in the (1,0) gauge structure. For our purposes, it suffices to restrict ourselves to so-called semistrict Lie 3-algebras. These arise from categorifying twice the notion of a Lie algebra and imposing antisymmetry on the higher products. For simplicity, we will often drop the label `semistrict' in the following. 

\subsection{Semistrict Lie 3-algebras}

Semistrict Lie 3-algebras are categorically equivalent to 3-term $L_\infty$- or strong homotopy Lie algebras \cite{Baez:2003aa}, see appendix \ref{app:SH_Lie_algebras} for the general definition of $L_\infty$-algebras. A 3-term $L_\infty$-algebra\footnote{also known as an $L_3$-algebra} is a graded vector space $L=L_{-2}\oplus L_{-1}\oplus L_0$, where $L_i$ has grading $i$, together with multilinear, totally graded antisymmetric maps
\begin{equation}\label{eq:products_Lie_3}
 \begin{aligned}
  &\mu_1:L_{-2}\rightarrow L_{-1}~,~~~&&\mu_1:L_{-1}\rightarrow L_{0}~,\\
  &\mu_2:L_{0}^{\wedge 2}\rightarrow L_0~,~~~&&\mu_2:L_{0}\wedge L_{-1}\rightarrow L_{-1}~,~~~&&\mu_2:L_{0}\wedge L_{-2}\rightarrow L_{-2}~,\\
  &\mu_2:L_{-1}^{\wedge 2}\rightarrow L_{-2}~,\\
  &\mu_3:L_0^{\wedge 3}\rightarrow L_{-1}~,~~~&&\mu_3:L_{-1}\wedge L_0^{\wedge 2}\rightarrow L_{-2}~,\\
  &\mu_4:L_0^{\wedge 4}\rightarrow L_{-2}~.
 \end{aligned}
\end{equation}
These maps satisfy a number of higher Jacobi or homotopy relations, which we list in the following. The map $\mu_1$ is a differential:
\begin{subequations}\label{eq:homotopy_relations}
\begin{equation}
  \mu_1^2(\lambda): =\mu_1(\mu_1(\lambda))=0~,\label{eq:homotopy_relation_a}
\end{equation}
and it is compatible with the products $\mu_2$:
\begin{eqnarray}
\mu_1(\mu_2(\gamma,\chi))&\!\!=\!\!&-\mu_2(\mu_1(\chi),\gamma)~,\label{eq:homotopy_relation_b}\\
\mu_1(\mu_2(\gamma,\lambda))&\!\!=\!\!&-\mu_2(\mu_1(\lambda),\gamma)~,\label{eq:homotopy_relation_c}\\
\mu_1(\mu_2(\chi_1,\chi_2))&\!\!=\!\!&\mu_2(\mu_1(\chi_1),\chi_2)+\mu_2(\mu_1(\chi_2),\chi_1)~.\label{eq:homotopy_relation_d}
\end{eqnarray}
The map $\mu_2$ satisfies a Jacobi identity up to correction terms given by $\mu_3$:
\begin{eqnarray}
 \mu_1(\mu_3(\gamma_1,\gamma_2,\gamma_3))&\!\!=\!\!&-\mu_2(\mu_2(\gamma_1,\gamma_2),\gamma_3)+\mu_2(\mu_2(\gamma_1,\gamma_3),\gamma_2)-\mu_2(\mu_2(\gamma_2,\gamma_3),\gamma_1)~,~~~\label{eq:homotopy_relation_e}\\
 \mu_1(\mu_3(\chi,\gamma_1,\gamma_2))&\!\!=\!\!&-\mu_3(\mu_1(\chi),\gamma_1,\gamma_2)-\mu_2(\mu_2(\gamma_1,\gamma_2),\chi)\nonumber\\
 &&-\mu_2(\mu_2(\chi,\gamma_1),\gamma_2)+\mu_2(\mu_2(\chi,\gamma_2),\gamma_1)~,\label{eq:homotopy_relation_f}\\
0&\!\!=\!\!&-\mu_3(\mu_1(\lambda),\gamma_1,\gamma_2)-\mu_2(\mu_2(\gamma_1,\gamma_2),\lambda)\nonumber\\
&&-\mu_2(\mu_2(\lambda,\gamma_1),\gamma_2)+\mu_2(\mu_2(\lambda,\gamma_2),\gamma_1) ~,\label{eq:homotopy_relation_g}\\
0&\!\!=\!\!&-\mu_3(\mu_1(\chi_1),\chi_2,\gamma)-\mu_3(\mu_1(\chi_2),\chi_1,\gamma)-\mu_2(\mu_2(\chi_1,\chi_2),\gamma)\nonumber\\
&&+\mu_2(\mu_2(\chi_1,\gamma),\chi_2)+\mu_2(\mu_2(\chi_2,\gamma),\chi_1)~.\label{eq:homotopy_relation_h}
\end{eqnarray}
The map $\mu_3$ is compatible with the map $\mu_2$ in the obvious way up to correction terms given by $\mu_4$:
\begin{equation}\label{eq:homotopy_relation_i}
 \begin{aligned}
 \mu_1(\mu_4(\gamma_1,&\gamma_2,\gamma_3,\gamma_4))+\mu_2(\mu_3(\gamma_1,\gamma_2,\gamma_3),\gamma_4)-\mu_2(\mu_3(\gamma_1,\gamma_2,\gamma_4),\gamma_3)\\
 &+\mu_2(\mu_3(\gamma_1,\gamma_3,\gamma_4),\gamma_2)-\mu_2(\mu_3(\gamma_2,\gamma_3,\gamma_4),\gamma_1)=\\
 &\mu_3(\mu_2(\gamma_1,\gamma_2),\gamma_3,\gamma_4))+\mu_3(\mu_2(\gamma_2,\gamma_3),\gamma_1,\gamma_4))+\mu_3(\mu_2(\gamma_3,\gamma_4),\gamma_1,\gamma_2))\\
 &+\mu_3(\mu_2(\gamma_1,\gamma_4),\gamma_2,\gamma_3))
 -\mu_3(\mu_2(\gamma_1,\gamma_3),\gamma_2,\gamma_4))-\mu_3(\mu_2(\gamma_2,\gamma_4),\gamma_1,\gamma_3))~.\\
  \end{aligned}
\end{equation}
\begin{equation}\label{eq:homotopy_relation_j}
 \begin{aligned}
\mu_2(\mu_3(\gamma_1,&\gamma_2,\gamma_3),\chi)-\mu_2(\mu_3(\chi,\gamma_1,\gamma_2),\gamma_3)+\mu_2(\mu_3(\chi,\gamma_1,\gamma_3),\gamma_2)\\
  &-\mu_2(\mu_3(\chi,\gamma_2,\gamma_3),\gamma_1)-\mu_4(\mu_1(\chi),\gamma_1,\gamma_2,\gamma_3)=\\
 &-\mu_3(\mu_2(\gamma_1,\gamma_2),\chi,\gamma_3))-\mu_3(\mu_2(\gamma_2,\gamma_3),\chi,\gamma_1))-\mu_3(\mu_2(\chi,\gamma_3),\gamma_1,\gamma_2))\\
 &-\mu_3(\mu_2(\chi,\gamma_1),\gamma_2,\gamma_3))
 +\mu_3(\mu_2(\gamma_1,\gamma_3),\chi,\gamma_2))+\mu_3(\mu_2(\chi,\gamma_2),\gamma_1,\gamma_3))~.
 \end{aligned}
\end{equation}
Finally, the map $\mu_4$ satisfies the following compatibility relation\footnote{Note that the total antisymmetrization is here equivalent to merely considering unshuffles in definition \eqref{eq:homotopyJacobi}.}
\begin{equation}\label{eq:homotopy_relation_k}
 \begin{aligned}
 \tfrac{1}{2}\mu_2(\mu_4(\gamma_{[1},\gamma_2,\gamma_3,\gamma_4),\gamma_{5]})+\mu_3(\mu_3(\gamma_{[1},&\gamma_2,\gamma_3),\gamma_4,\gamma_{5]})+\mu_4(\mu_2(\gamma_{[1},\gamma_2),\gamma_3,\gamma_4,\gamma_{5]})=0~.
 \end{aligned}
\end{equation}
\end{subequations}

A simple example of a Lie 3-algebra is that of the {\em Chern-Simons Lie 3-algebra} $\frcs_k(\frg)$ of a simple Lie algebra $\frg$, where $k\in\FR$ denotes the {\em level}. The graded vector space is $L=\FR[-2]\oplus(\FR\oplus \frg)[-1]\oplus \frg$, and we will denote elements of these spaces by $\lambda,\binom{\lambda}{\gamma}$ and $\gamma$, respectively. The non-vanishing higher products are defined as
\begin{equation}
 \begin{aligned}
  \mu_1(\lambda):=\binom{\lambda}{0}&~,~~\mu_1\binom{\lambda}{\gamma}:=\gamma~,~~  \mu_2(\gamma_1,\gamma_2):=[\gamma_1,\gamma_2]~,\\
\mu_2\left(\gamma_1,\binom{\lambda}{\gamma_2}\right)&:=\binom{k\langle\gamma_1,\gamma_2\rangle}{[\gamma_1,\gamma_2]}~,~~
  \mu_2\left(\binom{\lambda_1}{\gamma_1},\binom{\lambda_2}{\gamma_2}\right):=2k\langle\gamma_1,\gamma_2\rangle~,\\
  \mu_3(\gamma_1,\gamma_2,\gamma_3)&:=k\langle\gamma_1,[\gamma_2,\gamma_3]\rangle~,
 \end{aligned}
\end{equation}
where $\langle\cdot,\cdot\rangle$ denotes the Killing form on $\frg$. In the following, we will discuss some special Lie 3-algebras that will later serve as examples for the gauge structure of the (1,0)-model.

\subsection{Semistrict Lie 2-algebras and string Lie 2-algebras}\label{ssec:Lie_2_algebras}

General semistrict Lie 2-algebras are obtained by considering Lie 3-algebras with trivial $L_{-2}$. This reduces the non-trivial products \eqref{eq:products_Lie_3} to the following ones:
\begin{equation}\label{eq:products_Lie_2}
 \begin{aligned}
  \mu_1:L_{-2}\rightarrow L_{-1}~,~~~\mu_1:L_{-1}\rightarrow L_{0}~,\hspace{2cm}\\
  \mu_2:L_{0}^{\wedge 2}\rightarrow L_0~,~~~\mu_2:L_{0}\wedge L_{-1}\rightarrow L_{-1}~,~~~
  \mu_3:L_0^{\wedge 3}\rightarrow L_{-1}~,
 \end{aligned}
\end{equation}
while the higher Jacobi relations reduce in an obvious manner. 

Let us specialize a little further. A semistrict Lie 2-algebra is called {\em skeletal}, if isomorphic objects are equivalent. This amounts to setting $\mu_1=0$. A nice class of skeletal semistrict Lie 2-algebras is obtained from a Lie algebra $\frg$, a vector space $V$ carrying a representation $\rho$ of $V$ and a Lie algebra cocycle with values in $V$, $c=H^3(\frg,V)$ \cite{Baez:2003aa}. As products on the 2-term complex $V\rightarrow \frg$, we define $\mu_1:=0$, $\mu_2:\frg \times \frg\rightarrow \frg$ as the Lie bracket, $\mu_2:\frg\times V\rightarrow V$ as the action of $\frg$ onto $V$ in the representation $\rho$ and $\mu_3:\frg\times\frg\times\frg\rightarrow V$ is given by the Lie algebra cocycle $c$. 

It is shown in \cite{Baez:2003aa} that isomorphism classes of such data $(\frg,V,\rho,c)$ defining semistrict Lie 2-algebras are equivalent to isomorphism classes of general skeletal semistrict Lie 2-algebras. Moreover, any general semistrict Lie 2-algebras is categorically equivalent to a skeletal one, and therefore the data $(\frg,V,\rho,c)$ can be used to classify semistrict Lie 2-algebras. 

Particularly interesting is the {\em string Lie 2-algebra} of a simple Lie algebra $\frg$, which is defined by the data $(\frg,\FR,\rho,c)$, where $\rho$ is the trivial representation and $c(g_1,g_2,g_3):=k\left<{\rm ad}(g_1),{\rm ad}([g_2,g_3])\right>$, for $k\in\FR$, is a Lie algebra cocycle arising from the Killing form $\left<\cdot,\cdot\right>$ of $\frg$.

\subsection{(1,0) gauge structures and semistrict Lie 3-algebras}\label{ssec:(1,0)fromLie3}

Consider a (1,0) gauge structure with $\sfg=\sfb=0$. As we saw before in section \ref{ssec:Courant_Dorfman_algebras}, such a (1,0) gauge structure is equivalent to a Courant-Dorfman algebra. It is easy to verify that a Courant-Dorfman algebra $(\CR,\CE,\llbracket\cdot,\cdot\rrbracket)$ gives rise to a semistrict Lie 2-algebra with
\begin{equation}
 L_{-1}=\CR=\frh\eand L_0=\CE=\frg
\end{equation}
as well as higher products
\begin{equation}
\begin{aligned}
\mu_1(r)&:=\CD r=\sfh(r)~,\\\mu_2(e_1,e_2)&:=\llbracket e_1,e_2\rrbracket=-\sff(e_1,e_2)~,\\
\mu_2(e,r)&:=\tfrac{1}{2}\langle e,\CD r\rangle=\sfd(e,\sfh(r))~,\\\mu_3(e_1,e_2,e_3)&:=-\tfrac{1}{2}\langle e_{[1},\llbracket e_2,e_{3]}\rrbracket\rangle=\sfd(e_{[1},\sff(e_2,e_{3]}))~,
\end{aligned}
\end{equation}
where $e,e_1,e_2,e_3\in\CE$ and $r\in\CR$. In the special case of Courant algebroids, this observation was already made in \cite{Roytenberg:1998vn}.\footnote{As a side remark, note that a Courant-Dorfman algebra with the Dorfman bracket, which is not antisymmetric but satisfies the Jacobi identity, can be regarded as a {\em hemistrict} Lie 2-algebra, cf.\ \cite{Roytenberg:0712.3461}.}

Inversely, many interesting Lie 2-algebras do not form (1,0) gauge structures. For example, consider the Lie 2-algebra based on the octonions with $L_{-1}=L_0=\mathbbm{O}$, where $\mu_2$ is given by the commutator and $\mu_3$ is given by the Jacobiator. In this case, the Jacobiator cannot be written as $\sfd(\cdot,[\cdot,\cdot])$ for any symmetric map $\sfd:\mathbbm{O}\odot\mathbbm{O}\rightarrow\mathbbm{O}$.

For (1,0) gauge structures with $\sfg$ and $\sfb$ nontrivial, the situation is more involved. We evidently start from the chain complex
\begin{equation}\label{eq:chain_complex2}
 L_{-2}=\frg^*\ \xrightarrow{~~\mu_1:=\sfg~~}\ L_{-1}=\frh\ \xrightarrow{~~\mu_1:=\sfh~~}\ L_0=\frg
\end{equation}
together with the maps
\begin{equation}
 \begin{aligned}
 \mu_1(\lambda):=\sfg(\lambda)~,~~~\mu_1(\chi):=\sfh(\chi)\eand \mu_2(\gamma_1,\gamma_2):=-\sff(\gamma_1,\gamma_2)~.
 \end{aligned}
\end{equation}
The higher homotopy relations \eqref{eq:homotopy_relation_a}-\eqref{eq:homotopy_relation_f} then define the remaining products up to terms in the kernels of $\sfg$ and $\sfh$, where the latter turn out to lie in the image of $\sfg$:
\begin{equation}
 \begin{aligned}
  \mu_2(\gamma,\chi)&=\sfd(\gamma,\sfh(\chi))+\sfg(\phi_1(\gamma,\chi))~,\\
  \mu_2(\gamma,\lambda)&=\phi_1(\gamma,\sfg(\lambda))+\phi_2(\gamma,\lambda)~,&\phi_2(\gamma,\lambda)&\in\kernel\,\sfg\\
  \mu_2(\chi_1,\chi_2)&=\sfb(\chi_{(1},\sfh(\chi_{2)}))+2\phi_1(\sfh(\chi_{(1}),\chi_{2)})+\phi_3(\chi_1,\chi_2)~,&\phi_3(\chi_1,\chi_2)&\in\kernel\,\sfg\\
  \mu_3(\gamma_1,\gamma_2,\gamma_3)&=\sfd(\gamma_{[1},\sff(\gamma_2,\gamma_{3]}))+\sfg(\phi_4(\gamma_1,\gamma_2,\gamma_3))~,\\
  \mu_3(\chi,\gamma_1,\gamma_2)&=-\tfrac{2}{3}\sfb(\sfd(\gamma_{[1},\sfh(\chi)),\gamma_{2]})+2\phi_1(\gamma_{[1},\sfd(\gamma_{2]},\sfh(\chi)))\\
  &\hspace{1.5cm}+2\phi_1(\gamma_{[1},\sfg(\phi_1(\gamma_{2]},\chi)))+\phi_1(\sff(\gamma_1,\gamma_2),\chi)\\
  &\hspace{1.5cm}-\phi_4(\sfh(\chi),\gamma_1,\gamma_2)+\phi_5(\chi,\gamma_1,\gamma_2)~,&\phi_5(\chi,\gamma_1,\gamma_2)&\in\kernel\,\sfg
 \end{aligned}
\end{equation}
Equation \eqref{eq:homotopy_relation_i} defines $\mu_4(\gamma_1,\gamma_2,\gamma_3,\gamma_4)$ in a similar way. The challenge is now to fix the $\phi_i$ such that the remaining homotopy relations \eqref{eq:homotopy_relation_g}, \eqref{eq:homotopy_relation_h}, \eqref{eq:homotopy_relation_j} and \eqref{eq:homotopy_relation_k} are satisfied.

A detailed analysis using a computer algebra program suggests that in general, there are no such $\phi_i$ and one has to impose additional constraints onto the (1,0) gauge structure. We understand these constraints as a hint that the (1,0) gauge structure needs to be extended, and there are two possibilities for such extensions. First, the extensions discussed briefly in section \ref{ssec:Bianchi_Extension}, which result in an extended (1,0) gauge structure forming a Lie $n$-algebra with $n>3$. Second, one can extend the chain complex \eqref{eq:chain_complex} to an exact sequence, leading to a Lie 4-algebra. We will discuss this extension briefly in the next section.

But first, let us try to turn the (1,0) gauge structure into a Lie 3-algebra. There is a large number of possible constraints that do this, many of which involve the shifted-graded Jacobi identity for $\sfb$ and $\sfd$ given in equation \eqref{eq:angular_graded_jacobi}. Here we only want to study one. Because we considered the extreme case where $\sfg=0$ (as well as $\sfb=0$) before, let us now turn to the opposite extreme and impose the condition that the kernel of $\sfg$ is trivial. In this case, the maps $\phi_2,\phi_3$ and $\phi_5$ are trivial, and we put
\begin{equation}
 \phi_1(\gamma,\chi):=\alpha_1\sfb(\chi,\gamma)~,~~~\alpha_1\in\FR\eand \phi_4(\gamma_1,\gamma_2,\gamma_3)=0~.
\end{equation}
The map $\mu_4(\gamma_1,\gamma_2,\gamma_3,\gamma_4)$ is given by
\begin{equation}
 \mu_4(\gamma_1,\gamma_2,\gamma_3,\gamma_4)=-2(1+2\alpha_1)\sfb(\sfd(\gamma_{[1},\sff(\gamma_2,\gamma_3)),\gamma_{4]})~.
\end{equation}
If the kernel of $\sfg$ is trivial, these maps satisfy all the homotopy relations \eqref{eq:homotopy_relations} and thus form a semistrict Lie 3-algebra. 

There are two interesting choices for $\alpha_1$. First, the choice  $\alpha_1=-\tfrac{1}{2}$ gives 
\begin{equation}
\mu_2(\gamma,\chi)=\tfrac{1}{2}\rho(\gamma)\acton \chi~,~~\mu_2(\gamma,\lambda)=\tfrac{1}{2}\rho(\gamma)\acton \lambda\eand\mu_4=0~.
\end{equation}

Second, with the choice $\alpha_1=-1$ the curvatures $\CF$ and $\CH$ defined in \eqref{eq:curvatures} can be rewritten in the form 
\begin{equation}
\begin{aligned}
\CF&=\dpar  A+\tfrac{1}{2}\mu_2(A,A)+\mu_1(B)~,\\
\CH&=\dpar  B+ \mu_2(A,B)+\tfrac{1}{6}\mu_3(A,A,A)+\mu_1(C)~,
\end{aligned}
\end{equation}
provided we assume that the {\em fake curvature condition} $\CF=0$ is satisfied. This condition is very natural from the point of view of higher gauge theory, and we will return to it in section \ref{ssec:highergaugetheory}. Note that the Chern-Simons term in $\CH$ collapsed into Lie 3-algebra products. The above form for $\CH$ has been suggested in the context of semistrict higher gauge theory in \cite{Zucchini:2011aa}.

Moreover, demanding that both fake curvatures $\CF$ and $\CH$ vanish and that the graded Jacobi identity \eqref{eq:angular_graded_jacobi} is satisfied, we find that all products in the gauge transformations \eqref{eq:shiftqauge} can be written in terms of Lie 3-algebra products as follows:
\begin{equation}
\begin{aligned}
\delta A=&~\dpar  \alpha+\mu_2(A,\alpha)-\mu_1(\Lambda)~,\\
\delta B=&~\dpar  \Lambda+\mu_2(B,\alpha)+\mu_2(A,\Lambda)+\tfrac{1}{2}\mu_3(A,A,\alpha)-\mu_1(\Xi)~,\\
\delta  C=&~\dpar  \Xi+\mu_2(C,\alpha)+\mu_2(B,\Lambda)+\mu_2(A,\Xi)-\tfrac{1}{2}\mu_3(A,A,\Lambda)+\mu_3(B,A,\alpha)\\&\hspace{1cm}+\tfrac{2}{3}\mu_4(A,A,A,\alpha).
\end{aligned}
\end{equation}
We regard this as a good starting point for studying semistrict higher gauge theory based on Lie 3-algebras. As far as we are aware, this has yet to be developed.

Note however that several terms remain in the supersymmetry transformations and equations of motion which are not of the form of Lie 3-algebra products.

\subsection{Strong homotopy Lie algebras from resolutions of Lie algebras}

Demanding that $\sfg$ is injective is a first step towards turning the chain complex \eqref{eq:chain_complex} underlying the (1,0) gauge structure into an exact sequence. On such sequences, there is a canonical construction of strong homotopy Lie structures \cite{Barnich:1997ij}, as we briefly review in the following. Consider a resolution of a vector space $\frg_0$. That is, consider an exact sequence of vector spaces
\begin{equation}\label{eq:resolution}
 \cdots \xrightarrow{~\mu_1~} L_{-2} \xrightarrow{~\mu_1~} L_{-1} \xrightarrow{~\mu_1~} L_0\xrightarrow{~\mu_1~} \frg_0\xrightarrow{~\mu_1~} 0~.
\end{equation}
Because the sequence is exact, we can decompose $L_0=\frb\oplus \frg_0'$ where $\frb={\rm ker}(\mu_1)$ and $\frg_0'\cong \frg_0$. Assume now that there is a skew-symmetric bilinear map
\begin{equation}
 \mu_2:L_0\times L_0 \rightarrow L_0~,
\end{equation}
which satisfies for all $\ell\in L_0$ and $b\in\frb$ the following two properties:
\begin{conditions}
 \item[(i)] $\mu_2(\ell,b)\in\frb$,
 \item[(ii)] $\mu_2(\mu_2(\ell_1,\ell_2),\ell_3)-\mu_2(\mu_2(\ell_1,\ell_3),\ell_2)+\mu_2(\mu_2(\ell_2,\ell_3),\ell_1)\in \frb$.
\end{conditions}
	  
Then, as shown in \cite{Barnich:1997ij}, the map $\mu_2$ can be extended to a Lie bracket on $\frg_0$ and further to a strong homotopy Lie algebra on all of $L=L_\bullet$. First, one extends $\mu_2$ to all of $L_\bullet$ by showing that
\begin{equation}\label{eq:ext_cond_1}
 \mu_1(\mu_2(\mu_1(\ell_1\otimes \ell_2)))=0~,~~\mbox{for}~ \ell_1,\ell_2\in L_\bullet~.
\end{equation}
As the complex \eqref{eq:resolution} is exact, this equation implies $\mu_2(\mu_1(\ell_1\otimes \ell_2))=\mu_1(\ell_3)$ for some $\ell_3$, and we can define $\mu_2(\ell_1,\ell_2):=\ell_3$. Starting from $\mu_2$ on $L_0\times L_0$, one can iteratively define $\mu_2$ for all higher $L_n$. Note that for $\ell_1,\ell_2\in L_0$, \eqref{eq:ext_cond_1} follows from axiom (i), otherwise one can calculate it using the iteratively defined $\mu_2$. 

For higher products, we use the same method, applied to the corresponding higher Jacobi relations. For example, to define $\mu_3$, we use that
\begin{equation}\label{eq:ext_cond_2}
 \mu_1\big(\mu_3(\mu_1(\ell_1),\ell_2,\ell_3)\pm\mu_2(\mu_2(\ell_2,\ell_3),\ell_1)\pm\mu_2(\mu_2(\ell_1,\ell_2),\ell_3)\pm\mu_2(\mu_2(\ell_1,\ell_3),\ell_2)\big)=0~,
\end{equation}
where the signs are to be chosen according to the gradings of $\ell_1,\ell_2$ and $\ell_3$. Again, for $\ell_1,\ell_2,\ell_3\in L_0$, \eqref{eq:ext_cond_2} follows from axiom (ii), otherwise one can calculate it using the iteratively defined $\mu_3$. Together with the exactness of \eqref{eq:resolution} we thus have
\begin{equation}
 \mu_3(\mu_1(\ell_1),\ell_2,\ell_3)\pm\mu_2(\mu_2(\ell_2,\ell_3),\ell_1)\pm\mu_2(\mu_2(\ell_1,\ell_2),\ell_3)\pm\mu_2(\mu_2(\ell_1,\ell_3),\ell_2)=\mu_1(\ell_4)~,
\end{equation}
for some $\ell_4$, which leads us to define $\mu_3(\ell_1,\ell_2,\ell_3):=\ell_4$.

For a (1,0) gauge structure with $\sfb$ and $\sfg$ trivial, we consider the exact sequence
\begin{equation}\label{eq:chain_complex_3}
 0 \longrightarrow \frh \xrightarrow{~~\sfh~~} \frg \xrightarrow{~~{\rm proj}~~} \frg_0\longrightarrow 0~,
\end{equation}
which induces a splitting $\frg={\rm im}\sfh\oplus \frg_0$. As shown e.g.\ in \cite[sec. 3]{Samtleben:2012mi}, $\frg_0$ forms a Lie algebra with Lie bracket given by $-\sff|_{\frg_0}$. If we now follow the above construction, we recover precisely the Lie 2-algebra structure of a (1,0) gauge structure with $\sfb$ and $\sfg$ trivial: besides $\mu_1(\chi)=\sfh(\chi)$, we have the following higher products:
\begin{equation}
 \mu_2(\gamma_1,\gamma_2)=-\sff(\gamma_1,\gamma_2)~,~~\mu_2(\gamma,\chi)=\sfd(\gamma,\sfh(\chi))~~\mbox{and}~~ \mu_3(\gamma_1,\gamma_2,\gamma_3)=\sfd(g_1,\sff(g_2,g_3))~.
\end{equation}

Assuming that $\sfg$ has trivial kernel and that $\im(\sfg)=\kernel(\sfh)$, we can extend the exact sequence \eqref{eq:chain_complex_3} to 
\begin{equation}\label{eq:chain_complex_3}
 0 \longrightarrow \frg^* \xrightarrow{~~\sfg~~} \frh \xrightarrow{~~\sfh~~} \frg \xrightarrow{~~{\rm proj}~~} \frg_0\longrightarrow 0~.
\end{equation}
The above construction then recovers the Lie 3-algebra that we derived in the previous section with $\alpha_1=0$.

Note that more generally, if $\im(\sfg)=\kernel(\sfh)$, we obtain the exact sequence
\begin{equation}\label{eq:chain_complex_4}
 0 \longrightarrow \ker(\sfg)\longhookrightarrow \frg^* \xrightarrow{~~\sfg~~} \frh \xrightarrow{~~\sfh~~} \frg \xrightarrow{~~{\rm proj}~~} \frg_0\longrightarrow 0~,
\end{equation}
and correspondingly a Lie 4-algebra via the above construction. 

Finally, even if $\im(\sfg)\varsubsetneq\kernel(\sfh)$, we can construct an extension of the map $\sfg:\frg^*\rightarrow \frh$ to a map $\tilde{\sfg}:\frg^*\oplus \fra \rightarrow \frh$ for some vector space $\fra$ such that $\im(\tilde{\sfg})=\kernel(\sfh)$. Then the exact sequence
\begin{equation}\label{eq:chain_complex_4}
 0 \longrightarrow \ker(\tilde{\sfg})\longhookrightarrow \frg^*\oplus\fra \xrightarrow{~~\tilde{\sfg}~~} \frh \xrightarrow{~~\sfh~~} \frg \xrightarrow{~~{\rm proj}~~} \frg_0\longrightarrow 0
\end{equation}
yields again a Lie 4-algebra.

Since higher gauge theory has not been developed for Lie 4-algebras, our subsequent discussion has to remain restricted to (1,0) gauge structures that form Lie 3-algebras.

\section{Examples}

\subsection{Abelian gerbe} 

Our first example is the simplest, that of an abelian gerbe, cf.\ \cite{Hitchin:1999fh}. If we take the vector spaces 
\begin{equation}
0\longrightarrow\fru(1)\longrightarrow0~, 
\end{equation}
and set all the maps to zero, we are left with just the (1,0) tensor multiplet $(\phi,\chi,B)$ satisfying the equations of motion
\begin{equation}
\CH=\dpar  B=*\CH~,~~\slasha{\dpar}\chi=0\eand\Box \phi=0~,
\end{equation}
and transforming under the usual gauge transformation for an abelian gerbe 
\begin{equation}
\delta B=\dpar  \Lambda~.
\end{equation}

The supersymmetry transformations become
\begin{equation}\label{eq:susy2}
\begin{aligned}
\delta \phi&=\epsb\chi~,~~~
\delta \chi^i&=\tfrac{1}{8}\slasha{\CH}\eps^i+\tfrac{1}{4}\slasha{\dpar}\phi~\eps^i~,~~~
\delta B&=-\epsb\gamma^{(2)}\chi~,
\end{aligned}
\end{equation}
which match the full $(2,0)$ supersymmetry transformations for a single M5-brane \cite{Howe:1997fb} when reduced to a contained (1,0) multiplet.

\subsection{Field redefinitions for higher gauge theory}\label{ssec:highergaugetheory}

We will briefly perform some field redefinitions for the equations describing higher gauge theory in the previous chapter. We will also take the infinitesimal form of the gauge transformations. This will allow us to make contact with the gauge transformations of the (1,0)-gauge structure.

We redefine the fields $B$ and $H$ with factors of $-1$ to give
\begin{equation}\label{eq:DefOfFH}
 F := \dpar  A+\tfrac{1}{2}[A, A]\eand H:=\nabla B:=\dpar  B+A\acton B~.
\end{equation}
with the fake curvature condition
\begin{equation}\label{eq:fake_curvature}
 \CF:=\ F + \sft(B)=0~.
\end{equation}

For the infinitesimal gauge transformations we will use $g=\de^\alpha$ and ignore higher order terms. The gauge transformations become
\begin{equation}\label{eq:Space-time-GT}
\begin{aligned}
\delta A&=\dpar  \alpha+[A,\alpha]-\sft(\Lambda)~,\\
\delta B&=\dpar  \Lambda +A\acton \Lambda-\alpha\acton B~.
\end{aligned}
\end{equation}
Note that the curvature \eqref{eq:curvatureF10} of the (1,0) model has to be identified with the fake curvature $\CF$, as it is the only two-form curvature built from $A$ and $B$ that transforms covariantly.

In the case of principal 3-bundles, we also redefine $\Xi$ and $\CH$ with factors of $-1$, to give
\begin{equation}\label{eq:fake_curvature_3}
 \CF :=F+\sft(B)=0\eand \CH:=H+\sft(C)=0~,
\end{equation}
with gauge transformations
\begin{equation}\label{eq:Space-time-GT2}
\begin{aligned}
\delta A&=\dpar  \alpha+[A,\alpha]-\sft(\Lambda)~,\\
\delta B&=\dpar  \Lambda+A\acton \Lambda-\alpha\acton B -\sft(\Xi)~,\\
\delta C&=\dpar \Xi+A\acton\Xi-\alpha\acton C-\{B,\Lambda\}-\{\Lambda,B\}~.
\end{aligned}
\end{equation}

Let us stress here that the fake curvature condition $\CF=0$ is {\em not} stable under supersymmetry transformations \eqref{eq:susy} in general. Therefore, whenever we impose the fake curvature condition in the following, we implicitly break supersymmetry. A way out of this problem would be to impose, in addition, the equations arising from a supersymmetry variation of the fake curvature condition, as well as further equations arising from supersymmetry variations of the latter.

Note that in the models arising from twistor constructions, the fake curvature condition is indeed invariant under the corresponding supersymmetry transformations. 

\subsection{Principal 2-bundles}

To obtain differential crossed modules from a (1,0) gauge structure, we set $\sfg=\sfb=0$ and assume 
\begin{equation}\label{eq:Jacobi}
\begin{aligned}
\sfd(\sff(\gamma_{[1},\gamma_2),\gamma_{3]})=0~.
\end{aligned}
\end{equation}
This ensures that $\sff(\cdot,\cdot)$ is a Lie bracket on $\frg$ by \eqref{eq:algebra_relation_e} and corresponds to setting $\mu_3=0$ on the Lie 2-algebra level, making it a strict Lie 2-algebra. Nontrivial such (1,0) gauge structures are very restricted, but can indeed be constructed, e.g.\ by using the analysis in \cite[sec.\ 3]{Samtleben:2012mi}.

Now to obtain a differential crossed module we define 
\begin{equation}
\begin{aligned}
\sft:=\sfh~,~~~~[\gamma_1,\gamma_2]:=-\sff(\gamma_1,\gamma_2)\eand\gamma\acton\chi:=\sfd(\gamma,\sfh(\chi))~.
\end{aligned}
\end{equation}
Note that this is a differential crossed module with abelian $\frh$ since $[\chi_1,\chi_2]=\sft(\chi_{[1})\acton \chi_{2]}=\sfd(\sfh(\chi_{[1}),\sfh(\chi_{2]}))=0$, by \eqref{eq:Pfeif} and the symmetry of $\sfd$.

Note also that for $\sfg=0$ the vector multiplet equations of motion \eqref{eq:eom2} become trivial, and we can therefore eliminate the degrees of freedom by enforcing the fake curvature condition \eqref{eq:fake_curvature} of higher gauge theory:
\begin{equation}
\begin{aligned}
\CF=\dpar  A-\tfrac{1}{2}\sff(A,A)+\sfh(B)=0~.
\end{aligned}
\end{equation}
Using \eqref{eq:Jacobi}, the shifted form of the (1,0) gauge transformations \eqref{eq:shiftqauge} becomes
\begin{equation}
\begin{aligned}
\delta A&=\dpar  \alpha-\sff(A,\alpha)-\sfh(\Lambda)~,\\
\delta B&=\dpar  \Lambda+\sfd(A,\sfh(\Lambda))-\sfd(\alpha,\sfh(B))~,
\end{aligned}
\end{equation}
which matches exactly the higher gauge theory gauge transformations \eqref{eq:Space-time-GT}.

One of the most interesting classes of differential crossed modules is that of the 3-algebras appearing in the context of M2-brane models, cf.\ chapter \ref{ch:high}. However these are not included in the above discussion since they have a trivial map $\sft=0$ and a non trivial action $\acton$. Since the maps above were defined by $\gamma\acton\chi:=-\sfd(\gamma,\sfh(\chi))$ and $\sft:=\sfh$, a trivial map $\sft$ implies a trivial action. Luckily 3-algebras can be treated separately, and we will come back to them shortly.

\subsection{Principal 3-bundles}

Higher gauge theory has been developed not only for principal 2-bundles but also for principal 3-bundles, which have differential 2-crossed modules as underlying structure Lie 3-algebras. For this section we assume first that the products corresponding to Lie 3-algebra products $\mu_3$ and $\mu_4$ are zero:  
\begin{equation}\label{eq:Jacobi2}
 \begin{aligned}
\sfd(\gamma_{[1},\sff(\gamma_2,\gamma_{3]}))=0~,~~~\sfb(\sfd(\gamma_{[1},\sff(\gamma_2,\gamma_3)),\gamma_{4]})=0~&,\\
\sfb(\chi,\sff(\gamma_1,\gamma_2))-\tfrac{4}{3}\sfb(\sfd(\gamma_{[1},\sfh(\chi)),\gamma_{2]})+2\sfb(\sfg(\sfb(\gamma_{[1},\chi)),\gamma_{2]})&=0~,\\
 \end{aligned}
\end{equation}
and second that terms of the form $\sfb(\sfg(\cdot),\sfh(\cdot))$ vanish. These terms are in the kernel of $\sfg$ and are therefore expected to vanish, as discussed in section \ref{ssec:(1,0)fromLie3}. 

To obtain differential 2-crossed modules from (1,0) gauge structures we define
\begin{equation}
\begin{aligned}
\sft&(\lambda):=\sfg(\lambda),~\sft(\chi):=\sfh(\chi),~[\gamma_1,\gamma_2]:=-\sff(\gamma_1,\gamma_2),~\gamma\acton\chi:=\sfd(\gamma,\sfh(\chi))-\sfg(\sfb(\chi,\gamma))~,\\
[&\chi_1,\chi_2]=[\lambda_1,\lambda_2]=0,~\{\chi_1,\chi_2\}:=\tfrac{1}{2}\sfb(\chi_{1},\sfh(\chi_{2}))\eand \gamma\acton\lambda:=-\sfb(\sfg(\lambda),\gamma)~.
\end{aligned}
\end{equation}
Note that this is a differential 2-crossed module with abelian $\frl$ and $\frh$.

To reduce to principal 3-bundles, we have to impose the vanishing of the fake curvatures
\begin{equation}
\begin{aligned}
\CF&=\dpar  A-\tfrac{1}{2}\sff(A,A)+\sfh(B)=0~,\\
\CH&=\dpar  B+ 2\sfd(A,\sfh(B))-\sfg(\sfb(B,A))+\sfd(A,\dpar  A-\tfrac{1}{3}\sff(A,A))+\sfg(C)\\&=\dpar  B+ \sfd(A,\sfh(B))-\sfg(\sfb(B,A))+\sfg(C)=0~.
\end{aligned}
\end{equation}

This simplifies the shifted gauge transformations \eqref{eq:shiftqauge} to
\begin{equation}
\begin{aligned}
\delta A&=\dpar  \alpha-\sff(A,\alpha)-\sfh(\Lambda)~,\\
\delta B&=\dpar  \Lambda+\sfd(A,\sfh(\Lambda))+\sfg(\sfb(\Lambda,A))-\sfd(\alpha,\sfh(B))+\sfg(\sfb(B,\alpha))-\sfg(\Xi)~,\\
\delta  C&=\dpar  \Xi-\sfb(\sfg(\Xi),A)+\sfb(\sfg(C),\alpha)-\sfb(B,\sfh(\Lambda)) +\dots~,
\end{aligned}
\end{equation}
which match exactly the higher gauge theory transformations \eqref{eq:Space-time-GT2}.

The constraints \eqref{eq:Jacobi2} are again very restrictive. One admissible example is the Chern-Simons Lie 3-algebra of $\fru(1)$, which we will discuss in section \ref{ssec:chern}. If we are just interested in the algebraic structure and not in matching the gauge transformations to higher gauge theory, we can discuss many more interesting examples. 

\subsection{Representations of Lie algebras and M2-brane model 3-algebras}\label{ssec:onezero3algebras}

Let $\fra$ be a semi-simple Lie algebra with a representation $\rho$ acting on a vector space $V$. There are three types of models based on this information, as discussed in \cite{Samtleben:2012mi}; here we will just discuss the simplest one. An action is not possible for this type, however the type admitting an action is closely related.

We take the complex
\begin{equation}
0\longrightarrow V\longrightarrow V\times\fra~,
\end{equation}
and choose the maps 
\begin{equation}
 \begin{aligned}
\sfg=\sfb=&0~,~~\sfh(v)=\binom{v}{ 0}~,\\
\sfd\left(\binom{v_1}{ g_1},\binom{v_2}{ g_2}\right)&=\tfrac{1}{2}(\rho(g_1)\acton v_2+\rho(g_2)\acton v_1)~,~~\\
\sff\left(\binom{v_1}{ g_1},\binom{v_2}{ g_2}\right)&=\binom{\tfrac{1}{2}(\rho(g_2)\acton v_1-\rho(g_1)\acton v_2)}{[g_1,g_2]}~,
 \end{aligned}
\end{equation}
for $v\in V,~\binom{v_i}{g_i}\in V\times\fra$.

Recall that metric 3-algebras are obtained from metric Lie algebras with faithful orthogonal representations via the Faulkner construction \cite{deMedeiros:2008zh}, where the representation space is  the 3-algebra itself, $V=\CA$, and the Lie algebra is the associated Lie algebra of inner derivations 
 $\fra=\frg_\CA$. 

In order to use this relation we need to endow the (1,0) gauge structure with metrics on the spaces $\fra$ and $V$ which are invariant under the action of $\fra$. Explicitly, this construction gives the triple bracket
\begin{equation}\label{eq:triple}
[v_1,v_2,v_3]=\sfd(\sfm^*_\fra( \sfd^*(\sfm_\frh(v_1),\sfh(v_2))),\sfh(v_3))~,
\end{equation}
where $\sfm^*_\fra:\fra^*\rightarrow\fra$  and $\sfm_\frh:\frh\rightarrow\frh^*$ are maps induced from the metrics on $\fra$ and $\frh$, respectively. 

The simplest non-trivial example is that of $A_4$. We choose the fundamental representation of $\fra=\aso(4)$ acting on $V=\FR^4$, along with the standard euclidean metric on $\FR^4$ and a split signature metric on $\aso(4)$, explicitly:\begin{equation}
\begin{aligned}
\sfm_{\aso(4)}(A^{\pm})=\pm \big(A^{\pm}\big)^T~&,~~\sfm_{\FR^4}(v)=v^T~,\\
\sfd\left(\binom{v}{A},\binom{w}{B}\right)=\tfrac{1}{2}(A.w+B.v)~&,~~\sfd^*\left(v^T,\binom{w}{A}\right)=\tfrac{1}{2}\binom{v^T. A}{w v^T- vw^T}~,
\end{aligned}
\end{equation}
for $v,w\in \FR^4,~A,B\in\aso(4)$ and where $v^T$ denotes the transpose of $v$ and $A^\pm$ denote the selfdual and anti-selfdual parts of $A$. Then \eqref{eq:triple} gives the triple bracket on the basis vectors $e^\mu\in\FR^4$ as
\begin{equation}
[e^\mu,e^\nu,e^\rho]=\eps^{\mu\nu\rho\sigma}e^\sigma~.
\end{equation}

Similarly, the 3-algebra describing $N$ M2-branes in the ABJM model corresponds to the choice $\fra=\fru(N)\times\fru(N)$, with split signature metric
\begin{equation}
\sfm^*_\fra\binom{A_L}{A_R}=\binom{A_L^\dagger}{-A_R^\dagger}~,
\end{equation}
and where $V=\agl(N,\FC)$ is the bi-fundamental representation with the standard Hilbert-Schmidt metric $\sfm_\frh(A)=A^\dagger$. The triple bracket then becomes
\begin{equation}
[A,B;C]=\sfd(\sfm^*_\fra( \sfd^*(\sfm_\frh(A),\sfh(C))),\sfh(B))=AC^\dagger B-B C^\dagger A~.
\end{equation}
For $N=2$ this essentially coincides with the 3-Lie algebra $A_4$. 

We can now rewrite equations \eqref{eq:eom} in terms of the products appearing in 3-algebras. Note however a crucial difference here to the M2-brane models: the gauge field of M2-brane models lives only in $\fra$ and not in $V\times \fra$ and also that the gauge transformations have only one ($\fra$-valued) parameter. 

There is also a reason for which the algebraic structure describing M5-branes should be different from M2-branes. As we have seen in chapter \ref{ch:background}, when describing infinitely many M2-branes, the space of functions on a three-manifold was used. Similarly, the space of functions on a two manifold was used to describe infinitely many D-branes. The D-p-branes then merge into a single D-(p+2)-brane, with two scalar fields being redefined as two gauge fields. Similarly, the BLG model with the 3-algebra based on functions on a three-manifold gives the action for a single M5-brane \cite{Ho:2008nn}. If M5-branes had the same algebraic structure as M2-branes, then one would expect that functions on a three-manifold could again be used, which would then describe an M8-brane, which we know from supergravity does not exist. I would like to speculate that the algebraic structure relevant to M5-branes may then be related to a quantization of $S^5$. 

\subsection{Vectors in $\sG\times \sG$} 

Another example is found in \cite{Chu:2011fd}, where the $\sG\times \sG$-model is conjectured to describe the gauge sector of M5-brane dynamics. This conjecture passes many consistency checks, including selfdual string profiles which match gravity dual predictions \cite{Chu:2013hja}. One key difference between the $\sG\times \sG$-model and the (1,0) model is that in the former, the vector fields are on shell and that they are related to the tensor fields in a way reminiscent of the fake curvature condition \eqref{eq:fake_curvature}. Nevertheless, the algebraic structure is an example of a $(1,0)$ gauge structure with matching gauge transformations. In our notation, the vector spaces present are
\begin{equation}
0\longrightarrow\frg\longrightarrow\frg\times\frg~,
\end{equation}
where $\frg$ is a Lie algebra with Lie bracket $[\cdot,\cdot]$. We will use the notation $A=\binom{A_L}{A_R}$ and $\alpha=\binom{\alpha_L}{\alpha_R}$ to denote one-forms and functions taking values in $\frg\times\frg$. The gauge transformations take the following form:
\begin{equation}\label{eq:chu}
 \begin{aligned}
\delta A&=\dpar  \alpha+\binom{[A_L,\alpha_L+\alpha_R]}{[A_R,\alpha_L+\alpha_R]}+\binom{\Lambda}{-\Lambda}~,\\
\delta B&=\dpar  \Lambda+\tfrac{1}{2}[A_L+A_R,\Lambda]+\tfrac{1}{2}([A_R,\dpar \alpha_L]-[A_L,\dpar \alpha_R])+[B,\alpha_L+\alpha_R]~,
 \end{aligned}
\end{equation}
where, as before, wedge products are implied, e.g. $[A_L,\Lambda]=[A_L{}_{\mu},\Lambda_{\nu}]\dd x^\mu\wedge\dd x^\nu$ .

To make contact with the (1,0) gauge structure transformations \eqref{eq:qauge} we set $\sfg=\sfb=0$ and introduce the new shift of gauge parameters $(\alpha,\Lambda)\rightarrow(\alpha,\Lambda+2\sfd(\alpha,A))$. Using \eqref{eq:algebra_relation_a}, we obtain
\begin{equation}\label{eq:chu2}
\begin{aligned}
\delta A&=\dpar  \alpha -\sff(A,\alpha)-\sfh(\sfd(A,\alpha))-\sfh(\Lambda)~,\\
\delta B&=\dpar  \Lambda+\sfd(A,\sfh(\Lambda)-\dpar  \alpha)-2\sfd(\alpha,\sfh(B))~.
\end{aligned}
\end{equation}

With the following choice of maps\footnote{A slightly different set of maps, which satisfy the constraints \eqref{eq:algebra_relations}, was given in \cite{Samtleben:2011fj}. These, however, do not lead to the gauge transformations of \cite{Chu:2011fd}.}
\begin{equation}
 \begin{aligned}
\sfh(g)&=\binom{-g}{ g}~,~~\sfd\left(\binom{g_1}{ g_2},\binom{g_3}{ g_4}\right)=\tfrac{1}{2}([g_1,g_4]+[g_3,g_2])~,~~\\&\sff\left(\binom{g_1}{ g_2},\binom{g_3}{ g_4}\right)=\binom{-[g_1,g_3]-\tfrac{1}{2}([g_1,g_4]-[g_3,g_2])}{-[g_2,g_4]-\tfrac{1}{2}([g_1,g_4]-[g_3,g_2])}~,
 \end{aligned}
\end{equation}
the shifted gauge transformations \eqref{eq:chu2} match \eqref{eq:chu} exactly.

Since this $\sff$ does not satisfy the Jacobi identity, this is not a differential crossed module. However since the above (1,0) gauge structure has trivial maps $\sfg$ and $\sfb$, it is an example of a (semistrict) Lie 2-algebra.

\subsection{String Lie 2-algebras}

Another interesting Lie 2-algebra related to M-theory dynamics is $\mathfrak{string}$, or the string Lie 2-algebra \cite{Baez:2003aa}, defined in section \ref{ssec:Lie_2_algebras}. A Lie algebra $\frg$ is put into the complex

\begin{equation}
0\longrightarrow\FR\longrightarrow\frg~,
\end{equation}
and the Lie bracket and Killing form $\left<\cdot,\cdot\right>$ correspond to the maps
\begin{equation}
\begin{aligned}
 \sfg=\sfb=\sfh=0~,~~\sff(\gamma_1,\gamma_2):=-[\gamma_1,\gamma_2]\eand\sfd(\gamma_1,\gamma_2)
=\left<\gamma_1,\gamma_2\right>~.
\end{aligned}
\end{equation}

This model describes an abelian tensor multiplet sourced by a non-abelian vector multiplet. It was originally found in \cite{Bergshoeff:1996qm} and it provided crucial inspiration for the development of the (1,0) superconformal models of \cite{Samtleben:2011fj}. The equations of motion \eqref{eq:eom} now read as
\begin{equation}
\begin{aligned}
\CH^-&=-\left<\lambdab,\gamma^{(3)}\lambda\right>~,\\
\slasha{\dpar} \chi^i&=\left<\slasha{\CF},\lambda^i\right>+2\left<Y^{ij},\lambda_j\right>~,\\
\dpar^2 \phi&=2\left<Y^{ij},Y_{ij}\right>-*2\left<\CF,*\CF\right>-4\left<\lambdab,\slasha{\dpar} \lambda\right>~,
\end{aligned}
\end{equation}
where the field strengths are
\begin{equation}
\begin{aligned}
\CF&=&\dpar  A+\tfrac{1}{2}[A,A]\eand
\CH&=&\dpar  B+\left<A,\dpar  A+\tfrac{1}{3}[A,A]\right>~.
\end{aligned}
\end{equation}
The gauge and supersymmetry transformations can be easily read off from \eqref{eq:qauge} and \eqref{eq:susy}.

\subsection{Chern-Simons Lie 3-algebra}\label{ssec:chern}

In the Chern-Simons Lie 3-algebra $\frcs_k(\frg)$ of a simple Lie algebra $\frg$, the map $\mu_1:L_{-1}\rightarrow L_0$ is surjective. This map should be identified with the map $\sfh$ in a (1,0) gauge structure, and because of \eqref{eq:algebra_relation_d}, this implies that $\sff=0$. We therefore have to restrict ourselves to abelian $\frg$. The Chern-Simons Lie 3-algebra $\frcs_k(\FR)$ consists of the complex 
\begin{equation}\label{eq:chain_complex_chern_simons_3_algebra}
\FR\longrightarrow\FR\times\FR\longrightarrow\FR~,
\end{equation}
with the following non trivial products
\begin{equation}
  \mu_2\left(\gamma_1,\binom{\lambda}{\gamma_2}\right):=\binom{k\gamma_1\gamma_2}{0}\eand
  \mu_2\left(\binom{\lambda_1}{\gamma_1},\binom{\lambda_2}{\gamma_2}\right):=2k\gamma_1\gamma_2~.
\end{equation}
Note that the chain complex \eqref{eq:chain_complex_chern_simons_3_algebra} forms an exact sequence. By the identification of (1,0) gauge structures and Lie 3-algebras based on exact sequences we set
\begin{equation}
\begin{aligned}
 \sfg(\lambda):=\binom{\lambda}{0}~,~~~&\sfh\binom{\lambda}{\gamma}:=\gamma~,~~~\sff=0~,\\
 \sfd(\gamma_1,\gamma_2):=\binom{k\gamma_1\gamma_2}{0}~,~~~&\sfb\left(\binom{\lambda_1}{\gamma_1},\gamma_2\right):=2k\gamma_1\gamma_2~.
\end{aligned}
\end{equation}

The field strengths of $A$ and $B=\binom{B_L}{B_R}$ then read explicitly as
\begin{equation}
\begin{aligned}
\CF&=&\dpar  A+B_R\eand
\CH&=&\dpar  B+\binom{kA\wedge\dpar  A+C}{0}~.
\end{aligned}
\end{equation}
The gauge and supersymmetry transformations become
\begin{equation}
\begin{aligned}
\delta A&=\dpar  \alpha -\Lambda_R~,\\
\delta B&=\dpar  \Lambda+\binom{k A\wedge(\dpar  \alpha -\Lambda_R)-2k\alpha\CF-\Xi}{0}~,\\
\delta  C&=\dpar  \Xi+2k(\dpar  \alpha\wedge B_R+\Lambda_R\wedge\dpar  A+\alpha\CH)~,
\end{aligned}
\end{equation}
and
\begin{equation}\nonumber
\begin{aligned}
\delta \phi&=\epsb\chi~,~~~&\delta  Y^{ij}&=-\epsb^{(i}\slasha{\dpar} \lambda^{j)}+2 \epsb^{(i}\chi^{j)}_R~,\\
\delta \chi^i&=\tfrac{1}{8}\slasha{\CH}\eps^i+\tfrac{1}{4}\slasha{\dpar}\phi~\eps^i-\tfrac{k}{2}\binom{*(\gamma\lambda^i\wedge*\epsb\gamma\lambda)}{0}~,~~~&\delta \lambda^i&=\tfrac{1}{4}\slasha{\CF}\eps^i-\tfrac{1}{2}Y^{ij}\eps_j+\tfrac{1}{4}\phi_R\eps^i~,\\
\delta B&=-k\binom{A\wedge\epsb \gamma \lambda}{0}-\epsb\gamma^{(2)}\chi~,~~~&\delta A&=-\epsb \gamma \lambda~,
\end{aligned}
\end{equation}
\begin{equation}
 \delta C~=~-2k~(~B_R\wedge\epsb \gamma\lambda+\phi_R\epsb\gamma^{(3)}\lambda~)~,
\end{equation}
while the equations of motion read as
\begin{equation}
\begin{aligned}
\CH^-&=-\binom{k\lambdab\gamma^{(3)}\lambda}{0}~,\\
\slasha{\dpar} \chi^i&=k\binom{\slasha{\CF}\lambda^i+2Y^{ij}\lambda_j-3\phi_R\lambda^i}{0}~,\\
D^2 \phi&=2k\binom{Y^{ij}Y_{ij}-*(\CF\wedge*\CF)-2\lambdab\slasha{\dpar} \lambda+2\chib_R\lambda+8\lambdab\chi_R-\tfrac{3}{2}\phi_R^2}{0}~,
\end{aligned}
\end{equation}
\begin{equation}
\begin{aligned}
\phi_R Y^{ij}+2\chib_R^{(i}\lambda^{j)}&=0~,\\
4k(\phi_R\CF+2\chib_R\gamma^{(2)}\lambda)&=*\CH^{(4)}~,\\
\phi_R\slasha{\dpar} \lambda_i+\tfrac{1}{2}\slasha{\dpar}\phi_R \lambda_i&=\tfrac{1}{2}\slasha{\CF}\chi_{Ri}+\tfrac{1}{4}\slasha{\CH}_R\lambda_i-\chi_R^jY_{ij}+\tfrac{3}{2}\phi_R\chi_R~,
\end{aligned}
\end{equation}
where we used the notiation $\phi=\binom{\phi_L}{\phi_R}$ for fields $\phi\in\FR\times \FR$. Note that the field equations all remain interacting.

\subsection{Extreme Courant-Dorfman algebras}

Finally, let us briefly comment on the example of extreme Courant-Dorfman algebras with either $\sfh=0$ or $\sfd=0$ for which $\frg$ is a Lie algebra. In the first case, $\frg$ is a Lie algebra endowed with an invariant quadratic form over $\frh$. Here, we obtain a free (1,0) vector multiplet together with a tensor multiplet in the background of this vector multiplet. Furthermore, the tensor multiplet fields do not interact among each other; all interactions arise from source terms containing exclusively fields of the vector multiplet.

In the second case $\sfd=0$, $\frg$ is a Lie algebra over $\frh$ and $\sfh$ is a derivation with values in the center of $\frg$. The definitions of $\CF$ and $\CH$ correspond to the fake curvature and the curvature 3-form of a principal 2-bundle with strict structure 2-group. The action of the covariant derivative becomes trivial on $\frh$, and we obtain an abelian free tensor multiplet together with a free vector multiplet.

\chapter{Selfdual string and higher instanton solutions}
\label{ch:SDS}

In this chapter, explicit solutions to the selfdual string equation are presented in the context of higher gauge theory. In particular, we consider a spherically symmetric ansatz that is a rather straightforward generalizations of the 't Hooft-Polyakov monopole. As we expect a close link to M2-brane models, we base our ansatz on the differential crossed module corresponding to $A_4$. It turns out that this ansatz can be solved, and the scalar field of the selfdual string configuration can be classified by integer winding numbers, just as the scalar field of the $\sSU(2)$ monopole.

Motivation for the study of elementary selfdual string solutions stems from our goal to establish an ADHMN-like construction of selfdual strings. The related twistor constructions were given in \cite{Saemann:2012uq,Saemann:2013pca}, making it reasonable to expect the existence of such a construction. Given a potential ADHMN-like construction of selfdual strings, it is only natural to ask for an analogue of the ADHM construction, which would yield solutions to the selfduality equations in six dimensions. For lack of a better name, we will call such solutions higher instantons. Again, a twistor description of higher instantons was given in \cite{Saemann:2012uq,Saemann:2013pca}. To develop an ADHM-like construction, a good understanding of the elementary solutions to the higher instanton equation is crucial.

Using an ansatz closely related to the BPST instanton, we manage to find explicit higher instanton solutions which can be continued to solutions on a large region in the conformal compactification of six-dimensional Minkowski space. In fact, our solutions are invariant under an action of $\sSO(1,5)$ and share many of the properties of the BPST instanton.

\section{Selfdual strings}

There are various proposals for a non-abelian generalization of the selfdual string equation \eqref{eq:SelfDualString}, which should describe configurations involving $N\ge 2$ M5-branes. In this section, we will review the equations arising in the context of higher gauge theory and compare them to other recent proposals.

We start from a pair of Lie algebras $\frh$ and $\frg$ forming a differential crossed module. The non-abelian selfdual string equation then reads as 
\begin{equation}\label{eq:NonAbSelfDualString}
H:=\dd B+A\acton B=\star(\dd\Phi+A\acton \Phi)~,
\end{equation}
which was first suggested in \cite{Saemann:2012uq}, where also a construction mechanism for solutions was developed using a twistor approach. In the canonical description of higher gauge theory, the so-called \emph{fake curvature condition}
\begin{equation}\label{eq:fc}
\begin{aligned}
\CF:=\dd A+\tfrac{1}{2}[A,A]-\sft(B)=0
\end{aligned}
\end{equation}
is imposed. This equation guarantees that the parallel transport is consistent and it eliminates additional degrees of freedom from the potential one-form. In the following, we will not impose the fake curvature condition, but recall from section \ref{ssec:2crossmod} that we can embed into an inner derivation 2-crossed module such that the fake curvature conditions \eqref{eq:fake_curvature_conditions} hold.

Differential crossed modules are equivalent to strict Lie 2-algebras. Generalizing to the semistrict case, we obtain 2-term $L_\infty$-algebras, see $\cite{Baez:2002jn}$. The effect of this for the selfdual string equation would be an additional term 
\begin{equation}\label{eq:semi}
H:=\dd B+\mu_2(A, B)+\tfrac{1}{3!}\mu_3(A,A,A)=\star(\dd\Phi+\mu_2(A, \Phi))~,
\end{equation}
where $\mu_i$ are antisymmetric maps satisfying the homotopy Jacobi identities of the $L_\infty$-algebra.

For the special case of a differential crossed module corresponding to a 3-Lie algebra, the selfdual string equation \eqref{eq:NonAbSelfDualString} also arises as the BPS equation in the Lambert-Papageorgakis $\CN=(2,0)$ model in section \ref{ssec:tensormultiplet}. This model came with an additional vector field $C^\mu$. A selfdual string solution for this model should also satisfy the equations of motion
\begin{equation}\label{eq:(2,0)eom}
D^\mu D_\mu\Phi=0,~D(H_{\mu\nu\kappa},C^{\kappa})=F_{\mu\nu},~D_\mu C^\nu=D(C^\mu,C^\nu)=C^\mu D_\mu\Phi=C^\mu D_\mu H_{\nu\kappa\lambda}=0~.
\end{equation}

Another equation arises from the (1,0) superconformal models of the previous chapter. For this discussion we will set the gauge potential three-form $C$ and the maps $\sfb$ and $\sfg$ to zero and identify the map $\sfh=\sft$. The BPS equations in the (1,0) superconformal model reduced to $\FR^4$ then read as

 \begin{align}
H:=\dd B+ 2\sfd(A,\sft(B))+\sfd(A,\dd A-\tfrac{1}{3}\sff(A,A))&=\star(\dd\Phi+2\sfd(A,\sft(\Phi))~,\label{eq:(1,0)BPS}\\
\CF=\star\CF~,~~~~
\sft(\Phi)&=0~.
 \end{align}

The equation of motion, which is not implied by \eqref{eq:(1,0)BPS} alone, is 
\begin{equation}\label{eq:(1,0)eom}
D^2\Phi=\star\sfd(\CF,\star\CF)~,
\end{equation}
where $D^2:=\star D\star D$ with $D:=\dd+2\sfd(A,\cdot)$ and $\CF=\dd A-\tfrac{1}{2}\sff(A,A)+\sft(B)$. Note that the selfdual string equation \eqref{eq:(1,0)BPS} implies $D^2\Phi=\star\sfd(\CF,\CF)$.

Due to the large overlap with higher gauge theory, equation \eqref{eq:(1,0)BPS} agrees with \eqref{eq:NonAbSelfDualString} or \eqref{eq:semi} for the right choice of vector spaces and maps. The solutions presented below therefore also yield solutions to \eqref{eq:(1,0)BPS}. We will come back to this in section \ref{ssec:comments}. 

In another approach \cite{Chu:2012um}, one direction is singled out (as is common in many descriptions of M5-branes) and an additional relation connecting the curvature and potential two-forms is imposed, which is strongly reminiscent of the fake curvature condition. In the case of selfdual strings, this reads as
\begin{equation}\label{eq:GGconstraint}
F_{ij}=c\int\dd x^4 \dpar_4 B_{ij}~,
\end{equation}
where $i,j=1,\ldots,3$ and $c\in\FR$ is some fixed constant.

All fields live in the same Lie algebra\footnote{or a differential crossed module of the form $\frg\overset{\sft}{\rightarrow}\frg$} and the selfdual string equation reads
\begin{equation}\label{eq:GGeqn}
H:=\dd B+[A, B]=\star(\dd\Phi+[A, \Phi])~.
\end{equation}
This equation is invariant under the gauge transformations
\begin{equation}
\delta A=\dd  \alpha+[A,\alpha]~,~~~\delta B=\Sigma-[\alpha, B]~,~~~\delta \Phi=-[\alpha, \Phi]~,
\end{equation}
where $\Sigma$ is a two-form satisfying $\dd\Sigma+[A,\Sigma]=0$. 

\subsection{Previously constructed solutions}

Before presenting the solutions, we will briefly comment on solutions to the equations \eqref{eq:(1,0)BPS} and \eqref{eq:GGeqn} given previously.

In \cite{Akyol:2012cq,Akyol:2013ana}, solutions to the tensor hierarchy BPS equations \eqref{eq:(1,0)BPS} had been constructed. In the solutions corresponding to selfdual strings, however, the $B$-field was always put to zero. The explicit solution given in \cite{Akyol:2012cq} contains a $\au(1)$-valued scalar field $\Phi$ and an $\asu(2)$-valued one-form potential $A$. The solution is $\sSO(4)$-invariant and everywhere regular. Also, similarly to the 't Hooft-Polyakov monopole, the potential one-form $A$ can be gauged away at large radius by turning on a potential two-form $B$ and leaving the abelian Howe-Lambert-West selfdual string \cite{Howe:1997ue} with Higgs field: $\Phi=\frac{\di}{|x|^2}$.

Solutions to \eqref{eq:GGeqn} similar to Wu-Yang monopoles were constructed in \cite{Chu:2012rk}. These solutions were interpreted as corresponding to $N=2$ M2-branes and were generalized to the case $N>2$ in \cite{Chu:2013hja}, where all fields took values in $\asu(N)$. This class of solutions passes certain consistency checks, in particular the M2-brane spike profiles match supergravity predictions \cite{Chu:2012rk,Chu:2013hja}. These solutions, however, remain singular at the position of the selfdual string.

In \cite{Chu:2013gra}, a construction algorithm was given that turned an $\asu(N)$ monopole solution into a solution to the equations \eqref{eq:GGeqn}. The solution constructed from the 't Hooft-Polyakov monopole is a unit charge non-singular selfdual string, but lacks $\sSO(4)$ invariance. The construction also involved choosing a function with certain asymptotic behavior. In this sense the solution is not unique. This situation is similar to our non-singular and $\sSO(4)$-invariant selfdual string solution presented in the following section.

\subsection{$A_4$ non-singular selfdual strings}\label{ssec:A4solSDS}

We now come to a generalization of the 't Hooft-Polyakov monopole to a selfdual string solution based on differential crossed modules. The first issue here is to find the pair of Lie algebras describing our solution. The scalar field of the 't Hooft-Polyakov solution itself, $\Phi=e_i x^i f(r)$, where $f$ is some radial function, suggests a four-dimensional vector space with basis $e_\mu$, allowing for a scalar field $\Phi\propto e_\mu x^\mu$ for the selfdual string. This already leads us to the use of the 3-Lie algebra $A_4$. Now, having fixed the gauge structure, it remains to make an $\sSO(4)$-invariant ansatz for a solution to the selfdual string equation \eqref{eq:NonAbSelfDualString}. Inspired by the $\sSO(3)$-invariant 't Hooft-Polyakov monopole solution \eqref{eq:thooft}, we set
\begin{equation}\label{eq:SDS}
\begin{aligned}
\Phi&=\frac{ e_\mu x^\mu }{|x|^3}~f(\xi)~,\\
B_{\mu\nu}&=\eps_{\mu\nu\kappa\lambda}\frac{e_\kappa x^\lambda}{|x|^3}~g(\xi)~,\\
A_\mu&= \eps_{\mu\nu\kappa\lambda}  D(e_\nu,e_\kappa) ~\frac{x^\lambda}{|x|^2} ~h(\xi)~,
\end{aligned}
\end{equation}
where $\xi:=v|x|^2$ is a dimensionless parameter, $e_\mu$ are the generators of $\frh=A_4$ and $D(e_\mu,e_\nu)\in \frg=\frg_{A_4}$ are inner derivations. We will now seek solutions with non-singular $|\Phi(x)|$ and asymptotic behavior $|\Phi|\sim v-|x|^{-2}$.

\begin{figure}[h!]
\center
\begin{picture}(400,100)
\includegraphics[width=40mm]
{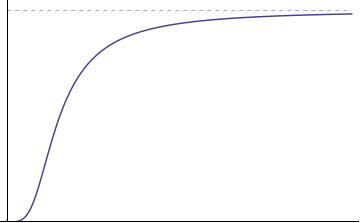}~~~~~~~~\includegraphics[width=40mm]{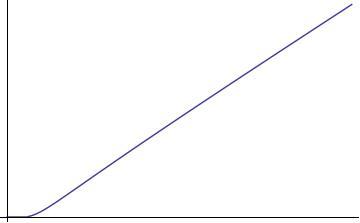}
~~~~~~~~\includegraphics[width=40mm]{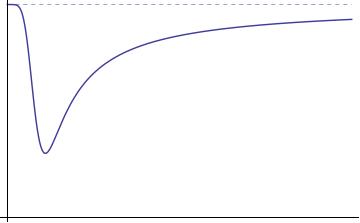}
\put(-158.0,-7.0){\makebox(0,0)[c]{$|x|$}}
\put(-408.0,67.0){\makebox(0,0)[c]{$v$}}
\put(-116.0,69.0){\makebox(0,0)[c]{$\tfrac{1}{2}$}}
\put(-13.0,-7.0){\makebox(0,0)[c]{$|x|$}}
\put(-303.0,-7.0){\makebox(0,0)[c]{$|x|$}}
\put(-323.0,57.0){\makebox(0,0)[c]{$|\Phi|$}}
\put(-181.0,57.0){\makebox(0,0)[c]{$g$}}
\put(-22.0,54.0){\makebox(0,0)[c]{$h$}}
\end{picture}
 \caption{The qualitative radial behavior of the scalar field $\Phi$ and the functions $g$ and $h$ appearing in the potentials $B$ and $A$ for a selfdual string solution \eqref{eq:SDS} with any of the $f(\xi)$ in \eqref{eq:choices_f}. Note that $g(\xi)\sim \sqrt{\xi}$ for large $\xi$.}
 \label{fig:SDSA4solution}
\end{figure}
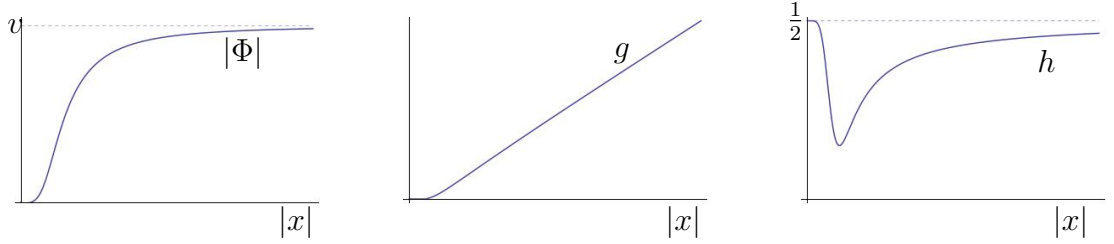

The above ansatz reduces the selfdual string equation to the following ODEs:
\begin{equation}
\begin{aligned}
f(\xi)+\tfrac12 g(\xi)-g(\xi)h(\xi)-\xi f'(\xi)&=0~,\\
f(\xi)-2f(\xi)h(\xi)-\tfrac23\xi g'(\xi)&=0~.
\end{aligned}
\end{equation}
Note that $h(\xi)$ appears only algebraically. Assuming that $g(\xi)$ vanishes only at isolated points, we can combine the above equations into a single ODE for $f(\xi)$ and $g(\xi)$:
\begin{equation}
 f(\xi)^2-\xi f(\xi)f'(\xi)+\tfrac{1}{3}\xi g(\xi)g'(\xi)=0~.
\end{equation}
The fact that we arrive at a single ODE for two functions shows that our ansatz was underconstraint. This gives us the freedom to choose a function $f$ such that $\Phi$ has the correct asymptotic behavior $|\Phi|\sim v-|x|^{-2}$, which implies $f(\xi)\sim \xi$ at infinity. Convenient choices satisfying this property are e.g.\
\begin{equation}\label{eq:choices_f}
\begin{aligned}
f(\xi)&=\xi\ {\rm coth}(\xi)-1~,\\
f(\xi)&=\xi-1+\frac{2}{\pi  }\tan^{-1}\left(\frac{2}{\pi \xi}\right)~,\\
f(\xi)&= \xi\left(1-\frac{1}{1+\xi}\right)~.
\end{aligned}
\end{equation}
Moreover, we can choose an initial value for $g$ such that $g(0)=0$. The analytical expressions for $g(\xi)$ and $h(\xi)$ can be computed, but their analytical form does not provide further insight. For example, for the third choice in \eqref{eq:choices_f}, we have
\begin{equation}
 \begin{aligned}
  g(\xi)&=\sqrt{15+6\left(\xi-\frac{5+6\xi}{2(1+\xi)^2}-3 \log(1+\xi)\right)}~,\\
  h(\xi)&=\frac{1}{2}-\frac{\xi^2}{3(1+\xi)\sqrt{\tfrac{1}{3}\xi(6+\xi(9+2\xi))-2(1+\xi)^2\log(1+\xi)}}~.
 \end{aligned}
\end{equation}
The qualitative behavior resulting from any of the choices for $f(\xi)$ is displayed in figure \ref{fig:SDSA4solution}.
 
\subsection{Matrix representation of $A_4$ and hermitian 3-algebras}

The differential crossed module $A_4\rightarrow \frg_{A_4}$ can be represented in terms of matrices in the following way:
\begin{equation}
e_\mu:=\frac{1}{\sqrt{2}}\left(\begin{array}{cc} 0 & \sigma_\mu\\ 0 & 0\end{array}\right)\eand
D(e_\mu,e_\nu):=\frac{1}{2}\gamma_5\gamma_{\mu\nu}=\frac{1}{2}\left(\begin{array}{cc}\sigma_{\mu\nu} & 0\\ 0 & -\sigmab_{\mu\nu} \end{array}\right)~,
\end{equation}
where $\sigma_{\mu\nu}:=\sigma_{[\mu}\sigma^\dagger_{\nu]}$ and $\sigmab_{\mu\nu}:=\sigma^\dagger_{[\mu}\sigma_{\nu]}$. Note that we always use weighted antisymmetrization of indices.  The commutator in $\frg_{A_4}$ and the action $\acton$ of $\frg_{A_4}$ onto $A_4$ are just the matrix commutator.

In this notation, the solution \eqref{eq:SDS} becomes
\begin{equation}\label{eq:SDSsol2}
\begin{aligned}
\Phi&=\frac{1}{\sqrt{2}}\left(\begin{array}{cc} 0 & \frac{x}{|x|^3}\\ 0 & 0\end{array}\right)f(\xi)~,\\
B&=-\frac{1}{\sqrt{2}^3|x|^3}\left(\begin{array}{cc} 0 & x\,\dd \xb\wedge\dd x+\dd x\wedge \dd \xb\, x\\ 0 & 0\end{array}\right)g(\xi)~,\\
A&= \frac{1}{2|x|^2}\Im\left(\begin{array}{cc}x\,\dd \xb & 0\\ 0 & -\xb\,\dd x \end{array}\right)h(\xi)~,
\end{aligned}
\end{equation}
and the selfdual string equation becomes
\begin{equation}
H:=\dd B+[A, B]=\star(\dd\Phi+[A, \Phi])~.
\end{equation}

Interestingly, we see that the gauge potential, up to its radial behavior, is a combination of an instanton and an anti-instanton for gauge group $\sSU(2)$. 

\subsection{Topological charges}\label{sec:top}

Recall that the topological charge of an $\sSU(2)$ monopole can be computed in three different ways. Firstly, the asymptotic behavior of $\Phi$ is
\begin{equation}\label{eq:fall-off}
||\Phi||=v-\frac{n}{|x|}+\CO\left(\frac{1}{|x|^{2}}\right) ~~~\mbox{as}~~~|x|\rightarrow\infty\ewith ||\Phi||:=\sqrt{\tfrac{1}{2}\tr(\Phi^\dagger \Phi)}~,
\end{equation}
for some $v\in\FR$ and $n\in\RZ$. 

Secondly, we can use the asymptotic behavior $|\Phi|\sim v$ to impose the following asymptotic gauge condition on $\Phi$:
\begin{equation}\label{eq:su(2)Higgs}
\Phi\sim g^{-1}\left(\begin{array}{cc} v & 0\\ 0& -v \end{array}\right) g~.
\end{equation}
The elements $g\in\sSU(2)$ which leave this expression invariant form the stabilizing group $\sU(1)$. Solutions are therefore classified by an integer topological charge
\begin{equation}
\pi_2(\sSU(2)/\sU(1))\cong\RZ~.
\end{equation}
Thirdly, this charge can be computed as
\begin{equation}\label{eq:top_charge_monopole}
 2\pi n=\tfrac{1}{2}\int_{S^2_\infty}\frac{\tr(F^\dagger \Phi)}{||\Phi||}~,
\end{equation}
where the integral is taken over the sphere at infinity, $S^2_\infty$.

Similarly to the case of magnetic monopoles we may set the asymptotic value of the scalar field $\Phi$ for an $A_4$ selfdual string to a specific matrix, up to a gauge transformation
\begin{equation}\label{eq:asymptotic}
\begin{aligned}
\Phi\sim g\acton\frac{1}{\sqrt{2}}\left(\begin{array}{cccc} 0 &0&\di v&0\\0&0& 0 & \di v\\0&0&0&0\\0&0&0&0\end{array}\right)~~\mbox{as}~|x|\rightarrow\infty~,
\end{aligned}
\end{equation}
where $g\in\sSU(2)\times\sSU(2)$ defines a map: $S^3_\infty\rightarrow \sSU(2)\times\sSU(2)/\sSU(2)$. The stabilizing group $\sSU(2)$ in the denominator is the unbroken symmetry group which leaves the form of $\Phi$ invariant. Since 
\begin{equation}
\begin{aligned}
\pi_3\left( \sSU(2)\times\sSU(2)/\sSU(2)\right)\cong\RZ~,
\end{aligned}
\end{equation}
we find that $A_4$ selfdual strings are indeed classified by an integer charge.

The element $g\in\sSU(2)\times\sSU(2)/\sSU(2)$ in \eqref{eq:asymptotic} that produces solution \eqref{eq:SDSsol2} can be represented by
\begin{equation}
 g=\left(\begin{array}{cc} x & 0 \\ 0 & 1 \end{array}\right)~,
\end{equation}
and we see that our solution has indeed charge 1, as the map $x:\sSU(2)\rightarrow \sSU(2)$ has winding number 1.

A charge formula analogue to \eqref{eq:top_charge_monopole} reads as
\begin{equation}
 (2\pi)^3 n=\tfrac{1}{2}\int_{S^3_\infty}\frac{(H,\Phi)}{||\Phi||}\ewith ||\Phi||:=\sqrt{\tfrac{1}{2}(\Phi,\Phi)}~,
\end{equation}
where $(\cdot,\cdot)$ denotes the Euclidean inner product on $A_4\cong \FR^4$. As the solutions arising from any of our choices \eqref{eq:choices_f} have all the same asymptotic behavior, they all yield the same result $n=1$.

\subsection{Comments on the solution}\label{ssec:comments}

To view this solution as a solution to the $\CN=(1,0)$ BPS equation \eqref{eq:(1,0)BPS}, we embed the gauge field $A$ taking values in $\frg_{A_4}$ into $A_4\oplus\frg_{A_4}$ and set the map $\sfd:(A_4\oplus\frg_{A_4})\odot(A_4\oplus\frg_{A_4})\rightarrow A_4$ to 
\begin{equation}
\sfd\left(\binom{a_1}{b_1},\binom{a_2}{b_2}\right)=\tfrac{1}{2}(b_1\acton a_2+b_2\acton a_1)~,
\end{equation}
for $a_{1,2}\in A_4,~b_{1,2}\in\frg_{A_4}$. If $A$ has no components in $A_4$, equation \eqref{eq:(1,0)BPS} reduces to the selfdual string equation \eqref{eq:NonAbSelfDualString}.

We may now look at the equation of motion
\begin{equation}
D^2\Phi=\star\sfd(\CF,\star\CF)~.
\end{equation}
Our ansatz alone implies $F\acton\star B=0$ and so the equation of motion reduces to $D^2\Phi=0$, which also appears in the Lambert-Papageorgakis equation of motion \eqref{eq:(2,0)eom}. Unfortunately, imposing $D^2\Phi=0$ yields
\begin{equation}
 f(\xi)=\xi~,~~~g(\xi)=0~,~~~h(\xi)=\tfrac{1}{2}~.
\end{equation}
This solution does not have the desired behavior at $\xi=0$ and $\xi=\infty$. Moreover, the field strength $H=\star D \Phi$ vanishes.

We also note that the gauge transformations of \cite{Samtleben:2011fj} allow for components of $A$ to be turned on in $A_4$, which in turn introduces the terms in $H$ involving only $A$ in \eqref{eq:(1,0)BPS}. Exactly the same structure appears in the modified non-abelian gerbes of \cite{Ho:2012nt}, where the fake curvature condition was also dropped.

To solve this issue, note that from the higher gauge theory point of view, the condition $D^2\Phi=0$ should in fact be ``categorified'' to 
\begin{equation}
 D^2\Phi=\sft(\star\{B,B\})~,
\end{equation}
which is implied by the selfdual string equation if we embed the solution into an inner derivation 2-crossed module, cf.\ section \ref{ssec:2crossmod}. 

Starting from a solution $(A_0, B_0)$ to the selfdual string equation based on the differential crossed module $A_4\stackrel{\sft}{\rightarrow}\frg_{A_4}$, we obtain a solution based on the differential 2-crossed module $\frder(A_4\stackrel{\sft}{\rightarrow}\frg_{A_4})$ by letting $A=A_0$, $B=B_0+\dd A+\tfrac{1}{2}[A,A]$ and $C=\dd B_0+[A,B_0]$. Then both the fake curvature conditions \eqref{eq:fake_curvature_conditions} are automatically satisfied.

Note, however, that the selfdual string equation is not invariant under the general gauge transformations \eqref{eq:gauge_transformations}. According to the results of \cite{Saemann:2012uq,Saemann:2013pca}, analogues of adjoint scalar fields in higher gauge theory such as $\Phi$ transform in the same way as the three-form curvature $H$. Moreover, for a covariant derivative to make sense, the possible gauge transformations have to restrict to $H\rightarrow \tilde{H}:=g^{-1}\acton H$. This breaks the gauge symmetries \eqref{eq:gauge_transformations} to a residual symmetry given by triples $(g,\Lambda,\Sigma)$ with
\begin{equation}\label{eq:residual_gauge}
 -(\tilde D+\sft(\Lambda)\acton)\Sigma+\{\tilde B+\tfrac12\tilde{D}\Lambda+\tfrac12[\Lambda,\Lambda],\Lambda\}+\{\Lambda,\tilde B-\tfrac12\tilde{D}\Lambda-\tfrac12[\Lambda,\Lambda]\}=0~.
\end{equation}

This observation is crucial: Because of the simple structure of our differential crossed and 2-crossed module, the solution would be gauge trivial if the gauge symmetries \eqref{eq:gauge_transformations} were not broken. The fact that equations of motion break the general gauge symmetries of higher gauge theory\footnote{which one might regard as the larger gauge symmetries of a flat connective structure} seems not unusual and has been observed previously in \cite{Baez:2012bn}. 

We thus arrive at a solution of the selfdual string equation based on a differential 2-crossed module satisfying both fake curvature conditions \eqref{eq:fake_curvature_conditions}. Such a solution should now have a twistor description in terms of holomorphic 3-bundles as described in \cite{Saemann:2013pca}. This procedure can also be applied to higher instantons, defined in the following section.

\section{Higher instantons}

We will first review ordinary Yang-Mills instantons before discussing higher instantons.

\subsection{Instantons}

Instantons on $\FR^4$ are defined as solutions to the selfduality equation
\begin{equation}\label{eq:instanton}
F=\star F~,
\end{equation}
where the non-abelian curvature $F:=\dd A+\tfrac{1}{2}[A,A]$ takes values in the Lie algebra $\frg$ of some gauge Lie group $\sG$ and vanishes sufficiently rapidly as $|x|\rightarrow\infty$. That is, the curvature becomes pure gauge
\begin{equation}
A\sim g^{-1}\dd g
\end{equation}
as $|x|\rightarrow\infty$ for some $g\in \CC^\infty(\FR^4\backslash \{0\},\sG)$. The function $g$ then defines a map $S^3_\infty \rightarrow \sSU(2)$ with an integer winding number
\begin{equation}
\pi_3(\sSU(2))\cong\RZ~.
\end{equation}
This integer is the {\em instanton number}, which is given by the second Chern number
\begin{equation}\label{eq:instanton_number}
 q=\frac{1}{8\pi^2}\int_{\FR^4} \tr(F^\dagger\wedge F)~.
\end{equation}

Just like monopoles, instanton solutions find a nice interpretation in terms of D-brane configurations. A $q$-instanton with gauge group $\sU(N)$ corresponds to a BPS-configuration of $q$ D0-branes bound to $N$ D4-branes:
\begin{equation}\label{diag:D0D4}
\begin{tabular}{rcccccccc}
& 0 & 1 & 2 & 3 & 4 &  \ldots\\
D0 & $\times$ & & & & \\
D4 & $\times$ & $\times$ & $\times$ & $\times$ &  $\times$
\end{tabular}
\end{equation}
Note that the Bogomolny monopole equation arises from the instanton equation via dimensional reduction. Analogously, this D-brane configuration yields the monopole D-brane configuration \eqref{diag:D1D3} via a T-duality along $x^4$.

\subsection{Basic instanton solutions}\label{ssec:instsols}

There are no abelian instantons on $\FR^4$. This is due to the fact that the fall-off conditions on the gauge potential correspond to a continuation of the instanton configuration from $\FR^4$ to $S^4$. The gauge potential then is the local description of the connection on a principal fiber bundle over $S^4$. Such a bundle is characterized by transition functions on the overlap of the two standard patches on $S^4$, which is contractible to an $S^3$. The transition functions are therefore given by elements of $\CC^\infty(S^3,\sU(1))$ or $\pi_3(S^1)$, which are all trivial. Alternatively, one can readily show that the instanton number \eqref{eq:instanton_number} for an abelian instanton necessarily vanishes.

Let us therefore turn to gauge group $\sSU(2)$. Just as the two-sphere $S^2\cong \CPP^1$ is conveniently described by the usual complex stereographic coordinates, the four-sphere $S^4\cong \FH P^1$ is described by analogous quaternionic stereographic coordinates. In the following, we use the notation
\begin{equation}
\begin{aligned}
x=x^i\sigma_i-\di x^4\unit_2\eand\xb=x^\mu \sigma^\dagger_\mu=x^i \sigma_i+\di x^4\unit_{2}~,
\end{aligned}
\end{equation}
where besides their interpretation as quaternion generators, $\sigma_\mu=(\sigma_i,-\di \unit_{2})$ are the van-der-Waerden symbols appearing in the Clifford algebra of $\FR^4$, which is generated by
\begin{equation}
\begin{aligned}
\gamma_\mu=\left(\begin{array}{cc} 0 &  \sigma^\dagger_\mu\\ \sigma_\mu & 0 \end{array}\right)~.
\end{aligned}
\end{equation}

The BPST instanton \cite{Belavin:1975:85,Atiyah:1979iu}, in regular Landau gauge, reads as
\begin{subequations}\label{eq:BPST_solution}
\begin{equation}
\begin{aligned}
A=\Im\left(\frac{x~\dd \xb}{\rho^2+|x|^2}\right)=\frac{1}{2}\left(\frac{x~\dd \xb}{\rho^2+|x|^2}-\frac{\xb~\dd x}{\rho^2+|x|^2}\right)~,
\end{aligned}
\end{equation}
where $\rho$ is a parameter corresponding to the distance of the D0-brane from the D4-brane\footnote{Taking the D0-brane infinitely far away gives a singular configuration known as the `small instanton'. Inversely, bringing the D0-brane into the worldvolume of the D4-brane yields vanishing curvature and thus no instanton.} and $\Im(M)$ denotes the antihermitian part of a matrix $M$. This gauge potential has the $\asu(2)$-valued curvature
\begin{equation}
\begin{aligned}
F=\rho^2\frac{\dd x\wedge\dd \xb}{(\rho^2+|x|^2)^2}~.
\end{aligned}
\end{equation}
\end{subequations}
Similarly the basic anti-instanton, with charge $q=-1$, has curvature
\begin{equation}
\begin{aligned}
F=\rho^2\frac{\dd \xb\wedge\dd x}{(\rho^2+|x|^2)^2}
\end{aligned}
\end{equation}
and satisfies $F=-\star F$.

Note that the formulas for the gauge potential and its curvature are related to those of the Dirac monopole \eqref{eq:Dirac} by setting $\rho=1$ and replacing quaternionic coordinates by complex stereographic coordinates.

\subsection{Higher instantons}

We define a \emph{higher instanton} as a solution to the six-dimensional selfduality equation
\begin{equation}\label{eq:selfduality_6d}
H=\star H~,~~~H:\dd B+A \acton B~,
\end{equation}
on $\FR^{1,5}$, where $A$ and $B$ are potential one- and two-forms taking values in the Lie algebras of a differential crossed module as before. We furthermore require that the curvature $H$ vanishes as $|x|\rightarrow\CCI$, implying that the solution extends to the conformal compactification of $\FR^{1,5}$. Here, $\CCI$ denotes the boundary of Minkowski space also known as {\em conformal infinity}, consisting of space-like, time-like and light-like infinity, see e.g.\ \cite{Penrose:1986ca} for more details. Comparing with the BPST instanton, we therefore expect that $H$ comes with a coefficient $\frac{1}{(\rho^2+|x|^2)^n}$ with $n\geq 2$.

Because we are dealing with a space with indefinite signature, we cannot expect our solutions to be regular everywhere. The fall-off behavior requires to include the norm of $x\in \FR^{1,5}$, and the expected coefficient $\frac{1}{(\rho^2+|x|^2)^n}$ therefore will yield divergences on a hyperboloid in $\FR^{1,5}$. In a neighborhood of the origin, however, the solutions will remain non-singular. In principle, we could apply a Wick rotation to $\FR^6$, but this would yield complex solutions of $H=\star\di H$.

Solutions to equations closely related to \eqref{eq:selfduality_6d} were previously constructed in \cite{Chu:2013joa}. These equations were interpreted as M-waves and the curvature of the solution's gauge potential one-form was given by an instanton solution.

We will now follow our strategy for selfdual strings and try to find as close an analogue to the BPST solution of instantons as possible. 

\subsection{Elementary higher instanton}

In section \ref{ssec:instsols}, we saw how the expression
\begin{equation}
\frac{\dd x\wedge\dd \xb}{(1+|x|^2)^2}
\end{equation}
appears both in the radially independent part of the Dirac monopole, where $x$ is the complex coordinate on one patch of $\CPP^1$, as well as in the basic instanton, where $x$ is a quaternionic coordinate on one patch of $S^4\cong \FH P^1$. This expression also describes a so-called octonionic instanton on $\FR^8$ when $x$ is an octonion \cite{Fubini:1985jm}. In this section, we will use the analogous selfdual three-forms on $\FR^{1,5}$ to find solutions to the higher instanton equations.

We denote the van-der-Waerden symbols appearing in the Clifford algebra of $\FR^{1,5}$ by $\sigma_M$, $M=0,\ldots,5$. We use the representation given implicitly by
\begin{equation}
  x_{AB}\ =\ x_M\sigma^M_{AB}\ =\ \left(\begin{array}{cccc}
			  0 & x_0+x_5 & -x_3-\di x_4 & -x_1+\di x_2\\
			  -x_0- x_5 & 0 & -x_1-\di x_2 & x_3-\di x_4\\
			  x_3+\di x_4 & x_1+\di x_2 & 0 & -x_0+x_5\\
			  x_1-\di x_2 & -x_3+\di x_4 & x_0-x_5 & 0
			\end{array}\right)~.
\end{equation}
We also define
\begin{equation}
 \xh=(\xh^{AB}):=(\tfrac{1}{2}\eps^{ABCD}x_{CD})~.
\end{equation}
We then have $x^\dagger=-\xh$ and $\xh^\dagger=-x$ and the norm of the vector $x$ is given by
\begin{equation}
 |x|^2=-\tfrac{1}{4}\tr(\xh x)=\sqrt{\det(x)}=\sqrt{\det(\xh)}~.
\end{equation}
Note also that
\begin{equation}
 x^{-1}=\frac{-\xh}{|x|^2}\eand \xh^{-1}=\frac{-x}{|x|^2}~.
\end{equation}
With this convention, the three-forms $\dd\xh\wedge\dd x\wedge\dd \xh$ and $\dd x\wedge\dd\xh\wedge\dd x$ are selfdual and anti-selfdual, respectively. 

As differential crossed module, we consider $\frh\stackrel{\sft}{\rightarrow}\frg$ with $\frh\cong \FR^{1,15}\supset \FR^{1,5}$ and $\frg=\aspin(1,5)$. We use a matrix representation similar to that for $A_4$. That is, we work with block matrices
\begin{equation}
 \left(\begin{array}{cc}M_1 & M_2\\ 0 & M_3\end{array}\right)~,
\end{equation}
where the $M_i$ are $4\times 4$-dimensional complex matrices. Elements of $\frg$ have $M_2=0$ and elements of $\frh$ have $M_1=M_3=0$. 

A first abelian solution of the selfduality equation \eqref{eq:selfduality_6d}, which is singular at the origin $x=0$, is given by the following fields:
\begin{equation}\label{eq:sing}
A=0~,~~~B=\tfrac{\rho^3}{|x|^6}x~\dd  \xh\wedge\dd x~,~~~H=\dd B = \tfrac{\rho^3}{|x|^8}x~ \dd \xh\wedge\dd x\wedge\dd \xh~x~.
\end{equation}

To find true non-abelian solutions of the form $H\sim \dd\xh\wedge\dd x\wedge\dd \xh$ with the right fall-off behavior, we make the following ansatz for the $B$-field:
\begin{subequations}
\begin{equation}
 B=\frac{1}{(\rho^2+|x|^2)^\frac{3}{2}}\left(\begin{array}{cc} 0 &\xh\,\dd x\wedge\dd \xh-\dd\xh\, x\wedge\dd \xh+\dd\xh\wedge \dd x\, \xh\\ 0 & 0\end{array}\right)~.
\end{equation}
Here, the power of the fall-off coefficient $\frac{1}{(\rho^2+|x|^2)}$ is determined by the fact that $B$ has to be dimensionless. Together with the instanton-inspired gauge potential
\begin{equation}
 A=\frac{1}{4(\rho^2+|x|^2)}\left(\begin{array}{cc} \dd \xh\,x-\xh\, \dd x & 0\\ 0 & \dd x\,\xh-x\, \dd \xh\end{array}\right)~,
\end{equation}
we obtain the selfdual three-form curvature
\begin{equation}
 H:= \frac{\rho^2}{(\rho^2+|x|^2)^\frac{5}{2}}\left(\begin{array}{cc} 0 &\dd \xh\wedge\dd x\wedge\dd \xh\\ 0 & 0\end{array}\right)=\star H
\end{equation}
as well as the two-form curvature
\begin{equation}
 F=-\frac{1}{(\rho^2+|x|^2)^2}\left(\begin{array}{cc} \rho^2\,\dd \xh\wedge \dd x+\tfrac{1}{2} \dd \xh\, x\wedge \dd \xh\, x & 0\\ 0 & \rho^2\,\dd x\wedge\dd\xh+\tfrac{1}{2}\dd x\,\xh\wedge \dd x\, \xh\end{array}\right)~.
\end{equation}
\end{subequations}

\subsection{Comments on the higher instanton solution}

As the coefficients controlling the fall-off appear with non-integer powers in $B$ and in particular in $H$, the above solution is only defined for $\rho^2+|x|^2>0$, i.e.\ in the region of $\FR^{1,5}$ containing the origin, which is bounded by the hyperboloid $\rho^2+|x|^2=0$. On the hyperboloid itself, the solution blows up, as expected. Outside of the hyperboloid, the above solution is purely imaginary. Multiplying it by an appropriate root of $-1$ then turns it again into a real solution. 

Note that because of the fall-off behavior of our solution, it extends to the region of the conformal compactification of Minkowski space that consists of the interior of the hyperboloid $\rho^2+|x|^2=0$.

Imposing less stringent conditions on the shapes of $A$, $B$, $H$ and $F$, many more general solutions can be found. In particular, one can replace the antisymmetrizations in the potential one- and two-forms, such as $\dd \xh\,x-\xh\,\dd x$, by more general terms, such as $\alpha_1\dd \xh\,x-\alpha_2\xh\,\dd x$ with constants $\alpha_{1,2}\in \FC$. Selfduality of $H$ then does not fix all the arising constants. The resulting curvatures $H$ and $F$, however, look less natural or symmetric.

Moreover, one easily realizes that our solutions can be `conjugated' to anti-higher instanton solutions satisfying $H=-\star H$. Explicitly, one needs to take the conjugate transpose and apply time-reversal on the fields.

Similarly to the selfdual string solutions, the higher instanton solutions can be embedded into an inner derivation 2-crossed module such that the fake curvature conditions \eqref{eq:fake_curvature_conditions} hold. 
\chapter{Conclusions}
\label{ch:conclusion}

At the start of this thesis, we saw a Nahm-like transform for selfdual strings using loop space. In spite of this success, a description of M5-branes not involving loop space seems more likely. One reason for this is that the relationship with even the well known theory for a single M5-brane remains unclear. In particular, the field $\Phi$ cannot be transgressed to a field on spacetime. Furthermore, much progress has been made in this field which does not use loop space.

Indeed, a transform for infinitely many selfdual strings was demonstrated in chapter \ref{ch:background}, which involved only ordinary spacetime. This was possible since the double-scaling limit rendered the field strength abelian and furthermore, the algebraic structure was simply the 3-algebra of functions on a three-manifold.

The challenge for the scientific community now is to find a full description of multiple M5-branes. Hopefully higher gauge theory will lead to an answer. In chapter \ref{ch:high}, we saw a higher gauge theory based on the Lambert-Papageorgakis $(2,0)$ model. The original model had the problems of only admitting 3-Lie algebras and not having a $B$ field, whereas the higher gauge theory had the problem of having the wrong BPS selfdual string equation. We also reformulated the 3-algebras of M2-brane models as differential crossed modules and differential 2-crossed modules. The differential crossed module corresponding to a 3-algebra $\CA$ with Lie algebra of inner derivations $\frg_\CA$ was
\begin{equation}
\CA\ \rightarrow\ \frg_\CA~,
\end{equation}
whereas the differential 2-crossed module was 
\begin{equation}
\CA\ \rightarrow\ \frg_\CA \ltimes \CA\ \rightarrow\ \frg_\CA~.
\end{equation}
The latter case allowed for the M2-brane models to be written in a way in which the fake curvature conditions held.

In chapter \ref{ch:design}, we analyzed the $\CN=(1,0)$ models of \cite{Samtleben:2011fj}. These models did not impose the fake curvature conditions and the vector spaces used were not Lie algebras. However, under certain conditions, these algebraic structures formed $L_\infty$-algebras or, under further conditions, differential crossed and 2-crossed modules. Perhaps most interestingly, under the conditions that the map $\sfg$ has trivial kernel and that the fake curvature conditions hold, the gauge transformations, although not the supersymmetry transformations, could be written entirely in terms of $L_\infty$-algebra products:
\begin{equation}
\begin{aligned}
\delta A=&~\dpar  \alpha+\mu_2(A,\alpha)-\mu_1(\Lambda)~,\\
\delta B=&~\dpar  \Lambda+\mu_2(B,\alpha)+\mu_2(A,\Lambda)+\tfrac{1}{2}\mu_3(A,A,\alpha)-\mu_1(\Xi)~,\\
\delta  C=&~\dpar  \Xi+\mu_2(C,\alpha)+\mu_2(B,\Lambda)+\mu_2(A,\Xi)-\tfrac{1}{2}\mu_3(A,A,\Lambda)+\mu_3(B,A,\alpha)\\&\hspace{1cm}+\tfrac{2}{3}\mu_4(A,A,A,\alpha)~,
\end{aligned}
\end{equation}
with vanishing fake curvatures
\begin{equation}
\begin{aligned}
\CF&=\dpar  A+\tfrac{1}{2}\mu_2(A,A)+\mu_1(B)=0~,\\
\CH&=\dpar  B+ \mu_2(A,B)+\tfrac{1}{6}\mu_3(A,A,A)+\mu_1(C)=0~.
\end{aligned}
\end{equation}

Finally, chapter \ref{ch:SDS} contained explicit solutions for a selfdual string similar to the 't Hooft-Polyakov monopole based on the 3-Lie algebra $A_4$. We also saw configurations similar to Yang-Mills instantons, called higher instantons. The solutions, however, were not unique, which suggests that these solutions are not the correct descriptions. Perhaps this is not surprising, due to the argument in section \ref{ssec:onezero3algebras}, that the algebraic structures describing M5-branes should not be the same as those describing M2-branes. If they were the same, the trick of using the Nambu 3-algebra of functions on a three-sphere, could possibly be used to describe an infinite number of M5-branes forming an M8-brane, which we know does not exist. This trick works for going from infinitely many M2-branes to an M5-brane \cite{Ho:2008nn,Bagger:2007vi} as well as going from infinitely many D-p-branes to a D-(p+2)-brane, for $-1\le p\le7$, using the Lie algebra of functions on a two-sphere. 

I would like to end this thesis with a conjecture, that the correct algebraic structure describing multiple M5-branes is related to the quantization of $S^4$. This would imply that the functions on a four-sphere could be used to describe infinitely many M5-branes forming an M9-brane, see \cite{deRoo:1997gq} for details on M9-branes. 

Furthermore, if this is the case, perhaps configurations of M5-branes suspended between M9-branes \cite{Bergshoeff:2006bs} could give rise to magnetic domains in five dimensions. This would involve an equation of the form $g=*\dd\phi$ on $\FR^5$, dual to an equation describing multiple M5-branes involving a quaternary bracket.
\appendix
\renewcommand{\chaptermark}[1]{\markboth{Appendix \thechapter.\ #1}{}}
\chapter{3-algebras}\label{app:real3algebras}

In the M-theory generalization of the Nahm equation proposed by Basu and Harvey \cite{Basu:2004ed}, Filippov's 3-Lie algebras \cite{Filippov:1985aa} play a prominent role. A {\em 3-Lie algebra} is a real vector space $\CA$ endowed with a totally antisymmetric, trilinear map $[\cdot,\cdot,\cdot]:\CA^{\wedge 3}\rightarrow \CA$, which satisfies the so-called {\em fundamental identity}:
\begin{equation}\label{eq:fun}
 [a_1,a_2,[b_1,b_2,b_3]]=[[a_1,a_2,b_1],b_2,b_3]+[b_1,[a_1,a_2,b_2],b_3]+[b_1,b_2,[a_1,a_2,b_3]]
\end{equation}
for all $a_1,a_2,b_1,b_2,b_3\in \CA$. Due to this identity, the span of the operators $D(a,b)$, $a,b\in\CA$, which act on $c\in\CA$ according to
\begin{equation}
 D(a,b)\acton c:=[a,b,c]~,
\end{equation}
forms a Lie algebra. We will call this Lie algebra the {\em associated Lie algebra} of $\CA$ and denote it $\frg_\CA$. We can turn $\CA$ into a {\em metric 3-Lie algebra} by introducing a positive definite, non-degenerate symmetric bilinear form $(\cdot,\cdot)$ on $\CA$, which is invariant under actions of $\frg_\CA$:
\begin{equation}
 ([a_1,a_2,b_1],b_2)+(b_1,[a_1,a_2,b_2])=0~.
\end{equation}

Because the only 3-Lie algebras with positive definite metric are $A_4$ (which is the four-dimensional 3-Lie algebra ${\rm span}(e^\mu)$, $\mu=1,\ldots,4$, with 3-bracket $[e^\mu,e^\nu,e^\kappa]=\eps^{\mu\nu\kappa\lambda}e^\lambda$) and direct sums thereof \cite{Nagy:2007aa}, generalizations of the above 3-Lie algebras were soon proposed. Here, we will drop the total antisymmetry (and the word Lie from 3-Lie algebra) and focus on the real and hermitian 3-algebras introduced in \cite{Cherkis:2008qr} and \cite{Bagger:2008se}. Real (and hermitian) 3-algebras are (complex) vector spaces endowed with 3-brackets which are anti-symmetric in their first two slots. The 3-brackets are linear in all their slots except for the third slot of hermitian 3-algebras, which is antilinear. They are required to satisfy the fundamental identity and the metric compatibility condition, which can be written in an intuitive form if we define
\begin{equation}\label{eq:3bra}
[a_1,a_2,a_3]:=D(a_1,a_2)\acton a_3~~\mbox{and}~~[a_3,a_1;a_2]:=D(a_1,a_2)\acton a_3~,
\end{equation}
for real and hermitian 3-algebras, respectively. We also define a complex conjugation $\overline{D(a_1,a_2)}:=-D(a_2,a_1)$, which leaves the inner derivations of real 3-algebras invariant. For both real and hermitian 3-algebras, we can then write the fundamental identity as
\begin{equation}\label{eq:GenFundIdent}
[D(a_1,a_2),D(b_1,b_2)]\acton c=D(D(a_1,a_2)\acton b_1,b_2)\acton c+D(b_1,\overline{D(a_1,a_2)}\acton b_2)\acton c
\end{equation}
and the metric compatibility condition as
\begin{equation}
 (D(a_1,a_2)\acton b_1,b_2)+(b_1,\overline{D(a_1,a_2)}\acton b_2)=0~.
\end{equation}

Using the 3-bracket and the metric on the 3-algebra, a nondegenerate invariant metric on the associated Lie algebra is induced by defining
\begin{equation}
 \lbr D(a_1,a_2),D(a_3,a_4)\rbr:=-(D(a_1,a_2)\acton a_4,a_3)~.
\end{equation}
Note that here, we started from a 3-bracket on a metric 3-algebra $\CA$ and constructed the map $D$ and a metric on $\frg_\CA$. In section \ref{ssec:3fromDCM}, the inverse operation is performed. A differential crossed module is used to construct a map $D$ and hence a 3-bracket.

\chapter{Strong homotopy Lie algebras}\label{app:SH_Lie_algebras}

Recall that a {\em strong homotopy Lie algebra} or {\em $L_\infty$-algebra} is a graded vector space $L=\oplus_n L_n$, equipped with graded antisymmetric multilinear maps
\begin{equation}
 \mu_i:L^{\wedge i}\rightarrow L~,~~~i\geq 1~,
\end{equation}
of degree $2-i$, such that the following higher Jacobi relations are satisfied for each\footnote{Sometimes, a zero-bracket is introduced in addition and $L_\infty$-algebras for which this bracket vanishes (as in our definition) are called `strict'. This nomenclature unfortunately collides with that of a strict $n$-category and we will not use it here.} $m\geq 1$ and homogeneous elements $\ell_1,\ldots,\ell_m$:
\begin{equation}\label{eq:homotopyJacobi}
 \sum_{i+j=m}\sum_\sigma\chi(\sigma;\ell_1,\ldots,\ell_m)(-1)^{i\cdot j}\mu_{j+1}(\mu_i(\ell_{\sigma(1)},\cdots,\ell_{\sigma(i)}),\ell_{\sigma(i+1)},\cdots,\ell_{\sigma(m)})=0~.
\end{equation}
Here, the sum over $\sigma$ is taken over all $(i,j)$ unshuffles. Recall that a permutation $\sigma$ of $i+j$ elements is called an {\em $(i,j)$-unshuffle}, if the first $i$ and the last $j$ images of $\sigma$ are ordered: $\sigma(1)<\cdots<\sigma(i)$ and $\sigma(i+1)<\cdots<\sigma(i+j)$. Moreover, $\chi(\sigma;\ell_1,\ldots,\ell_n)$ is the skew-symmetric Koszul sign defined implicitly via
\begin{equation}
 \ell_1\wedge \ldots \wedge \ell_m=\chi(\sigma;\ell_1,\ldots,\ell_m)\ell_{\sigma(1)}\wedge \ldots \wedge\ell_{\sigma(m)}~,
\end{equation}
where $\wedge$ is seen as a graded anticommutative operation. 

We will only be interested in $L_\infty$-algebras which consist of graded vector spaces with non-positive gradings. If the degrees of the vector spaces $L_n$ are further truncated and the $L_\infty$-algebra is concentrated in degrees $-n+1$ to $0$, we call the resulting $L_\infty$-algebra a (semistrict\footnote{General (or {\em weak}) Lie $n$-algebras arise as categorifications of the notion of a Lie algebra, see e.g.\ \cite{Roytenberg:0712.3461}. In this thesis, however, we only needed semistrict Lie $n$-algebras.}) {\em Lie $n$-algebra} or {\em $L_n$-algebra}.  

There is an elegant alternative definition of an $L_\infty$-algebra that makes use of a nilpotent differential. First, note that if we shift the grading of an $L_\infty$-algebra $L$ by $-1$ and consider $L[-1]=\oplus_n L_n[-1]$, where $L_n[-1]$ has now grading $n-1$, the degree of all brackets $\mu_i$ becomes +1. After the shift, we can define an $L_\infty$-algebra as a $\RZ^{<0}$-graded vector space $L$ equipped with a differential $\CD:\wedge^\bullet L\rightarrow \wedge^\bullet L$ of degree 1, which satisfies $\CD^2=0$. The connection to the previous definition is made by decomposing
\begin{equation}
 \CD=\CD_1+\CD_2+\CD_3+\cdots
\end{equation}
and demanding that $\CD_i$ acts on elements of $\wedge^i L$ as $\mu_i$, and otherwise it is extended to a coderivation via
\begin{equation}
 \mu_i(\ell_1\wedge\cdots\wedge \ell_m)=\sum_\sigma \chi(\sigma;\ell_1,\ldots,\ell_m)(-1)^{i\cdot (m-i)} \mu_i(\ell_{\sigma(1)},\cdots,\ell_{\sigma(i)})\wedge \ell_{\sigma(i+1)}\wedge \cdots\wedge \ell_{\sigma(m)}~.
\end{equation}
From here, it is rather obvious that the condition $\CD^2=0$ on $\wedge^\bullet L$ translates into the higher Jacobi relations \eqref{eq:homotopyJacobi}.

If all the homogeneously graded vector subspaces $L_n$ of $L$ are finite-dimensional, we can dualize this construction and obtain the {\em Chevalley-Eilenberg algebra} $\CEa(L)=(\wedge^\bullet L^*,Q)$ of $L$, where $Q$ is the dual of $\CD$. The Chevalley-Eilenberg algebra can be regarded as the polynomials on the space $L[-1]$ and $Q:\CEa(L)\rightarrow \CEa(L)$ becomes a homological vector field of degree 1. Altogether, we thus reinterpreted an $L_\infty$-algebra in terms of a {\em $Q$-manifold} as defined in \cite{Alexandrov:1995kv}.

\chapter{Jacobi elliptic functions and generalizations}\label{app:Jacobi}

An elliptic function is a doubly-periodic, meromorphic\footnote{Note that any doubly-periodic, holomorphic function must be constant.} function and any such function can be expressed in terms of {\em Jacobi} (or {\em Weierstra\ss}) {\em elliptic functions}. The Jacobi functions satisfy the relations\footnote{Many more relations can be found at \href{http://functions.wolfram.com}{functions.wolfram.com}.} 
\begin{equation}\label{C1}
\begin{aligned}
{\rm sn}_0z=\sin z\ ,\ \ {\rm cn}_0z=\cos z\ ,\ \ {\rm dn}_0z=1\ ,\  \ &{\rm cn}_k^2z+{\rm sn}_k^2z=1\ ,\ \ {\rm dn}_k^2z+k^2{\rm sn}_k^2z=1~,\\
{\rm sn}_kz={\rm sn}_k(z+4K(k))={\rm sn}_k(z+2\di K(k'))=-&{\rm sn}_k(z+2K(k))=\frac{{\rm sn}_{k^{-1}}kz}{k}=\frac{-\di {\rm sn}_{k'}\di z}{{\rm cn}_{k'}\di z}~,\\
{\rm cn}_k0={\rm dn}_k0=1\ ,\ \ {\rm sn}_k0=0\ ,\ \ {\rm sn}_k(z+K(k))&=\frac{{\rm cn}_kz}{{\rm dn}_kz}\  ,\ \ \ {\rm sn}_k(z+\di K(k'))=\frac{1}{k\ {\rm sn}_kz}~,\\
{\rm cn}_k({\rm sn}_k^{-1}s)=\sqrt{1-s^2}\ ,\ \ {\rm dn}_k({\rm sn}_k^{-1}s)=&\sqrt{1-k^2s^2}\ ,\ \ \dder{s}{\rm sn}_ks={\rm cn}_ks\ {\rm dn}_ks~,
\end{aligned}
\end{equation}
where $K(k)={\rm sn}_k^{-1}(1)$ and $k'^2=1-k^2$. 

They can be defined in terms of theta functions (which are not doubly-periodic) or in terms of integrals. Since the Jacobi functions are related, it suffices to define
\begin{equation}
{\rm sn}_k^{-1}(s)=\int_0^s\frac{\dd t}{\sqrt{(1-t^2)(1-k^2t^2)}}~.
\end{equation}

A {\em generalized Jacobi elliptic function} \cite{Pawellek:2009er} is given by 
\begin{equation}\label{genjacdefinition}
S^{-1}(s,k_1,k_2)=\int_0^s\frac{\dd t}{\sqrt{(1-t^2)(1-k_1^2t^2)(1-k^2_2t^2)}}~.
\end{equation}
The function $S(s,k_1,k_2)$ is hyperelliptic but can be viewed as a single-valued meromorphic function on a Riemann surface of genus two \cite{Pawellek:2009er}. It has been shown to be related to the Jacobi elliptic functions by
\begin{equation}\label{genjac}
S(s,k_1,k_2)=\frac{{\rm sn}_{\kappa}(k'_2s)}{\sqrt{k'^2_2+k_2^2{\rm sn}^2_{\kappa}(k'_2s)}}~,
\end{equation}
where $\kappa^2=\frac{k_1^2-k_2^2}{1-k_2^2}$ and $k'^2_2=1-k_2^2$.

\chapter{Ends of manifolds and volume types of volume forms}\label{app:A}

In this appendix, we briefly review the notion of an end of a manifold and its volume.

Consider a topological space $M$ together with an ascending sequence $K_i\subset K_{i+1}$, $i\in \NN$, of compact subsets whose interiors cover $M$. Then $M$ has an end for every sequence $U_i\supset U_{i+1}$, where $U_i$ is a connected component of $M\backslash K_i$. For example, the real line $\FR$ has two ends, which are obtained from the sequence $K_i=[-i,i]$ with $U_i=(i,\infty)$ and $U'_i=(-\infty,-i)$. 

More generally, one defines an end of a manifold $M$ as an element of the inverse limit system $\{K, \mbox{components of}~M\backslash K\}$ indexed by compact subsets $K$ of $M$, cf.\ \cite{Greene:1979aa}. 

If $M$ is orientable and endowed with a volume form, we say that an end has a finite volume, if there is a compact set $K$ such that the volume of the component of $M\backslash K$ containing the end is finite. Otherwise, we say that the volume is infinite.

\chapter{BLG Supersymmetry transformations}\label{app:blgsusy}

In the hopes that this might be useful to someone, below is a detailed calculation showing the closure of the supersymmetry algebra in the BLG model, as outlined in \cite{Bagger:2007jr}. Our starting point contains a free parameter $\kappa$, which will turn out to be fixed to $\kappa=-\tfrac{1}{6}$. The supersymmetry transformations are
\begin{equation*}
\begin{aligned}
\delta X^I&=i \epsb \Gamma^I \Psi\\
\delta\Psi&=D_\mu X^I\Gamma^\mu\Gamma^I\eps +\kappa [X^I,X^J,X^K]\Gamma^{IJK}\eps\\
\delta A_\mu&=i\epsb\Gamma_\mu\Gamma^ID(X^I,\Psi)
\end{aligned}
\end{equation*}
where $\Gamma_{012}\eps=\eps~,~~\Gamma_{012}\Psi=-\Psi$. 

\section{Fierz Identities}
When $\eps_1,\eps_2,\chi$ are the same $\Gamma_{012}$ chirality we have
\begin{equation*}
\begin{aligned}
&(\epsb_2\chi)\eps_1-(\epsb_1\chi)\eps_2=\\
&-\frac{1}{16}\left(2(\epsb_2\Gamma_\mu\eps_1)\Gamma^\mu\chi-(\epsb_2\Gamma_{IJ}\eps_1)\Gamma^{IJ}\chi+\frac{1}{4!}(\epsb_2\Gamma_\mu\Gamma_{IJKL}\eps_1)\Gamma^\mu\Gamma^{IJKL}\chi\right)
\end{aligned}
\end{equation*}

Note that $\Gamma^\mu\chi$ preserves $\Gamma_{012}$ chirality while $\Gamma^I\chi$ flips it.

\section{Charge conjugation}

The charge conjugation matrix satisfies
\begin{equation*}
\begin{aligned}
C^T=-C~, ~~C\Gamma^\mu C^{-1}=-(\Gamma^\mu)^T~, ~~C\Gamma^I C^{-1}=-(\Gamma^I)^T
\end{aligned}
\end{equation*} 
and since $\epsb=\eps^T C$ we have 
\begin{equation}\label{charge}
\begin{aligned}
\epsb_{[2}M\eps_{1]}=\eps_2^\alpha\eps_1^\beta((CM)_{\alpha\beta}+(CM)_{\beta\alpha})
\end{aligned}
\end{equation}

\section{$[\delta_1,\delta_2]X^I$}

We may now begin the calculation. We take antisymmetrization in 1,2 implicitly.

\begin{equation*}
\begin{aligned}
~[\delta_1,\delta_2]X^I=i\epsb_2\Gamma^I (D_\mu X^J \Gamma^\mu\Gamma^J\eps_1+\kappa [X^J,X^K,X^L]\Gamma^{JKL}\eps_1)
\end{aligned}
\end{equation*}
then \eqref{charge} helps us calculate $-\epsb_2\Gamma^I \Gamma^J\Gamma^\mu\eps_1$, we just need
\begin{equation*}
\begin{aligned}
(C\Gamma^I\Gamma^J\Gamma^\mu)^T=-(\Gamma^\mu)^T(\Gamma^J)^T(\Gamma^I)^TC=C\Gamma^\mu\Gamma^J\Gamma^I =C\Gamma^J\Gamma^I\Gamma^\mu
\end{aligned}
\end{equation*}
so we get $\epsb_2\Gamma^I \Gamma^\mu\Gamma^J\eps_1=-2\epsb_2 \Gamma^\mu\delta^{IJ}\eps_1$ .

Similarly 
\begin{equation*}
\begin{aligned}
(C\Gamma^I\Gamma^{JKL})^T=-(\Gamma^{JKL})^T(\Gamma^I)^T C=C\Gamma^{JKL}\Gamma^I 
\end{aligned}
\end{equation*}
and now we need $\{\Gamma^I,\Gamma^{JKL}\}=2\delta^{IJ}\Gamma^{KL}-2\delta^{IK}\Gamma^{JL}+2\delta^{IL}\Gamma^{JK}$ but this gets antisymmetrized in $JKL$ for the BLG model, so we get
\begin{equation*}
\begin{aligned}
~[\delta_1,\delta_2]X^I=-2i\epsb_2 (D_\mu X^I \Gamma^\mu\eps_1+6\kappa [X^J,X^K,X^I]\Gamma^{JK}\eps_1)~,
\end{aligned}
\end{equation*}
which is a translation by 
\begin{equation*}
\begin{aligned}
v^\mu=-2i\epsb_2\Gamma^\mu\eps_1
\end{aligned}
\end{equation*}
and a gauge transformation by
\begin{equation*}
\begin{aligned}
g=6i\kappa\epsb_2\Gamma_{JK}\eps_1D(X^J,X^K)~.
\end{aligned}
\end{equation*}

\section{$[\delta_1,\delta_2]\Psi$}
Take antisymmetrization in 1,2 implicitly.

\begin{equation*}
\begin{aligned}
~[\delta_1,\delta_2]\Psi=~&i\epsb_1\Gamma^I D_\mu \Psi \Gamma^\mu\Gamma^I\eps_2-i\epsb_1\Gamma_\mu\Gamma^J[X^J,\Psi,X^I]\Gamma^\mu\Gamma^I\eps_2\\&+	3\kappa i [\epsb_1\Gamma^I \Psi, X^J,X^K]\Gamma^{IJK}\eps_2
\end{aligned}
\end{equation*}

\subsection{First term $i\epsb_1\Gamma^I D_\mu \Psi \Gamma^\mu\Gamma^I\eps_2$}
\begin{equation*}
\begin{aligned}
i\epsb_1\Gamma^I D_\mu \Psi \Gamma^\mu\Gamma^I\eps_2=\frac{i}{16}(2(\epsb_2\Gamma_\nu\eps_1)\Gamma^\mu\Gamma^I\Gamma^\nu\Gamma^ID_\mu\Psi-(\epsb_2\Gamma_{JK}\eps_1)\Gamma^\mu\Gamma^I\Gamma^{JK}\Gamma^ID_\mu\Psi
\end{aligned}
\end{equation*}
plus a term which vanishes due to the identity $\Gamma^I\Gamma^{JKLM}\Gamma^I=0$
\begin{equation}\label{first}
\begin{aligned}
i\epsb_1\Gamma^I D_\mu \Psi \Gamma^\mu\Gamma^I\eps_2=-2i(\epsb_2\Gamma^\mu\eps_1)D_\mu\Psi+i(\epsb_2\Gamma_\nu\eps_1)\Gamma^\nu\Gamma^\mu D_\mu\Psi-\frac{i}{4}(\epsb_2\Gamma_{JK}\eps_1)\Gamma^{JK}\Gamma^\mu D_\mu\Psi
\end{aligned}
\end{equation}
\subsection{Second term $-i\epsb_1\Gamma_\mu\Gamma^J[X^J,\Psi,X^I]\Gamma^\mu\Gamma^I\eps_2$}
\begin{equation*}
\begin{aligned}
(\epsb_1\Gamma_\mu\Gamma^J\Psi)\Gamma^\mu\Gamma^I\eps_2 =\frac{1}{16}&\left(2(\epsb_2\Gamma_\nu\eps_1)\Gamma^\mu\Gamma^I\Gamma^\nu\Gamma_\mu\Gamma^J\Psi-(\epsb_2\Gamma_{LM}\eps_1)\Gamma^\mu\Gamma^I\Gamma^{LM}\Gamma_\mu\Gamma^J\Psi\vphantom{\int_1^2} \right. \\
&~~~~~~~~~~\left.  +\frac{1}{4!}(\epsb_2\Gamma_\nu\Gamma_{KLMN}\eps_1)\Gamma^\mu\Gamma^I\Gamma^\nu\Gamma^{KLMN}\Gamma_\mu\Gamma^J\Psi\right)
\end{aligned}
\end{equation*}
use the antisymmetrization of the first and third slot to antisymmetrize I,J. Now $\Gamma^\mu\Gamma_\mu=3$ so $\Gamma^\mu\Gamma^\nu\Gamma_\mu=-\Gamma^\nu$
\begin{equation*}
\begin{aligned}
&=\frac{1}{16}\left(-2(\epsb_2\Gamma_\nu\eps_1)\Gamma^\nu\Gamma^{IJ}\Psi+3(\epsb_2\Gamma_{LM}\eps_1)\Gamma^I\Gamma^{LM}\Gamma^J\Psi
\vphantom{\int}\right.\\
&~~~~~~~~\left.-\frac{1}{4!}(\epsb_2\Gamma_\nu\Gamma_{KLMN}\eps_1)\Gamma^\nu\Gamma^I\Gamma^{KLMN}\Gamma^J\Psi\right)
\\&=\frac{1}{16}(-2(\epsb_2\Gamma_\nu\eps_1)\Gamma^\nu\Gamma^{IJ}\Psi+3(\epsb_2\Gamma_{LM}\eps_1)(\Gamma^{LM}\Gamma^{IJ}-\Gamma^{LJ}\delta^{IM}\\
&~~~+\Gamma^{LI}\delta^{JM}-\Gamma^{MI}\delta^{JL}-\Gamma^{MJ}\delta^{IL}-2\delta^{LJ}\delta^{IM}+2\delta^{LI}\delta^{JM})\Psi
\\&~~~-\frac{1}{4!}(\epsb_2\Gamma_\nu\Gamma_{KLMN}\eps_1)\Gamma^\nu\Gamma^I\Gamma^{KLMN}\Gamma^J\Psi)
\end{aligned}
\end{equation*}

\subsection{Third term $3\kappa i [\epsb_1\Gamma^I \Psi, X^J,X^K]\Gamma^{IJK}\eps_2$}

Use $\Gamma^{IJK}\Gamma^I=6\Gamma^{JK}$
\begin{equation*}
\begin{aligned}
&(\epsb_1\Gamma^I \Psi)\Gamma^{IJK}\eps_2\\
&=\frac{1}{16}\left(2(\epsb_2\Gamma_\mu\eps_1)\Gamma^{IJK}\Gamma^\mu\Gamma^I\Psi-(\epsb_2\Gamma_{LM}\eps_1)\Gamma^{IJK}\Gamma^{LM}\Gamma^I\Psi
\vphantom{\int}\right.\\
&~~~\left.+\frac{1}{4!}(\epsb_2\Gamma_\mu\Gamma_{LMNP}\eps_1)\Gamma^{IJK}\Gamma^\mu\Gamma^{LMNP}\Gamma^I\Psi\right)
\\
&=-(\epsb_2\Gamma_\mu\eps_1)\Gamma^{JK}\Gamma^\mu\Psi\\
&~~~-\frac{1}{16}(\epsb_2\Gamma_{LM}\eps_1)(2\Gamma^{LM}\Gamma^{JK}-6\Gamma^{LK}\delta^{JM}\\
&~~~+6\Gamma^{LJ}\delta^{KM}-6\Gamma^{MJ}\delta^{KL}+6\Gamma^{MK}\delta^{JL}+4\delta^{LK}\delta^{JM}-4\delta^{LJ}\delta^{KM})\Psi
\\
&~~~-\frac{1}{4!16}(\epsb_2\Gamma_\mu\Gamma_{LMNP}\eps_1)\Gamma^\mu(\Gamma^K\Gamma^{LMNP}\Gamma^J-\Gamma^J\Gamma^{LMNP}\Gamma^K )\Psi 
\\
&=-(\epsb_2\Gamma_\mu\eps_1)\Gamma^{JK}\Gamma^\mu\Psi\\
&~~~-\frac{1}{16}\left((\epsb_2\Gamma_{LM}\eps_1)2\Gamma^{LM}\Gamma^{JK}-(\epsb_2\Gamma_{LJ}\eps_1)24\Gamma^{LK}\Psi+(\epsb_2\Gamma_{KJ}\eps_1)8\Psi\right)
\\
&~~~-\frac{1}{4!16}(\epsb_2\Gamma_\mu\Gamma_{LMNP}\eps_1)\Gamma^\mu(\Gamma^K\Gamma^{LMNP}\Gamma^J-\Gamma^J\Gamma^{LMNP}\Gamma^K )\Psi 
\end{aligned}
\end{equation*}

\subsection{Putting the second and third terms together}

We can cancel the two terms proportional to $(\epsb_2\Gamma_\mu\Gamma_{LMNP}\eps_1)$ by setting $\kappa=-\frac{1}{6}$ .

We also get.
\begin{equation*}
\begin{aligned}
(\epsb_2\Gamma_\nu\eps_1)\Gamma^\nu\Gamma^{IJ}[X^I,X^J,\Psi](-\frac{3\kappa i}{8} +\frac{i}{8})
\\
+(\epsb_2\Gamma_{LM}\eps_1)\Gamma^{LM}\Gamma^{IJ}[X^I,X^J,\Psi](-3\kappa i\frac{2}{16}-\frac{3i}{16})
\end{aligned}
\end{equation*}

with $\kappa=-\frac{1}{6}$ everything's right :)  They combine with the non-translation terms of \eqref{first} to give two copies of the fermion eom. We also get terms which give the gauge transformation 
\begin{equation*}
\begin{aligned}
(\epsb_2\Gamma_{IJ}\eps_1)[X^I,X^J,\Psi](\frac{3\kappa i}{2}-\frac{3i}{4})
\\
\end{aligned}
\end{equation*}
and some other terms proportional $(\epsb_2\Gamma_{IL}\eps_1)\Gamma^{JL}[X^I,X^J,\Psi]$ which cancel.

\section{$[\delta_1,\delta_2]A_\mu$}
Take $D(\cdot,\cdot)$ on the RHS implicitly
\begin{equation*}
\begin{aligned}
~[\delta_1,\delta_2]A_\mu=&-(\epsb_1\Gamma^I\Psi)(\epsb_2\Gamma_\mu\Gamma^I\Psi)\\&+i\epsb_2\Gamma_\mu\Gamma_I X^I(D_\nu X^J\Gamma^\nu\Gamma^J\eps_1+\kappa[X^J,X^K,X^L]\Gamma^{JKL}\eps_1)
\end{aligned}
\end{equation*}

The last term vanishes, first of all \eqref{charge} gives us $\epsb_{[2}\Gamma_\mu\Gamma_I\Gamma^{JKL}\eps_{1]}=\epsb_{[2}\Gamma_\mu\Gamma^{IJKL}\eps_{1]}$ and then the fundamental identity gives us $D(X^{[I},[X^J,X^K,X^{L]}])=0$. 

Let's look at the first term
\begin{equation*}
\begin{aligned}
(\epsb_1\Gamma^I\Psi)(\epsb_2\Gamma_\mu\Gamma^I\Psi)&=-(\epsb_1\Gamma^I\Psi)(\bar\Psi\Gamma^I\Gamma_\mu\eps_2)\\
&=-\frac{1}{16}\left(2(\epsb_2\Gamma^\nu\eps_1)\bar\Psi\Gamma_I\Gamma_\mu\Gamma_\nu\Gamma^I\Psi-(\epsb_2\Gamma_{JK}\eps_1)\bar\Psi\Gamma_I\Gamma_\mu\Gamma^{JK}\Gamma^I\Psi\right)
\end{aligned}
\end{equation*}
plus a term which vanishes due to the identity $\Gamma^I\Gamma^{JKLM}\Gamma^I=0$. Now the $\Psi$'s are in $D(\Psi,\Psi)$ so they get antisymmetrised and we can use \eqref{charge} again, the last term vanishes and we are left with 
\begin{equation*}
\begin{aligned}
(\epsb_1\Gamma^I\Psi)(\epsb_2\Gamma_\mu\Gamma^I\Psi)&=-(\epsb_2\Gamma^\nu\eps_1)(\bar\Psi\Gamma_{\mu\nu}\Psi)=-(\epsb_2\Gamma^\nu\eps_1)(\bar\Psi\Gamma^{\lambda}\Psi)\eps_{\mu\nu\lambda}
\end{aligned}
\end{equation*}
since $\Gamma_{012}\Psi=-\Psi$

The second term was 
\begin{equation*}
\begin{aligned}
i\epsb_2\Gamma_\mu\Gamma_I X^ID_\nu X^J\Gamma^\nu\Gamma^J\eps_1=2i\epsb_2\Gamma^\nu X^ID^\lambda X^I\eps_1 \eps_{\mu\nu\lambda}-2i\epsb_2\Gamma^{IJ} X^ID_\mu X^J\eps_1
\end{aligned}
\end{equation*}
so all together we get 
\begin{equation*}
\begin{aligned}
~[\delta_1,\delta_2]A_\mu=&2i(\epsb_2\Gamma^\nu\eps_1)(X^ID^\lambda X^I+\frac{i}{2}\bar\Psi\Gamma^{\lambda}\Psi)\eps_{\mu\nu\lambda}\\
&-2i\epsb_2\Gamma^{IJ} X^ID_\mu X^J\eps_1\\
=&v^\nu F_{\mu\nu}+D_\mu\Lambda
\end{aligned}
\end{equation*}
if the eom is $F_{\mu\nu}=-\eps_{\mu\nu\lambda}(X^ID^\lambda X^I+\frac{i}{2}\bar\Psi\Gamma^{\lambda}\Psi)$

\section{Possible extensions}

Perhaps these supersymmetry transformations can be generalized to involve loop space, double geometry, one compact direction with $\nabla_\mu=\partial_\mu+A_\mu+\frac{B_{\mu2}}{R}$, Jacobiators or terms involving the map $\sft$ such as
\begin{equation*}
\begin{aligned}
\delta X^I&=i \epsb \Gamma^I \Psi\\
\delta\Psi&=D_\mu X^I\Gamma^\mu\Gamma^I\eps +\kappa [X^I,X^J,X^K]\Gamma^{IJK}\eps+\kappa_1 [[X^I,X^J],X^J]\Gamma^{I}\eps\\
\delta A_\mu&=i\epsb\Gamma_\mu\Gamma^I(D(X^I,\Psi)+t([X^I,\Psi]))~.
\end{aligned}
\end{equation*}
Supersymmetry transformations for a $B$ field might also be possible, such as
\begin{equation*}
\begin{aligned}
\delta B_{\mu\nu}&=i\epsb\Gamma_\mu\Gamma^I D_\nu [X^I,\Psi]~.
\end{aligned}
\end{equation*}

\chapter{Comments on M5-brane supersymmetry transformations}\label{app:m5susy}

The supersymmetry transformations for the M5-brane model in section \ref{ssec:tensormultiplet} are
\begin{equation}
 \begin{aligned}
  \delta X^I&=\di \epsb \Gamma^I\Psi~,\\
  \delta \Psi&=\Gamma^\mu\Gamma^I\nabla_\mu X^I\eps+\tfrac{1}{2\times 3!}\Gamma_{\mu\nu\lambda}h^{\mu\nu\lambda}\eps-\tfrac{1}{2}\Gamma^{IJ}\Gamma [X^I,X^J]\eps~,\\
  \delta B_{\mu\nu}&=3\di \epsb\Gamma_{[\mu\nu}c^\lambda\nabla_{\lambda]}\Psi~,\\
  \delta A_\mu&=\di\epsb\Gamma_{\mu\lambda}c^\lambda \sft(\Psi)~,\\
  \delta c^\mu&=0~.
 \end{aligned}
\end{equation}
where $\Gamma_{012345}\eps=\eps~,~~\Gamma_{012345}\Psi=-\Psi~,~~\Gamma:=\Gamma^\mu c_\mu$ and $\Gamma_{\mu\nu\lambda}h^{\mu\nu\lambda}\eps=\frac{1}{|c|^2}\Gamma_{\mu\nu\lambda}B^{\mu\nu}c^{\lambda}\eps$ .

\section{Fierz Identities}
When $\eps_1,\eps_2$ have the opposite $\Gamma_{012345}$ chirality  as $\chi$ we have
\begin{equation*}
\begin{aligned}
&(\epsb_2\chi)\eps_1-(\epsb_1\chi)\eps_2=\\
&-\frac{1}{16}\left(2(\epsb_2\Gamma_\mu\eps_1)\Gamma^\mu\chi-2(\epsb_2\Gamma_{\mu}\Gamma^I\eps_1)\Gamma^{\mu}\Gamma^I\chi+\frac{1}{12}(\epsb_2\Gamma_{\mu\nu\lambda}\Gamma_{IJ}\eps_1)\Gamma^{\mu\nu\lambda}\Gamma_{IJ}\chi\right)
\end{aligned}
\end{equation*}

Note that $\Gamma^\mu\chi$ swaps $\Gamma_{012345}$ chirality while $\Gamma^I\chi$ preserves it.

\section{Charge conjugation}

The charge conjugation matrix satisfies

$C^T=C^{-1}=-C=-\Gamma_0~, ~~C\Gamma^\mu C^{-1}=-(\Gamma^\mu)^T~, ~~C\Gamma^I C^{-1}=-(\Gamma^I)^T$ 
\begin{equation}\label{charge2}
\begin{aligned}
\epsb_{[2}M\eps_{1]}=\eps_2^\alpha\eps_1^\beta((CM)_{\alpha\beta}+(CM)_{\beta\alpha})
\end{aligned}
\end{equation}

\section{$[\delta_1,\delta_2]X^I$}

Take antisymmetrization in 1,2 implicitly.

\begin{equation*}
\begin{aligned}
~[\delta_1,\delta_2]X^I=i\epsb_2\Gamma^I (\Gamma^\mu\Gamma^J\nabla_\mu X^J\eps_1+\tfrac{1}{2\times 3!}\frac{1}{|c|^2}\Gamma_{\mu\nu\lambda}B^{\mu\nu}c^{\lambda}\eps_1-\tfrac{1}{2}\Gamma^{JK}\Gamma [X^J,X^K]\eps_1)
\end{aligned}
\end{equation*}
so \eqref{charge2} helps us calculate $-\epsb_2\Gamma^I \Gamma^J\Gamma^\mu\eps_1$ we just need
\begin{equation*}
\begin{aligned}
(C\Gamma^I\Gamma^J\Gamma^\mu)^T=-(\Gamma^\mu)^T(\Gamma^J)^T(\Gamma^I)^TC=C\Gamma^\mu\Gamma^J\Gamma^I =C\Gamma^J\Gamma^I\Gamma^\mu
\end{aligned}
\end{equation*}
so we get $\epsb_2\Gamma^I \Gamma^\mu\Gamma^J\eps_1=-2\epsb_2 \Gamma^\mu\delta^{IJ}\eps_1$

Similarly 
\begin{equation*}
\begin{aligned}
(C\Gamma^I\Gamma^{\mu\nu\lambda})^T=-(\Gamma^{\mu\nu\lambda})^T (\Gamma^I)^TC=-C\Gamma^I \Gamma^{\mu\nu\lambda} 
\end{aligned}
\end{equation*}
so the second term vanishes.
For the third term 
\begin{equation*}
\begin{aligned}
(C\Gamma^I \Gamma^{JK}\Gamma^{\mu})^T=- C\Gamma^{\mu}\Gamma^{KJ}\Gamma^{I}=-C\Gamma^{JK}\Gamma^{I}\Gamma^{\mu}
\end{aligned}
\end{equation*}
then $[\Gamma^I , \Gamma^{JK}]=2(\delta^{IJ}\Gamma^K-\delta^{IK}\Gamma^J)$ giving
\begin{equation*}
\begin{aligned}
~[\delta_1,\delta_2]X^I=-2i\epsb_2 (\nabla_\mu X^I \Gamma^\mu\eps_1- [X^J,X^I]\Gamma^{J}\Gamma\eps_1)
\end{aligned}
\end{equation*}

which is a translation by 
\begin{equation}
\begin{aligned}
v^\mu=-2i\epsb_2\Gamma^\mu\eps_1
\end{aligned}
\end{equation}
and a gauge transformation 
\begin{equation}\label{eq:ginimt}
\begin{aligned}
g\acton X^I:=\sft(2i\epsb_2\Gamma_{J}\Gamma\eps_1X^J)\acton X^I~.
\end{aligned}
\end{equation}

The other terms work out analogously.

\section{Preserving $H=*H$}

To examine the preservation of $H=*H$ we will look at
\begin{equation*}
\begin{aligned}
\delta H_{\rho\mu\nu}= 3\di \epsb\Gamma_{[\mu\nu}c^\lambda F_{\rho\lambda]}\acton\Psi+ \di\epsb\Gamma_{[\rho\lambda}c^\lambda \sft(\Psi) \acton B_{\mu\nu]}~,
\end{aligned}
\end{equation*}
which simplifies to $3\di \epsb\Gamma_{[\mu\nu}c^\lambda (F_{\rho\lambda]}-\sft(B_{\rho\lambda]}))\acton\Psi$ and therefore vanishes, or is of the form of a (non-ample) gauge transformation $(F-\sft(B))\acton a$, meaning these susy transformations don't require vanishing fake curvature themselves. 

This fits with 
\begin{equation*}
\begin{aligned}
~[\delta_1,\delta_2]H=0=v^\mu\nabla_\mu H +g\acton H
\end{aligned}
\end{equation*}
since $\nabla_\mu H =0~,~\sft(H)=0$ and here \eqref{eq:ginimt} tells us that $g$ is in the image of $\sft$ and therefore $g\acton H=0$.

\bibliographystyle{abbrv}

\end{document}